\documentclass[journal,onecolumn]{IEEEtran}

\usepackage{amssymb}
\usepackage{amsfonts}
\usepackage[intlimits,sumlimits]{amsmath}
\usepackage[bbgreekl]{mathbbol}
\usepackage{framed}
\usepackage[boxruled,vlined,longend]{algorithm2e}
\usepackage{rotating}
\usepackage{hyperref}
\usepackage[sans]{dsfont}
\usepackage{yfonts}
\usepackage{pstcol} 

\newtheorem{theorem}{Theorem}

\newtheorem{condition}[theorem]{Condition}

\newtheorem{definition}[theorem]{Definition}

\newtheorem{lemma}[theorem]{Lemma}

\newtheorem{problem}[theorem]{Problem}
\newtheorem{proposition}[theorem]{Proposition}
\newtheorem{remark}[theorem]{Remark}

\newcommand{\dbigcup}{\bigcup}  
\newcommand{\dbigcap}{\bigcap}
\newcommand{\dprod}{\prod}
\newcommand{\dsum}{\sum}  

\begin{document}

%
\title{A precise bare simulation approach to 
the minimization of some distances. I. Foundations}
%
%
%

\author{Michel~Broniatowski
and Wolfgang~Stummer
\thanks{M. Broniatowski is with the 
LPSM, Sorbonne Universit\'{e}, 4 place Jussieu, 75252 Paris, France.
ORCID 0000-0001-6301-5531.}
\thanks{W. Stummer is with the Department of Mathematics, University of Erlangen--N\"{u}rnberg (FAU),
Cauerstrasse $11$, 91058 Erlangen, Germany; e-mail: stummer@math.fau.de.
ORCID 0000-0002-7831-4558. Corresponding author.}
}

%
%

\markboth{\ }
{Broniatowski \& Stummer: 
A bare simulation approach to minimization}
%



\maketitle

\begin{abstract}
In information theory --- as well as in the adjacent fields of 
statistics, machine learning, artificial intelligence, 
signal processing 
and pattern recognition ---
many \textit{flexibilizations} of the omnipresent Kullback-Leibler information 
distance (relative entropy) 
and of the closely related Shannon entropy
have become frequently used tools. To tackle corresponding constrained 
minimization (respectively maximization) problems by 
a newly developed \textit{dimension-free bare (pure) simulation} method,
is the main goal of this paper. Almost no assumptions (like convexity) on the set 
of constraints are needed, within our 
discrete setup of arbitrary dimension, and our method is precise (i.e., converges in the limit).
As a side effect, we also derive an innovative way of constructing new 
useful distances/divergences.  
To illustrate the core of our approach, we present numerous 
solved cases.
The potential for wide-spread applicability
is indicated, too; in particular, we deliver many recent references for uses of
the involved distances/divergences and entropies in various different
research fields (which may also serve as an interdisciplinary interface).
\end{abstract}

\begin{IEEEkeywords}
f-divergences of Csiszar-Ali-Silvey-Morimoto type, power divergences, 
Kullback-Leibler information distance, relative entropy, 
Renyi divergences, Bhattacharyya distance, Jensen-Shannon divergence/distance,
alpha-divergences, Shannon entropy, Renyi entropies, Bhattacharyya coefficient, 
Tsallis (cross) entropies,
Cressie-Read measures, Hellinger distance, 
Euclidean norms,
generalized maximum entropy method, importance sampling.
\end{IEEEkeywords}

%
\IEEEpeerreviewmaketitle

 
\hfill 2nd November 2022


%
%

\section{Introduction}
%
%
%
%

\IEEEPARstart{T}{his}
paper develops a new approach to perform the nonlinear constrained optimization of directed
distances --- and connected quantities --- based on a random simulation method. 
Given a set $\mathbf{\Omega} \subset \mathbb{R}^{K}$ 
with mild
regularity, the problem is to find the $\inf \left\{ \phi (\mathbf{Q}),
\mathbf{Q}\in \mathbf{\Omega }\right\} $ or the $\sup \left\{ \phi (\mathbf{Q
}),\mathbf{Q}\in \mathbf{\Omega }\right\} $ (depending on what the
optimization entails), where $\phi $ is a general objective function.
Such $\phi$ that satisfy the assumptions which allow the
proposed method to work, are named as \textquotedblleft bare-simulation
optimizable\textquotedblright .

In particular, this paper motivates the approach by considering $\phi 
(\mathbf{Q})=D_{\varphi }(\mathbf{Q},\mathbf{P})$ where $\mathbf{P} \in 
\mathbb{R}^{K}$ with positive entries and $D_{\varphi}(\cdot,\cdot)$ is a
Csiszar-Ali-Silvey-Morimoto-type 
(\cite{Csi:63},\cite{Ali:66},\cite{Mori:63})
\textit{$\varphi-$divergence} 
of specific form for which a basic link between $\varphi$ and an instrumental
probability distribution can be stated; simulation of random variables under this
distribution is the cornerstone for the optimization procedure. Most
commonly used $\varphi-$divergences can enter this scheme, and the wide range of
applicability of our proposal is presented through numerous 
solved cases.

\enlargethispage{0.5cm}

\vspace{0.2cm}
\noindent
Let us present an outlook of the core steps of the present approach.
The \textit{first step} to perform the distance (divergence) minimization is to
normalize the vector\footnote{in this paper, vectors are taken to be row vectors} 
$\mathbf{P}$ into a probability vector
$\mathbf{P}$ \footnote{with a slight abuse of notation; see the main text for a
more comprehensive notation} 
(e.g. the $\varphi-$entropy triggering case $\mathbf{P}=\left( 1,\ldots,1\right)$
is converted into the uniform-probability vector $\mathbf{P} = \left( 1/K,...,1/K\right)$).
The \textit{second step} follows from expressing the function $\varphi$ in form
of the Fenchel-Legendre transform of the 
cumulant (i.e., log moment) generating function of some
random variable $W$; a probabilistic construction based on i.i.d. copies
of $W$ allows to interpret $\inf \left\{ D_{\varphi }(\mathbf{Q},\mathbf{P}),
\mathbf{Q}\in \mathbf{\Omega }\right\} $ as an asymptotic characteristic for
some explicitly constructable scheme involving both $\mathbf{P}$ and the 
$W_{i}$'s.
The \textit{third and final step} consists in the construction of this
probabilistic scheme, and it differs for the specific problem context. 

In
general, for a deterministic setup where the (transformed) probability vector $\mathbf{P}
=\left( p_{1},\ldots,p_{K}\right)$ is completely known
and $\mathbf{\Omega}$ has non-void interior, one can construct 
the integer part
$n_{i} := \lfloor np_{i} \rfloor$, 
partition the index set $\left\{ 1,\ldots,n\right\} $
into $K$ sets of size $n_{1},\ldots,n_{K}$ and build a $K$-component vector;
each component of this vector is an ad hoc weighted empirical mean of the 
$W_{i}$ 's ; up to standard transformations the empirical count of the visits
of this vector in $\mathbf{\Omega }$ approximates the solution of the
optimization problem $\inf \left\{ D_{\varphi }(\mathbf{Q},\mathbf{P}),
\mathbf{Q}\in \mathbf{\Omega }\right\}$. Therefore, the resulting approximation 
can be performed straightforwardly:
the (typically) very complicated minimization task is replaced by a 
much more comfortable 
--- nevertheless convergent ---
random count procedure
which can be based on a fast and accurate 
--- pseudo, true, natural, quantum  
--- random number generator.

\enlargethispage{0.5cm}

In case of the statistical problem one has instead of a \textit{known} 
probability vector $\mathbf{P}$ 
a data-describing 
\textit{sample} $X_{1},\ldots,X_{n}$ of $n$ i.i.d. (and even more general) copies of a discrete random variable $X$ 
with \textit{unknown} distribution (described by an unknown probability vector) $\mathbf{P}$, 
and $\mathbf{\Omega}$ is now a subset of the probability simplex 
$\mathbb{S}^{K}$ in $\mathbb{R}^K$,
where $\mathbb{S}^{K}$ obviously has void interior but nevertheless a very useful special structure.
For such contexts, we can adapt the above-described bare-simulation method
by basically using the corresponding (vectorized)
sample-based \textit{empirical probability mass
function} as $\mathbf{P}$ and accordingly calculate a specific scheme which
also makes (in a slightly different way) use of the random variables $W_{i}$ associated with the function 
$\varphi$,
and estimate $\inf \left\{ D_{\varphi }(\mathbf{Q},\mathbf{P}),\mathbf{Q}\in 
\mathbf{\Omega }\right\}$. 

To work out the above-sketched program in detail, requires 
many known and newly developed results from
a number of different topics 
from various fields of information theory, applied probability, computer simulation and
analysis. This explains the need for a considerable number of different sections. 
We also found it 
necessary
to present \textit{explicit} 
solutions 
for the optimization
procedure in a number of cases which are of common use in the corresponding domains; the range of applications of the outcoming directed distances 
covers many areas in
engineering and in natural sciences, for which we selected recent prominently placed contributions with specific focus in relation with the aims of this paper. All this results in a 
paper of very substantial length.

\vspace{0.2cm}
\noindent
Let us first review the class of directed (i.e. not necessarily symmetric)
distances (also called divergences) $D(\mathbf{P},\mathbf{Q})$ between
two finite discrete\footnote{
for reasons of technicality, 
in this paper we only deal with such kind of distributions;
for instance, these can be also 
achieved from more involved systems
by quantizations of observations represented by finite partitions of the observation/data space,
or by making use of the dual representation for CASM $\varphi-$divergences
(cf.~\cite{Lie:06},~\cite{Bro:09}). 
}
(probability) distributions $\mathbf{P},\mathbf{Q}$ or between two 
general 
Euclidean 
vectors $\mathbf{P},\mathbf{Q}$
which are proved to be bare simulation optimizable; 
those serve as important (dis)similarity measures, proximity measures and discrepancy measures
in various different research areas such as
information theory, statistics, artificial intelligence, machine learning, signal processing,
pattern recognition, physics, finance, etc.\footnote{
since there exists a vast literature on divergences and connected entropies in these fields,
for the sake of brevity we will give in this introduction only some basic references;
many corresponding concrete 
applications will be mentioned in
the following sections.
}.
A major class are the above-mentioned \textit{$\varphi-$divergences 
$D_{\varphi}( \mathbf{P},\mathbf{Q})$
of Csiszar-Ali-Silvey-Morimoto} (CASM); 
this covers 
--- with corresponding choices of 
$\varphi$ --- 
e.g. the omnipresent \textit{Kullback-Leibler information distance/divergence} \cite{Kul:51} 
(also known as relative entropy),
the \textit{Jensen-Shannon distance/divergence}, as well as the 
\textit{power divergences} 
(also known as alpha-divergences, Cressie-Read measures, and Tsallis
cross-entropies).
For some comprehensive overviews on CASM $\varphi-$divergences,
the reader is referred to the insightful
books~\cite{Lie:87}--\cite{Lie:08}, 
the survey articles~\cite{Lie:06},\cite{Vaj:10}--\cite{Bass:13}
and the references therein;
an imbedding of CASM $\varphi-$divergences to more general frameworks 
can be found e.g.
in~\cite{Stu:12}--\cite{Bro:22}. 

Frequently used special cases of the above-mentioned power divergences
are e.g. the (squared) \textit{Hellinger distance}, the \textit{Pearson chi-square divergence},
and the \textit{Neyman chi-square divergence}.
Moreover, several deterministic transformations of power divergences 
are also prominently used in research,
most notably 
the \textit{Bhattacharyya distance}~\cite{Bha:43} and 
the more general
\textit{Renyi divergences}~\cite{Ren:61} 
(also known as Renyi cross-entropies); a 
comprehensive exposition of the latter is given e.g. 
in~\cite{VanErv:14}.
Some other important deterministic transformations of power divergences
include the 
\textit{Bhattacharyya coefficient}
(cf.~\cite{Bha:43},\cite{Bha:46},\cite{Bha:47})
 --- which is also called 
\textit{affinity} (cf.~\cite{Mat:51}) 
and \textit{
fidelity similarity} (cf. e.g.~\cite{Dez:16}) 
--- as well as 
the \textit{Bhattacharyya arccos distance 
} (cf.~\cite{Bha:47}) and the \textit{Fisher distance} 
(also known as \textit{Rao distance}, \textit{geodesic distance}, 
cf. e.g.~\cite{Dez:16}).
As shown below, by further explicit 
transformations
we can also recover Sundaresan\textquoteright s divergence~\cite{Sun:02}\cite{Sun:07}.

\vspace{0.2cm}
\noindent
Let us mention that
from CASM $\varphi-$divergences one can also derive
the widely used \textit{$\varphi-$entropies} $\mathcal{E}_{\varphi}(\mathbf{Q})$
of a distribution $\mathbf{Q}$
(and non-probability versions thereof)
in the sense of~\cite{Bur:82} (see also~\cite{Csi:72}--\cite{Vaj:07});
these entropies can be constructed from $D_{\varphi}(\mathbf{Q},\mathbf{P}^{unif})$
where $\mathbf{P}^{unif}$ denotes the uniform distribution.
Moreover, by use of certain deterministic transformations $h$ 
one can also deduce the more general
\textit{$(h,\varphi)-$entropies}
(and non-probability versions thereof)
in the sense of~\cite{Sal:93} (see also e.g.~\cite{Par:06}). 
As will be worked out in detail 
below, from this one can deduce as special cases a variety of 
prominently used quantities in research, such as for instance: 

\begin{itemize}

\item 
the omnipresent \textit{Shannon entropy} \cite{Sha:48}, 
the \textit{$\gamma-$order Renyi entropy} \cite{Ren:61},
the \textit{$\gamma-$order entropy of Havrda-Charvat} \cite{Hav:67} (also called 
non-additive \textit{$\gamma-$order Tsallis entropy} \cite{Tsa:88} in statistical physics),
the \textit{$\widetilde{\gamma}-$order entropy of Arimoto} \cite{Ari:71},
Vajda's quadratic entropy \cite{Vaj:89},
\textit{Sharma-Mittal entropies} \cite{Shar:75}, 

\item the Euclidean \textit{$\gamma-$norms}, as well as

\item measures of diversity, heterogeneity and unevenness, 
like the \textit{Gini-Simpson diversity index},
the \textit{diversity index of Hill} \cite{Hil:73},
the \textit{Simpson-Herfindahl index}
(which is also known as \textit{index of coincidence}, cf.~\cite{Har:01} 
and its generalization in~\cite{Har:08}), 
the \textit{diversity index of
Patil \& Taillie}~\cite{Pati:82}, 
the \textit{$\gamma-$mean heterogeneity index} (see e.g.~\cite{VanDerLub:86});
see also~\cite{Nay:85}~and~\cite{Jost:06}
for some interrelations with the above-mentioned entropies.

\end{itemize}

\vspace{0.2cm}
\noindent
Given that the constraint set $\mathbf{\Omega}$ 
reflects some \textit{incomplete/partial} information about a system (e.g. moment constraints),
the maximization over $\mathbf{Q}\in\mathbf{\Omega}$
of the above-mentioned 
entropies, norms and diversity indices (and the more general \textit{$(h,\varphi)-$entropies})
is important for many research topics, most notably manifested
in Jaynes\textquoteright s \cite{Jay:57a},\cite{Jay:57b}
omnipresent, 
\textquotedblleft  universally applicable\textquotedblright\ 
\textit{maximum entropy principle} 
(which employs the Shannon entropy), and 
its generalizations (see e.g. 
the books~\cite{Kap:89}--\cite{Gzy:18}
for comprehensive surveys). 

\vspace{0.2cm}
In the \textit{statistical} context, 
the \textit{minimization} 
$\inf_{\mathbf{Q}\in\mathbf{\Omega}} D( \mathbf{Q}, \mathbf{P} )$
of divergences 
from one distribution (respectively, its equivalent vector of frequencies) $\mathbf{P}$ 
to an appropriate set $\mathbf{\Omega}$ of distributions (frequency vectors)
appears in a natural way,
as indicated in the following.
For instance, 
let $\mathbf{P} = \mathbf{P}_{true}$ be the \textit{true}  
distribution of a mechanism which generates 
non-deterministic data
and $\mathbf{\Omega}$ be a pregiven \textit{model} in the sense of a
(parametric or non-parametric) family of distributions 
which serves as an 
\textquotedblleft approximation\textquotedblright\ 
(in fact, a collection of approximations) of the 
\textquotedblleft  truth\textquotedblright\ $\mathbf{P}_{true}$.
If $\mathbf{P}_{true} \notin \mathbf{\Omega}$ --- e.g.  
since $\mathbf{\Omega}$ reflects some simplifications of $\mathbf{P}_{true}$ 
which is in line with the general scientific procedure ---
then the positive quantity 
$\Phi_{\mathbf{P}_{true}}(\mathbf{\Omega}) := \inf_{\mathbf{Q}\in\mathbf{\Omega}} 
D( \mathbf{Q}, \mathbf{P}_{true} )$
can be used as an \textit{index of model adequacy} 
in the sense of a degree of departure between the model and the truth
(cf.~\cite{Lind:04}, see also 
e.g.~\cite{Lind:08}--\cite{Mark:19});
small index values should indicate
high adequacy.
If $\mathbf{P}_{true} \in \mathbf{\Omega}$, then
$\Phi_{\mathbf{P}_{true}}(\mathbf{\Omega}) = 0$
which corresponds to full adequacy.
This index of model adequacy $\Phi_{\mathbf{P}_{true}}(\mathbf{\Omega})$
can also be seen as 
\textit{index of goodness/quality of approximation to the truth}
or as \textit{model misspecification error},
and it can be used for model assessment as well as 
for model search (model selection, model hunting)
by comparing the indices $\Phi_{\mathbf{P}_{true}}(\mathbf{\Omega_{1}}),
\Phi_{\mathbf{P}_{true}}(\mathbf{\Omega_{2}}), \ldots$
of competing models $\mathbf{\Omega_{1}}, \mathbf{\Omega_{2}}, \ldots$ and choosing
the one with the smallest index;
this idea can be also used for classification (e.g. analogously
to~\cite{Bil:12}  
who deal with continuous (rather than discrete) distributions)
where the $\mathbf{\Omega_{i}}$ are interpreted as (possibly data-derived but fixed) classes
which are disjoint and non-exhaustive.

Typically, in statistical analyses the true distribution $\mathbf{P}_{true}$ is unknown 
and is either replaced by a hypothesis-distribution $\mathbf{P}_{hyp}$
or by a distribution $\mathbf{P}_{data}$ derived from
data (generated by $\mathbf{P}_{true}$) which converges
to $\mathbf{P}_{true}$ as the data/sample size tends to infinity
(e.g. $\mathbf{P}_{data}$ may be the well-known empirical distribution or a conditional distribution).
Correspondingly, $\Phi_{\mathbf{P}_{data}}(\mathbf{\Omega})
=\min_{\mathbf{Q}\in\mathbf{\Omega}} D( \mathbf{Q}, \mathbf{P}_{data})$
reflects a data-derived 
approximation (estimate)
of the index of model adequacy
(resp. of the model misspecification error) 
from which one can
cast corresponding model-adequacy tests and related goodness-of-fit tests.
Accordingly, our new above-described procedure is well fitted for model
choice. (For reasons of efficiency, especially in high dimension $K$).
\textit{After} choosing the most adequate model (say) $\mathbf{\Omega}_{i_0}$,
one can then tackle the problem of finding the corresponding 
(not necessarily existent or unique)
best-model-member/element choice (i.e., the \textit{minimizer})
$\arg \min_{\mathbf{Q}\in \mathbf{\Omega}_{i_0}}D(\mathbf{Q},\mathbf{P}_{data})$ 
which amounts to the well-known corresponding \textit{minimum distance estimator}
(for comprehensive surveys on divergence-based
statistical testing and estimation, the reader is referred to e.g. the
references in \cite{Bro:22});
for the sake of brevity, this will be treated in a follow-up paper.

\vspace{0.2cm}
\noindent
Besides the above-mentioned principal overview, let us now briefly discuss
some existing \textit{technical issues} 
for the minimization of CASM $\varphi-$divergences
$\Phi_{\mathbf{P}}(\mathbf{\Omega}) := 
\inf_{\mathbf{Q}\in\mathbf{\Omega}} D_{\varphi}( \mathbf{Q}, \mathbf{P} )$.
For (not necessarily discrete) probability distributions/measures $\mathbf{P}$ 
and sets $\mathbf{\Omega}$ of probability distributions/measures satisfying
a finite set of linear equality constraints,
$\Phi_{\mathbf{P}}(\mathbf{\Omega})$ 
has been characterized in~\cite{Csi:75} and more
recently 
in~\cite{Csi:03}--\cite{Pel:11}
among others, in various contexts; 
those results extend to inequality constraints. 
Minimizations of $\gamma-$order Renyi divergences on $\gamma-$convex sets $\mathbf{\Omega}$
are studied e.g. in~\cite{Kum:16}, whereas~\cite{Kum:15a} \cite{Kum:15b}
investigate minimizations of Sundaresan\textquoteright s divergence on
certain convex sets $\mathbf{\Omega}$.

To our knowledge, no general
representation for 
$\Phi_{\mathbf{P}}(\mathbf{\Omega})$ for a positive distribution/measure 
$\mathbf{P}$ (respectively, for a Euclidean vector with positive components)
and a general set $\mathbf{\Omega}$ of signed measures 
(respectively, of Euclidean vectors with components of arbitrary sign)
exists. At the contrary, many \textit{algorithmic} approaches for such minimization problems 
have been proposed; they mostly aim at finding minimizers  
more than at the evaluation of the minimum divergence itself, which is
obtained as a by-product. Moreover, it is well-known that 
\textit{such kind} of
CASM $\varphi-$divergence minimization 
approaches
may be hard to tackle or even intractable via usual methods such as the omnipresent
gradient descent method and versions thereof,
especially for non-parametric or semi-parametric $\mathbf{\Omega}$ in 
sufficiently high-dimensional situations.
For instance, $\mathbf{\Omega}$ may consist (only) of constraints on 
moments or on L-moments (see e.g.~\cite{Bro:16}); 
alternatively, $\mathbf{\Omega}$ may be e.g. a
tubular neighborhood 
of a parametric model
(see e.g.~\cite{Liu:09},\cite{Gho:18}).
The same intractability problem holds for the above-mentioned
$(h,\varphi)-$entropy maximization problems.
In the light of this, the goals of this paper are:

\begin{itemize}

\item 
to solve constrained minimization problems 
of a large range of CASM $\varphi-$divergences and deterministic transformations thereof
(respectively constrained maximization problems of $(h,\varphi)-$entropies 
including Euclidean norms and diversity indices),
by means of a newly developed \textit{dimension-free bare (pure) simulation} method
which is precise (i.e., converges in the limit)
and which needs almost no assumptions (like convexity) on the set 
$\mathbf{\Omega}$ of constraints;
in doing so, for the sake of brevity 
we concentrate on finding/computing the minimum divergences themselves rather than the 
corresponding minimizers
(to achieve the latter, e.g. dichotomous search 
could be used in a subsequent step, however);

\item to derive a method of constructing new 
useful distances/divergences;

\item to present numerous solved cases
in order to illuminate
our method and its potential for wide-spread applicability;
as we go along, we also deliver many recent references for uses of
the outcoming distances/divergences and entropies 
(covering in particular all the above-mentioned ones).
 
\end{itemize}

\noindent This agenda is achieved in the following way. In the next 
Section \ref{SectMain}, we briefly introduce the principal idea of our new
bare-simulation optimization paradigm. After manifesting 
the fundamentally employed class of CASM $\varphi -$divergences in 
Section \ref{SectDirDist},
the correspondence between the function $\varphi$ and the
distribution of the random variable $W$ is stated in 
Section \ref{Sect Prelude}.
The random simulation scheme aiming at bare-simulation minimizability for deterministic
problems is presented in Section \ref{Sect Minimization},
and Section \ref{Subsect Approx emp risk}
adapts this scheme to the statistical context, together with some deterministic
simplex-constraints variant,
also providing estimations or approximations for bounds of the
solutions on the minimization/maximization problem at hand; 
the asymptotics which justify the simulation scheme as providing approximations for the
optimization problem is discussed.
Sections \ref{SectRenyi},\ref{SectMaxEnt} and \ref{SectFurther} present
specific optimization schemes for the important R\'{e}nyi family of
divergences, for the constrained optimization of entropies, as well as a
number of prominent deterministic optimization problems which can benefit from the wide
applicability of this approach (high dimension and highly disconnected
constraint sets, linear assignment problem with side constraints, etc). 
Section \ref{Sect Estimators}
presents the construction of estimators together with some importance-sampling 
procedures and explicit algorithms.
Section \ref{SectFind} provides
explicit constructions for the distribution of the instrumental random
variable $W$ for wide classes of functions $\varphi $; this section is based
on a new theorem which allows for an easy bridge between moment generating
functions and their Fenchel-Legendre transform. Finally,
Section \ref{Sect Cases} provides
explicitly solved cases which cover both deterministic and statistical
important bare-simulation amenable problems.
Two important proofs are given in 
the Appendices A and B of the paper; the
remaining technical proofs and discussions are presented in the Supplementary
Material. 

\vspace{0.2cm}
\noindent
Let us finally mention that a first simulation-based algorithm in vein with the present proposal has been developed by~\cite{Bro:17}
in the restricted setup of
risk estimation for power divergences.
The present paper extends this very considerably by (amongst other things) treating
\textit{general} CASM $\varphi-$divergences \textit{as well as} related entropies,
by dealing with corresponding \textit{general} optimization problems
of \textit{both} deterministic and stochastic type, respectively, 
and by developing new \textit{types} of more sophisticated simulation algorithms.

%
%

\section{A new minimization paradigm \label{SectMain}}

\noindent
We concern with minimization problems of the following type, where $\mathcal{
M}$ is a topological space 
and $\mathcal{T}$ is the Borel $\sigma-$field
over a given base on $\mathcal{M}$; e.g. take $\mathcal{M} = \mathbb{R}^{K}$
to be the $K-$dimensional Euclidean space equipped with the Borel $\sigma-$
field $\mathcal{T}$. 

\vspace{0.2cm}

\begin{definition}
\label{brostu3:def.1} 
A measurable function $\Phi: \mathcal{M} \mapsto
\mathbb{R} \cup \{-\infty, \infty\}$ 
and measurable set $\Omega \subset \mathcal{M}$ 
\footnote{
i.e. $\Omega \in \mathcal{T}$} are called 
\textquotedblleft bare-simulation minimizable\textquotedblright\ 
(BS-minimizable) respectively 
\textquotedblleft bare-simulation maximizable\textquotedblright\ 
(BS-maximizable)
if for 
\begin{equation}
\Phi(\Omega ):=\inf_{Q\in\Omega}\left\{ \Phi(Q) \right\} 
\in \, ]-\infty, \infty[
\qquad \textrm{respectively} \qquad 
\Phi(\Omega ):=\sup_{Q\in\Omega}\left\{ \Phi(Q) \right\} 
\in \, ]-\infty, \infty[
\label{brostu3:fo.1}
\end{equation}
there exists a measurable function $G: [0,\infty[ \mapsto 
\mathbb{R}$ as well as 
a sequence $\left((\mathfrak{X}_{n},\mathcal{A}_{n},\mathbb{\Pi}_{n})\right)_{n \in \mathbb{N}}$ of probability spaces and on them a
sequence $(\xi_{n})_{n\in \mathbb{N}}$ 
\footnote{
in order to emphasize the dependence on $\Phi$, one should use the notations 
$(\xi_{\Phi,n})_{n\in \mathbb{N}}$, $\mathbb{\Pi}_{\Phi,n}$, etc.; this is avoided
for the sake of a better readability.} of $\mathcal{M}-$valued random
variables such that 
\begin{eqnarray}
& &
G\Big(-\lim_{n\rightarrow\infty}\frac{1}{n}\log \mathbb{\Pi}_{n}\negthinspace \left[
\xi_{n}\in\Omega\right] \Big) =\inf_{Q\in\Omega}\Phi(Q)=\Phi(\Omega)
\label{brostu3:fo.2} \\
&&
\hspace{-2.0cm} \textrm{respectively} \quad
G\Big(-\lim_{n\rightarrow\infty}\frac{1}{n}\log \mathbb{\Pi}_{n}\negthinspace \left[
\xi_{n}\in\Omega\right] \Big) =\sup_{Q\in\Omega}\Phi(Q) =\Phi(\Omega);
\label{brostu3:fo.2b}
\end{eqnarray}
in situations where $\Phi$ is fixed and different $\Omega$\textquoteright s are considered,  
we say that 
\textquotedblleft  $\Phi$ is bare-simulation minimizable (BS-minimizable)
on $\Omega$\textquotedblright\ 
respectively 
\textquotedblleft  $\Phi$ is bare-simulation maximizable (BS-maximizable)
on $\Omega$\textquotedblright.
\end{definition}

\vspace{0.4cm}
\noindent

\noindent The basic idea/incentive of 
this new approach is: if a minimization problem \eqref{brostu3:fo.1} has no
explicit solution and is computationally intractable (or unfeasible) but can
be shown to be BS-minimizable with concretely constructable $(\xi_{n})_{n\in 
\mathbb{N}}$ and $(\mathbb{\Pi}_{n})_{n\in \mathbb{N}}$, then one can basically
simulate the log-probabilities $-\frac{1}{n} \log \mathbb{\Pi}_{n}\negthinspace \left[
\xi_{n}\in\Omega\right]$ for large enough integer $n \in \mathbb{N}$ 
to obtain an approximation of \eqref{brostu3:fo.1}
without having to evaluate the corresponding (not necessarily unique) 
minimizer, where the latter is typically time-costly.
Finding minimizers can be performed e.g. through dichotomic search, once
an algorithm leading to the minimal value of the divergence on
adequate families of sets $\Omega$ is at hand;
for the sake of brevity, this is omitted in the current paper.

\vspace{0.2cm} 
\noindent 
For reasons of transparency, we \textit{start} to demonstrate this approach for the
following important/prominent
class of \textit{deterministic} constrained minimization problems with the following components:

\begin{enumerate}
\item[(i)] 
$\mathcal{M}$ is the $K-$dimensional
Euclidean space $\mathbb{R}^{K}$, i.e. $\Omega$ is a set of vectors $Q$ with 
a number of $K$ components
(where $K$ may be huge, as it is e.g. the case in big data contexts); 

\item[(ii)] $\Phi(\cdot) := \Phi_{P}(\cdot)$ 
depends on some known vector $P$ in $\mathbb{R}^{K}$ 
with $K$ nonnegative components;

\item[(iii)] $\Phi_{P}(\cdot)$ is a \textquotedblleft directed distance\textquotedblright\ (divergence) from $P$ into 
$\Omega$ in the sense of $\Omega \ni Q \mapsto \Phi_{P}(Q)
:= D(Q, P)$, where $D(\cdot,\cdot)$ has the the two properties \textquotedblleft $D(Q, P)
\geq 0$\textquotedblright\ and \textquotedblleft $D(Q, P) = 0$ if and only if $Q=P$\textquotedblright. In particular, 
$D(\cdot,\cdot)$ needs neither satisfy the symmetry $D(Q,P)= D(P,Q)$ nor the
triangular inequality.
\end{enumerate}

\noindent In other words, (1) together with (i)-(iii) constitutes a
\textit{deterministic}
constrained
distance/divergence-minimization problem; we design a \textquotedblleft
universal\textquotedblright\ method to solve such problems by constructing
appropriate (cf.\eqref{brostu3:fo.2}) sequences $(\xi _{n})_{n\in \mathbb{N}}
$ of $\mathbb{R}^{K}-$valued random variables, for all directed distances $D(\cdot ,\cdot )$ from a
large subclass of the important omnipresent Csiszar-Ali-Silvey-Morimoto
$\varphi-$divergences (also called $f-$divergences) given in
Definition \ref{def div} below. 

%
%

\section{Directed distances \label{SectDirDist}}

\noindent To begin with, concerning the above-mentioned point (i) we take the $K-$
dimensional Euclidean space $\mathcal{M}=\mathbb{R}^{K}$, 
denote from now on --- as usual --- its elements (i.e. vectors) in boldface letters,
and also employ the subsets 
\begin{eqnarray}
&& 
\mathbb{R}_{\ne 0}^{K} := \{\mathbf{Q}:= (q_{1},\ldots,q_{K}) \in \mathbb{R}^K: \,
q_{i} \ne 0 \ \text{for all} \ i=1,\ldots,K \},  
\notag \\
&& 
\mathbb{R}_{> 0}^{K} := \{\mathbf{Q}:= (q_{1},\ldots,q_{K}) \in \mathbb{R}^K: \,
q_{i} > 0 \ \text{for all} \ i=1,\ldots,K \},  \notag \\
&& \mathbb{R}_{\geq 0}^{K} := \{\mathbf{Q}:= (q_{1},\ldots,q_{K}) \in \mathbb{R}^K
: \, q_{i} \geq 0 \ \text{for all} \ i=1,\ldots,K \},  \notag \\
&& \mathbb{R}_{\gneqq 0}^{K} := \mathbb{R}_{\geq 0}^{K}\backslash \{\boldsymbol{0}\} 
 := \{\mathbf{Q}:= (q_{1},\ldots,q_{K}) \in \mathbb{R}_{\geq 0}^{K}
: \, q_{i} \ne 0 \ \text{for some} \ i=1,\ldots,K \},
\notag \\
&& 
\textstyle
\mathbb{S}^{K} := \{\mathbf{Q} := (q_{1},\ldots,q_{K}) \in \mathbb{R}_{\geq 0}^{K}:
\, \sum_{i=1}^{K} q_{i} =1 \} \quad \text{(simplex of probability vectors, probability simplex)},
\notag \\
&& 
\textstyle
\mathbb{S}_{> 0}^{K} := \{\mathbf{Q}:= (q_{1},\ldots,q_{K}) \in \mathbb{R}_{>
0}^{K}: \, \sum_{i=1}^{K} q_{i} =1 \}.  \notag
\end{eqnarray}
Concerning the directed distances $D(\cdot,\cdot)$ in (ii) and (iii), we
deal with the following

\vspace{0.2cm}

\begin{definition}
\label{def div}
(a) \, Let the \textquotedblleft divergence generator\textquotedblright\ be a 
lower semicontinuous convex function $\varphi: \, ]-\infty,\infty[ \rightarrow [0,\infty]$ 
satisfying $\varphi(1)=0$.
Furthermore, for the effective domain $dom(\varphi) := \{ t \in \mathbb{R} : \varphi(t) < \infty \}$
we assume that its interior $int(dom(\varphi))$ is non-empty
which implies that $int(dom(\varphi)) = \, ]a,b[$ for
some $-\infty \leq a < 1 < b \leq \infty$.
Moreover, we suppose that $\varphi$ is strictly convex
in a non-empty neighborhood $]t_{-}^{sc},t_{+}^{sc}[ \, \subseteq \, ]a,b[$ of one
($t_{-}^{sc} < 1 < t_{+}^{sc}$).
Also, we set $\varphi(a) := \lim_{t \downarrow a} \varphi(t)$ 
and $\varphi(b) := \lim_{t \uparrow b} \varphi(t)$ 
(these limits always exist).
The class of all such functions $\varphi$ will be denoted
by $\widetilde{\Upsilon}(]a,b[)$. A frequent choice is e.g. $]a,b[ \, = \, ]0,\infty[$ or 
$]a,b[ \, = \, ]-\infty,\infty[$.\\
(b) \, For $\varphi \in \widetilde{\Upsilon}(]a,b[)$, $\mathbf{P} := (p_{1},\ldots,p_{K}) \in 
\mathbb{R}_{\gneqq 0}^{K}$ and $\mathbf{Q} := (q_{1},\ldots,q_{K}) \in \mathbf{\Omega} \subset \mathbb{R}^{K}$, we define the
Csiszar-Ali-Silvey-Morimoto (CASM) $\varphi-$divergence 
\begin{equation}
\Phi_{\mathbf{P}} \left(\mathbf{Q}\right) := D_{\varphi}( \mathbf{Q}, \mathbf{P} ) := 
\sum_{k=1}^{K} p_{k} \cdot
\varphi \left( \frac{q_{k}}{p_{k}}\right) \, \geq 0.  
\label{brostu3:fo.div}
\end{equation}
As usual, in \eqref{brostu3:fo.div} we employ the three conventions that $p
\cdot \varphi \left( \frac{0}{p}\right) = p \cdot \varphi(0) >0$ for all $p > 0$, 
and $0 \cdot \varphi \left( \frac{q}{0}\right) = q \cdot \lim_{x
\rightarrow \infty} \frac{\varphi(x \cdot \textrm{sgn}(q))}{x\cdot \textrm{sgn}(q)} >0$ 
for $q \neq 0$ (employing the sign of $q$), and $0 \cdot
\varphi \left( \frac{0}{0}\right) :=0$. Throughout the paper, we only
consider constellations $(\varphi,\mathbf{P},\mathbf{\Omega})$ 
for which the very mild condition 
\ $\Phi_{\mathbf{P}}(\Omega) := \inf_{\mathbf{Q}\in\mathbf{\Omega}} 
D_{\varphi}( \mathbf{Q}, \mathbf{P} ) \neq \infty$ \ holds.
\end{definition}

\vspace{0.3cm}
\noindent For probability vectors $\mathds{P}$ and $\mathds{Q}$ in 
$\mathbb{S}^{K}$, the 
$\varphi-$divergences $D_{\varphi}( \mathds{Q}, \mathds{P} )$ were introduced by 
Csiszar~\cite{Csi:63}, Ali \& Silvey~\cite{Ali:66} and 
Morimoto~\cite{Mori:63}
(where the first two references even deal with more general probability distributions); 
for some comprehensive overviews 
--- including statistical applications to goodness-of-fit testing and
minimum distance estimation ---
the reader is referred to the insightful 
books~\cite{Lie:87}--\cite{Lie:08},
the survey articles~\cite{Lie:06},\cite{Vaj:10}--\cite{Bass:13},
and the references therein.
Some exemplary recent studies and applications
of CASM $\varphi-$ divergences
appear e.g. in~\cite{Qiao:10}--\cite{Stu:21}. 
For the setup of $D_{\varphi}( \mathbf{Q}, \mathbf{P} )$ for vectors $\mathbf{P}$, 
$\mathbf{Q}$ with non-negative components
the reader is referred to e.g.~\cite{Stu:10} 
(who deal with even more general nonnegative measures
and give some statistical as well as information-theoretic applications)
and~\cite{Gie:17} 
(including applications
to iterative proportional fitting). 
The case of $\varphi-$divergences for vectors with arbitrary components
can be extracted from e.g.~\cite{Bro:06} 
who actually deal with finite \textit{signed} measures. For a comprehensive technical
treatment, see also~\cite{Bro:19b}. 

\vspace{0.2cm} 
\noindent 
Clearly, from \eqref{brostu3:fo.div} it is obvious
that in general $D_{\varphi}( \mathbf{Q}, \mathbf{P} ) \ne 
D_{\varphi}(\mathbf{P}, \mathbf{Q})$ (non-symmetry).
Moreover, it is straightforward to deduce that $D_{\varphi}(\mathbf{Q}, \mathbf{P}) = 0$ if
and only if $\mathbf{Q}=\mathbf{P}$ (reflexivity). By appropriate choice of 
$\varphi$, one can get as special cases many very prominent divergences  
which are frequently used in information theory and its applications to e.g.
statistics, artificial intelligence, and machine learning. We shall address them 
later on as we go along.

\vspace{0.3cm}

\begin{remark}
Since, in general, our methods work also for \textit{non-}probability vectors
$\mathbf{Q}$ and $\mathbf{P}$, we can also deal with --- plain versions and transformations of --- 
\textit{weighted $\varphi-$divergences} 
of the form
\begin{equation}
D_{\varphi}^{wei}( \mathbf{Q}, \mathbf{P} ) := 
\sum_{k=1}^{K} c_{k} \cdot p_{k} \cdot
\varphi \left( \frac{q_{k}}{p_{k}}\right) \, \geq 0  
\label{brostu3:fo.div.weighted1}
\end{equation}
where $c_{k} >0$ \, ($k=1,\ldots,K$) are weights which not necessarily add up to one.
Indeed, we can formally rewrite 
\begin{equation}
\inf_{\mathbf{Q}\in\mathbf{\Omega}}
D_{\varphi}^{wei}( \mathbf{Q}, \mathbf{P} ) = 
\inf_{\mathbf{Q}^{wei}\in\mathbf{\Omega}^{wei}}
D_{\varphi}( \mathbf{Q}^{wei}, \mathbf{P}^{wei} )
\nonumber
\end{equation}
where $\mathbf{P}^{wei} := (c_{1} \cdot p_{1},\ldots,c_{K} \cdot p_{K})$, 
$\mathbf{Q}^{wei} := (c_{1} \cdot q_{1},\ldots,c_{K} \cdot q_{K})$
and $\mathbf{\Omega}^{wei}$ is the corresponding rescaling of $\mathbf{\Omega}$.
Of course, all the necessary technicalities for the $\varphi-$divergences
(see below) have to be adapted to the weighted $\varphi-$divergences;
for the sake of brevity, this will not be discussed in detail.
Notice that $\mathbf{P}^{wei}$, $\mathbf{Q}^{wei}$ are generally not
probability vectors anymore, even if $\mathbf{Q},\mathbf{P}$ are
probability vectors. 
In the latter case, and under the assumption $\sum_{k=1}^{K} c_{k} = 1$,
the divergences \eqref{brostu3:fo.div.weighted1} 
coincide with the discrete versions of the \textit{($\mathbf{c}-$)local divergences} of
Avlogiaris et al. \cite{Avl:16a}, \cite{Avl:16b} who also
give absolutely-continuous versions and beyond (see also \cite{Bro:19b} 
for an imbedding in a general divergence framework).

\end{remark}

%
%

\section{Construction principles: 
the cornerstone \label{Sect Prelude}}

\noindent
For the divergence-minimization $\Phi_{\mathbf{P}}(\Omega) := \inf_{\mathbf{Q}\in\mathbf{\Omega}} 
D_{\varphi}( \mathbf{Q}, \mathbf{P} )$ (and its variants),
in order to obtain the \eqref{brostu3:fo.2}-conform construction of the 
desired sequences $(\boldsymbol{\xi}_{n})_{n\in \mathbb{N}}$
of $\mathbb{R}^{K}-$valued random variables 
and $(\mathbb{\Pi}_{n})_{n \in \mathbb{N}}$
of probability distributions, we will assume 
(directly or after some multiplication) that the divergence generator 
$\varphi \in \widetilde{\Upsilon} (]a,b[)$ 
has the additional property that it can be 
represented as
\begin{equation}
\varphi (t)=\sup_{z\in \mathbb{R}}\Big( z\cdot t-\log \int_{\mathbb{R}}
e^{z \cdot y}d\mathbb{\bbzeta} (y)\Big), \qquad t\in \mathbb{R},\ \ 
\label{Phi Legendre of mgf(W)}
\end{equation}
for some probability distribution/measure $\mathbb{\bbzeta} $ 
on the real line $\mathbb{R}$ such that the
function $z\mapsto MGF_{\mathbb{\bbzeta} }(z):=
\int_{\mathbb{R}}e^{z \cdot y}d\mathbb{\bbzeta} (y)$ is
finite on some open interval containing zero \footnote{
in particular, this implies 
that $\mathbb{\bbzeta} $ has light tails;}.
From this, we shall construct 
a sequence $(W_{n})_{n\in \mathbb{N}}$ of i.i.d. copies of a 
random variable $W$ whose distribution 
is $\mathbb{\bbzeta}$
\big(i.e. $\mathbb{\Pi}[W \in \cdot \, ] = \mathbb{\bbzeta}[ \, \cdot \,]$
under some $\mathbb{\Pi}$\big),
from which the desired $(\boldsymbol{\xi}_{n})_{n\in \mathbb{N}}$ 
and $(\mathbb{\Pi}_{n})_{n \in \mathbb{N}}$ will be constructed.
The class of functions $\varphi \in \widetilde{\Upsilon} (]a,b[)$
satisfying the representability \eqref{Phi Legendre of mgf(W)} will be denoted by $\Upsilon (]a,b[)$.
Notice that $\Upsilon (]a,b[)$ contains many divergence
generators; this and $\varphi -$construction principles will be developed in 
Section \ref{SectFind} below.

\vspace{0.2cm}
\noindent
The representability \eqref{Phi Legendre of mgf(W)} is \textit{the} cornerstone for
our approach, and opens the gate to make use of 
simulation methods in appropriate contexts. We first
develop this approach for \textit{deterministic} minimization problems
in the following Section \ref{Sect Minimization}
(where we retransform the
generator $\varphi $ into $\widetilde{c}\cdot \varphi $ for 
strictly positive scales $\widetilde{c}$ and where $\mathbb{\Pi}_{n} \equiv \mathbb{\Pi}$). 
Thereafter, in Section \ref{Subsect Approx emp risk},
we deal with the setup 
where $\mathbf{P}$ is identified with an unknown probability vector in the simplex $\mathbb{S}^{K}$
which is supposed to
be the limit (as $n$ tends to infinity) of the data-based empirical 
distribution pertaining to a collection of observations $\mathbf{X}_{n}$ $:=\left(
X_{1},..,X_{n}\right) $; 
this amounts to the estimation of $\Phi _{\mathbf{P}}\left( \Omega \right) $ based on 
$\mathbf{X}_{n}$, leading to the important \textquotedblleft minimization-distance
estimation problem\textquotedblright\ in statistics, artificial intelligence
and machine learning.

%
%

\section{BS-minimizability/amenability: \, 
deterministic minimization problems}
\label{Sect Minimization}

\noindent

\begin{problem}
\label{det Problem}
For pregiven $\varphi \in \widetilde{\Upsilon}(]a,b[)$, 
positive-entries vector $\mathbf{P}:=\left( p_{1},..,p_{K}\right) \in \mathbb{R}_{>0}^{K}$
(or from some subset thereof), and subset $\mathbf{\Omega} \subset \mathbb{R}^{K}$
(also denoted in boldface letters, with a slight abuse of notation) 
with regularity properties
\begin{equation}
cl(\mathbf{\Omega} )=cl\left( int\left( \mathbf{\Omega} \right) \right) ,  \qquad int\left( \mathbf{\Omega} \right) \ne \emptyset,
\label{regularity}
\end{equation}
find 
\begin{equation}
\Phi_{\mathbf{P}}(\mathbf{\Omega}) := \inf_{\mathbf{Q}\in \mathbf{\Omega} } D_{\varphi }(\mathbf{Q},\mathbf{P}),  
\label{min Pb}
\end{equation}
provided that 
\begin{equation}
\inf_{\mathbf{Q}\in \mathbf{\Omega} } D_{\varphi }(\mathbf{Q},\mathbf{P}) < \infty .
\label{def fi wrt Omega}
\end{equation}
An immediate consequence thereof is --- for pregiven $\varphi \in \widetilde{\Upsilon}(]a,b[)$ --- 
the treatment of the more flexible problem
\begin{equation}
\Phi_{\mathbf{P},h}(\mathbf{\Omega}) := \inf_{\mathbf{Q}\in \mathbf{\Omega} } 
h\Big(D_{\varphi }(\mathbf{Q},\mathbf{P}) \Big)
= h\Big(\inf_{\mathbf{Q}\in \mathbf{\Omega} }  D_{\varphi }(\mathbf{Q},\mathbf{P}) \Big)  
\label{min Pb with h 1}
\end{equation}
for any continuous strictly increasing function $h: \mathcal{H} \, \mapsto \mathbb{R}$
with $\mathcal{H} := [0,\infty[$ 
and extension $h(\infty) := \sup_{y \in \mathcal{H}}(y)$
(depending on the problem, a sufficiently 
large $\mathcal{H} \subset [0,\infty[$ may be enough),
respectively of 
\begin{equation}
\sup_{\mathbf{Q}\in \mathbf{\Omega} } 
h\Big(D_{\varphi }(\mathbf{Q},\mathbf{P}) \Big)
= h\Big(\inf_{\mathbf{Q}\in \mathbf{\Omega} }  D_{\varphi }(\mathbf{Q},\mathbf{P}) \Big)  
\label{min Pb with h 2}
\end{equation}
for any continuous strictly decreasing function $h: \mathcal{H} \, \mapsto \mathbb{R}$
and extension $h(\infty) := \inf_{y \in \mathcal{H}}(y)$.

\end{problem}

\vspace{0.4cm}

\begin{remark} \ 
\label{after det Problem}
(a) \ By the basic properties of $\varphi $, it follows that for all $c>0$
the level sets $\mathbf{\varphi }_{c}:=\left\{ x\in \mathbb{R}:\varphi
(x)\leq c\right\} $ are compact and so are the level sets 
$\Gamma _{c}:=\left\{ \mathbf{Q}\in \mathbb{R}^{K}:D_{\varphi }(\mathbf{Q},
\mathbf{P})\leq c\right\}$
for all $c>0$ .\\
(b) \ When $\mathbf{\Omega }$ is not closed but merely satisfies \eqref{regularity}, 
then the infimum in \eqref{min Pb} may not be reached in $\mathbf{\Omega }$ 
although being finite; however we aim for finding the 
\textit{infimum/minimum} in \eqref{min Pb}. Finding the \textit{minimizers}
in \eqref{min Pb} is another question. For instance, this can be solved
whenever, additionally, $\mathbf{\Omega }$ is a closed set which implies the
existence of minimizers in $\mathbf{\Omega }$. In this case, and when the
number of such minimizers is finite, those can be e.g. approximated by dichotomic
search. For the sake of brevity, this will not be addressed in this paper.\\ 
(c) \thinspace\ The purpose of condition \eqref{regularity} is to get
rid of the $\lim \sup$ type and $\lim \inf$ type results in our
below-mentioned \textquotedblleft bare-simulation\textquotedblright\
approach and to obtain \textit{limit}-statements which motivate our construction. In
practice, it is enough to verify $\mathbf{\Omega }\subseteq cl\left(
int\left( \mathbf{\Omega }\right) \right) $, which is equivalent to 
the left-hand part of \eqref{regularity}. Clearly, any open set 
$\mathbf{\Omega }\subset \mathbb{R}^{K}$ 
satisfies the left-hand part of \eqref{regularity}.
In the subsetup where $\mathbf{\Omega }$ is a closed convex set 
and $int(\mathbf{\Omega })\neq \emptyset $, 
\eqref{regularity} is satisfied and the minimizer $\mathbf{Q}_{min}\in 
\mathbf{\Omega }$ in 
\eqref{min Pb} is attained and even unique. When $\mathbf{\Omega }
$ is open and satisfies \eqref{regularity}, then the
infimum in (\ref{min Pb}) exists but is reached at some generalized
projection of $\mathbf{P}$ on $\mathbf{\Omega }$ 
(see \cite{Csi:84} 
for the Kullback-Leibler divergence case of probability measures, 
which extends to any $\varphi-$divergence in our framework).
\\
(d) Without further mentioning,
the regularity condition \eqref{regularity} is supposed to hold in the \textit{full}
topology. Of course, 
$int\left( \mathbb{S}^{K} \right) = \emptyset$
and thus, for the important probability-vector setup $\mathbf{\Omega} \subset \mathbb{S}^{K}$
the condition \eqref{regularity} is violated which requires extra refinements
(cf. Section \ref{Subsect Approx emp risk} below).
The same is needed for $\mathbf{\Omega} \subset A \cdot \mathbb{S}^{K}$
for some $A \ne 1$, since obviously $int\left( A \cdot\mathbb{S}^{K} \right) = \emptyset$;
such a context appears naturally e.g. in connection with mass transportation problems 
(cf. \eqref{Monge1a} below) and with distributed energy management (cf. the
paragraph after \eqref{quadopt2}).
\\
(e) \ Often, $\mathbf{\Omega }$ will present a (discrete) model
\footnote{recall that an alternative naming also used in literature is to call $\mathbf{\Omega}$
a model class (rather than model), and each $\mathbf{Q} \in \mathbf{\Omega}$ a model (rather than model element)
}.
Since $\mathbf{\Omega }$ is assumed to have a non-void interior (cf. the
right-hand part of \eqref{regularity}), this will exclude parametric models 
$\mathbf{\Omega }:=\{\mathbf{Q}_{\theta }:\theta \in \Theta \}$ for some 
$\Theta \subset \mathbb{R}^{d}$ ($d<K-1$), for which $\theta \mapsto 
\mathbf{Q}_{\theta }$ constitutes a curve/surface in 
$\mathbb{R}^{K}$;
however, for such a situation, one
can employ standard minimization principles. Our approach is 
predestined
for \textit{non- or semiparametric} models, instead. For instance, \eqref{regularity} 
is valid for appropriate \textit{tubular neighborhoods} of parametric models
or for more general non-parametric settings such as e.g. shape constraints. 

\end{remark}

\vspace{0.4cm}
\noindent
Let us now present our \textit{new bare-simulation approach}
(cf. Definition \ref{brostu3:def.1})
for solving the distance-optimization Problem 
\ref{det Problem}:

\vspace{0.2cm}

\begin{enumerate}

\item[(BS1)] Step 1: equivalently rewrite \eqref{min Pb} 
such that the vector $\mathbf{P}$ \textquotedblleft turns into\textquotedblright\ a probability vector $\widetilde{\mathds{P}}$. 
More exactly, define $M_{\mathbf{P}}:=\sum_{i=1}^{K}p_{i}>0$ and let 
$\widetilde{\mathds{P}}:=\mathbf{P}/M_{\mathbf{P}},$ and
for $\mathbf{Q}$ in $\mathbf{\Omega}$, let $\widetilde{\mathbf{Q}}:=\mathbf{Q}/M_{\mathbf{P}}$ 
(notice that $\widetilde{\mathbf{Q}}$ 
may be a non-probability vector).
With the function 
$\widetilde{\varphi} \in \widetilde{\Upsilon}(]a,b[)$ defined through $\widetilde{\varphi }:=M_{\mathbf{P}} \cdot \varphi $, 
we obtain
\begin{equation}
D_{\varphi }(\mathbf{Q},\mathbf{P})=\sum_{k=1}^{K}p_{k}\cdot \varphi \Big( \frac{q_{k}}{p_{k}}
\Big) =\sum_{k=1}^{K}M_{\mathbf{P}}\cdot \widetilde{p_{k}}\cdot \varphi
\Big( \frac{M_{\mathbf{P}}\cdot \widetilde{q_{k}}}{M_{\mathbf{P}}\cdot \widetilde{p_{k}}}
\Big)=D_{\widetilde{\varphi }}(\widetilde{\mathbf{Q}},\widetilde{\mathds{P}}).
\label{min Pb prob1}
\end{equation}
It follows that the solution of \eqref{min Pb} coincides with the one
of the problem of finding
\begin{equation}
\widetilde{\Phi}_{\widetilde{\mathds{P}}}(\widetilde{\mathbf{\Omega}}) := \inf_{\widetilde{\mathbf{Q}}\in \widetilde{\mathbf{\Omega}} }
D_{\widetilde{\varphi} }(\widetilde{\mathbf{Q}},\widetilde{\mathds{P}}),  
\qquad \textrm{with } \widetilde{\mathbf{\Omega}}:=\mathbf{\Omega} /M_{\mathbf{P}};
\label{min Pb prob2}
\end{equation}
as a side remark, one can see that in such a situation the rescaling of the divergence generator $\varphi$
is important, which is one incentive that we incorporate multiples of $\varphi$ below.

\vspace{0.2cm}
\noindent
As an important special case we get for the choice $\mathbf{P} := (1, \ldots, 1) := \mathbf{1}$ that
the \textquotedblleft prominent/frequent\textquotedblright\ separable nonlinear optimization problem 
of finding the optimal value
$\inf_{\mathbf{Q} \in \mathbf{\Omega} } \sum_{k=1}^{K}\varphi (q_{k})$
--- with objective (e.g. cost, energy, purpose) function $\varphi \in \widetilde{\Upsilon}(]a,b[)$
and constraint set (choice set, search space) $\mathbf{\Omega}$
--- can be imbedded into our BS-approach by
\begin{equation}
\inf_{\mathbf{Q} \in \mathbf{\Omega} } \sum_{k=1}^{K}\varphi (q_{k}) = 
\inf_{\mathbf{Q} \in \mathbf{\Omega} } D_{\varphi }(\mathbf{Q},\mathbf{1})  =
\inf_{\widetilde{\mathbf{Q}} \in \mathbf{\Omega}/K } 
D_{K \cdot \varphi}(\widetilde{\mathbf{Q}},\mathds{P}^{unif}),   
\label{min Pb one}
\end{equation}
with $\mathds{P}^{unif} := (\frac{1}{K}, \ldots, \frac{1}{K})$ being the probability vector 
of frequencies of the uniform distribution on $\{1, \ldots, K\}$.

\end{enumerate}
 
\vspace{0.2cm}

\begin{remark}
\label{inequality}
(a) Since $\mathbf{1}$ can be seen 
as a reference vector with (normalized)
equal components, 
$\inf_{\mathbf{Q} \in \mathbf{\Omega} } D_{\varphi }(\mathbf{Q},\mathbf{1})$
in \eqref{min Pb one} 
can be interpreted as an \textquotedblleft  index/degree of (in)equality of
the set $\mathbf{\Omega}$\textquotedblright ,
respectively as an \textquotedblleft  index/degree of diversity of
the set $\mathbf{\Omega}$\textquotedblright .\\
(b) The quantity $\sum_{k=1}^{K}\varphi (q_{k})$ in \eqref{min Pb one}
can be interpreted as (non-probability extension of a)
$\varphi-$entropy in the sense of Burbea \& Rao \cite{Bur:82}
(see also~\cite{Csi:72}--\cite{Vaj:07});
for applications to scalar quantization for lossy coding of information sources
see e.g.~\cite{Gyo:00}.
\end{remark}

\vspace{0.2cm}
\noindent
Returning to the original distance-minimizing Problem \ref{det Problem},
after the first step \eqref{min Pb prob1} and \eqref{min Pb prob2},
we proceed as follows:

\vspace{0.2cm}

\begin{enumerate}

\item[(BS2)] Step 2: construct an appropriate sequence 
$(\boldsymbol{\xi}_{n})_{n\in \mathbb{N}}$ 
of $\mathbb{R}^{K}-$valued random variables
(cf. \eqref{brostu3:fo.2} in Definition \ref{brostu3:def.1}):

\vspace{0.3cm}
\noindent
The following condition transposes the minimization problem \eqref{min Pb prob2} 
(and thus the equivalent problem \eqref{min Pb})
into a 
\textit{BS minimizable/amenable problem} in the sense of Definition 
\ref{brostu3:def.1}.
The connection of this condition with \eqref{Phi Legendre of mgf(W)}
will be discussed in Proposition \ref{ID1} and its surroundings, 
see Section \ref{SectFind} below.

\vspace{0.3cm}

\begin{description}
\item 
\begin{condition}
\label{Condition  Fi Tilda in Minimization}
With $M_{\mathbf{P}} =\sum_{i=1}^{K}p_{i}>0$, the divergence generator $\varphi$ 
in \eqref{min Pb}
(cf. also \eqref{min Pb prob1}) satisfies
$\widetilde{\varphi} := M_{\mathbf{P}} \cdot \varphi \in \Upsilon(]a,b[)$, \, 
i.e. $\widetilde{\varphi} \in \widetilde{\Upsilon}(]a,b[)$ 
(which is equivalent to $\varphi \in \widetilde{\Upsilon}(]a,b[)$)
and there holds the representation 
\begin{equation}
\widetilde{\varphi}(t) = 
\sup_{z \in \mathbb{R}} \Big( z\cdot t - \log \int_{\mathbb{R}} e^{zy} d\widetilde{\mathbb{\bbzeta}}(y) \Big),
\qquad t \in \mathbb{R},  
\label{brostu3:fo.link.var}
\end{equation}
for some probability measure $\widetilde{\mathbb{\bbzeta}}$ on the real line $\mathbb{R}$
such that the function $z \mapsto MGF_{\widetilde{\mathbb{\bbzeta}}}(z) := \int_{\mathbb{R}} e^{zy} d\widetilde{\mathbb{\bbzeta}}(y)$ is finite
on some open interval containing zero
\footnote{in particular, this implies that
$\int_{\mathbb{R}} y d\widetilde{\mathbb{\bbzeta}}(y) =1$ 
and that $\widetilde{\mathbb{\bbzeta}}$
has light tails. 
}. Notice that $\widetilde{\mathbb{\bbzeta}}$ may depend on $M_{\mathbf{P}}$
in a highly non-trivial way (see e.g. Section \ref{Sect Cases} below).
\end{condition}
\end{description}

\vspace{0.4cm}
\noindent
Next, we explain the above-mentioned Step 2 in detail: \ for any $n \in \mathbb{N}$ and 
any $k \in \left\{ 1, \ldots ,K\right\}$, let $n_{k}:=\lfloor n \cdot \widetilde{p}_{k}\rfloor $ 
where $\lfloor x \rfloor$ denotes the integer part
of $x$. We assume $\mathbf{P} \in \mathbb{R}_{> 0}^{K}$, 
and since thus none of the $\widetilde{p}_{k}$'s is zero, one has 
\begin{equation}
\lim_{n\rightarrow \infty} \frac{n_{k}}{n} = \widetilde{p}_{k}.
\label{fo.freqlim}
\end{equation}
Moreover, we assume that $n \in \mathbb{N}$ is large enough, 
namely
$n \geq \max_{k \in \{1, \ldots, K\}} \frac{1}{\widetilde{p}_{k}}$,
and decompose the set $\{1, \ldots, n\}$ of all integers from $1$ to $n$
into the following disjoint blocks: $I_{1}^{(n)}:=\left\{
1,\ldots ,n_{1}\right\} $, $I_{2}^{(n)}:=\left\{ n_{1}+1,\ldots
,n_{1}+n_{2}\right\} $, and so on until the last block 
$I_{K}^{(n)} := \{ \sum_{k=1}^{K-1} n_{k} + 1, \ldots, n \}$ which
therefore contains all integers from $n_{1}+ \ldots +n_{K-1}+1$ to $n$.
Clearly, $I_{k}^{(n)}$ has $n_{k} \geq 1$ elements (i.e. $card(I_{k}^{(n)}) = n_{k}$
where $card(A)$ denotes the number of elements in a set $A$) 
for all $k \in \{1, \ldots, K-1\}$, and the last block $I_{K}^{(n)}$
has \, $n- \sum_{k=1}^{K-1} n_{k} \geq 1$ \, elements which anyhow
satisfies $\lim_{n\rightarrow \infty } card(I_{K}^{(n)})/n=\widetilde{p}_{K}$
\footnote{
if all $\widetilde{p}_{k}$ ($k=1,\ldots,K$) are rational numbers in $]0,1[$ with 
$\sum_{k=1}^{K} \widetilde{p}_{k} =1$ 
and $N$ is the (always existing) smallest integer such that all
$N \cdot \widetilde{p}_{k}$ ($k=1,\ldots,K$) are integers (i.e. $\in \mathbb{N}$),
then for any multiple $n= \ell \cdot N$ ($\ell \in \mathbb{N}$) 
one gets that all $n_{k} = n \cdot \widetilde{p}_{k}$ are integers
and that $card(I_{K}^{(n)}) = n_{K}$.
}.
Furthermore, consider a vector $\mathbf{\widetilde{W}}:=\left( \widetilde{W}
_{1},\ldots ,\widetilde{W}_{n}\right) $ where the $\widetilde{W}_{i}$'s are
i.i.d. copies of the random variable $\widetilde{W}$ whose distribution is
associated with the divergence-generator $\widetilde{\varphi }:=
M_{\textbf{P}} \cdot \varphi $ through 
\eqref{brostu3:fo.link.var}, in the sense that 
$\mathbb{\Pi }[\widetilde{W}\in \cdot \,]=\widetilde{\mathbb{\bbzeta}}
[\,\cdot \,]$. We group the $\widetilde{W}_{i}$'s according to the
above-mentioned blocks and sum them up blockwise, in order to build the
following $K-$ component random vector 
\begin{equation}
\boldsymbol{\xi }_{n}^{\mathbf{\widetilde{W}}}:=\Big(\frac{1}{n}\sum_{i\in
I_{1}^{(n)}}\widetilde{W}_{i},\ldots ,\frac{1}{n}\sum_{i\in I_{K}^{(n)}}
\widetilde{W}_{i}\Big);
\label{Xi_n^W vector}
\end{equation}
notice that the signs of 
its components
may be negative, depending on the
nature of the $\widetilde{W}_{i}$'s; moreover, the expectation of its $k-$th
component converges to $\widetilde{p}_{k}$ as $n$ tends to infinity 
(since the expectation of $\widetilde{W}_{1} $ is $1$), 
whereas the $n-$fold of the corresponding variance converges to $\widetilde{p}_{k}$ times
the variance of $\widetilde{W}_{1}$. 

\end{enumerate}

\vspace{0.3cm}
\noindent
For such a context, we obtain the following assertion
on BS-minimizability (which will be proved in in Appendix A):

\vspace{0.3cm}

\begin{theorem}
\label{brostu3:thm.divW.var}
Let $\mathbf{P} \in \mathbb{R}_{> 0}^{K}$, 
$M_{\mathbf{P}}:=\sum_{i=1}^{K}p_{i}>0$,
and suppose that the divergence generator $\varphi$ 
satisfies Condition \ref{Condition  Fi Tilda in Minimization}
above, with $\widetilde{\mathbb{\bbzeta}}$ (cf. \eqref{brostu3:fo.link.var}). 
Additionally, let $\widetilde{W}:=(\widetilde{W}_{i})_{i\in \mathbb{N}}$ be
a sequence of random variables where the $
\widetilde{W}_{i}$'s are i.i.d. copies of the random variable $\widetilde{W}$
whose distribution 
is $\mathbb{\Pi }[\widetilde{W}\in \cdot \,]=\widetilde{\mathbb{\bbzeta}}[\,\cdot \,]$ \footnote{
and thus, $E_{\mathbb{\Pi }}[\widetilde{W}_{i}]=1$}. 
Then, in terms of 
the random vectors $\boldsymbol{\xi }_{n}^{\mathbf{\widetilde{W}}}$
(cf. \eqref{Xi_n^W vector})
there holds 
\begin{equation}
-\lim_{n\rightarrow \infty }\frac{1}{n}\log \,\mathbb{\Pi }\negthinspace \left[
\boldsymbol{\xi }_{n}^{\mathbf{\widetilde{W}}}
\in \mathbf{\Omega} /M_{\mathbf{P}}
\right]
=\inf_{Q\in \mathbf{\Omega }}D_{\varphi }(\mathbf{Q},\mathbf{P})
\label{LDP Minimization}
\end{equation}
for any $\mathbf{\Omega }\subset \mathbb{R}^{K}$ with regularity properties 
\eqref{regularity} and finiteness property \eqref{def fi wrt Omega}.
In particular, for each $\mathbf{P} \in \mathbb{R}_{> 0}^{K}$ the function 
$\Phi_{\mathbf{P}} \left( \cdot \right) := D_{\varphi}( \cdot , \mathbf{P} )$
(cf. \eqref{brostu3:fo.div}) is bare-simulation minimizable (BS-minimizable, 
cf. \eqref{brostu3:fo.2}) on 
any such $\mathbf{\Omega }\subset \mathbb{R}^{K}$.

\end{theorem}

\vspace{0.2cm} 
\noindent
\begin{remark}
\label{dist of components}
(i)  For some contexts, one can \textit{explicitly} give the
distribution of each of the independent (non-deterministic parts of the) components
$\Big(\sum_{i\in I_{k}^{(n)}}\widetilde{W}_{i}\Big)_{k=1,\ldots,K}$
of the vector $\boldsymbol{\xi }_{n}^{\mathbf{\widetilde{W}}}$;
this will ease the corresponding concrete simulations.
For instance, we shall give those in the solved-cases Section \ref{Sect Cases}
below.\\
(ii)  Let us emphasize that we have assumed $\mathbf{P} \in \mathbb{R}_{> 0}^{K}$
in Theorem \ref{brostu3:thm.divW.var} which excludes $\mathbf{P}$
from having zero components. 
However, in cases
where $\lim_{x\rightarrow \infty }\left\vert \frac{\varphi \left(
x \cdot sgn(q)\right) }{x \cdot sgn(q)}\right\vert =+\infty $ for $q\neq 0$, then if 
$p_{k_{0}}=0$ for some $k_{0\text{ }}$ it follows that $q_{k_{0}}=0$,
which proves that $\mathbf{P} \in \mathbb{R}_{>0}^{K}$ imposes no restriction
in Theorem \ref{brostu3:thm.divW.var},
since the projection of $\mathbf{P}$ in $\mathbf{\Omega }$ then belongs
to the subspace of $\mathbb{R}^{K}$ generated by the non-null components of $\mathbf{P}$;
such a situation appears e.g. for power divergence generators $\varphi_{\gamma }$ with $\gamma > 2$. 
So there is no loss of generality assuming $\mathbf{P} \in \mathbb{R}_{>0}^{K}$ in this case.

\end{remark}

\vspace{0.3cm}
\noindent
As examples for the applicability of Theorem \ref{brostu3:thm.divW.var}, 
one can e.g. combine \textit{each} of the divergence generators $\varphi$  
of Section \ref{Sect Cases}
(except for the one in Subsection \ref{Subsect Case9})
with \textit{any} of the optimization
problems \eqref{min Pb},\eqref{min Pb with h 1},\eqref{min Pb with h 2},\eqref{min Pb one} 
as well as the below-mentioned 
\eqref{min Pb one flex1},\eqref{min Pb one flex2};
the needed distributions $\mathbb{\Pi }[\widetilde{W}\in \cdot \,]=\widetilde{\mathbb{\bbzeta}}[\,\cdot \,]$
correspond to the entry of the corresponding Subsection \ref{Sect Cases}$-$
with the choice $\widetilde{c} \cdot M_{\mathbf{P}}$ instead of $\widetilde{c}$.
By taking $\zeta := - \varphi$ instead, one can solve the
below-mentioned problems \eqref{min Pb one flex3} and \eqref{min Pb one flex4}.

\vspace{0.5cm} 
\noindent
Returning to the general context, the limit statement \eqref{LDP Minimization} provides the principle for the approximation of the solution of Problem \eqref{min Pb}. Indeed,
by replacing the left-hand side in \eqref{LDP Minimization} by its finite
counterpart, we deduce for given large $n$  
\begin{equation}
- \frac{1}{n}\log \mathbb{\Pi} \negthinspace \left[ \boldsymbol{\xi}_{n}^{\mathbf{\widetilde{W}}}\in 
\mathbf{\Omega}/M_{\mathbf{P}} \right] 
\approx \inf_{Q\in \mathbf{\Omega} }D_{\varphi }(\mathbf{Q},\mathbf{P});
\label{fo.approx.1} 
\end{equation}
it remains to estimate the left-hand side of \eqref{fo.approx.1}.
The latter can be performed either by a \textit{naive estimator} of the
frequency of those replications of $\boldsymbol{\xi}_{n}^{\mathbf{\widetilde{W}}}$  
which hit $\mathbf{\Omega}/M_{\mathbf{P}}$, or more efficiently by some improved estimator; 
for details, see Section \ref{Sect Estimators} below.

\vspace{0.3cm}

\begin{remark} \ 
\label{computingGeneral}
According to 
\eqref{LDP Minimization} of Theorem \ref{brostu3:thm.divW.var} as well as \eqref{fo.approx.1},
we can principally tackle 
the (approximative) computation of the minimum value
$\inf_{Q\in \mathbf{\Omega}}
D_{\varphi }(\mathbf{Q},\mathbf{P}) =
\inf_{Q\in \mathbf{\Omega}}
\sum_{k=1}^{K}p_{k}\cdot \varphi \left( \frac{q_{k}}{p_{k}}\right)$
and in particular of
$\inf_{\mathbf{Q} \in \mathbf{\Omega} } \sum_{k=1}^{K}\varphi (q_{k}) = 
\inf_{\mathbf{Q} \in \mathbf{\Omega} } D_{\varphi }(\mathbf{Q},\mathbf{1})  
\  \textrm{(cf. \eqref{min Pb one})}$
by basically \textit{only employing a fast and accurate 
--- pseudo, true, natural, quantum  
--- random number generator}\footnote{
see e.g.~\cite{Tuc:13}--\cite{Sto:21} 
},
provided that the constraint set $\mathbf{\Omega}$ 
satisfies the mild assumptions \eqref{regularity} and \eqref{def fi wrt Omega}.
Notice that \eqref{regularity} also covers 
(e.g. high-dimensional) constraint sets $\mathbf{\Omega}$
which are \textit{non-convex} and even \textit{highly disconnected}, and for which
other minimization methods (e.g. pure enumeration, gradient or steepest descent methods, etc.
\footnote{ a detailed
discussion and comparisons are beyond the scope of this paper,
given its current length 
}
) may be problematic or intractable.
For instance, \eqref{regularity} covers 
kind of \textquotedblleft  $K-$dimensional 
(not necessarily regular)
polka dot (leopard skin) pattern type\textquotedblright\ relaxations
$\mathbf{\Omega} := \dot{\bigcup}_{i=1}^{N} \mathcal{U}_{i}(\mathbf{Q}_{i}^{dis})$ of finite 
discrete constraint sets $\mathbf{\Omega}^{dis} := \{\mathbf{Q}_{1}^{dis}, \ldots, \mathbf{Q}_{N}^{dis}\}$ 
of high cardinality $N$ 
(e.g. being exponential or factorial in a large $K$),
where each $K-$dimensional vector $\mathbf{Q}_{i}^{dis}$ (e.g. having pure integer components only)
is surrounded by some small (in particular, non-overlapping/disjoint) 
neighborhood $\mathcal{U}_{i}(\mathbf{Q}_{i}^{dis})$;
in such a context, e.g. $\inf_{\mathbf{Q} \in \mathbf{\Omega} } \sum_{k=1}^{K}\varphi (q_{k})$
can be regarded as a 
\textit{\textquotedblleft  BS-tractable\textquotedblright\
relaxation} of the nonlinear discrete (e.g. integer, combinatorial
\footnote{
see e.g.~\cite{Schr:03}--\cite{Wol:21}
for comprehensive books 
on discrete, integer and combinatorial programming and their vast applications
}
) optimization program
$\inf_{\mathbf{Q} \in \mathbf{\Omega}^{dis} } \sum_{k=1}^{K}\varphi (q_{k})$.
\end{remark}

\vspace{0.3cm}
\noindent
Returning to the general context, notice that Theorem \ref{brostu3:thm.divW.var}
does not cover cases where $\mathbf{\Omega}$ consists of $\mathbf{Q}$ satisfying the
additional constraint $\sum_{i=1}^{K} q_{i} =A$ for some fixed $A>0$
(and thus $int\left( \mathbf{\Omega} \right) = \emptyset$ violating \eqref{regularity}).
However, such situations can be still handled with an adaption of 
the above-described BS method, see Remark \ref{remark divnormW}(v),(vi), 
Lemma \ref{Lemma Indent Rate finite case_new} and Section \ref{Sect Cases} below.

\enlargethispage{1.0cm}


\subsection{Generalizations}

Recalling \eqref{min Pb one}, let us point out that with our new BS approach one may even tackle more general optimization problems of the form 
$\inf_{\breve{\mathbf{Q}} \in \breve{\mathbf{\Omega}}} \sum_{k=1}^{K} \breve{\varphi} (\breve{q}_{k})$
where basically $\breve{\varphi}$ is some function which is finite and convex in a
non-empty neighborhood (say, $]t_{0} + a- 1, t_{0} + b- 1[$ with $a < 1 < b$) 
of some point $t_{0} \in \mathbb{R}$ as well as differentiable and strictly convex in a non-empty
sub-neighborhood of $t_{0}$; for this, the function
$\varphi(t) := \breve{\varphi}(t+t_{0}-1) - \breve{\varphi}^{\prime}(t_{0})
\cdot \Big((t+t_{0}-1) - t_{0} \Big) - \breve{\varphi}(t_{0}), 
\ t \in ]a,b[,$
(which corresponds to 
argument-shifting and adding an affine-linear function) 
should be such that $K \cdot \varphi \in \Upsilon (]a,b[)$, and from the corresponding minimization problem
\begin{eqnarray}
& &
\inf_{\widetilde{\mathbf{Q}} \in \mathbf{\Omega}/K } 
D_{K \cdot \varphi}(\widetilde{\mathbf{Q}},\mathds{P}^{unif}) =
\inf_{\mathbf{Q} \in \mathbf{\Omega} } \sum_{k=1}^{K} \varphi (q_{k})
= \inf_{\mathbf{Q} \in \mathbf{\Omega} } \sum_{k=1}^{K}
\Big( 
\breve{\varphi}(q_{k}+t_{0}-1) - \breve{\varphi}^{\prime}(t_{0})
\cdot ((q_{k}+t_{0}-1) -t_{0}) - \breve{\varphi}(t_{0})
\Big)
\nonumber \\
& & = \inf_{\breve{\mathbf{Q}} \in \mathbf{\Omega} + t_{0} -1} \sum_{k=1}^{K} \ 
\Big( 
\breve{\varphi}(\breve{q}_{k}) - \breve{\varphi}^{\prime}(t_{0})
\cdot (\breve{q}_{k} -t_{0}) - \breve{\varphi}(t_{0})
\Big)
\nonumber \\
& & = K \cdot \Big(t_{0} \cdot \breve{\varphi}^{\prime}(t_{0}) - \breve{\varphi}(t_{0})\Big)
+ \inf_{\breve{\mathbf{Q}} \in \breve{\mathbf{\Omega}} } \left(
\sum_{k=1}^{K} \breve{\varphi}(\breve{q}_{k})
-  \breve{\varphi}^{\prime}(t_{0}) \cdot \sum_{k=1}^{K} \breve{q}_{k}
\right),
\qquad \textrm{with } \breve{\mathbf{\Omega}} :=  \mathbf{\Omega} + t_{0} -1, 
\label{min Pb one shifted}
\end{eqnarray}
the term $\inf_{\breve{\mathbf{Q}} \in \breve{\mathbf{\Omega}}} 
\sum_{k=1}^{K} \breve{\varphi} (\breve{q}_{k})$ 
should be recoverable; for instance, later on 
we shall employ constraints sets
$\breve{\mathbf{\Omega}}$ which particularly include $\sum_{k=1}^{K} \breve{q}_{k} = A > 0$,
whereas another possibility would be to use a $\breve{\varphi}$ 
which satisfies $\breve{\varphi}^{\prime}(t_{0}) =0$. 
As a different line of flexibilization of \eqref{min Pb one},
we can also deal with the problem 
$\inf_{\mathbf{Q} \in \mathbf{\Omega} } h\Big(\sum_{k=1}^{K}\varphi (q_{k})\Big)$
through
\begin{equation}
\inf_{\mathbf{Q} \in \mathbf{\Omega} } 
h\Big(\sum_{k=1}^{K}\varphi (q_{k})\Big) =  
h\Big( \inf_{\widetilde{\mathbf{Q}} \in \mathbf{\Omega}/K }
D_{K \cdot \varphi}(\widetilde{\mathbf{Q}},\mathds{P}^{unif}) \Big)   
\label{min Pb one flex1}
\end{equation}
for any $\varphi$ with $K \cdot \varphi \in \Upsilon (]a,b[)$ and any 
continuous strictly increasing function $h: \mathcal{H} \, \mapsto \mathbb{R}$
with $\mathcal{H} := [0,\infty[$ (or a sufficiently large subset thereof),
and with the problem 
$\sup_{\mathbf{Q} \in \mathbf{\Omega} } h\Big(\sum_{k=1}^{K}\varphi (q_{k})\Big)$ through
\begin{equation}
\sup_{\mathbf{Q} \in \mathbf{\Omega} } 
h\Big(\sum_{k=1}^{K}\varphi (q_{k})\Big) =  
h\Big( \inf_{\widetilde{\mathbf{Q}} \in \mathbf{\Omega}/K }
D_{K \cdot \varphi}(\widetilde{\mathbf{Q}},\mathds{P}^{unif}) \Big)   
\label{min Pb one flex2}
\end{equation}
for any $\varphi$ with $K \cdot \varphi \in \Upsilon (]a,b[)$ and any 
continuous strictly decreasing function $h: \mathcal{H} \, \mapsto \mathbb{R}$.
As a continuation of Remark \ref{inequality}(b),
the quantity $h\Big(\sum_{k=1}^{K}\varphi (q_{k})\Big)$
in \eqref{min Pb one flex1} can be seen as 
(non-probability extension of a) 
$(h,\varphi)-$entropy in the sense of
Salicru et al.~\cite{Sal:93} (see also 
e.g.~\cite{Par:06},\cite{Vaj:85}
as well as~\cite{Che:07},\cite{Ren:15} for exemplary applications);
important special cases will be discussed in more detail, below.
Combining \eqref{min Pb one shifted} with \eqref{min Pb one flex1}
(respectively, with \eqref{min Pb one flex2}) leads to a further flexibilization.
Of course, we can also apply our BS method to the maximization
$\sup_{\mathbf{Q} \in \mathbf{\Omega} } h\Big(\sum_{k=1}^{K}\zeta (q_{k})\Big)$
for any concave function $\zeta$ with $- K \cdot \zeta \in \Upsilon(]a,b[)$ and any 
continuous strictly increasing function $h: \mathcal{H} \, \mapsto \mathbb{R}$
with $\mathcal{H} := - [\infty,0]$ (or a sufficiently large subset thereof),
via 
\begin{equation}
\sup_{\mathbf{Q} \in \mathbf{\Omega} } 
h\Big(\sum_{k=1}^{K}\zeta (q_{k})\Big) =  
h\Big( - \inf_{\widetilde{\mathbf{Q}} \in \mathbf{\Omega}/K }
D_{- K \cdot \zeta}(\widetilde{\mathbf{Q}},\mathds{P}^{unif}) \Big) ,   
\label{min Pb one flex3}
\end{equation}
and to $\inf_{\mathbf{Q} \in \mathbf{\Omega} } h\Big(\sum_{k=1}^{K}\zeta (q_{k})\Big)$
for any concave $\zeta$ with $-K \cdot \zeta \in \Upsilon(]a,b[)$ and any 
continuous strictly decreasing 
$h: \mathcal{H} \, \mapsto \mathbb{R}$, via 
\begin{equation}
\inf_{\mathbf{Q} \in \mathbf{\Omega} } 
h\Big(\sum_{k=1}^{K}\zeta (q_{k})\Big) =  
h\Big( - \inf_{\widetilde{\mathbf{Q}} \in \mathbf{\Omega}/K }
D_{- K \cdot \zeta}(\widetilde{\mathbf{Q}},\mathds{P}^{unif}) \Big) .   
\label{min Pb one flex4}
\end{equation} 
Moreover, we can tackle  
$\sup_{\breve{\mathbf{Q}} \in \breve{\mathbf{\Omega}}} \sum_{k=1}^{K} \breve{\zeta} (\breve{q}_{k})$
where $\breve{\zeta}$ is some function which is finite and concave in a
non-empty neighborhood $]t_{0} + a- 1, t_{0} + b- 1[$ (with $a < 1 < b$) 
of some point $t_{0} \in \mathbb{R}$ as well as differentiable and strictly concave in a non-empty
sub-neighborhood of $t_{0}$; for this, the function
\begin{equation}
- \zeta(t) := - \breve{\zeta}(t+t_{0}-1) + \breve{\zeta}^{\prime}(t_{0})
\cdot \Big((t+t_{0}-1) - t_{0} \Big) + \breve{\zeta}(t_{0}), \qquad t \in ]a,b[,
\nonumber
\end{equation}
should be such that $- K \cdot \zeta \in \Upsilon (]a,b[)$, and from the corresponding minimization problem
\begin{eqnarray}
& &
- \inf_{\widetilde{\mathbf{Q}} \in \mathbf{\Omega}/K }
D_{- K \cdot \zeta}(\widetilde{\mathbf{Q}},\mathds{P}^{unif})
= \sup_{\mathbf{Q} \in \mathbf{\Omega} } \sum_{k=1}^{K} \zeta(q_{k})
= \sup_{\mathbf{Q} \in \mathbf{\Omega} } \sum_{k=1}^{K}
\Big( 
\breve{\zeta}(q_{k}+t_{0}-1) - \breve{\zeta}^{\prime}(t_{0})
\cdot ((q_{k}+t_{0}-1) -t_{0}) - \breve{\zeta}(t_{0})
\Big)
\nonumber \\
& & = \sup_{\breve{\mathbf{Q}} \in \mathbf{\Omega} + t_{0} -1} \sum_{k=1}^{K} \ 
\Big( 
\breve{\zeta}(\breve{q}_{k}) - \breve{\zeta}^{\prime}(t_{0})
\cdot (\breve{q}_{k} -t_{0}) - \breve{\zeta}(t_{0})
\Big)
\nonumber \\
& & = K \cdot \Big(t_{0} \cdot \breve{\zeta}^{\prime}(t_{0}) - \breve{\zeta}(t_{0})\Big)
+ \sup_{\breve{\mathbf{Q}} \in \breve{\mathbf{\Omega}} } \left(
\sum_{k=1}^{K} \breve{\zeta}(\breve{q}_{k})
-  \breve{\zeta}^{\prime}(t_{0}) \cdot \sum_{k=1}^{K} \breve{q}_{k}
\right),
\qquad \textrm{with } \breve{\mathbf{\Omega}} :=  \mathbf{\Omega} + t_{0} -1, 
\label{min Pb one shifted 2}
\end{eqnarray}
the term $\sup_{\breve{\mathbf{Q}} \in \breve{\mathbf{\Omega}}}
\sum_{k=1}^{K} \breve{\zeta} (\breve{q}_{k})$ 
should be recoverable; the left-hand side of \eqref{min Pb one shifted 2} corresponds to the special case 
$h(x) := x$ of the BS-minimizable \eqref{min Pb one flex3}.
A combination of \eqref{min Pb one shifted 2} with 
\eqref{min Pb one flex3}
(respectively, with \eqref{min Pb one flex4}) leads to a further flexibilization.

%
%

\section{
Stochastic minimum distance/risk estimation and deterministic simplex cases
\label{Subsect Approx emp risk}}

\subsection{General stochastic construction}

\noindent
In contrast to the previous Section \ref{Sect Minimization}, 
we now work out our BS method for the important setup where basically $P$ is a 
\textit{random} (unknown) element of the simplex $\mathbb{S}^{K}$ 
of $K-$component probability (frequency) vectors 
and $\mathbf{\Omega} \subset \mathbb{S}^{K}$ (which violates \eqref{regularity}
since $int\left( \mathbf{\Omega} \right) = \emptyset$, cf. 
Remark \ref{after det Problem}(d), and thus requires a different treatment). 
To begin with, in the statistics of discrete data --- and in the adjacent research fields of
information theory, artificial intelligence and machine learning --- one
often encounters the following \textit{minimum distance estimation (MDE)
problem} which is often also named as 
\textit{estimation of the empirical risk}:

\vspace{0.4cm}

\begin{enumerate}

\item[(MDE1)] for index $i\in \mathbb{N}$, let the generation of the $i-$th
(uncertainty-prone) data point be represented by the random variable $X_{i}$
which takes values in the discrete set $\mathcal{Y}:=\left\{ d_{1},\cdots ,d_{K}\right\}$ 
of $K$ distinct values \textquotedblleft of any
kind\textquotedblright. It is assumed that there exists a probability
measure $\mathbb{P}[\cdot \,]$ on $\mathcal{Y}$ which is the a.s. limit 
(as $n$ tends to infinity) of
the empirical measures $\mathbb{P}_{n}^{emp}$ defined by the collection 
$\left( X_{1},..,X_{n}\right)$, in
formula  
\begin{equation}
\lim_{n\rightarrow \infty }\mathbb{P}_{n}^{emp}:=\lim_{n\rightarrow \infty }
\frac{1}{n}\sum_{i=1}^{n}\delta _{X_{i}}=\mathbb{P} \qquad \text{a.s.}
\label{cv emp measure X to P}
\end{equation}
where $\delta _{y}$ denotes the one-point distribution (Dirac mass) at point 
$y$ 
\footnote{
notice that $\mathbb{P}_{n}^{emp}$ a probability measure 
(on the data space
$\mathcal{Y}$), which is random due to its dependence on the $X_{i}$\textquoteright s 
}. We will assume that none of the entries of $\mathbb{P}$ bears zero mass
so that $\mathbb{P}$ is identified with a point in the interior of $\mathbb{S}^{K}$ 
(see below).
The underlying probability space (say, $(\mathfrak{X},\mathcal{A},\mathbb{\Pi})$) 
where the above a.s. convergence holds, pertains 
to the random generation of the sequence $\left( X_{i}\right)_{i\in \mathbb{N}}$, 
of which we do not need to know but for \eqref{cv emp measure X to P}. 
Examples 
include the i.i.d. case (where the $X_{i}$\textquoteright s
are independent and have common distribution $\mathbb{P}$), 
ergodic Markov chains on $\mathcal{Y}$ with
stationary distribution $\mathbb{P}$, 
more globally autoregressive chains
with stationary measure $\mathbb{P}$, etc.

\vspace{0.2cm}
\noindent 
Let us briefly discuss our assumption
\eqref{cv emp measure X to P}
(resp. its vector form \eqref{cv emp measure X to P vector} below)
on the limit behavior of the empirical distribution 
of the observed sample $\mathbf{X}_{n}:=\left(X_{1},..,X_{n}\right)$ 
as $n$ tends to infinity.
In the \textquotedblleft  basic\textquotedblright\ statistical context, 
the sample $\mathbf{X}_{n}$ consists
of i.i.d. replications of a generic random variable $X$ with 
probability distribution $\mathds{P}$.
However, our approach captures many other sampling schemes, where the 
distribution $\mathds{P}$ is defined implicitly through 
\eqref{cv emp measure X to P}
for which we aim at some estimate of $\Phi_{\mathds{P}}\left( 
\mathbb{\Omega} \right)$
of a family $\mathbb{\Omega}$  
of probability distributions on $\mathcal{Y}$.
Sometimes the sequence of samples $\left( \mathbf{X}_{n}\right) _{n\in \mathbb{N}}$
may stem from a triangular array so that $\mathbf{X}_{n}=\left(
X_{1,n},..,X_{k_{n},n}\right) $ with $k_{n}\rightarrow \infty$ and
\eqref{cv emp measure X to P} 
is substituted by 
\vspace{-0.2cm}
\[
\lim_{n\rightarrow \infty }\frac{1}{k_{n}}\sum_{i=1}^{k_{n}}\delta_{X_{i,n}} =
\mathds{P} \text{ \ a.s.} 
\]
\vspace{-0.2cm}
\noindent
which does not alter the results of this paper by any means.

\item[(MDE2)] given a \textit{model} $\mathbb{\Omega}$, i.e. a 
family $\mathbb{\Omega}$  
of probability distributions $\mathbb{Q}$ on $\mathcal{Y}$
each of which serves as a potential description of
the underlying (unknown) data-generating mechanism $\mathbb{P}$, one would like to find
\begin{equation}
\Phi_{\mathbb{P}}(\mathbb{\Omega}) := \inf_{\mathbb{Q}\in\mathbb{\Omega}} D_{\varphi}( \mathbb{Q}, \mathbb{P} )
\label{inf proba}
\end{equation}
which quantifies the \textit{adequacy} of the model
$\mathbb{\Omega}$ for modeling $\mathbb{P}$, 
\textit{via} the minimal distance/dissimilarity of $\mathbb{\Omega}$ to $\mathbb{P}$;
a lower $\Phi_{\mathbb{P}}-$value means 
a better adequacy
(in the sense of a lower departure between the model and the truth,
cf.~\cite{Lind:04}--\cite{Mark:19}).
Hence, especially in the context of \textit{model selection} 
within complex big-data contexts, for the \textit{search of appropriate 
models} $\mathbb{\Omega}$ and
model elements/members therein,
the (fast and efficient) computation of $\Phi_{\mathbb{P}}(\mathbb{\Omega})$ constitutes
a decisive first step, since
if the latter is \textquotedblleft  too large\textquotedblright\ (respectively 
\textquotedblleft  much larger than\textquotedblright\ 
$\Phi_{\mathbb{P}}(\overline{\mathbb{\Omega}})$ for some competing
model $\overline{\mathbb{\Omega}}$), then the model 
$\mathbb{\Omega}$ is \textquotedblleft  not adequate enough\textquotedblright\ (respectively
\textquotedblleft  much less adequate than\textquotedblright\ 
 $\overline{\mathbb{\Omega}}$);
in such a situation, the effort of computing the (not necessarily unique) 
best model
element/member 
$\arg \inf_{\mathbb{Q}\in\mathbb{\Omega}} D_{\varphi}( \mathbb{Q}, \mathbb{P} )$
within the model $\mathbb{\Omega}$ 
is \textquotedblleft  not very useful\textquotedblright\ and is thus
a \textquotedblleft waste of computational time\textquotedblright.
Because of such considerations, we concentrate 
on finding the infimum
\eqref{inf proba} rather than finding the corresponding minimizer(s).
Variants of \eqref{inf proba} are of interest, too.

\end{enumerate}

\vspace{0.3cm}
\noindent
Since $int(\mathbb{\Omega})$ is required to be a non-empty set 
(in the relative topology)
in the space of 
probability distributions on $\mathcal{Y}$, the present procedure is fitted for
semi-parametric models $\mathbb{\Omega}$, 
e.g. defined through
moment conditions (as extensions of the Empirical Likelihood paradigm, see
e.g. \cite{Bro:12}), 
or through L-moment conditions
(i.e. moment conditions pertaining to quantile measures, see~\cite{Bro:16}), 
or even more involved non-parametric models
where the geometry of $\mathbb{\Omega }$ does not allow for ad-hoc
procedures. In such setups, there is typically no closed form of
the divergence with respect to any 
probability distribution
available.

\vspace{0.2cm}
\noindent
The measurement or the estimation of $\Phi _{\mathbb{P}}(\mathbb{\Omega})$ is a tool
for the choice of pertinent putative models $\mathbb{\Omega}$ among a class of
specifications. The case when  $\Phi _{\mathbb{P}}(\mathbb{\Omega })>0$ \,  is 
interesting in its own, 
since it is quite common in engineering modelling to argue in
favor of misspecified models (or (non-void) neighborhoods of such models
for sake of robustness issues), due to quest for conservatism; the choice
between them is a widely open field e.g. in the practice of reliability.
This also opens the question of the choice of the divergence generator $\varphi$;
although this will not be discussed in this paper, 
as a motivating running example the reader may keep in mind 
the generator 
$\varphi_{2}(x):=(x-1)^{2}/2$ 
which induces the divergence 
$D_{\varphi_{2}}(\mathbb{Q},\mathbb{P})$
(see \eqref{brostu3:fo.powdiv.new} below for details)
which quantifies the 
expected square relative error when
substituting the true distribution $\mathbb{P}$ by the model $\mathbb{Q}$. 

\vspace{0.2cm}
\noindent
An estimate of $\Phi_{\mathbb{P}}(\mathbb{\Omega})$ can be used as a statistics for some
test of fit, and indeed the likelihood ratio test adapted to some semi-parametric models
has been generalized to the divergence setting (see~\cite{Bro:12}). 
The statement of the limit distributions of our estimate, under
the model and under misspecification, is postponed to future work.

\vspace{0.2cm}
\noindent
In the following, we compute/approximate \eqref{inf proba} --- and some variants thereof --- by our 
\textit{bare simulation (BS)} method,
by \textquotedblleft mimicking\textquotedblright\ the deterministic minimization 
problem \eqref{min Pb} respectively 
\eqref{min Pb prob2}. Let us first remark that, as usual, each probability
distribution (probability measure) $\mathbb{P}$ on $\mathcal{Y}=\left\{d_{1},\ldots,d_{K}\right\}$
can be uniquely identified with the (row) vector $\mathds{P} := (p_{1}, \ldots, p_{K}) \in \mathbb{S}^{K}$ 
of the corresponding probability masses (frequencies) 
$p_{k} = \mathbb{P}[\{ d_{k} \}]$ via 
$\mathbb{P}[A] = \sum_{k=1}^{K} p_{k} \cdot \textfrak{1}_A(d_{k}) $ for each $A \subset \mathcal{Y}$,
where $\textfrak{1}_A(\cdot)$ denotes the indicator function on the set $A$.
In particular, the probability distribution $\mathbb{P}$ in (MDE1) can be identified with 
$(p_{1}, \ldots, p_{K})$ in terms of $p_{k} = \mathbb{P}[\{ d_{k} \}]$
(which in the i.i.d. case turns into $p_{k} = \mathbb{\Pi}[X_{1} = d_{k}]$).
Along this line, the family
$\mathbb{\Omega}$ of probability distributions in (MDE2)
can be identified with a subset $\boldsymbol{\Omega}$\hspace{-0.23cm}$\boldsymbol{\Omega} 
 \subset  \mathbb{S}^{K}$ of probability vectors 
(viz. of vectors of probability masses).
Analogously, each finite nonnegative measure $Q$ on $\mathcal{Y}$
can be uniquely identified with a vector $\mathbf{Q} := (q_{1}, \ldots, q_{K}) \in \mathbb{R}_{\geq 0}^{K}$,
and each finite signed measure $Q$ with a vector 
$\mathbf{Q} := (q_{1}, \ldots, q_{K}) \in \mathbb{R}^{K}$.
The corresponding divergences between distributions/measures are then, as usual, defined 
through the divergences between their respective masses/frequencies:

\begin{equation}
D_{\varphi }(Q,\mathbb{P}) : = D_{\varphi }(\mathbf{Q},\mathds{P}).
\label{measure divergence}
\end{equation}

\vspace{0.2cm}
\noindent
In particular, $\mathbb{P}_{n}^{emp}$ can be identified with 
the vector $\mathds{P}_{n}^{emp} := (p_{n,1}^{emp}, \ldots, p_{n,K}^{emp})$ where
\begin{equation}
p_{n,k}^{emp} \ := \ 
\frac{1}{n} \cdot n_{k} \ := \ \frac{1}{n} \cdot card(\bigl\{ i \in \{ 1, \ldots, n\}:  
\ X_{i} = d_{k} \bigr\})
\ =: \ \frac{1}{n} \cdot card(I_{k}^{(n)}) , \quad k \in \{1, \ldots, K\},
\label{I^(n)_k for stat case}
\end{equation}
and accordingly the required limit behaviour \eqref{cv emp measure X to P}
is equivalent to the vector-convergence
\begin{equation}
\lim_{n\rightarrow \infty } \Big( \frac{n_{1}}{n}, \ldots, \frac{n_{K}}{n} \Big) 
= (p_{1}, \ldots, p_{K}) 
\qquad 
\textrm{a.s.}
\label{cv emp measure X to P vector}
\end{equation}

\noindent
Notice that, in contrast to 
the above Section \ref{Sect Minimization},
the sets $I_{k}^{(n)}$ of indexes introduced in \eqref{I^(n)_k for stat case} 
and their numbers $n_{k} = card(I_{k}^{(n)})$ of elements are now
\textit{random} (due to their dependence on the $X_{i}$\textquoteright s)
and $M_{\mathds{P}_{n}^{emp}}=1$.
In a \textit{batch procedure}, when 
$D_{\varphi}(\textrm{$\boldsymbol{\Omega}$\hspace{-0.23cm}$\boldsymbol{\Omega}$},\mathds{P}_{n}^{emp})
:= \inf_{\mathds{Q}\in \textrm{$\boldsymbol{\Omega}$\hspace{-0.19cm}$\boldsymbol{\Omega}$} }
D_{\varphi}(\mathds{Q},\mathds{P}_{n}^{emp})$ 
is estimated once the sample  $\left(X_{1},..,X_{n}\right)$ is observed, 
we may reorder this sample by putting the $n_{1}$ sample points $X_{i}$ which are equal to $d_{1}$ 
in the first places, and so on; accordingly one ends up with index sets $I_{k}^{(n)}$ as defined in 
Section \ref{Sect Minimization}. When the \textit{online acquisition} of the data $X_{i}$\textquoteright s
is required, then we usually do not reorder the sample, and the $I_{k}^{(n)}$'s do
not consist in consecutive indexes, which does not make any change with
respect to the resulting construction nor to the estimator.

\vspace{0.2cm}
\noindent
The above considerations open the gate to our desired 
\textquotedblleft mimicking\textquotedblright\ of \eqref{min Pb} and \eqref{min Pb prob2}
to achieve \eqref{inf proba} (and some variants thereof) by our bare simulation (BS) 
method.
To proceed, we employ
a family of random variables $(W_{i})_{i \in  \mathbb{N}}$  
of independent and identically distributed $\mathbb{R}-$valued random variables
with probability distribution $\mathbb{\bbzeta}[ \cdot \, ] := \mathbb{\Pi}[W_{1} \in \cdot \, ]$
--- being connected with the divergence generator $\varphi \in \Upsilon(]a,b[)$ via the representability
\eqref{Phi Legendre of mgf(W)} --- 
such that $(W_{i})_{i \in  \mathbb{N}}$ is independent of $(X_{i})_{i \in  \mathbb{N}}$
\footnote{
on the common underlying probability space $(\mathfrak{X},\mathcal{A},\mathbb{\Pi})$
}.

\vspace{0.2cm}
\noindent 
As a next step, notice that the \textquotedblleft natural candidate\textquotedblright\ 
\begin{equation*}
\xi_{n,\mathbf{X}}^{\mathbf{W}}:= \frac{1}{n} \cdot \sum_{k=1}^{K}\bigg(
\sum_{i \in I_{k}^{(n)}}W_{i}\bigg) \cdot \delta _{d_{k}}
\ = \ \frac{1}{n}\sum_{i=1}^{n} W_{i} \cdot \delta_{X_{i}}
\end{equation*}
is not a probability measure since its total mass is not $1$ in general, since in terms of
its equivalent vector version
\begin{equation}
\boldsymbol{\xi}_{n,\mathbf{X}}^{\mathbf{W}} := \Big(\frac{1}{n} 
\sum_{i\in I_{1}^{(n)}} W_{i}, \ldots, \frac{1}{n} \sum_{i\in I_{K}^{(n)}} W_{i} \Big) 
\label{brostu3:fo.weiemp.var2}
\end{equation}
the sum $\sum_{k=1}^{K} \frac{1}{n} \sum_{i\in I_{k}^{(n)}} W_{i} = 
\frac{1}{n} \sum_{j=1}^{n} W_{i}$ of the $K$ 
vector components of \eqref{brostu3:fo.weiemp.var2} is typically 
not equal to 1;
this implies that no limit result of the form \eqref{LDP Minimization}
with finite limit can hold, since $\boldsymbol{\xi}_{n,\mathbf{X}}^{\mathbf{W}}$ 
takes values in $\mathbb{R}^{K}$ and $\textrm{$\boldsymbol{\Omega}$\hspace{-0.23cm}$\boldsymbol{\Omega}$}$ 
is a subset in the probability simplex $\mathbb{S}^{K}$ which
has \textit{void} interior in $\mathbb{R}^{K}$ 
causing a violation of condition \eqref{regularity}
(cf. Remark \ref{after det Problem}(d));
moreover, depending on the concrete form of the generator $\varphi$,
the corresponding weights may take \textit{negative values}.
Therefore, we need some \textquotedblleft rescaling\textquotedblright .  Indeed, 
let us introduce the \textit{normalized weighted empirical measure} 
\begin{eqnarray}
\xi_{n,\mathbf{X}}^{w\mathbf{W}} &:=&
\begin{cases}
\frac{1}{\sum_{k=1}^{K}\sum_{i \in I_{k}^{(n)}}W_{i}} \cdot 
\sum_{k=1}^{K}\left(
\sum_{i \in I_{k}^{(n)}}W_{i}\right) \cdot \delta _{d_{k}}
 = \sum_{i=1}^{n}  \frac{W_{i}}{\sum_{j=1}^{n} W_{j}} \cdot \delta _{X_{i}},
\qquad \textrm{if } \sum_{j=1}^{n} W_{j} \ne 0, \\
\infty \cdot \sum_{k=1}^{K} \delta _{d_{k}} =: \underline{\infty}, \hspace{7.6cm} 
\textrm{if } \sum_{j=1}^{n} W_{j} = 0,
\end{cases}
\label{brostu3:fo.norweiemp} 
\end{eqnarray}
which will substitute $\xi _{n,\mathbf{X}}^{\mathbf{W}}$
and which may belong to $\textrm{$\boldsymbol{\Omega}$\hspace{-0.23cm}$\boldsymbol{\Omega}$}$
 with positive probability.
The equivalent vector version of $\xi _{n,\mathbf{X}}^{w\mathbf{W}}$ is 
\begin{eqnarray}
\boldsymbol{\xi}_{n,\mathbf{X}}^{w\mathbf{W}} &:=&
\begin{cases}
\left(\frac{\sum_{i \in I_{1}^{(n)}}W_{i}}{\sum_{k=1}^{K}\sum_{i \in I_{k}^{(n)}}W_{i}},
\ldots, \frac{\sum_{i \in I_{K}^{(n)}}W_{i}}{\sum_{k=1}^{K}\sum_{i \in I_{k}^{(n)}}W_{i}} \right) ,
\qquad \textrm{if } \sum_{j=1}^{n} W_{j} \ne 0, \\
\ (\infty, \ldots, \infty) =: \boldsymbol{\infty}, \hspace{4.0cm} \textrm{if } \sum_{j=1}^{n} W_{j} = 0,
\end{cases}
\label{brostu3:fo.norweiemp.vec} 
\end{eqnarray}
a point in the linear subset of $\mathbb{R}^{K}$ spanned by $\mathbb{S}^{K}$
at infinity.

\vspace{0.4cm}

\begin{remark} 
(i) \ 
(Concerning e.g. computer-program command availability) 
In case of $\sum_{j=1}^{n}W_{j}=0$, in \eqref{brostu3:fo.norweiemp} 
we may 
equivalently
assign to $\xi _{n,\mathbf{X}}^{w\mathbf{W}}$ 
instead of $\underline{\infty}$
any measure (e.g. probability distribution) which does not belong to $\mathbb{\Omega}$,
respectively, in \eqref{brostu3:fo.norweiemp.vec} we may 
equivalently
choose for 
$\boldsymbol{\xi}_{n,\mathbf{X}}^{w\mathbf{W}}$ any vector outside of
$\boldsymbol{\Omega}$\hspace{-0.23cm}$\boldsymbol{\Omega}$
instead of $\boldsymbol{\infty}$.\\
(ii) \ By construction, in case of $\sum_{j=1}^{n} W_{j} \ne 0$, the sum of the random 
$K$ vector components of \eqref{brostu3:fo.norweiemp.vec} is now automatically equal to 1, 
but --- as (depending on $\varphi$) the $W_{i}$\textquoteright s may take both positive 
and negative values\footnote{see e.g. the below-mentioned solved Case 4 of
Subsection \ref{Subsect Case4}} --- 
these random components may be negative with probability strictly 
greater than zero (respectively nonnegative with probability strictly less than 1);
in the framework of \eqref{brostu3:fo.norweiemp} this means that $\xi _{n,\mathbf{X}}^{w\mathbf{W}}$ is in general a random \textit{signed} measure with total mass $1$, 
in case of $\sum_{j=1}^{n} W_{j} \ne 0$.
However, $\mathbb{\Pi} [\boldsymbol{\xi}_{n,\mathbf{X}}^{w\mathbf{W}}\in \mathbb{S}_{>0}^{K}]$ 
converges to $1$ as $n$ tends to infinity, since all the (identically distributed) random variables $W_{i}$ have expectation 1 
(as a consequence of the assumed representability
\eqref{Phi Legendre of mgf(W)});
in case of $\mathbb{\Pi}[W_{1}>0]=1$ one has even 
$\mathbb{\Pi}[\boldsymbol{\xi}_{n,\mathbf{X}}^{w\mathbf{W}}\in \mathbb{S}_{>0}^{K}]=1$
for all $n \in \mathbb{N}$.\\
(iii) By generalizing the terminology of e.g. 
\cite{Vaj:96},
through the right-hand side of \eqref{brostu3:fo.norweiemp}
one can interpret (for $\sum_{j=1}^{n} W_{j} \ne 0$) 
the normalized weighted empirical measure  $\xi_{n,\mathbf{X}}^{w\mathbf{W}}$
as response of an output neuron
in a random perceptron consisting of random inputs $\mathbf{X}$, 
a layer with $n$ units having one-point-distribution-valued responses
$\delta_{X_{1}}, \ldots, \delta_{X_{n}}$, and independent random synaptic
weights $\left(\frac{W_{1}}{\sum_{j=1}^{n} W_{j}}, \ldots, \frac{W_{n}}{\sum_{j=1}^{n} W_{j}}\right)$.
By our below-mentioned methods, we can approximate 
$\mathbb{\Pi}[\boldsymbol{\xi}_{n,\mathbf{X}}^{w\mathbf{W}}\in 
\textrm{$\boldsymbol{\Omega}$\hspace{-0.23cm}$\boldsymbol{\Omega}$}]$
for nearly any model $\textrm{$\boldsymbol{\Omega}$\hspace{-0.23cm}$\boldsymbol{\Omega}$}$,
and therefore propose proxies of Bayesian rules associated with hidden layers
in neural networks, as e.g. suggested in \cite{Vaj:96}.

\end{remark}

\enlargethispage{0.5cm}

\vspace{0.5cm}
\noindent
With the above-mentioned ingredients, we are now in the position to tackle
a variant of the distance minimization problem \eqref{inf proba},
by our bare simulation method 
through \textquotedblleft mimicking\textquotedblright\ the deterministic minimization problem \eqref{min Pb} 
respectively \eqref{min Pb prob2}.
For this, we also employ the \textit{conditional} distributions
$\mathbb{\Pi}_{n}[\, \cdot \, ] := \mathbb{\Pi}_{X_{1}^{n}}[\, \cdot \, ]  := \mathbb{\Pi}[ \, \cdot \, | \, 
X_{1}, \ldots, X_{n} ]$
and obtain the following

\vspace{0.4cm}

\begin{theorem}
\label{brostu3:thm.divnormW.new} 
Suppose that $(X_{i})_{i\in \mathbb{N}}$ is a sequence of random variables
with values in $\mathcal{Y}:=\left\{ d_{1},\cdots ,d_{K}\right\}$ 
such that \eqref{cv emp measure X to P} 
holds for some probability
measure $\mathbb{P}[\cdot \,]$ on $\mathcal{Y}$ having no zero-mass frequencies
(or equivalently, \eqref{cv emp measure X to P vector} holds for some
probability vector $\mathds{P} \in \mathbb{S}_{> 0}^{K}$).
Moreover, let $(W_{i})_{i \in  \mathbb{N}}$  be  a family
of independent and identically distributed $\mathbb{R}-$valued random variables
with probability distribution $\mathbb{\bbzeta}[ \cdot \, ] := \mathbb{\Pi}[W_{1} \in \cdot \, ]$
being connected with the divergence generator $\varphi \in \Upsilon(]a,b[)$ via the representability
\eqref{Phi Legendre of mgf(W)},
such that $(W_{i})_{i \in  \mathbb{N}}$ is independent of $(X_{i})_{i \in  \mathbb{N}}$. 
Then there holds 
\begin{align}
-\lim_{n\rightarrow \infty }\frac{1}{n}\log \, 
\mathbb{\Pi}_{X_{1}^{n}}\negthinspace \left[\xi _{n,\mathbf{X}}^{w\mathbf{W}}\in 
\mathbb{\Omega}\right]& =\inf_{\mathbb{Q}\in \mathbb{\Omega} }\ \inf_{m\neq
0}D_{\varphi }(m\cdot \mathbb{Q},\mathbb{P})\,  
\label{LDP Normalized} \\
& =\inf_{m\neq 0}\ \inf_{\mathbb{Q}\in \mathbb{\Omega} }D_{\varphi }(m\cdot \mathbb{Q},\mathbb{P})  
\notag
\\
& =\inf_{m\neq 0}\ \inf_{\mathds{Q}\in \textrm{$\boldsymbol{\Omega}$\hspace{-0.19cm}$\boldsymbol{\Omega}$} }
D_{\varphi }(m\cdot \mathds{Q},\mathds{P}) 
\label{LDP Normalized link}
\\
& = \inf_{\mathds{Q}\in \textrm{$\boldsymbol{\Omega}$\hspace{-0.19cm}$\boldsymbol{\Omega}$} }
\ \inf_{m\neq 0} 
D_{\varphi }(m\cdot \mathds{Q},\mathds{P})  
= -\lim_{n\rightarrow \infty }\frac{1}{n}\log \, 
\mathbb{\Pi}_{X_{1}^{n}}\negthinspace \left[\boldsymbol{\xi}_{n,\mathbf{X}}^{w\mathbf{W}}\in 
\textrm{$\boldsymbol{\Omega}$\hspace{-0.23cm}$\boldsymbol{\Omega}$}\right]
\label{LDP Normalized Vec}
\end{align}
for all sets $\mathbb{\Omega}$ of probability distributions such that
their equivalent probability-vector form 
$\boldsymbol{\Omega}$\hspace{-0.23cm}$\boldsymbol{\Omega}$ satisfies the 
regularity properties \eqref{regularity} 
\textit{in the relative topology}
and the finiteness property \eqref{def fi wrt Omega};
notice that for 
the equality \eqref{LDP Normalized link}
we have used the \textquotedblleft divergence link\textquotedblright\ \eqref{measure divergence}.
In particular, 
for each $\mathds{P} \in \mathbb{S}_{>0}^{K}$
(respectively, its equivalent probability-distribution $\mathbb{P}$)
the function 
$\mathds{Q} \mapsto \inf_{m\neq 0} 
D_{\varphi }(m\cdot \mathds{Q},\mathds{P}) $
(respectively, the function $\mathbb{Q} \mapsto \inf_{m\neq 0}D_{\varphi }(m\cdot \mathbb{Q},\mathbb{P})$) is
BS-minimizable (cf. \eqref{brostu3:fo.2}) on all sets 
$\textrm{$\boldsymbol{\Omega}$\hspace{-0.23cm}$\boldsymbol{\Omega}$} \subset \mathbb{S}^{K}$ 
satisfying \eqref{regularity}
\textit{in the relative topology} and
\eqref{def fi wrt Omega}
(respectively, on their probability-distribution-equivalent $\mathbb{\Omega}$).
\end{theorem}

\vspace{0.4cm}
\noindent
The proof of Theorem \ref{brostu3:thm.divnormW.new} will be given in Appendix B.
Analogous to Remark \ref{dist of components}(ii),
let us emphasize that we have assumed $\mathds{P} \in \mathbb{S}_{> 0}^{K}$
in Theorem \ref{brostu3:thm.divnormW.new}.
Henceforth, for sets  
$\textrm{$\boldsymbol{\Omega}$\hspace{-0.23cm}$\boldsymbol{\Omega}$} \subset \mathbb{S}^{K}$
of probability vectors we deal with \eqref{regularity} only in
the relative topology; thus, the latter will be unmentioned for the sake of brevity.
Remark \ref{after det Problem}(a),(b),(c),(e) applies accordingly.

\vspace{0.4cm}

\begin{remark}
\label{remark divnormW} 
(i) \ In strong contrast to Theorem \ref{brostu3:thm.divW.var},
the above result does not provide a
direct tool for the solution of Problem \eqref{inf proba} since the limit in 
\eqref{LDP Normalized} bears no \textit{direct} information on the minimum divergence 
$D_{\varphi }\left( \mathbb{\Omega},\mathbb{P} \right) :=  
\inf_{\mathbb{Q}\in\mathbb{\Omega}} D_{\varphi}( \mathbb{Q}, \mathbb{P})$;
the link between the corresponding quantities
can be emphasized and exploited e.g. in the case of power type divergences,
which leads to explicit minimization procedures as shown in 
Subsection \ref{Sub Contruction Princ Power case} below. 
For general divergences, Theorem \ref{brostu3:thm.divnormW.new} 
allows for the estimation of upper and lower
bounds of $D_{\varphi }\left( \mathbb{\Omega},\mathbb{P} \right)$, 
as developed in Subsection \ref{Sub Bounds General div} below.\\
(ii) \ Notice that $\breve{D}_{\varphi}(\mathbb{Q},\mathbb{P}):=
\inf_{m\neq 0} D_{\varphi }(m\cdot \mathbb{Q},\mathbb{P})$ satisfies the axioms of a
divergence, that is,
$\breve{D}_{\varphi}(\mathbb{Q},\mathbb{P}) \geq 0$, as well as 
$\breve{D}_{\varphi}(\mathbb{Q},\mathbb{P}) = 0$ if
and only if $\mathbb{Q} = \mathbb{P}$ (reflexivity). 
Hence, in Theorem \ref{brostu3:thm.divnormW.new} we are still within our
framework of bare simulation of a divergence minimum w.r.t. 
its first component (however, notice the difference to (i)).\\
(iii) \ Viewed from a \textquotedblleft reverse\textquotedblright\ angle, 
Theorem \ref{brostu3:thm.divnormW.new} gives
a crude approximation for the probability for
$\xi _{n,\mathbf{X}}^{w\mathbf{W}}$ to belong to $\mathbb{\Omega}$, 
conditionally upon $\mathbf{X} = (X_{1}, \ldots, X_{n})$.\\
(iv) \ In the same spirit as Remark \ref{dist of components}(i), for some contexts
one can \textit{explicitly} give the
distribution of each of the independent components
$\Big(\sum_{i\in I_{k}^{(n)}}W_{i}\Big)_{k=1,\ldots,K}$
of the vector $\boldsymbol{\xi}_{n,\mathbf{X}}^{w\mathbf{W}}$ given $\mathbf{X} =\mathbf{x}$;
this will ease the corresponding concrete simulations in a batch procedure.
For instance, we shall give some of those in 
the solved-cases Section \ref{Sect Cases}
below.\\
(v) \ Consider the special \textquotedblleft  degenerate\textquotedblright\ case
where all the data observations are \textit{certain}
and thus $(X_{i})_{i \in  \mathbb{N}}$ is nothing but a
\textit{purely deterministic} sequence, say $(\widetilde{x}_{i})_{i\in \mathbb{N}}$, 
of elements $\widetilde{x}_{i}$ from the arbitrary set $\mathcal{Y}:=\left\{d_{1},\ldots ,d_{K}\right\}$ of 
$K$ distinct values \textquotedblleft of any kind\textquotedblright\ (e.g., $\mathcal{Y}$ 
may consist of $K$ distinct numbers);
then the corresponding empirical distribution $\mathbb{P}_{n}^{emp}$ can be identified with 
the vector $\mathds{P}_{n}^{emp} := (p_{n,1}^{emp}, \ldots, p_{n,K}^{emp})$ where
\begin{equation}
p_{n,k}^{emp} \ := \ 
\frac{1}{n} \cdot n_{k} \ := \ \frac{1}{n} \cdot card(\bigl\{ i \in \{ 1, \ldots, n\}:  \ \widetilde{x}_{i} = d_{k} \bigr\})
\ =: \ \frac{1}{n} \cdot card(I_{k}^{(n)}) , \quad k \in \{1, \ldots, K\},
\nonumber
\end{equation}
and accordingly the required limit behaviour \eqref{cv emp measure X to P}
is equivalent to the vector-convergence
\begin{equation}
\lim_{n\rightarrow \infty } \Big( \frac{n_{1}}{n}, \ldots, \frac{n_{K}}{n} \Big) = (p_{1}, \ldots, p_{K})
\quad 
\textrm{for some $p_{1} >0$, \ldots, $p_{K} >0$ such that $\sum_{k=1}^{K} p_{k} = 1$.}
\nonumber
\end{equation}

\noindent
Correspondingly, with the notations $\mathds{P} := (p_{1}, \ldots, p_{K})$
and $\mathbf{\widetilde{x}} := (\widetilde{x}_{1}, \ldots, \widetilde{x}_{n})$,
the vector-form part of the assertion \eqref{LDP Normalized}
of Theorem \ref{brostu3:thm.divnormW.new} becomes
\begin{equation}
-\lim_{n\rightarrow \infty }\frac{1}{n}\log \, 
\mathbb{\Pi}\negthinspace \left[\boldsymbol{\xi}_{n,\mathbf{\widetilde{x}}}^{w\mathbf{W}}\in 
\textrm{$\boldsymbol{\Omega}$\hspace{-0.23cm}$\boldsymbol{\Omega}$}\right] 
\ = \ \inf_{\mathds{Q}\in \textrm{$\boldsymbol{\Omega}$\hspace{-0.19cm}$\boldsymbol{\Omega}$} }
\ \inf_{m\neq 0} 
D_{\varphi }(m\cdot \mathds{Q},\mathds{P})  
\ = \ \inf_{m\neq 0}\ \inf_{\mathds{Q}\in \textrm{$\boldsymbol{\Omega}$\hspace{-0.19cm}$\boldsymbol{\Omega}$} }
D_{\varphi }(m\cdot \mathds{Q},\mathds{P}) 
\nonumber
\end{equation}
for all subsets $\boldsymbol{\Omega}$\hspace{-0.23cm}$\boldsymbol{\Omega} \subset \mathbb{S}^{K}$ 
satisfying the regularity properties \eqref{regularity} and the 
finiteness property \eqref{def fi wrt Omega};
notice that the conditional probability $\mathbb{\Pi}_{X_{1}^{n}}[\, \cdot\, ]$ has degenerated to the ordinary
probability $\mathbb{\Pi}[\, \cdot \, ]$. \\
(vi) \ In a similar fashion to the proof of (the special degenerate case (v) of)
Theorem \ref{brostu3:thm.divnormW.new}, one can show
\begin{equation}
-\lim_{n\rightarrow \infty }\frac{1}{n}\log \, 
\mathbb{\Pi}\negthinspace \left[\boldsymbol{\xi}_{n}^{w\mathbf{W}}
\in \textrm{$\boldsymbol{\Omega}$\hspace{-0.23cm}$\boldsymbol{\Omega}$}\right] 
\ = \ \inf_{\mathds{Q}\in \textrm{$\boldsymbol{\Omega}$\hspace{-0.19cm}$\boldsymbol{\Omega}$} }
\ \inf_{m\neq 0} 
D_{\varphi }(m\cdot \mathds{Q},\mathds{P})  
\ = \ \inf_{m\neq 0}\ \inf_{\mathds{Q}\in \textrm{$\boldsymbol{\Omega}$\hspace{-0.19cm}$\boldsymbol{\Omega}$} }
D_{\varphi }(m\cdot \mathds{Q},\mathds{P}) 
\label{LDP Normalized det 2}
\end{equation}
for all subsets $\boldsymbol{\Omega}$\hspace{-0.23cm}$\boldsymbol{\Omega} \subset \mathbb{S}^{K}$ 
with regularity properties \eqref{regularity} and the finiteness property \eqref{def fi wrt Omega}, where 
\begin{eqnarray}
\boldsymbol{\xi}_{n}^{w\mathbf{W}} &:=&
\begin{cases}
\left(\frac{\sum_{i \in I_{1}^{(n)}}W_{i}}{\sum_{k=1}^{K}\sum_{i \in I_{k}^{(n)}}W_{i}},
\ldots, \frac{\sum_{i \in I_{K}^{(n)}}W_{i}}{\sum_{k=1}^{K}\sum_{i \in I_{k}^{(n)}}W_{i}} \right) 
= \frac{n \cdot \boldsymbol{\xi}_{n}^{\mathbf{W}}}{\sum_{i=1}^{n}W_{i}},
\qquad \textrm{if } \sum_{j=1}^{n} W_{j} \ne 0, \\
\ (\infty, \ldots, \infty) =: \boldsymbol{\infty}, \hspace{5.6cm} \textrm{if } \sum_{j=1}^{n} W_{j} = 0,
\end{cases}
\label{brostu3:fo.norweiemp.vec det} 
\end{eqnarray}
with
$I_{1}^{(n)}:=\left\{
1,\ldots ,n_{1}\right\} $, $I_{2}^{(n)}:=\left\{ n_{1}+1,\ldots
,n_{1}+n_{2}\right\} $, \ldots, 
$I_{K}^{(n)} := \{ \sum_{k=1}^{K-1} n_{k} + 1, \ldots, n \}$
and $n_{k}:=\lfloor n \cdot p_{k}\rfloor $ ($k \in \{1,\ldots,K\}$) for some pregiven
\textit{known} probability vector $\mathds{P} := (p_{1}, \ldots, p_{K})$. 
Recall the definition of $\boldsymbol{\xi}_{n}^{\mathbf{W}}$ in \eqref{Xi_n^W vector}
(with $\mathbf{W}$ instead of $\mathbf{\widetilde{W}}$).
The limit behaviour 
\eqref{LDP Normalized det 2}
contrasts to the one
of Theorem \ref{brostu3:thm.divW.var}, where
\begin{equation}
-\lim_{n\rightarrow \infty }\frac{1}{n}\log \,\mathbb{\Pi }\negthinspace \left[
\boldsymbol{\xi }_{n}^{\mathbf{\widetilde{W}}}
\in \mathbf{\Omega} /M_{\mathbf{P}}\right]
=\inf_{Q\in \mathbf{\Omega}}D_{\varphi }(\mathbf{Q},\mathbf{P})
\hspace{4.0cm} \text{(cf. \eqref{LDP Minimization})}
\nonumber
\end{equation}
for any $\mathbf{\Omega }\subset \mathbb{R}^{K}$ with regularity properties \eqref{regularity}
and the finiteness property \eqref{def fi wrt Omega};
recall that $(\widetilde{W}_{i})_{i \in  \mathbb{N}}$  
are i.i.d. random variables
with probability distribution $\widetilde{\mathbb{\bbzeta}}$
(being connected with the divergence generator 
$\widetilde{\varphi} := M_{\mathbf{P}} \cdot \varphi 
$ via the representability
\eqref{brostu3:fo.link.var}), 
whereas $(W_{i})_{i \in  \mathbb{N}}$  
are i.i.d. random variables
with probability distribution $\mathbb{\bbzeta}$
(being connected with the divergence generator $\varphi 
$ via the representability
\eqref{Phi Legendre of mgf(W)}).
Indeed, the construction leading to Theorem \ref{brostu3:thm.divW.var} does not hold any longer 
when 
$\mathbf{\Omega} \subset \mathbb{S}^{K}$
is a set of vectors within the probability simplex $\mathbb{S}^{K}$ and 
$\mathbf{P} \in \mathbb{S}_{>0}^{K}$  
is a known vector in this simplex with no zero entries. In such a case, one
has to use \eqref{LDP Normalized det 2} and \eqref{brostu3:fo.norweiemp.vec det} instead.
Notice that for each constant $A >0$, \eqref{LDP Normalized det 2} can be rewritten as
\begin{eqnarray}
& & 
\hspace{-1.5cm}
-\lim_{n\rightarrow \infty }\frac{1}{n}\log \, 
\mathbb{\Pi}\negthinspace \left[ \boldsymbol{\xi}_{n}^{w\mathbf{W}}
\in \textrm{$\boldsymbol{\Omega}$\hspace{-0.23cm}$\boldsymbol{\Omega}$} \right]
= \inf_{\mathbf{Q}\in A \cdot \textrm{$\boldsymbol{\Omega}$\hspace{-0.19cm}$\boldsymbol{\Omega}$} }
\ \inf_{m\neq 0} 
D_{\varphi }\Big(\frac{m}{A} \cdot \mathbf{Q},\mathds{P}\Big) 
 = \inf_{\mathbf{Q}\in A \cdot \textrm{$\boldsymbol{\Omega}$\hspace{-0.19cm}$\boldsymbol{\Omega}$} }
\ \inf_{\widetilde{m}\neq 0} 
D_{\varphi }(\widetilde{m} \cdot \mathbf{Q},\mathds{P})  
= \inf_{\widetilde{m}\neq 0}\ \inf_{\mathbf{Q}\in A \cdot 
\textrm{$\boldsymbol{\Omega}$\hspace{-0.19cm}$\boldsymbol{\Omega}$} }
D_{\varphi }(\widetilde{m} \cdot \mathbf{Q},\mathds{P}) ; 
\label{LDP Normalized det 3}
\end{eqnarray}
therein, the constraint $\mathbf{Q}\in A \cdot \textrm{$\boldsymbol{\Omega}$\hspace{-0.23cm}$\boldsymbol{\Omega}$}$
means geometrically that the vector $\mathbf{Q}$ 
lives in a subset of a simplex which is parallel to the simplex $\mathbb{S}^{K}$ 
of probability vectors and which is cut off at the edges of the
first/positive orthant; 
in view of Remark \ref{after det Problem}(d) and \eqref{LDP Normalized det 3},
we can also handle such a situation.
Namely, in the light of the third expression in \eqref{LDP Normalized det 3}
in combination with \eqref{min Pb prob1} to \eqref{min Pb one}
for the special case of $\mathbf{\Omega} := \textrm{$\boldsymbol{\Omega}$\hspace{-0.23cm}$\boldsymbol{\Omega}$}$ 
lying in the probability simplex, it makes sense
to study e.g. functional relationships between 
$\inf_{\widetilde{m}\neq 0}D_{\widetilde{c}\cdot \varphi}(\widetilde{m} \cdot \mathbf{Q},\mathds{P})$
and $D_{\widetilde{c}\cdot \varphi}(\mathbf{Q},\mathds{P})$ \, ($\widetilde{c} >0$) \, 
for $\mathbf{Q} \in A \cdot \mathbb{S}^{K}$ with arbitrary $A >0$  not necessarily being equal to 1
(i.e. $\mathbf{Q} = A \cdot \mathds{Q}$ for some probability vector $\mathds{Q}$).
Indeed, such a context appears naturally e.g. in connection with mass transportation problems 
(cf. \eqref{Monge1a} below) and with distributed energy management (cf. the
paragraph after \eqref{quadopt2});
the special case $A=1/K$ of \eqref{LDP Normalized det 3} will also be used 
below for the application of our BS method to
solving \textit{(generalized) minimum/maximum entropy problems}
for probability 
vectors (and even for sub-/super-probability vectors)
$\mathds{Q}$ with constraints.

\end{remark}

\vspace{0.4cm}
\noindent
Let us proceed with the main context. 
As indicated in Remark \ref{remark divnormW}(i), in a number of important cases 
the limit in the above Theorem 
\ref{brostu3:thm.divnormW.new} can be stated in terms of an invertible function 
$G^{-1}$ (cf. \eqref{brostu3:fo.2}) of 
$\inf_{\mathds{Q}\in \textrm{$\boldsymbol{\Omega}$\hspace{-0.19cm}$\boldsymbol{\Omega}$}} 
D_{\varphi }(\mathds{Q},\mathds{P})$
by elimination of $m$; 
as explained above, for the degenerate case (cf. Remark \ref{remark divnormW}(v),(vi)) 
the search for $G^{-1}$ is even interesting for the more general infimum over non-probability vectors. 
This $m-$elimination is the scope of the development in the following 
Subsection \ref{Sub Contruction Princ Power case}. For cases where $m$ can not be 
(yet) explicitly eliminated, we deliver bounds in the second next
Subsection \ref{Sub Bounds General div}.


\subsection{Construction principle for the estimation of the minimum
divergence, the power-type case \label{Sub Contruction Princ Power case}} 

\noindent
Within the context of Theorem \ref{brostu3:thm.divnormW.new} respectively 
Remark \ref{remark divnormW}(v) and (vi),  
we obtain an explicit solution for the inner
(i.e. $m-$concerning) minimization in \eqref{LDP Normalized Vec}
for the important case of power-divergence generators $\varphi_{\gamma} : \mathbb{R} \mapsto [0,\infty]$
defined by
\begin{eqnarray}
\varphi_{\gamma}(t) \hspace{-0.2cm} &:=& \hspace{-0.2cm}
\begin{cases}
\frac{t^\gamma-\gamma \cdot t+ \gamma - 1}{\gamma \cdot (\gamma-1)}, \hspace{6.0cm} \textrm{if }  \gamma \in \, ]-\infty,0[ \ 
\textrm{and } t \in ]0,\infty[,  \\
- \log t + t - 1, \hspace{5.5cm} \textrm{if }  \gamma = 0 \ \textrm{and } t \in ]0,\infty[,
\\
\frac{t^\gamma-\gamma \cdot t+ \gamma - 1}{\gamma \cdot (\gamma-1)}, \hspace{6.0cm} \textrm{if }  \gamma \in \, ]0,1[ \ 
\textrm{and } t \in [0,\infty[,
 \\
t \cdot \log t + 1 - t, \hspace{5.5cm} \textrm{if }  \gamma = 1 \ \textrm{and } t \in [0,\infty[,
\\
\frac{t^\gamma-\gamma \cdot t+ \gamma - 1}{\gamma \cdot (\gamma-1)} \cdot \textfrak{1}_{]0,\infty[}(t)
+(\frac{1}{\gamma} - \frac{t}{\gamma-1}) \cdot \textfrak{1}_{]-\infty,0]}(t),
\hspace{1.0cm} \textrm{if }  \gamma \in \, ]1,2[ \ \textrm{and } t \in \, ]-\infty,\infty[,
\\
\frac{(t - 1)^2}{2}, \hspace{6.8cm} \textrm{if }  \gamma = 2 \ 
\textrm{and } t \in \, ]-\infty,\infty[,
\\
\frac{t^\gamma-\gamma \cdot t+ \gamma - 1}{\gamma \cdot (\gamma-1)} \cdot \textfrak{1}_{]0,\infty[}(t)
+(\frac{1}{\gamma} - \frac{t}{\gamma-1}) \cdot \textfrak{1}_{]-\infty,0]}(t),
\hspace{1.0cm} \textrm{if }  \gamma \in \, ]2,\infty[ \ \textrm{and } t \in \, ]-\infty,\infty[,
\\
\infty, \hspace{7.35cm} \textrm{else},
\end{cases}
\label{brostu3:fo.powdivgen} 
\end{eqnarray}
which for arbitrary multiplier $\widetilde{c} >0$ generate (the vector-valued form of) the 
\textit{generalized power divergences} given by
\begin{eqnarray}
D_{\widetilde{c} \cdot \varphi_{\gamma}}(\mathbf{Q},\mathbf{P}) \hspace{-0.2cm} &:=& \hspace{-0.2cm}
\begin{cases}
\widetilde{c} \cdot \Big\{\frac{ \sum\limits_{k=1}^{K} (q_{k})^{\gamma} \cdot (p_{k})^{1-\gamma}}{\gamma \cdot (\gamma-1)}
- \frac{1}{\gamma -1} \cdot \sum\limits_{k=1}^{K} q_{k} + \frac{1}{\gamma} \cdot \sum\limits_{k=1}^{K} p_{k} \Big\}, 
\hspace{2.0cm} \textrm{if }  \gamma \in \, ]-\infty,0[, \  
\mathbf{P} \in \mathbb{R}_{\gneqq 0}^{K} \ \textrm{and } \mathbf{Q} \in \mathbb{R}_{> 0}^{K}, 
 \\
\widetilde{c} \cdot \Big\{ \sum\limits_{k=1}^{K} p_{k} \cdot \log \Big(\frac{p_{k}}{q_{k}} \Big) 
+ \sum\limits_{k=1}^{K} q_{k} - \sum\limits_{k=1}^{K} p_{k} \Big\}, 
\hspace{3.1cm} \textrm{if }  \gamma = 0, \  
\mathbf{P} \in \mathbb{R}_{\gneqq 0}^{K} \ \textrm{and } \mathbf{Q} \in \mathbb{R}_{> 0}^{K},
\\
\widetilde{c} \cdot \Big\{\frac{ \sum\limits_{k=1}^{K} (q_{k})^{\gamma} \cdot (p_{k})^{1-\gamma}}{\gamma \cdot (\gamma-1)}
- \frac{1}{\gamma -1} \cdot \sum\limits_{k=1}^{K} q_{k} + \frac{1}{\gamma} \cdot \sum\limits_{k=1}^{K} p_{k} \Big\}, 
\hspace{2.1cm} \textrm{if }  \gamma \in \, ]0,1[, \  
\mathbf{P} \in \mathbb{R}_{\gneqq 0}^{K} \ \textrm{and } \mathbf{Q} \in \mathbb{R}_{\geq 0}^{K}, 
 \\
\widetilde{c} \cdot \Big\{ \sum\limits_{k=1}^{K} q_{k} \cdot \log \Big(\frac{q_{k}}{p_{k}} \Big)
- \sum\limits_{k=1}^{K} q_{k} + \sum\limits_{k=1}^{K} p_{k} \Big\}, 
\hspace{3.2cm} \textrm{if }  \gamma = 1, \  
\mathbf{P} \in \mathbb{R}_{> 0}^{K} \ \textrm{and } \mathbf{Q} \in \mathbb{R}_{\geq 0}^{K},
\\
\widetilde{c} \cdot \Big\{ \sum\limits_{k=1}^{K} \frac{(q_{k})^{\gamma} \cdot (p_{k})^{1-\gamma}}{\gamma \cdot (\gamma-1)} 
\cdot \textfrak{1}_{[0,\infty[}(q_{k})
- \frac{1}{\gamma -1} \cdot \sum\limits_{k=1}^{K} q_{k} + \frac{1}{\gamma} \cdot \sum\limits_{k=1}^{K} p_{k} \Big\},
\hspace{0.3cm}  \textrm{if }  \gamma \in \, ]1,2[, \  
\mathbf{P} \in \mathbb{R}_{>0}^{K} \ \textrm{and } \mathbf{Q} \in \mathbb{R}^{K}, 
\\
\widetilde{c}\cdot \sum\limits_{k=1}^{K}\frac{  (q_{k}-p_{k})^{2}}{2 \cdot p_{k}} , 
\hspace{7.1cm}  \textrm{if }  \gamma = 2, \  
\mathbf{P} \in \mathbb{R}_{>0}^{K} \ \textrm{and } \mathbf{Q} \in \mathbb{R}^{K},
\\
\widetilde{c} \cdot \Big\{ \sum\limits_{k=1}^{K} \frac{(q_{k})^{\gamma} \cdot (p_{k})^{1-\gamma}}{\gamma \cdot (\gamma-1)} 
\cdot \textfrak{1}_{[0,\infty[}(q_{k})
- \frac{1}{\gamma -1} \cdot \sum\limits_{k=1}^{K} q_{k} + \frac{1}{\gamma} \cdot \sum\limits_{k=1}^{K} p_{k} \Big\},
\hspace{0.4cm}  \textrm{if }  \gamma \in \, ]2,\infty[, \  
\mathbf{P} \in \mathbb{R}_{>0}^{K} \ \textrm{and } \mathbf{Q} \in \mathbb{R}^{K},
\\
\infty, \hspace{9.05cm} \textrm{else};
\end{cases}
\label{brostu3:fo.powdiv.new} 
\end{eqnarray}

\noindent
notice that one has the straightforward relationship $D_{\widetilde{c}\cdot
\varphi _{\gamma }}(\cdot ,\cdot )=\widetilde{c}\cdot D_{\varphi _{\gamma
}}(\cdot ,\cdot )$; however, as a motivation for the introduction of 
$\widetilde{c}>0$, we shall show in 
the solved-cases Section \ref{Sect Cases}
below that the corresponding
probability distribution $\mathbb{\mathbb{\bbzeta}}$ 
(cf. \eqref{Phi Legendre of mgf(W)})
of the $W_{i}$\textquoteright s depends on $\widetilde{c}$ in a
non-straightforward way (see also Remark \ref{remark divnormW}(vi) for another motivation for $\widetilde{c}$). 
In the course of this, it turns out that 
$\widetilde{c} \cdot \varphi_{\gamma} \in \Upsilon(]a_{\gamma},\infty[)$
with $a_{\gamma} =0$ for $\gamma \in ]-\infty,1]$ and $a_{\gamma} = -\infty$ for 
$\gamma \in [2,\infty[$.

\vspace{0.3cm}
\noindent
For $\widetilde{c}=1$ and probability vectors $\mathds{Q}$, $\mathds{P}$ in $\mathbb{S}^{K}$ respectively 
$\mathbb{S}_{>0}^{K}$, the divergences \eqref{brostu3:fo.powdiv.new}
simplify considerably, namely to the well-known
\textit{power divergences} $D_{\varphi_{\gamma}}(\mathds{Q},\mathds{P})$
in the scaling of e.g. Liese \& Vajda \cite{Lie:87} 
(in other scalings they are also called
\textit{Rathie \& Kannapan\textquoteright s non-additive directed divergences of order $\gamma$}
\cite{Rat:72}, \textit{Cressie-Read divergences} \cite{Cre:84} \cite{Rea:88}, 
\textit{relative Tsallis entropies or Tsallis cross-entropies} \cite{Tsa:98}
(see also \cite{Shi:98}), 
\textit{Amari\textquoteright s alpha-divergences} \cite{Ama:85});
for some comprehensive overviews on power divergences 
$D_{\varphi_{\gamma}}(\mathds{Q},\mathds{P})$
--- including statistical applications to goodness-of-fit testing and
minimum distance estimation ---
the reader is referred to the insightful 
books~\cite{Lie:87}--\cite{Lie:08},
the survey articles~\cite{Lie:06},\cite{Vaj:10},
and the references therein.
Prominent and widely used special cases of $D_{\varphi_{\gamma}}(\mathds{Q},\mathds{P})$
are the omnipresent \textit{Kullback-Leibler information divergence (relative entropy)}
where $\gamma=1$, the equally important
\textit{reverse Kullback-Leibler information divergence (reverse relative entropy)}
where $\gamma =0$,
the \textit{Pearson chi-square divergence} ($\gamma=2$), the 
\textit{Neyman chi-square divergence} ($\gamma=-1$),
the \textit{Hellinger divergence} ($\gamma=\frac{1}{2}$,
also called squared Hellinger distance, 
squared Matusita distance \cite{Mat:51} or squared Hellinger-Kakutani metric, 
see e.g. \cite{Dez:16} 
\footnote{in some literature, the (square root of the) Hellinger divergence (HD)
is misleadingly called Bhattacharyya distance; however, the latter is \textit{basically} some
rescaled logarithm of HD, namely $R_{1/2}(\mathds{Q},\mathds{P})$ (cf. \eqref{def Renyi}
with $\gamma=1/2$)
}).
Some exemplary (relatively) recent studies and applications
of power divergences $D_{\varphi_{\gamma}}(\mathds{Q},\mathds{P})$
--- aside from the vast statistical literature (including 
in particular maximum likelihood estimation and Pearson\textquoteright s chi-square test) ---
appear e.g. 
in 
\cite{Sas:19},\cite{Mat:03}--\cite{Wang2:21};
in \cite{Sig:20}~with $\gamma=2$;
in \cite{Ha:19}~with $\gamma=1$;
in \cite{Bek:20}~with $\gamma=-1$ and $\gamma=\frac{1}{2}$;
in \cite{Lup:19}--\cite{Che2:21}~with $\gamma=\frac{1}{2}$.

\vspace{0.3cm}
\noindent
For $\widetilde{c}=1$ and nonnegative-component vectors $\mathbf{Q}$, $\mathbf{P}$ in 
$\mathbb{R}_{\geq 0}^{K}$ respectively 
$\mathbb{R}_{>0}^{K}$ respectively $\mathbb{R}_{\gneqq 0}^{K}$, the generalized power divergences 
$D_{\varphi_{\gamma}}(\mathbf{Q},\mathbf{P})$ of \eqref{brostu3:fo.powdiv.new}
also (partially) simplify, and were treated by 
\cite{Stu:10} (for even more general probability measures, deriving
e.g. also generalized Pinsker\textquoteright s  inequalities);
for a more general comprehensive technical treatment see also e.g. \cite{Bro:19b}.

\vspace{0.4cm}
\noindent
Returning to the general context,
in Theorem \ref{brostu3:thm.divnormW.new}
we stated that for each $\mathds{P} \in \mathbb{S}_{>0}^{K}$ the function
$\mathds{Q} \mapsto \inf_{m\neq 0} 
D_{\varphi }(m\cdot \mathds{Q},\mathds{P}) $ is
BS-minimizable (cf. \eqref{brostu3:fo.2}) on all sets 
$\textrm{$\boldsymbol{\Omega}$\hspace{-0.23cm}$\boldsymbol{\Omega}$} \subset \mathbb{S}^{K}$ 
satisfying \eqref{regularity} and \eqref{def fi wrt Omega}. 
The (corresponding subsetup of the) following 
Lemma \ref{Lemma Indent Rate finite case_new} is the \textit{cornerstone} 
leading from this statement to BS-minimizability of the function 
$\mathds{Q} \mapsto D_{\varphi }(\mathds{Q},\mathds{P}))$ 
on those same sets, for the special divergences in \eqref{brostu3:fo.powdiv.new}.
To formulate this in a transparent way, we employ the following
three fundamental quantities $H_{\gamma}(\mathbf{Q},\mathds{P})$,
$I(\mathbf{Q},\mathds{P})$, $\widetilde{I}(\mathbf{Q},\mathds{P})$
and the arbitrary constant $A >0$ (where for $A=1$ all the following vectors $\mathbf{Q}$ 
will turn into probability vectors $\mathds{Q}$).
Indeed --- for any constellation $(\gamma , \mathds{P},\mathbf{Q}) \in 
\widetilde{\Gamma }\times \widetilde{\mathcal{M}}_{1}\times 
\widetilde{\mathcal{M}}_{2}$, where
$\widetilde{\Gamma }\times \widetilde{\mathcal{M}}_{1}\times 
\widetilde{\mathcal{M}}_{2}:= \, ]0,1[\times \, \mathbb{S}^{K}\times A \cdot \mathbb{S}^{K}$
\ or  \ $\widetilde{\Gamma }\times \widetilde{\mathcal{M}}_{1}\times 
\widetilde{\mathcal{M}}_{2}:= \, ]-\infty ,0[\times \, \mathbb{S}^{K}\times A \cdot \mathbb{S}_{>0}^{K} $
 \ or  \  $\widetilde{\Gamma }\times \widetilde{\mathcal{M}}_{1}\times 
\widetilde{\mathcal{M}}_{2}:= \, ]1,\infty \lbrack \times \, \mathbb{S}_{>0}^{K}\times 
A \cdot \mathbb{S}^{K}$ --- 
let 
\begin{eqnarray}
0 < H_{\gamma}(\mathbf{Q},\mathds{P}) := \sum\displaylimits_{k=1}^{K} (q_{k})^{\gamma} \cdot (p_{k})^{1-\gamma}
\ = \ 1 + \gamma \cdot (A-1) +
\gamma \cdot (\gamma-1) \cdot D_{\varphi_{\gamma}}(\mathbf{Q}, \mathds{P}), 
\qquad \gamma \in \mathbb{R}\backslash\{0,1\},
\label{brostu3:fo.divpow.hellinger1}
\end{eqnarray}
be the 
\textit{modified $\gamma-$order Hellinger integral
of $\mathbf{Q}$ and $\mathds{P}$}.
Furthermore, for any $\mathds{P} \in \mathbb{S}_{>0}^{K}$, $\mathbf{Q} \in A \cdot \mathbb{S}^{K}$, let 
\begin{eqnarray}
-1 \ < \ I(\mathbf{Q},\mathds{P}) :=  \sum\displaylimits_{k=1}^{K} q_{k} \cdot \log\left( \frac{q_{k}}{p_{k}} \right)
\ = \ D_{\varphi_{1}}(\mathbf{Q}, \mathds{P}) + A - 1
\label{brostu3:fo.divpow.Kull1}
\end{eqnarray}
be the \textit{modified Kullback-Leibler information (modified relative entropy)}.
Finally, for any $\mathds{P}\in \mathbb{S}^{K}$, $\mathbf{Q} \in A \cdot \mathbb{S}_{>0}^{K}$, let 
\begin{eqnarray}
1 - A \ \leq \ \widetilde{I}(\mathbf{Q},\mathds{P}) := \sum\displaylimits_{k=1}^{K} p_{k} \cdot \log\left( \frac{p_{k}}{q_{k}} \right)
\ = \ D_{\varphi_{0}}(\mathbf{Q}, \mathds{P}) + 1 - A
\label{brostu3:fo.divpow.RevKull1}
\end{eqnarray}
be the \textit{modified reverse Kullback-Leibler information (modified reverse relative entropy)}.
In terms of \eqref{brostu3:fo.divpow.hellinger1}, 
\eqref{brostu3:fo.divpow.Kull1} and
\eqref{brostu3:fo.divpow.RevKull1}  
we obtain the following assertions which will be proved in Appendix C. 

\vspace{0.2cm}

\begin{lemma}
\label{Lemma Indent Rate finite case_new}
\ Let $A >0$ be an arbitrary constant.\\ 
(a) Let $\widetilde{c}>0$ be
arbitrary and $(\gamma , \mathds{P},\mathbf{Q}) \in 
\widetilde{\Gamma }\times \widetilde{\mathcal{M}}_{1}\times 
\widetilde{\mathcal{M}}_{2}$
as above. Then one has \\[-0.4cm]
\begin{eqnarray}
\inf_{m\neq 0}D_{\widetilde{c}\cdot \varphi _{\gamma }}(m \cdot \mathbf{Q},\mathds{P})=
\inf_{m>0}D_{\widetilde{c}\cdot \varphi _{\gamma }}(m\cdot \mathbf{Q},\mathds{P})
\hspace{-0.2cm} &=& \hspace{-0.2cm}
\frac{\widetilde{c}}{\gamma }
\cdot \bigg[ 1- A^{\gamma/(\gamma-1)} \cdot 
\bigg[ 1+ \gamma \cdot (A-1) + \frac{\gamma \cdot \left( \gamma -1\right) }{
\widetilde{c}}\cdot D_{\widetilde{c}\cdot \varphi _{\gamma }}(\mathbf{Q},\mathds{P})\bigg]
^{-1/\left( \gamma -1\right) }\bigg]
\nonumber \\
& &  
\label{brostu3:fo.676}
\\
&=& \hspace{-0.2cm}
\frac{\widetilde{c}}{\gamma }
\cdot \left[ 1- A^{\gamma/(\gamma-1)} \cdot 
H_{\gamma}(\mathbf{Q},\mathds{P})^{-1/\left( \gamma -1\right) }\right]
\nonumber
\end{eqnarray}

\vspace{-0.2cm}
\noindent
and consequently for any subset 
$A \cdot \boldsymbol{\Omega}$\hspace{-0.23cm}$\boldsymbol{\Omega} \subset \widetilde{\mathcal{M}}_{2}$ 
\begin{eqnarray}
&&\hspace{-0.7cm}
\inf_{\mathbf{Q}\in A \cdot \textrm{$\boldsymbol{\Omega}$\hspace{-0.19cm}$\boldsymbol{\Omega}$}}
\ \inf_{m\neq 0} D_{\widetilde{c}\cdot
\varphi _{\gamma }}(m \cdot \mathbf{Q},\mathds{P})=\frac{\widetilde{c}}{\gamma }\cdot \bigg[ 1-
A^{\gamma/(\gamma-1)} \cdot \bigg[ 1+ \gamma \cdot (A-1) + 
\frac{\gamma \cdot \left( \gamma -1\right) }{\widetilde{c}}\cdot
\inf_{\mathbf{Q}\in A \cdot \textrm{$\boldsymbol{\Omega}$\hspace{-0.19cm}$\boldsymbol{\Omega}$} }
D_{\widetilde{c}\cdot \varphi _{\gamma }}(\mathbf{Q},\mathds{P})\bigg]
^{-1/\left( \gamma -1\right) }\bigg] ,\qquad \ 
\label{Inf in Lemma rate case finite 1} \\
&&\hspace{-0.7cm}\arg \inf {}_{\mathbf{Q}\in A \cdot \textrm{$\boldsymbol{\Omega}$\hspace{-0.19cm}$\boldsymbol{\Omega}$} }\inf_{m\neq 0}
\ D_{\widetilde{c}
\cdot \varphi _{\gamma }}(m \cdot \mathbf{Q},\mathds{P})=
\arg \inf {}_{\mathbf{Q}\in A \cdot \textrm{$\boldsymbol{\Omega}$\hspace{-0.19cm}$\boldsymbol{\Omega}$} } \ 
D_{\widetilde{c}
\cdot \varphi _{\gamma }}(\mathbf{Q},\mathds{P}),\qquad \ 
\label{equiv infima in lemma rate finite case 1} \\
&&\hspace{-0.7cm}
\inf_{\mathbf{Q}\in A \cdot \textrm{$\boldsymbol{\Omega}$\hspace{-0.19cm}$\boldsymbol{\Omega}$}}
\ \inf_{m\neq 0} D_{\varphi_{\gamma }}(m \cdot \mathbf{Q},\mathds{P})=
\frac{1}{\gamma }\cdot \bigg[ 1-
A^{\gamma/(\gamma-1)} \cdot 
\bigg[ \inf_{\mathbf{Q}\in A \cdot \textrm{$\boldsymbol{\Omega}$\hspace{-0.19cm}$\boldsymbol{\Omega}$} }
H_{\gamma}(\mathbf{Q},\mathds{P})\bigg]^{-1/\left( \gamma -1\right) }\bigg] ,\qquad 
\textrm{for $\gamma <0$ and $\gamma >1$}, 
\label{Inf in Lemma rate case finite 1b} \\
&&\hspace{-0.7cm}\arg \inf {}_{\mathbf{Q}\in A \cdot \textrm{$\boldsymbol{\Omega}$\hspace{-0.19cm}$\boldsymbol{\Omega}$} }\inf_{m\neq 0}
\ D_{\varphi_{\gamma }}(m \cdot \mathbf{Q},\mathds{P})=
\arg \inf {}_{\mathbf{Q}\in A \cdot \textrm{$\boldsymbol{\Omega}$\hspace{-0.19cm}$\boldsymbol{\Omega}$} } \ 
H_{\gamma}(\mathbf{Q},\mathds{P}),\hspace{3.4cm} \textrm{for $\gamma <0$ and $\gamma >1$},
\label{equiv infima in lemma rate finite case 1b} \\
&&\hspace{-0.7cm}
\inf_{\mathbf{Q}\in A \cdot \textrm{$\boldsymbol{\Omega}$\hspace{-0.19cm}$\boldsymbol{\Omega}$}}
\ \inf_{m\neq 0} D_{\varphi_{\gamma }}(m \cdot \mathbf{Q},\mathds{P})=
\frac{1}{\gamma }\cdot \bigg[ 1-
A^{\gamma/(\gamma-1)} \cdot 
\bigg[ \sup_{\mathbf{Q}\in A \cdot \textrm{$\boldsymbol{\Omega}$\hspace{-0.19cm}$\boldsymbol{\Omega}$} }
H_{\gamma}(\mathbf{Q},\mathds{P})\bigg]^{-1/\left( \gamma -1\right) }\bigg] , \hspace{0.8cm} 
\textrm{for $\gamma \in \, ]0,1[$}, 
\label{Inf in Lemma rate case finite 1c} \\
&&\hspace{-0.7cm}\arg \inf {}_{\mathbf{Q}\in A \cdot \textrm{$\boldsymbol{\Omega}$\hspace{-0.19cm}$\boldsymbol{\Omega}$} }\inf_{m\neq 0}
\ D_{\varphi_{\gamma }}(m \cdot \mathbf{Q},\mathds{P})=
\arg \sup {}_{\mathbf{Q}\in A \cdot \textrm{$\boldsymbol{\Omega}$\hspace{-0.19cm}$\boldsymbol{\Omega}$} } \ 
H_{\gamma}(\mathbf{Q},\mathds{P}), \hspace{3.4cm} \textrm{for $\gamma \in \, ]0,1[$},
\label{equiv infima in lemma rate finite case 1c} 
\end{eqnarray}

\vspace{-0.2cm}
\noindent
provided that the infimum on the right-hand side of \eqref{Inf in Lemma rate case finite 1} 
exists.\\[0.1cm]
\noindent (b) For any $\mathds{P} \in \mathbb{S}_{>0}^{K}$, $\mathbf{Q} \in A \cdot \mathbb{S}^{K}$, 
$\widetilde{c}>0$ one gets 
\begin{eqnarray}
\inf_{m\neq 0}D_{\widetilde{c}\cdot \varphi _{1}}(m \cdot \mathbf{Q},\mathds{P})=
\inf_{m>0}D_{\widetilde{c}\cdot \varphi _{1}}(m \cdot \mathbf{Q},\mathds{P})
&=&
\widetilde{c}\cdot \Big[ 1- A \cdot \exp
\Big( -\frac{1}{A \cdot \widetilde{c}}\cdot D_{\widetilde{c}\cdot \varphi_{1}}(\mathbf{Q},\mathds{P})
+ \frac{1}{A} -1 \Big) \Big]  
\label{brostu3:fo.677} \\
&=& 
\widetilde{c}\cdot \Big[ 1- A \cdot \exp
\Big( -\frac{1}{A}\cdot I(\mathbf{Q},\mathds{P})\Big) \Big]  
\nonumber
\end{eqnarray}

\vspace{-0.2cm}
\noindent
and consequently for any subset 
$A \cdot \boldsymbol{\Omega}$\hspace{-0.23cm}$\boldsymbol{\Omega} \subset A \cdot \mathbb{S}^{K}$ 
\begin{eqnarray}
& &
\inf_{\mathbf{Q}\in A \cdot \textrm{$\boldsymbol{\Omega}$\hspace{-0.19cm}$\boldsymbol{\Omega}$} } \ 
\inf_{m\neq 0}D_{\widetilde{c}\cdot
\varphi _{1}}(m \cdot \mathbf{Q},\mathds{P})
=\widetilde{c}\cdot \Big[ 1-
A \cdot \exp \Big( -\frac{1}{A \cdot \widetilde{c}}\cdot 
\inf_{Q\in A \cdot \textrm{$\boldsymbol{\Omega}$\hspace{-0.19cm}$\boldsymbol{\Omega}$} }D_{\widetilde{c}\cdot \varphi
_{1}}(\mathbf{Q},\mathds{P}) + \frac{1}{A} -1 \Big) \Big] ,  
\label{brostu3:fo.mmin-KL1} \\
& & \arg \inf {}_{\mathbf{Q}\in A \cdot \textrm{$\boldsymbol{\Omega}$\hspace{-0.19cm}$\boldsymbol{\Omega}$} } \ 
\inf_{m\neq 0}D_{\widetilde{c}\cdot \varphi _{1}}(m \cdot \mathbf{Q},\mathds{P}) =
\arg \inf {}_{\mathbf{Q}\in A \cdot \textrm{$\boldsymbol{\Omega}$\hspace{-0.19cm}$\boldsymbol{\Omega}$} } \ 
D_{\widetilde{c}\cdot \varphi _{1}}(\mathbf{Q},\mathds{P}),\qquad \   
\label{brostu3:fo.mmin-KL2} \\
& &\inf_{\mathbf{Q}\in A \cdot \textrm{$\boldsymbol{\Omega}$\hspace{-0.19cm}$\boldsymbol{\Omega}$} } \ 
\inf_{m\neq 0}D_{\varphi_{1}}(m \cdot \mathbf{Q},\mathds{P})
=\Big[ 1- A \cdot \exp \Big( -\frac{1}{A}\cdot 
\inf_{Q\in A \cdot \textrm{$\boldsymbol{\Omega}$\hspace{-0.19cm}$\boldsymbol{\Omega}$} }
I(\mathbf{Q},\mathds{P}) \Big) \Big] ,  
\label{brostu3:fo.mmin-KL1b} \\
& & \arg \inf {}_{\mathbf{Q}\in A \cdot \textrm{$\boldsymbol{\Omega}$\hspace{-0.19cm}$\boldsymbol{\Omega}$} } \ 
\inf_{m\neq 0}D_{\varphi_{1}}(m \cdot \mathbf{Q},\mathds{P}) =
\arg \inf {}_{\mathbf{Q}\in A \cdot \textrm{$\boldsymbol{\Omega}$\hspace{-0.19cm}$\boldsymbol{\Omega}$} } \ 
I(\mathbf{Q},\mathds{P}),\qquad \   
\label{brostu3:fo.mmin-KL2b}
\end{eqnarray}

\vspace{-0.1cm}
\noindent
provided that the infimum on the right-hand side of \eqref{brostu3:fo.mmin-KL1} 
exists. \\[0.1cm]
\noindent (c) For any $\mathds{P}\in \mathbb{S}^{K}$, $\mathbf{Q} \in A \cdot \mathbb{S}_{>0}^{K}$, 
$\widetilde{c}>0$ we obtain 
\begin{eqnarray}
\hspace{-0.7cm}
\inf_{m\neq 0}D_{\widetilde{c}\cdot \varphi _{0}}(m \cdot \mathbf{Q},\mathds{P})
\ = \
\inf_{m>0}D_{\widetilde{c}\cdot \varphi _{0}}(m \cdot \mathbf{Q},\mathds{P}) 
\ = \ D_{\widetilde{c}\cdot \varphi_{0}}(\mathbf{Q},\mathds{P})  
+ \widetilde{c} \cdot (1 - A + \log A) 
\label{brostu3:fo.678} 
\ = \ \widetilde{c} \cdot \Big( \widetilde{I}(\mathbf{Q},\mathds{P}) + \log A \Big)
\end{eqnarray}
and consequently for any set subset
$A \cdot \boldsymbol{\Omega}$\hspace{-0.23cm}$\boldsymbol{\Omega} \subset A \cdot \mathbb{S}_{>0}^{K}$ 
\begin{eqnarray}
&&\hspace{-0.7cm}
\inf_{\mathbf{Q} \in A \cdot 
\textrm{$\boldsymbol{\Omega}$\hspace{-0.19cm}$\boldsymbol{\Omega}$}} \ 
\inf_{m\neq 0} D_{\widetilde{c}\cdot
\varphi _{0}}(m \cdot \mathbf{Q},\mathds{P})
= \widetilde{c} \cdot (1 - A + \log A) +
\inf_{\mathbf{Q} \in A \cdot \textrm{$\boldsymbol{\Omega}$\hspace{-0.19cm}$\boldsymbol{\Omega}$}} \ 
D_{\widetilde{c}\cdot \varphi_{0}}(\mathbf{Q},\mathds{P}),  
\label{brostu3:fo.mmin-RKL1} \\
&&\hspace{-0.7cm}
\arg \inf {}_{\mathbf{Q}\in A \cdot 
\textrm{$\boldsymbol{\Omega}$\hspace{-0.19cm}$\boldsymbol{\Omega}$} } \ 
\inf_{m\neq 0}D_{\widetilde{c}\cdot 
\varphi_{0}}(m \cdot \mathbf{Q},\mathds{P})=
\arg \inf {}_{\mathbf{Q}\in A \cdot 
\textrm{$\boldsymbol{\Omega}$\hspace{-0.19cm}$\boldsymbol{\Omega}$} } \ 
D_{\widetilde{c}\cdot\varphi_{0}}(\mathbf{Q},\mathds{P}), \qquad \   
\label{brostu3:fo.mmin-RKL2} \\
&&\hspace{-0.7cm}
\inf_{\mathbf{Q} \in A \cdot 
\textrm{$\boldsymbol{\Omega}$\hspace{-0.19cm}$\boldsymbol{\Omega}$}} \ 
\inf_{m\neq 0} D_{\varphi_{0}}(m \cdot \mathbf{Q},\mathds{P})
= \log A +
\inf_{\mathbf{Q} \in A \cdot \textrm{$\boldsymbol{\Omega}$\hspace{-0.19cm}$\boldsymbol{\Omega}$}} \ 
\widetilde{I}(\mathbf{Q},\mathds{P}),  
\label{brostu3:fo.mmin-RKL1b} \\
&&\hspace{-0.7cm}
\arg \inf {}_{\mathbf{Q}\in A \cdot 
\textrm{$\boldsymbol{\Omega}$\hspace{-0.19cm}$\boldsymbol{\Omega}$} } \ 
\inf_{m\neq 0} D_{\varphi_{0}}(m \cdot \mathbf{Q},\mathds{P})=
\arg \inf {}_{\mathbf{Q}\in A \cdot 
\textrm{$\boldsymbol{\Omega}$\hspace{-0.19cm}$\boldsymbol{\Omega}$} } \ 
\widetilde{I}(\mathbf{Q},\mathds{P}), \qquad \   
\label{brostu3:fo.mmin-RKL2b}
\end{eqnarray}
\vspace{-0.2cm}
\noindent
provided that the infimum on the right-hand side of \eqref{brostu3:fo.mmin-RKL1}  exists. \\

\end{lemma}

\begin{remark}
\label{rem.negativity}
Notice that for $\mathds{P} \in \mathbb{S}_{>0}^{K}$ and $\mathbf{Q} \in A \cdot \mathbb{S}^{K}$,
the modified Kullback-Leibler information has the property
$I(\mathbf{Q},\mathds{P}) \geq 0$ if $A \geq 1$ (cf. \eqref{brostu3:fo.divpow.Kull1}); 
otherwise,
$I(\mathbf{Q},\mathds{P})$ may become negative, as
can be easily seen from the case where $\mathds{P} := \mathds{P}^{unif} := 
(\frac{1}{K}, \ldots, \frac{1}{K})$ is the probability vector 
of frequencies of the uniform distribution on $\{1, \ldots, K\}$,
and $\mathbf{Q} := (\frac{1}{K+1}, 0, \ldots, 0)$. Analogously,
for $\mathds{P}\in \mathbb{S}^{K}$ and $\mathbf{Q} \in A \cdot \mathbb{S}_{>0}^{K}$
one gets $\widetilde{I}(\mathbf{Q},\mathds{P}) \geq 0$ if $A \leq 1$
(cf. \eqref{brostu3:fo.divpow.RevKull1});
otherwise, $\widetilde{I}(\mathbf{Q},\mathds{P})$ may become negative
(take e.g. $\mathbf{Q} = (\frac{K+1}{K}, \ldots, \frac{K+1}{K})$
and $\mathds{P} := (1,0, \ldots, 0)$).

\end{remark}

\vspace{0.3cm}
\begin{remark}
\label{rem.inversions}
(a) In the context of Remark \ref{remark divnormW}(vi),
according to \eqref{LDP Normalized det 3} applied to 
$\varphi := \widetilde{c}\cdot \varphi_{\gamma}$, 
for all cases\\
 $\gamma \in \, ]-\infty,0[ \ \cup \ ]0,1[ \ \cup \ [2,\infty[$
the left-hand side of each of \eqref{Inf in Lemma rate case finite 1},
\eqref{Inf in Lemma rate case finite 1b}, \eqref{Inf in Lemma rate case finite 1c}
is independent of $A >0$ and equal to 
$-\lim_{n\rightarrow \infty }\frac{1}{n}\log \, 
\mathbb{\Pi}\big[\boldsymbol{\xi}_{n}^{w\mathbf{W}}
\in \textrm{$\boldsymbol{\Omega}$\hspace{-0.23cm}$\boldsymbol{\Omega}$}\big]$
where --- as will be shown in the solved-cases Section \ref{Sect Cases} below --- 
the corresponding $\mathbf{W}$'s
have probability distribution 
$\mathbb{\bbzeta}[\cdot \,]=
\mathbb{\Pi }[W_{1}\in \cdot \,]$ (cf. \eqref{Phi Legendre of mgf(W)}) 
which varies \textquotedblleft quite
drastically\textquotedblright\ with $\gamma$ (and the case $\gamma \in \, ]1,2[$
has to be even excluded for analytical difficulties
\footnote{because in this case there are some indications that the representation 
\eqref{Phi Legendre of mgf(W)} only holds for some \textit{signed}
probability distribution $\bbzeta$ (e.g. having a density with positive and negative 
values).
}). Analogously, each of the left-hand sides of 
\eqref{brostu3:fo.mmin-KL1},\eqref{brostu3:fo.mmin-KL1b},\eqref{brostu3:fo.mmin-RKL1},\eqref{brostu3:fo.mmin-RKL1b}
is also independent of $A >0$ and equal to 
$-\lim_{n\rightarrow \infty }\frac{1}{n}\log \, 
\mathbb{\Pi}\big[\boldsymbol{\xi}_{n}^{w\mathbf{W}}
\hspace{-0.15cm} \in \hspace{-0.1cm}
\textrm{$\boldsymbol{\Omega}$\hspace{-0.23cm}$\boldsymbol{\Omega}$}\big]$
for some $\mathbf{W}$ of respective distribution.
Hence, by inversion, 
all the extremum-describing target quantities 
$\inf_{\mathbf{Q}\in A \cdot \textrm{$\boldsymbol{\Omega}$\hspace{-0.19cm}$\boldsymbol{\Omega}$} }
D_{\widetilde{c}\cdot \varphi_{\gamma}}(\mathbf{Q},\mathds{P})$ 
($\gamma \in \mathbb{R}\backslash]1,2[$),
$\inf_{\mathbf{Q}\in A \cdot \textrm{$\boldsymbol{\Omega}$\hspace{-0.19cm}$\boldsymbol{\Omega}$} }
H_{\gamma}(\mathbf{Q},\mathds{P})$ 
($\gamma \in ]-\infty,0[ \, \cup \, [2,\infty[$), \, 
$\sup_{\mathbf{Q}\in A \cdot \textrm{$\boldsymbol{\Omega}$\hspace{-0.19cm}$\boldsymbol{\Omega}$} }
H_{\gamma}(\mathbf{Q},\mathds{P})$ ($\gamma \in \, ]0,1[$), \, 
$\inf_{Q\in A \cdot \textrm{$\boldsymbol{\Omega}$\hspace{-0.19cm}$\boldsymbol{\Omega}$} }
I(\mathbf{Q},\mathds{P})$ \, and \, 
$\inf_{Q\in A \cdot \textrm{$\boldsymbol{\Omega}$\hspace{-0.19cm}$\boldsymbol{\Omega}$} }
\widetilde{I}(\mathbf{Q},\mathds{P})$
can be expressed as 
$G\Big(-\lim_{n\rightarrow \infty }\frac{1}{n}\log \, 
\mathbb{\Pi}\negthinspace \Big[\boldsymbol{\xi}_{n}^{w\mathbf{W}}
\in \textrm{$\boldsymbol{\Omega}$\hspace{-0.23cm}$\boldsymbol{\Omega}$})\, \Big]\Big)$
for some explicitly known ($A-$dependent) function $G$.
This means that --- in the sense 
of Definition \ref{brostu3:def.1} ---
all the corresponding 
four \textquotedblleft  cornerstone quantities\textquotedblright\
$D_{\widetilde{c}\cdot \varphi_{\gamma}}(\mathbf{Q},\mathds{P})$,
$H_{\gamma}(\mathbf{Q},\mathds{P})$, $I(\mathbf{Q},\mathds{P})$,
$\widetilde{I}(\mathbf{Q},\mathds{P})$
are BS-minimizable, respectively BS-maximizable,
on $\mathbf{\Omega} = A \cdot \textrm{$\boldsymbol{\Omega}$\hspace{-0.23cm}$\boldsymbol{\Omega}$}$.
The above-mentioned inversions (i.e. constructions of $G(\cdot)$)
will be concretely carried out in Section \ref{Sect Cases} below
--- namely in the Propositions \ref{brostu3:cor.powdivexpl2} 
to \ref{brostu3:cor.powdivexpl5}.\\ 
\noindent
(b) The case $\varphi := \widetilde{c}\cdot \varphi_{\gamma}$ \,  
($\gamma \in \, ]-\infty,0[ \ \cup \ ]0,1[ \ \cup \ [ \ 2,\infty[$)
of Theorem \ref{brostu3:thm.divnormW.new} works analogously to (a), 
with the differences that we employ 
$A=1$ (instead of arbitrary $A>0$),
\eqref{LDP Normalized Vec} (instead of \eqref{LDP Normalized det 3}),
 $\mathbb{\Pi}_{X_{1}^{n}}[\cdot]$ (instead of $\mathbb{\Pi }[\cdot]$),
and $\xi _{n,\mathbf{X}}^{w\mathbf{W}}$
(instead of $\boldsymbol{\xi}_{n}^{w\mathbf{W}}$).\\
(c) As can be seen in the proof of Lemma \ref{Lemma Indent Rate finite case_new}, 
for the important case $\gamma=2$
the formulas \eqref{brostu3:fo.676} to \eqref{equiv infima in lemma rate finite case 1b}
also hold for $A<0$.

\end{remark}

\vspace{0.3cm}
\noindent
For the rest of the paper, in connection with the three points (a),(b),(c) 
of Remark \ref{rem.inversions}, we will always interpret (without explicit mentioning, for the sake of brevity)
the expression \textquotedblleft BS minimizable/maximizable\textquotedblright\ 
accordingly
in terms of $-\lim_{n\rightarrow\infty}\frac{1}{n}\log \mathbb{\Pi }\Big[\boldsymbol{\xi}_{n}^{w\mathbf{W}} \in \cdot \ \Big]$
respectively of $-\lim_{n\rightarrow\infty}\frac{1}{n}\log \mathbb{\Pi}_{X_{1}^{n}}\Big[\xi _{n,\mathbf{X}}^{w\mathbf{W}}\in \cdot \ \Big]$.

\vspace{0.6cm}
\noindent
Let us end this subsection with a comparison: 
suppose that we have a (sufficiently large) number $n$ 
of \textit{concrete} data observations $X_{i} = x_{i}$
($i=1,\ldots,n$) from the unknown probability distribution 
$\mathds{P}$ (in vector form), 
and from these we want to approximate/estimate 
the unknown distance $\inf_{\mathds{Q}\in \textrm{$\boldsymbol{\Omega}$\hspace{-0.19cm}$\boldsymbol{\Omega}$}} 
D_{\varphi_{\gamma}}( \mathds{Q}, \mathds{P} )$
from a family
of probability models (in vector form)\\
$\boldsymbol{\Omega}$\hspace{-0.23cm}$\boldsymbol{\Omega}$
(e.g. for model-adequacy
evaluations, for goodness-of-fit testing purposes):
by the elaborations in Section \ref{Sect Cases} below,
for the approximation of $\inf_{\mathds{Q}\in \textrm{$\boldsymbol{\Omega}$\hspace{-0.19cm}$\boldsymbol{\Omega}$}} 
D_{\varphi_{\gamma}}( \mathds{Q}, \mathds{P} )$) we can use
\begin{equation}
G\Big(-\frac{1}{n} \cdot \log \, 
\mathbb{\Pi}_{x_{1}^{n}}\negthinspace \left[
\boldsymbol{\xi}_{n,\mathbf{x}}^{w\mathbf{W}}\in 
\textrm{$\boldsymbol{\Omega}$\hspace{-0.23cm}$\boldsymbol{\Omega}$} \right]
\Big)
\label{approx 1}
\end{equation}
where $\mathbb{\Pi}_{x_{1}^{n}}[\, \cdot \, ]  := \mathbb{\Pi}[ \, \cdot \, | \, 
X_{1}=x_{1}, \ldots, X_{n} = x_{n}]$, $\mathbf{x} := (x_{1},\ldots,x_{n})$, 
and $G$ (cf. \eqref{brostu3:fo.2}) is e.g. chosen as follows:\\
$G(z) :=  -\frac{\widetilde{c}}{\gamma \cdot \left( \gamma -1\right) } 
\cdot \left\{ 1-\left( 1-\frac{\gamma }{\widetilde{c}} \cdot z \right) ^{1-\gamma } \right\}$
for the three cases $\gamma <0$, $\gamma \in \, ]0,1[$ and $\gamma \geq 2$,
$G(z) := z$ for $\gamma=0$ (reversed Kullback-Leibler divergence),
and $G(z):= - \widetilde{c} \cdot \log(1- \frac{1}{\widetilde{c}} \cdot z)$
for $\gamma=1$ (Kullback-Leibler divergence).
Notice that \eqref{approx 1} contrasts to the alternative approximation 
(of $\inf_{\mathds{Q}\in \textrm{$\boldsymbol{\Omega}$\hspace{-0.19cm}$\boldsymbol{\Omega}$}} 
D_{\varphi_{\gamma}}( \mathds{Q}, \mathds{P} )$) given by
\begin{equation}
\inf_{\mathds{Q}\in \textrm{$\boldsymbol{\Omega}$\hspace{-0.19cm}$\boldsymbol{\Omega}$}} 
D_{\varphi_{\gamma}}(\mathds{Q}, \mathbb{P}_{n}^{emp, co} ) 
\label{approx 2}
\end{equation}
which is used in the context of \textquotedblleft classical\textquotedblright\ statistical minimum distance estimation (MDE) 
with power divergences;
in \eqref{approx 2}, we have employed 
$\mathbb{P}_{n}^{emp, co} = \frac{1}{n} \cdot \sum_{i=1}^{n}  \delta_{x_{i}}$
to be the realization of the empirical distribution
$\mathbb{P}_{n}^{emp} = \frac{1}{n} \cdot \sum_{i=1}^{n}  \delta_{X_{i}}$.
Indeed, especially in complicated high-dimensional non-parametric or
semi-parametric big-data contexts,
we have substituted a quite difficult \textit{optimization problem} \eqref{approx 2} 
by a much easier solvable \textit{counting problem} \eqref{approx 1}. 
The same holds analogously for Renyi distances/divergences, etc.


\subsection{Construction principle for bounds of the minimum divergence
in the general case \label{Sub Bounds General div}} 

\noindent
Returning to Theorem 
\ref{brostu3:thm.divnormW.new}, we now consider
the general case when the divergence generator $\varphi \in \Upsilon(]a,b[)$ is \textit{not} of the power type \eqref{brostu3:fo.powdivgen}. 
Recall from \eqref{LDP Normalized link},\eqref{LDP Normalized Vec}
the crucial terms (with $\mathds{P} \in \mathbb{S}_{>0}$)

\begin{equation}
\inf_{m\neq 0} D_{\varphi }(m\cdot \textrm{$\boldsymbol{\Omega}$\hspace{-0.23cm}$\boldsymbol{\Omega}$},\mathds{P})
:= 
\inf_{m\neq 0}\ \inf_{\mathds{Q}\in \textrm{$\boldsymbol{\Omega}$\hspace{-0.19cm}$\boldsymbol{\Omega}$} }
D_{\varphi }(m\cdot \mathds{Q},\mathds{P})
= 
\inf_{\mathds{Q}\in \textrm{$\boldsymbol{\Omega}$\hspace{-0.19cm}$\boldsymbol{\Omega}$} }
\ \inf_{m\neq 0} 
D_{\varphi }(m\cdot \mathds{Q},\mathds{P}) \,  <  \, \infty 
\label{LDP Normalized 2}
\end{equation}
for all sets 
$\boldsymbol{\Omega}$\hspace{-0.23cm}$\boldsymbol{\Omega}$ satisfying the 
regularity properties \eqref{regularity} and the convenient,
more restrictive finiteness property 

\begin{equation}
\inf_{\mathds{Q}\in \textrm{$\boldsymbol{\Omega}$\hspace{-0.19cm}$\boldsymbol{\Omega}$}}\ 
\inf_{k=1,\ldots ,K}\frac{q_{k}}{p_{k}}\in dom(\varphi ),\quad 
\sup_{\mathds{Q}\in \textrm{$\boldsymbol{\Omega}$\hspace{-0.19cm}$\boldsymbol{\Omega}$}}\
\sup_{k=1,\ldots ,K}\frac{q_{k}}{p_{k}}\in dom(\varphi )
\label{def fi wrt Omega 2}
\end{equation}
which implies \eqref{def fi wrt Omega};
notice that $\inf_{k=1,\ldots ,K}\frac{q_{k}}{p_{k}} \leq 1$, 
$\sup_{k=1,\ldots ,K}\frac{q_{k}}{p_{k}} \geq 1$ with equalities if and only if
$\mathds{Q} = \mathds{P}$. Since 
$\textrm{$\boldsymbol{\Omega}$\hspace{-0.23cm}$\boldsymbol{\Omega}$} \ne \{ \mathds{P} \}$
(cf. the right-hand side of \eqref{regularity}),
the double infimum (supremum) in \eqref{def fi wrt Omega 2} is strictly smaller
(larger) than $1$.
In general, the inner minimization 
$\inf_{m\neq 0} D_{\varphi }(m\cdot \mathds{Q},\mathds{P})$ 
in \eqref{LDP Normalized 2} can not be performed in
explicit closed form, but e.g. in the specific case of power divergences
(cf. \eqref{brostu3:fo.powdivgen}, \eqref{brostu3:fo.powdiv.new})
the optimization
$\inf_{m\neq 0} D_{\widetilde{c} \cdot \varphi_{\gamma}}(m\cdot \mathds{Q},\mathds{P})$
produces an explicit
form, which in turn leads to a straightforward one-to-one correspondence between 
$D_{\widetilde{c} \cdot \varphi_{\gamma}}(\textrm{$\boldsymbol{\Omega}$\hspace{-0.23cm}$\boldsymbol{\Omega}$},
\mathds{P})$ and 
$\inf_{m\neq 0}\ D_{\widetilde{c} \cdot \varphi_{\gamma}}(m\cdot 
\textrm{$\boldsymbol{\Omega}$\hspace{-0.23cm}$\boldsymbol{\Omega}$},\mathds{P})$ 
(cf. Lemma \ref{Lemma Indent Rate finite case_new}). 
Under \eqref{regularity} and \eqref{def fi wrt Omega 2} one has
\begin{equation}
\inf_{m\neq 0} D_{\varphi }(m\cdot \textrm{$\boldsymbol{\Omega}$\hspace{-0.23cm}$\boldsymbol{\Omega}$},\mathds{P})
\ \leq \ D_{\varphi }(\textrm{$\boldsymbol{\Omega}$\hspace{-0.23cm}$\boldsymbol{\Omega}$},\mathds{P})
\ \leq \ 
D_{\varphi }(\mathds{Q},\mathds{P}) .
\nonumber
\end{equation}
For transparency, we first investigate the (widely useable) subsetup
where $dom(\varphi)= \, ]0,\infty[$ (and thus, $int(dom(\varphi)) = \, ]a,b[ \,  = \, ]0,\infty[$)
and 
$\textrm{$\boldsymbol{\Omega}$\hspace{-0.23cm}$\boldsymbol{\Omega}$} \subset \mathbb{S}_{> 0}^{K}$.
Let us start with the lower bound 
$\inf_{m\neq 0} D_{\varphi }(m\cdot \textrm{$\boldsymbol{\Omega}$\hspace{-0.23cm}$\boldsymbol{\Omega}$},\mathds{P})$.
We prove that the minimizer in $m$ is a well defined
constant, which belongs to a compact set in $\mathbb{R}_{> 0}$. To see this,
notice first that 
from \eqref{def fi wrt Omega 2} one can obtain
\begin{equation}
\Big[ 
\ \inf_{\mathbf{Q}\in \mathbf{\Omega }}\ \inf_{k=1,\ldots ,K}\frac{m \cdot q_{k}}{p_{k}}
\in dom(\varphi )\quad \textrm{and} \quad \sup_{\mathbf{Q}\in \mathbf{\Omega }}\
\sup_{k=1,\ldots ,K}\frac{m \cdot q_{k}}{p_{k}}\in dom(\varphi ) \Big]
\ \Longleftrightarrow \ m \in \, ]0,\infty[ .
\nonumber 
\end{equation}
Moreover, for any fixed 
$\mathds{Q}$ in 
\textrm{$\boldsymbol{\Omega}$\hspace{-0.23cm}$\boldsymbol{\Omega}$} 
there is a unique number $m=m(\mathds{Q}) > 0$ which satisfies the
first-order optimality condition 
\begin{eqnarray}
& &\psi _{\mathds{Q}}(m):=\frac{d}{dm} D_{\varphi }\left( m \cdot \mathds{Q},\mathds{P}\right)
=\sum_{k=1}^{K}q_{k} \cdot \varphi^{\prime }\left( \frac{m \cdot q_{k}}{p_{k}}\right) =0 
\quad \textrm{for $m \in \, ]0,\infty[$}
\label{equ m(Q) variant}
\\
\textrm{and thus } \quad
& & D_{\varphi }\left( m(\mathds{Q}) \cdot \mathds{Q},\mathds{P}\right) =
\inf_{m \neq 0} D_{\varphi }\left( m \cdot \mathds{Q},\mathds{P}\right) ; 
\label{equ m(Q) 2}
\end{eqnarray}
indeed, the mapping $ ]0,\infty[ \ \ni m \rightarrow 
D_{\varphi }\left( m \cdot \mathds{Q},\mathds{P}\right) $ is strictly convex
and infinitely differentiable, 
and the strictly increasing function $\psi_{\mathds{Q}}$ is such that
$\psi_{\mathds{Q}}(m)$ is strictly negative for all $m \in \, ]0,1[$ for which
$\sup_{k=1,\ldots ,K}\frac{m \cdot q_{k}}{p_{k}} <1$ whereas 
$\psi_{\mathds{Q}}(m)$ is strictly positive for all $m > 1$ for which
$\inf_{k=1,\ldots ,K}\frac{m \cdot q_{k}}{p_{k}} >1$ (recall the note right after 
\eqref{def fi wrt Omega 2} and $\varphi^{\prime }(1)=0$).
Hence, for any $\mathds{Q} \in \textrm{$\boldsymbol{\Omega}$\hspace{-0.23cm}$\boldsymbol{\Omega}$}$
the unique zero $m(\mathds{Q})$ of \eqref{equ m(Q) variant} 
(and hence, the unique minimizer in \eqref{equ m(Q) 2})
is in the compact interval
\begin{equation*}
\Big[
\frac{1}{\sup_{k=1,\ldots ,K}\frac{q_{k}}{p_{k}}}, \, \frac{1}{\inf_{k=1,\ldots ,K}\frac{q_{k}}{p_{k}}}
\Big]
\ \subseteq \ 
\Big[
\frac{1}{\sup_{\mathds{Q}\in \textrm{$\boldsymbol{\Omega}$\hspace{-0.19cm}$\boldsymbol{\Omega}$}}\ 
\sup_{k=1,\ldots ,K}\frac{q_{k}}{p_{k}}}, \, 
\frac{1}{\inf_{\mathds{Q}\in \textrm{$\boldsymbol{\Omega}$\hspace{-0.19cm}$\boldsymbol{\Omega}$}}\ 
\inf_{k=1,\ldots ,K}\frac{q_{k}}{p_{k}}}
\Big]
\ \subset \ \Big]\frac{1}{b}, \frac{1}{a}\Big[ \ = \ ]0,\infty[ .
\end{equation*} 
\noindent
When $\textrm{$\boldsymbol{\Omega}$\hspace{-0.23cm}$\boldsymbol{\Omega}$}$
is closed in $\mathbb{S}^{K}$, then by continuity of the function
$\mathds{Q} 
\mapsto D_{\varphi }\left( m\left( \mathds{Q}\right) \cdot \mathds{Q} ,\mathds{P}\right) $
there
exists  a $\mathds{Q}^{\ast }$ in 
$\textrm{$\boldsymbol{\Omega}$\hspace{-0.23cm}$\boldsymbol{\Omega}$}$
which achieves the infimum on 
$\textrm{$\boldsymbol{\Omega}$\hspace{-0.23cm}$\boldsymbol{\Omega}$}$.
When $\textrm{$\boldsymbol{\Omega}$\hspace{-0.23cm}$\boldsymbol{\Omega}$}$ 
is not closed but satisfies \eqref{regularity}, then the infimum exists anyway,
possibly on the boundary 
$\partial \textrm{$\boldsymbol{\Omega}$\hspace{-0.23cm}$\boldsymbol{\Omega}$}$. 
Anyhow, for such $\mathds{Q}^{\ast }$ there holds
\begin{equation}
D_{\varphi }\left( m(\mathds{Q}^{\ast }) \cdot \mathds{Q}^{\ast },\mathds{P}\right) \leq D_{\varphi }\left(
\textrm{$\boldsymbol{\Omega}$\hspace{-0.23cm}$\boldsymbol{\Omega}$} ,\mathds{P}\right) 
\leq D_{\varphi }\left( \mathds{Q}^{\ast },\mathds{P}\right) ,    
\label{estimator-general-divbounds-two}
\end{equation}
where we use the continuity of $\mathds{Q} 
\mapsto D_{\varphi }\left( \mathds{Q},\mathds{P}\right) $ and \eqref{regularity}
to obtain the last inequality above, even when 
$\mathds{Q}^{\ast}\in \partial \textrm{$\boldsymbol{\Omega}$\hspace{-0.23cm}$\boldsymbol{\Omega}$}$ 
and $\mathds{Q}^{\ast }\notin \textrm{$\boldsymbol{\Omega}$\hspace{-0.23cm}$\boldsymbol{\Omega}$}$. 
That \eqref{estimator-general-divbounds-two} provides sharp bounds can be seen 
through the case of power divergences.
Indeed, for the latter 
one basically gets (cf. Appendix C) 
$m(\mathds{Q}) = (1 + \frac{\gamma \cdot (\gamma-1)}{\widetilde{c}} \cdot 
D_{\widetilde{c} \cdot \varphi_{\gamma}}(\mathds{Q}, \mathds{P}) \, )^{1/(1-\gamma)}$
and $D_{\widetilde{c} \cdot \varphi_{\gamma}}\left( m(\mathds{Q}) \cdot \mathds{Q},\mathds{P}\right) = 
\frac{\widetilde{c}}{\gamma} (1- m(\mathds{Q}))$ for the case $\gamma \in \mathbb{R}\backslash\{0,1\}$,
respectively,   
$m(\mathds{Q}) = \exp(-\frac{1}{\widetilde{c}} \cdot D_{\widetilde{c} \cdot \varphi_{1}}(\mathds{Q}, \mathds{P}))$
and $D_{\widetilde{c} \cdot \varphi_{1}}\left( m(\mathds{Q}) \cdot \mathds{Q},\mathds{P}\right) = 
\widetilde{c} \cdot (1- m(\mathds{Q}))$
for the case $\gamma=1$,
respectively, 
$m(\mathds{Q}) = 1$ and $D_{\widetilde{c} \cdot \varphi_{0}}\left( m(\mathds{Q}) \cdot \mathds{Q},\mathds{P}\right) = 
D_{\widetilde{c} \cdot \varphi_{0}}\left(\mathds{Q},\mathds{P}\right)$
for the remaining case $\gamma=0$.
In all cases, $D_{\widetilde{c} \cdot \varphi_{\gamma}}\left( m(\mathds{Q}) \cdot \mathds{Q},\mathds{P}\right)$
is an increasing function of $D_{\widetilde{c} \cdot \varphi_{\gamma}}\left(\mathds{Q},\mathds{P}\right)$
and therefore, $\mathds{Q}^{\ast }\in \arg \inf_{\mathds{Q}\in 
\textrm{$\boldsymbol{\Omega}$\hspace{-0.19cm}$\boldsymbol{\Omega}$} } D_{\widetilde{c} \cdot \varphi_{\gamma}}\left(
m(\mathds{Q}) \cdot \mathds{Q},\mathds{P}\right)$ also satisfies 
$\mathds{Q}^{\ast }\in \arg \inf_{\mathds{Q}\in 
\textrm{$\boldsymbol{\Omega}$\hspace{-0.19cm}$\boldsymbol{\Omega}$}} 
D_{\widetilde{c} \cdot \varphi_{\gamma}}\left( \mathds{Q},\mathds{P}\right)$.
Hence, the right-hand side and the left-hand side of \eqref{estimator-general-divbounds-two}
coincide. Now due to \eqref{Phi Legendre of mgf(W)}, the LHS of 
\eqref{estimator-general-divbounds-two} can be estimated since by 
Theorem \ref{brostu3:thm.divnormW.new} for each $\mathds{P}\in \mathbb{S}_{>0}^{K}$
the divergence $\inf_{m\neq 0}D_{\varphi }\left( m \cdot \mathds{Q},\mathds{P}\right)$ is 
BS-minimizable on sets $\textrm{$\boldsymbol{\Omega}$\hspace{-0.23cm}$\boldsymbol{\Omega}$} \subset \mathbb{S}^{K}$.  
We shall propose in the below-treated Subsection \ref{subsub-estimator-GC2} 
an algorithm to handle the estimation of the RHS of
\eqref{estimator-general-divbounds-two}, whenever $\mathds{P}$ 
is known (as in Remark 13(v)) or when $\mathds{P}$ is
approximated by the empirical distribution of the data set $\left(X_{1},..,X_{n}\right)$.
Also note that \eqref{estimator-general-divbounds-two}
holds also for $\textrm{$\boldsymbol{\Omega}$\hspace{-0.23cm}$\boldsymbol{\Omega}$}$ 
substituted by  
$A \cdot \textrm{$\boldsymbol{\Omega}$\hspace{-0.23cm}$\boldsymbol{\Omega}$}$
for any $A\neq 0$. 


\vspace{0.3cm}
\noindent
Other cases of interest include when $dom\left( \varphi \right) $ is not 
$\left] 0,\infty \right[$. We list two cases which extend the above
discussion. Firstly, consider $\varphi $ with $dom\left( \varphi \right) =
\left[ 0,\infty \right[$. Then we may extend \eqref{estimator-general-divbounds-two} 
to cases when 
$\textrm{$\boldsymbol{\Omega}$\hspace{-0.23cm}$\boldsymbol{\Omega}$} \subset \mathbb{S}^{K}$ 
instead of $\textrm{$\boldsymbol{\Omega}$\hspace{-0.23cm}$\boldsymbol{\Omega}$} \subset \mathbb{S}_{>0}^{K}$,
hence allowing for possible null entries in $\textrm{$\boldsymbol{\Omega}$\hspace{-0.23cm}$\boldsymbol{\Omega}$}$.
When $dom\left( \varphi \right) =\left] a,b\right[ $ for some $a<0$, then clearly
the same argument leading to 
\eqref{estimator-general-divbounds-two}
holds; this case is of interest, for
instance, when extending a statistical model to signed measures  (see e.g.
\cite{Bro:19a} for the important task of testing the number of components
in a parametric probability mixture model).


\subsection{On the difference between minimization problems of 
deterministic nature and risk minimization} 

\noindent
In the context of minimization of the functional 
$\Phi_{\mathbf{P}}(\mathbf{Q})$ over $\mathbf{\Omega} \subset \mathbb{R}^{K}$
for \textit{known} vector $\mathbf{P}$, due to Theorem \ref{brostu3:thm.divW.var}
our bare simulation approach allows for the
\textit{approximate solution} for any divergence 
$D_{\varphi}$ satisfying the basic representation \eqref{brostu3:fo.link.var}.
Indeed, any proxy of 
$\mathbb{\Pi }\big[\boldsymbol{\xi }_{n}^{\mathbf{\widetilde{W}}}\in \mathbf{\Omega} /M_{\mathbf{P}}\big]$
yields a corresponding proxy for $\Phi_{\mathbf{P}}\negthinspace \left[\mathbf{\Omega}\right]$. 
This paves the way to the solution of numerous optimization problems, where the
divergence $D_{\varphi}$  is specifically suited to the problem at hand.
In the statistical context, when the \textit{probability distribution} 
(in its vector-form)
$\mathds{P}$ is \textit{unknown} up
to some indirect information provided by sampling or by any mean providing
a sequence $(X_{i})_{i\in \mathbb{N}}$ 
satisfying condition \eqref{cv emp measure X to P} 
(resp. \eqref{cv emp measure X to P vector}),
Theorem \ref{brostu3:thm.divnormW.new} adds a complementary
step of complexity; indeed, the \textit{estimation} of 
$\Phi_{\mathbf{P}}(\textrm{$\boldsymbol{\Omega}$\hspace{-0.23cm}$\boldsymbol{\Omega}$})$ 
over $\textrm{$\boldsymbol{\Omega}$\hspace{-0.23cm}$\boldsymbol{\Omega}$} 
\subset \mathbb{S}^{K}$
results as its subproduct through the optimization upon $m$ which can be performed explicitly only
in a number of specific divergences $D_{\varphi}$, e.g. 
the power divergences $D_{\varphi_{\gamma}}$, and which carries over also to their 
monotone transformations 
such as e.g. the Renyi divergences.
It is of relevance to mention that --- as already indicated above --- 
these divergences cover a \textit{very broad range} of
statistical criteria, indeed most of them, 
from the (maximum-likelihood estimation connected) likelihood divergence ($\gamma=0$) to
the Kullback-Leibler one ($\gamma=1$), the two standard Chi-square distances
($\gamma=2$, \, $\gamma=-1$), the Hellinger
distance ($\gamma=1/2$), etc.; in contrast with deterministic minimization problems, the choice
of a statistical criterion (or risk function) is not imposed by the modelling of
the problem at hand, but is dictated by the need for sharp measures of fit. 
Other divergences are more difficult to handle and our
general results in Section \ref{Sub Bounds General div} still prove some 
usefulness, since estimation
of upper and lower bounds for risk is of common use.

\vspace{0.2cm}
\noindent
As a \textquotedblleft  preparatory\textquotedblright\ remark, 
recall first that each probability
distribution (probability measure) $\mathbb{P}$ on $\mathcal{Y}=\left\{d_{1},\ldots,d_{K}\right\}$
has been uniquely identified with the vector $\mathds{P} := (p_{1}, \ldots, p_{K}) \in \mathbb{S}^{K}$ 
of the corresponding probability masses (frequencies) 
$p_{k} = \mathbb{P}[\{ d_{k} \}]$ via 
$\mathbb{P}[A] = \sum_{k=1}^{K} p_{k} \cdot \textfrak{1}_A(d_{k}) $ for each $A \subset \mathcal{Y}$;
from this, we have measured the distance/divergence between two probability
distributions $\mathbb{P}, \mathbb{Q}$ through the distance/divergence between their 
frequency vectors $\mathds{P}, \mathds{Q}$:
\begin{equation}
D_{\varphi }(\mathbb{Q},\mathbb{P}) : = D_{\varphi }(\mathds{Q},\mathds{P})
\qquad \textrm{(cf. \eqref{measure divergence})} .
\nonumber
\end{equation}
However, it has been noted in \cite{Kis:16} in a
context of even more general divergences $D(\mathbf{Q}, \mathbf{P})$
between vectors $\mathbf{P}, \mathbf{Q}$
that --- alternatively --- the latter two may consist of 
components  $p_{k} = \mathbb{P}[\{ E_{k} \}]$,  $q_{k} = \mathbb{Q}[\{ E_{k} \}]$
which are probabilities of only
some \textit{selected} (e.g.\ increasing) events $(E_{k})_{k \in \{1,\ldots,M\}} 
$ of \textit{application-based concrete} interest (within \textit{not necessarily discrete} 
probability models), e.g. related to cumulative distribution functions
(see~\cite{Bro:22}). Of course, we can apply our BS method to such a vector context.
As other alternatives, we can also deal with divergences between 
\textit{non-probabilistic} uncertainty quantifications, such as fuzzy sets and basic belief assignments (see our paper's full arXiv-version \cite{Bro:21}).

%
%

\section{Renyi divergences and friends \label{SectRenyi}}

\noindent
It is well known that Renyi divergences are widely used important tools 
in information theory as well as in the adjacent fields of 
statistics, machine learning, artificial intelligence, 
signal processing and pattern recognition. As a consequence of our
considerations in the above Subsection \ref{Sub Contruction Princ Power case}
in combination with the solved cases in Section \ref{Sect Cases} below, 
we can also apply our BS method
to the constrained optimization of Renyi divergences and closely 
related important quantities. To start with,
let us fix $\widetilde{c}=1$ and an arbitrary triple $(\gamma , \mathds{P},\mathbf{Q})$
which satisfies the assumptions of Lemma \ref{Lemma Indent Rate finite case_new}(a)
with $A := \sum_{k=1}^{K} q_{k} >0$.
For such a setup, we have obtained in   
\eqref{brostu3:fo.divpow.hellinger1}
the $\gamma-$order Hellinger integral (of $\mathbf{Q}$ and $\mathds{P}$)
$H_{\gamma}(\mathbf{Q},\mathds{P}) > 0$,
which is not a divergence;
as a terminology-concerning side remark, let us mention that 
$H_{\gamma}(\mathbf{Q},\mathds{P})$ ($\gamma \geq 1$) is called 
\textit{relative information generating function} in \cite{Gui:85}, 
see e.g. also \cite{Clark:20}; moreover,
$H_{\gamma}(\mathds{Q},\mathds{P})$
is sometimes termed \textit{($\gamma-$order) Chernoff coefficient} 
being a component of the Chernoff distances/informations \cite{Cher:52}.
In \cite{Tor:91} the name \textit{($\gamma-$order) Hellinger transform} is used.  
Notice that the special case $\gamma =\frac{1}{2}$ is nothing but
(a multiple of) the well-known important \textit{Bhattacharyya coefficient}
(cf.~\cite{Bha:43},\cite{Bha:46},\cite{Bha:47})
\begin{equation}
BC(\mathbf{Q},\mathds{P}) := 
H_{1/2}(\mathbf{Q},\mathds{P}) = 
\sum_{k=1}^{K} \sqrt{q_{k} \cdot p_{k}} =
1 + \frac{1}{2} \cdot (A-1) + \frac{1}{2} \cdot (\frac{1}{2}-1) \cdot D_{\varphi_{\frac{1}{2}}}(\mathbf{Q},\mathds{P})
\nonumber
\end{equation} 
which is also known as 
\textit{affinity} (cf. \cite{Mat:51}, see e.g. also \cite{Tou:78})
and \textit{(classic, non-quantum) fidelity similarity} (cf. e.g. \cite{Dez:16});
for non-probability vectors $\mathbf{P} \in \mathbb{R}_{\gneqq 0}^{K}$ 
with $M_{\mathbf{P}} >0$ one can simply retransform
$\mathds{P} := \frac{\mathbf{P}}{M_{\mathbf{P}}}$ and thus imbed
$BC(\mathbf{Q},\mathbf{P}) = \sqrt{M_{\mathbf{P}}} \cdot BC(\mathbf{Q},\mathds{P})$
into our BS context.
There is a vast literature on very recent applications of the Bhattacharyya coefficient,
for instance it appears exemplarily in~\cite{Peng:14}--\cite{Zhou:21}.
To proceed with the general context, for any 
$\gamma \in \, ]-\infty,0[ \ \cup \ ]0,1[ \ \cup \ [ \ 2,\infty[$
let the function
$h_{\gamma} \negthinspace : \, ]0,\infty[ \, \mapsto \, ]-\infty,\infty[$
be such that 
$x \mapsto h_{\gamma}(1 + \gamma \cdot (A-1) + \gamma \cdot (\gamma-1) \cdot x)$
is continuous and strictly increasing (respectively, strictly decreasing)
for all $x \geq 0$ with $1 + \gamma \cdot (A-1) + \gamma \cdot (\gamma-1) \cdot x > 0$;
since $D_{\varphi_{\gamma}}(\mathbf{Q},\mathds{P})$ is
BS-minimizable on
$\mathbf{\Omega} = A \cdot \textrm{$\boldsymbol{\Omega}$\hspace{-0.23cm}$\boldsymbol{\Omega}$}$,
then also the --- not necessarily nonnegative --- quantity 
$h_{\gamma}\Big(1 + \gamma \cdot (A-1) + \gamma \cdot (\gamma-1) \cdot 
D_{\varphi_{\gamma}}(\mathbf{Q},\mathds{P})\Big)
= h_{\gamma}\Big(H_{\gamma}(\mathbf{Q},\mathds{P})\Big)$
is BS-minimizable (respectively, BS-maximizable) on
$\mathbf{\Omega} = A \cdot \textrm{$\boldsymbol{\Omega}$\hspace{-0.23cm}$\boldsymbol{\Omega}$}$. 
If $h_{\gamma}$ satisfies additionally $h_{\gamma}(1)=0$ as well as
$h_{\gamma}(1 + \gamma \cdot (A-1) + \gamma \cdot (\gamma-1) \cdot x) \geq 0$
for all $x \geq 0$ with $1 + \gamma \cdot (A-1) + \gamma \cdot (\gamma-1) \cdot x > 0$,
then $D_{h_{\gamma}}(\mathbf{Q},\mathds{P}) := 
h_{\gamma}\Big(1 + \gamma \cdot (A-1) + \gamma \cdot (\gamma-1) \cdot 
D_{\varphi_{\gamma}}(\mathbf{Q},\mathds{P})\Big)
= h_{\gamma}\Big(H_{\gamma}(\mathbf{Q},\mathds{P})\Big)$
constitutes a divergence 
\footnote{in the usual sense that
$D_{h_{\gamma}}(\mathbf{Q},\mathds{P}) \geq 0 $ with equality iff
$\mathds{Q}=\mathds{P}$.}
which is BS-minimizable on
$\mathbf{\Omega} = A \cdot \textrm{$\boldsymbol{\Omega}$\hspace{-0.23cm}$\boldsymbol{\Omega}$}
$ (respectively, BS-maximizable on
$\mathbf{\Omega} = A \cdot \textrm{$\boldsymbol{\Omega}$\hspace{-0.23cm}$\boldsymbol{\Omega}$}
$).

\vspace{0.2cm}
Let us consider some important examples.
For the identity mapping $h_{\gamma}^{Id}(y) := y$ \, ($y>0$) the function
$x \mapsto 1 + \gamma \cdot (A-1) + \gamma \cdot (\gamma-1) \cdot x$
is strictly increasing for $\gamma <0$ and $\gamma >1$ (on the required domain of $x$), and
strictly decreasing for $\gamma \in ]0,1[$. Accordingly,
$H_{\gamma}(\mathbf{Q},\mathds{P})$ is BS-minimizable 
on
$\mathbf{\Omega} = A \cdot \textrm{$\boldsymbol{\Omega}$\hspace{-0.23cm}$\boldsymbol{\Omega}$}$
for $\gamma <0$ and $\gamma \geq 2$ 
and BS-maximizable on
$\mathbf{\Omega} = A \cdot \textrm{$\boldsymbol{\Omega}$\hspace{-0.23cm}$\boldsymbol{\Omega}$}$
for $\gamma \in\,  ]0,1[$ (this is consistent with
\eqref{Inf in Lemma rate case finite 1b},\eqref{Inf in Lemma rate case finite 1c});
in particular, the Bhattacharyya coefficient
$BC(\mathbf{Q},\mathds{P})$ is BS-maximizable on
$\mathbf{\Omega} = A \cdot \textrm{$\boldsymbol{\Omega}$\hspace{-0.23cm}$\boldsymbol{\Omega}$}$.
Some other important choices are 
\begin{eqnarray}
& & h_{\gamma}(y) := h_{c_{1},c_{2},c_{3}}(y) :=  c_{1} \cdot \Big(y^{c_{2}} - c_{3} \Big), 
\qquad y>0, \, c_{1}, c_{2} \in \mathbb{R}\backslash\{0\}, \, 
c_{3} \in \mathbb{R}, 
\label{fo.hgamma1}\\
& & h_{\gamma}(y) := h_{c_{4},f}^{R}(y) := \lim_{c_{2} \rightarrow 0} h_{c_{4}/f(c_{2}),c_{2},1}(y) =
\frac{c_{4}}{f^{\prime}(0)} \cdot \log(y),
\qquad y>0, \, c_{4} \in \mathbb{R}\backslash\{0\}, \, 
\label{fo.hgamma2}\\
& & h_{\gamma}(y) := h_{c_{5},c_{6}}^{GB2}(y) := c_{5} \cdot (\arccos(y))^{c_{6}}, 
\qquad \gamma \in \,]0,1[, \, y \in \, ]0,1], \, c_{5}>0,  \, c_{6}>0,
\label{fo.hgamma3}\\
& & h_{\gamma}(y) := h_{\nu,c_{7}}^{BB}(y) := c_{7} \cdot \frac{\log(1-\frac{1-y}{\nu})}{\log(1-\frac{1}{\nu})},
\qquad \gamma \in \,]0,1[, \, y \in \, ]0,1], \, c_{7}>0, \, \nu \in \, ]-\infty,0[ \, \cup \, ]1,\infty[,
\label{fo.hgamma4}
\end{eqnarray}
where the constants $c_{1}$ to $c_{7}$ may depend on $\gamma$, 
and $f$ is some (maybe $\gamma-$dependent) 
function which is differentiable in a neighborhood of $0$
and satisfies $f(0)=0$, $f^{\prime}(0) \ne 0$ \, (e.g. $f(z)=c_{8} \cdot z$
for some non-zero constant $c_{8}$).
Clearly, $h_{c_{1},c_{2},c_{3}}(\cdot)$ is strictly increasing (respectively,
strictly decreasing) if and only if $c_{1} \cdot c_{2} >0$
(respectively, $c_{1} \cdot c_{2} <0$). Moreover, $h_{c_{4},f}^{R}(\cdot)$ is
 strictly increasing (respectively,
strictly decreasing) if and only if $\frac{c_{4}}{f^{\prime}(0)} >0$ 
(respectively, $\frac{c_{4}}{f^{\prime}(0)} < 0$). 
Furthermore, both $h_{c_{5},c_{6}}^{GB2}(\cdot)$ and $h_{\nu,c_{7}}^{BB}(\cdot)$
are strictly decreasing.
For instance, the special case $h_{\gamma}(y) = h_{c_{4},Id}^{R}(y)$
with $c_{4} := \frac{1}{\gamma \cdot (\gamma-1)}$ 
(recall that $\gamma \in \, ]-\infty,0[ \ \cup \ ]0,1[ \ \cup \ [ \ 2,\infty[$)
and identity function $f := Id$ leads to the quantities
\begin{eqnarray}
\hspace{-1.5cm}
R_{\gamma}(\mathbf{Q},\mathds{P}) &: =&
D_{h_{c_{4},Id}^{R}}(\mathbf{Q},\mathds{P}) =
\frac{ 
\log \Big( 1 + \gamma \cdot (A-1) + 
\gamma \cdot (\gamma-1) \cdot D_{\varphi_{\gamma}}(\mathbf{Q},\mathds{P}) \Big)
}{\gamma \cdot (\gamma-1)}
= \frac{ 
\log \Big( H_{\gamma}(\mathbf{Q},\mathds{P}) \Big)
}{\gamma \cdot (\gamma-1)} 
\nonumber \\
&=& \frac{ 
\log \Big( \sum\displaylimits_{k=1}^{K} (q_{k})^{\gamma} \cdot (p_{k})^{1-\gamma} \Big)
}{\gamma \cdot (\gamma-1)} \, ,
\qquad \gamma \in \, ]-\infty,0[ \ \cup \ ]0,1[ \ \cup \ [ \ 2,\infty[,
\label{def Renyi}
\end{eqnarray}
(provided that all involved power divergences are finite),
which are thus BS-minimizable on
$\mathbf{\Omega} = A \cdot \textrm{$\boldsymbol{\Omega}$\hspace{-0.23cm}$\boldsymbol{\Omega}$}$; notice that $R_{\gamma}(\mathbf{Q},\mathds{P}) \geq 0$ if 
$\gamma \in [\, 2,\infty[$ together with $A \in [1,\infty[$,
and if $\gamma \in \, ]-\infty,0[ \ \cup \ ]0,1[$ together with $A \in \, ]0,1]$. 
The special subcase $A=1$ in \eqref{def Renyi} (and thus, $\mathbf{Q}$ is
a probability vector $\mathds{Q}$) corresponds to the prominent 
\textit{Renyi divergences/distances} \cite{Ren:61} (in the scaling
of e.g. Liese \& Vajda \cite{Lie:87} and in probability-vector form),
see e.g. \cite{VanErv:14} for
a comprehensive study of their properties; as a side remark,
$\gamma \cdot (\gamma-1) \cdot R_{\gamma}(\mathds{Q},\mathds{P})$
is also employed in the Chernoff distances/informations \cite{Cher:52}.
The special subcase $R_{1/2}(\mathds{Q},\mathds{P})$
(i.e. $\gamma =1/2$ and $A=1$ in \eqref{def Renyi})
corresponds to (a multiple of) the widely used
\textit{Bhattacharyya distance} (of type 1) between $\mathds{Q}$ and $\mathds{P}$, cf. \cite{Bha:43}
(see e.g. also \cite{Kai:67}). 
Sometimes, $\exp(R_{\gamma}(\mathbf{Q},\mathds{P}))$ is also called
\textit{Renyi divergence/distance}.
Some exemplary (relatively) recent 
studies and applications
of Renyi divergences $R_{\gamma}(\mathds{Q},\mathds{P})$
(respectively, their multiple or exponential)
--- aside from the substantial statistical literature ---
appear e.g. in~\cite{Sun:07},\cite{Zhao:18a}--\cite{Mao:21}.
There is vast literature on recent applications of the above-mentioned special case 
$R_{1/2}(\mathds{Q},\mathds{P})$ 
--- that is, the Bhattacharyya distance (of type 1); 
for instance, it appears 
in~\cite{Tar:16}--\cite{Xiahou:21}.
As a further example, consider (cf. \eqref{fo.hgamma3})
\begin{eqnarray}
\mathcal{B}_{\gamma,c_{5},c_{6}}(\mathds{Q},\mathds{P}) &: =&
D_{h_{c_{5},c_{6}}^{GB2}}(\mathds{Q},\mathds{P}) =
c_{5} \cdot \Big(\arccos\Big(
1 + \gamma \cdot (\gamma-1) \cdot D_{\varphi_{\gamma}}(\mathds{Q},\mathds{P})
\Big) \, \Big)^{c_{6}} 
= c_{5} \cdot \Big(\arccos\Big(
H_{\varphi_{\gamma}}(\mathds{Q},\mathds{P})
\Big) \, \Big)^{c_{6}}
\nonumber \\
&=&  c_{5} \cdot \Big(\arccos\Big(
\sum\displaylimits_{k=1}^{K} (q_{k})^{\gamma} \cdot (p_{k})^{1-\gamma}
\Big) \, \Big)^{c_{6}}
\geq 0 \, ,  
\qquad \gamma \in \, ]0,1[, \, c_5 >0, \, c_{6} >0,
\nonumber
\end{eqnarray}
which is BS-minimizable on
$\textrm{$\boldsymbol{\Omega}$\hspace{-0.23cm}$\boldsymbol{\Omega}$}$. 
The case $\mathcal{B}_{1/2,1,1}(\mathds{Q},\mathds{P})$
corresponds to the well-known \textit{Bhattacharyya arccos distance (Bhattacharyya distance of type 2)} 
in \cite{Bha:47}
(which is also called Wootters distance \cite{Woot:81}), and $\mathcal{B}_{1/2,1,2}(\mathds{Q},\mathds{P})$ to its variant in \cite{Bha:46};
the case $\mathcal{B}_{1/2,2,1}(\mathds{Q},\mathds{P})$ is known as \textit{Fisher distance}
or \textit{Rao distance} or \textit{geodesic distance} (see e.g. \cite{Dez:16});
a nice graphical illustration of the geometric connection between 
the Fisher distance $\mathcal{B}_{1/2,2,1}(\mathds{Q},\mathds{P})$
and the Hellinger distance/metric $\sqrt{\frac{1}{2} \cdot D_{\varphi_{1/2}}(\mathds{Q},\mathds{P})}$ 
can be found e.g. on p.35 in \cite{Ay:17}.
Some exemplary applications of the
Bhattacharyya arccos distance
$\mathcal{B}_{1/2,1,1}(\mathds{Q},\mathds{P})$
can be found e.g. in~\cite{Rao:77}--\cite{Gre:05},\cite{Che2:21}.
Let us give another example, namely (cf. \eqref{fo.hgamma4})
\begin{eqnarray}
\widetilde{\mathcal{B}}_{\gamma,\nu,c_{7}}(\mathds{Q},\mathds{P}) &: =&
D_{h_{\nu,c_{7}}^{BB}}(\mathds{Q},\mathds{P}) =
\frac{c_{7}}{\log(1-\frac{1}{\nu})} \cdot \log\bigg(1-\frac{1-
\Big( 1 + \gamma \cdot (\gamma-1) \cdot 
D_{\varphi_{\gamma}}(\mathds{Q},\mathds{P}) \Big) 
}{\nu}\bigg)
\nonumber \\
&=& \frac{c_{7}}{\log(1-\frac{1}{\nu})} \cdot \log\bigg(1-\frac{1-
H_{\varphi_{\gamma}}(\mathds{Q},\mathds{P})
}{\nu}\bigg)
= \frac{c_{7}}{\log(1-\frac{1}{\nu})} \cdot \log\bigg(1-\frac{1-
\sum\displaylimits_{k=1}^{K} (q_{k})^{\gamma} \cdot (p_{k})^{1-\gamma}
}{\nu}\bigg)
\in [0, c_{7}[ \, 
, \nonumber \\  
& & \hspace{8.0cm} \gamma \in \,]0,1[, \, \, c_{7}>0, \, \nu \in \, ]-\infty,0[ \, \cup \, ]1,\infty[,
\nonumber
\end{eqnarray}
which is BS-minimizable on
$
\textrm{$\boldsymbol{\Omega}$\hspace{-0.23cm}$\boldsymbol{\Omega}$}
$. The case $\widetilde{\mathcal{B}}_{1/2,\nu,1}(\mathds{Q},\mathds{P})$ corresponds to
the \textit{Bounded Bhattacharyya Distance Measures} of \cite{Jol:16}.
We can also employ divergences of the form
$\breve{R}_{\gamma}(\mathds{Q},\mathds{P}) := 
R_{\gamma}(T_{1}(\mathds{Q}),T_{2}(\mathds{P}))$
\footnote{
and analogously power divergences
$\breve{D}_{\widetilde{c} \cdot \varphi_{\gamma}}(\mathds{Q},\mathds{P}) := 
D_{\widetilde{c} \cdot \varphi_{\gamma}}(T_{1}(\mathds{Q}),T_{2}(\mathds{P}))$ etc.
}
where $T_{1}: \mathcal{D}_{1} \mapsto \mathcal{R}_{1}$, 
$T_{2}: \mathcal{D}_{1} \mapsto \mathcal{R}_{2}$
are (say) invertible functions on 
appropriately chosen 
subsets $\mathcal{D}_{1}, \mathcal{D}_{2}, \mathcal{R}_{1}, \mathcal{R}_{2}$
of the probability-vector simplex $\mathbb{S}^{K}$.
For instance, consider the following special case (with a slight abuse of notation):
\begin{equation}
\breve{R}_{\gamma}(\mathds{Q},\mathds{P}): = R_{\gamma}(\widetilde{\mathds{Q}},\widetilde{\mathds{P}})
= 
\frac{1}{\gamma \cdot (\gamma-1)} \cdot
\log \bigg( \sum\displaylimits_{k=1}^{K} 
\Big(
\frac{(q_{k})^{\nu_{1}}}{\sum_{j=1}^{K} (q_{j})^{\nu_{1}}}
\Big)^{\gamma} \cdot 
\Big(
\frac{(p_{k})^{\nu_{2}}}{\sum_{j=1}^{K} (p_{j})^{\nu_{2}}}
\Big)^{1-\gamma} 
\bigg)
\label{def Renyi 2}
\end{equation}
where (i) $\widetilde{\mathds{Q}} := (\widetilde{q}_{k})_{k=1}^{K}$ 
with $\widetilde{q}_{k} := \frac{(q_{k})^{\nu_{1}}}{\sum_{j=1}^{K} (q_{j})^{\nu_{1}}}$
is the \textit{escort probability distribution (in vector form)
associated with the probability distribution (in vector form)} 
$\mathds{Q} := (q_{k})_{k=1}^{K} \in \mathbb{S}_{> 0}^{K}$,
and (ii) $\widetilde{\mathds{P}} := (\widetilde{p}_{k})_{k=1}^{K}$ 
with $\widetilde{p}_{k} := \frac{(p_{k})^{\nu_{2}}}{\sum_{j=1}^{K} (p_{j})^{\nu_{2}}}$
is the escort probability distribution 
associated with the probability distribution 
$\mathds{P} := (p_{k})_{k=1}^{K} \in \mathbb{S}_{> 0}^{K}$, 
in terms of some fixed escort para\-meters $\nu_{1} >0$, $\nu_{2} >0$.
For the special choice $\nu_{1}= \nu_{2} >0$ and $\gamma := \frac{\nu}{\nu_{1}}$ with 
$\nu \in \, ]0,\nu_{1}[ \, \cup \, [2\nu_{1}, \infty[$ we obtain from \eqref{def Renyi 2}
\begin{eqnarray}
\hspace{-1.0cm}
&& 0 \leq \frac{\nu}{\nu_{1}} \cdot R_{\nu/\nu_{1}}(\widetilde{\mathds{Q}},\widetilde{\mathds{P}}) 
=
\frac{ 
\log \Big( \sum\displaylimits_{k=1}^{K} (\widetilde{q}_{k})^{\nu/\nu_{1}} \cdot 
(\widetilde{p}_{k})^{1-(\nu/\nu_{1})} \Big)
}{\frac{\nu}{\nu_{1}} -1} 
\nonumber\\
\hspace{-1.0cm}
&& =
\frac{\nu_{1}}{\nu - \nu_{1}} \cdot 
\log \Big( \sum\displaylimits_{k=1}^{K} (q_{k})^{\nu} \cdot 
(p_{k})^{\nu_{1} - \nu} \Big)
- \frac{\nu}{\nu - \nu_{1}} \cdot 
\log \Big( \sum\displaylimits_{k=1}^{K} (q_{k})^{\nu_{1}} \Big)
+ \log \Big( \sum\displaylimits_{k=1}^{K} (p_{k})^{\nu_{1}} \Big)
=: \breve{R}_{\nu/\nu_{1}}(\mathds{Q},\mathds{P})
\label{def.Renyi.escort}
\end{eqnarray}
which is BS-minimizable (in $\widetilde{\mathds{Q}}$) on
$\textrm{$\boldsymbol{\Omega}$\hspace{-0.23cm}$\boldsymbol{\Omega}$}$.
Our divergence $\breve{R}_{\nu/\nu_{1}}(\mathds{Q},\mathds{P})$ in \eqref{def.Renyi.escort} is 
basically a multiple of a divergence which has been very recently used in
\cite{Gho:21}. Moreover, $\breve{R}_{1/\nu_{1}}(\mathds{Q},\mathds{P})$
(i.e. the special case $\nu =1$ in \eqref{def.Renyi.escort}) is equal to  
\textit{Sundaresan\textquoteright s divergence} \cite{Sun:02} \cite{Sun:07}
(see also \cite{Lut:05}, \cite{Kum:15a}, \cite{Kum:15b}, \cite{Yagli:18});
for our BS-approach, we need 
the restriction $\nu_{1} \in \, ]0,\frac{1}{2}] \, \cup \, ]1,\infty[$.
Notice that Sundaresan\textquoteright s divergence can be employed in 
mismatch-cases of (i) Campbell\textquoteright s coding problem, 
(ii) Arikan\textquoteright s guessing problem, (iii) memoryless guessing,
and (iv) task partitioning problems; see e.g. \cite{Sun:07}, \cite{Bun:14}, \cite{Kum:19}.

\vspace{0.3cm}
\noindent
Returning to the general context,
functions of the modified Kullback-Leibler information $I(\mathbf{Q},\mathds{P})$ and 
the modified reverse Kullback-Leibler information 
$\widetilde{I}(\mathbf{Q},\mathds{P})$ can be treated analogously.
For the sake of brevity, we only deal with the former and fix arbitrary 
$\mathds{P} \in \mathbb{S}_{>0}^{K}$ and $\mathbf{Q} \in A \cdot \mathbb{S}^{K}$
with $A := \sum_{k=1}^{K} q_{k} >0$.
For this, in \eqref{brostu3:fo.divpow.Kull1} we have obtained 
$I(\mathbf{Q},\mathds{P})$ which is generally not a divergence 
(cf. Remark \ref{rem.negativity}). 
In the following, let the function
$h_{1} \negthinspace : \, ]-1,\infty[ \, \mapsto \, ]-\infty,\infty[$
be continuous and strictly increasing (respectively, strictly decreasing);
since $D_{\varphi_{1}}(\mathbf{Q},\mathds{P})$ is
BS-minimizable on
$\mathbf{\Omega} = A \cdot \textrm{$\boldsymbol{\Omega}$\hspace{-0.23cm}$\boldsymbol{\Omega}$}$,
also the quantity 
$h_{1}\Big( A - 1 +  D_{\varphi_{1}}(\mathbf{Q},\mathds{P})\Big)
= h_{1}\Big(I(\mathbf{Q},\mathds{P})\Big)$
is BS-minimizable on
$\mathbf{\Omega} = A \cdot \textrm{$\boldsymbol{\Omega}$\hspace{-0.23cm}$\boldsymbol{\Omega}$}$
(respectively, BS-maximizable on
$\mathbf{\Omega} = A \cdot \textrm{$\boldsymbol{\Omega}$\hspace{-0.23cm}$\boldsymbol{\Omega}$}$).
In particular, by using the negative identity mapping $h_{\gamma}^{-Id}(y) := -y$ \, ($y>-1$) 
we get that $-I(\mathbf{Q},\mathds{P})$ is BS-maximizable.
Another exemplary choice for $h_{1}$ is (cf. \cite{Shar:77}
in the scaling of e.g. \cite{Mor:94})
\begin{eqnarray}
& & h_{1}(y) := h_{s}^{SM}(y) :=  \frac{e^{(s-1) \cdot y} -1}{s-1},
\qquad y \in \mathbb{R}, \, s \in \, ]0,1[ \, \cup \, ]1,\infty[,
\label{fo.hKL1}
\end{eqnarray}
which is strictly increasing; hence, $h_{s}^{SM}(I(\mathbf{Q},\mathds{P}))$ 
(and also $h_{s}^{SM}(D_{\varphi_{1}}(\mathbf{Q},\mathds{P}))$) is
BS-minimizable on
$\mathbf{\Omega} = A \cdot \textrm{$\boldsymbol{\Omega}$\hspace{-0.23cm}$\boldsymbol{\Omega}$}$.

%
%

\section{Constrained entropy maximizations \label{SectMaxEnt}}

\noindent
Of course, entropies are extremely important tools 
in information theory, as well as in the adjacent fields of 
statistics, machine learning, artificial intelligence, 
signal processing and pattern recognition. As a consequence of our
considerations in the above Subsection \ref{Sub Contruction Princ Power case}
in combination with the solved cases in Section \ref{Sect Cases}, 
we can also apply our BS method
to the constrained optimization of a wide range of entropies and closely 
related diversity indices. To begin with,
let us fix any
$(\gamma , \mathbf{Q}) \in 
(\widetilde{\Gamma}\backslash]1,2[) \times \widetilde{\mathcal{M}}_{2}$
(cf. Lemma \ref{Lemma Indent Rate finite case_new}(a))
with $A := \sum_{k=1}^{K} q_{k} >0$.
Moreover, we take $\mathds{P} := \mathds{P}^{unif} := 
(\frac{1}{K}, \ldots, \frac{1}{K})$ to be the probability vector 
of frequencies of the uniform distribution on $\{1, \ldots, K\}$.
Then, for $\gamma \in \, ]-\infty,0[ \ \cup \ ]0,1[ \ \cup \ [ \ 2,\infty[$
one gets
$H_{\gamma}(\mathbf{Q},\mathds{P}^{unif}) = 
K^{\gamma -1} \cdot \sum_{k=1}^{K} q_{k}^{\gamma}$.
One can rewrite $K^{1- \gamma} \cdot H_{\gamma}(\mathbf{Q},\mathds{P}^{unif})
= \sum_{k=1}^{K} q_{k}^{\gamma}$; 
the latter is sometimes called \textit{heterogeneity index of type $\gamma$}, see e.g.
\cite{VanDerLub:86}, with $\gamma =2$ being the \textit{Simpson-Herfindahl index}
which is also known as \textit{index of coincidence} (cf. 
\cite{Har:01} and its generalization in \cite{Har:08}).
Alternatively, $\sum_{k=1}^{K} q_{k}^{\gamma}$ is also called 
Onicescu\textquoteright s information energy in case of $\gamma=2$ 
(cf. \cite{Oni:66}, see also \cite{Par:91} 
for comprehensive investigations) and in general
\textit{information energy of order $\gamma$} (cf. \cite{The:77}, 
see also e.g. \cite{Par:86}); 
for exemplary applications 
the reader may take (discretized versions of) 
e.g.~\cite{Liu4:15}--\cite{Rong:20}.
In some other literature (see e.g. 
\cite{Clark:20}), $\sum_{k=1}^{K} q_{k}^{\gamma}$
is alternatively called \textit{Golomb\textquoteright s \cite{Gol:66} 
information generating function (of a probability distribution $\mathds{Q}$)};
yet another name is  \textit{generalized information potential}
and for $\gamma=2$ \textit{information potential}
(cf. e.g. \cite{Pri:10}, \cite{Acu:21}).
From the above investigations, we obtain that $\sum_{k=1}^{K} q_{k}^{\gamma}$
is BS-minimizable on
$\mathbf{\Omega} = A \cdot \textrm{$\boldsymbol{\Omega}$\hspace{-0.23cm}$\boldsymbol{\Omega}$}$
for $\gamma <0$ and 
$\gamma \geq 2$,
and BS-maximizable on
$\mathbf{\Omega} = A \cdot \textrm{$\boldsymbol{\Omega}$\hspace{-0.23cm}$\boldsymbol{\Omega}$}
$ for $\gamma \in ]0,1[$.
More generally, by employing \eqref{fo.hgamma1} and \eqref{fo.hgamma2},
for the class of entropies (diversity indices)
\begin{eqnarray}
 \mathcal{E}_{\gamma,c_{1},c_{2},c_{3}}(\mathbf{Q}) &:=&
h_{c_{1},c_{2},c_{3}}\left(\sum_{k=1}^{K} q_{k}^{\gamma}\right) = 
c_{1} \cdot \bigg(\bigg(\sum_{k=1}^{K} q_{k}^{\gamma}\bigg)^{c_{2}} - c_{3} \bigg)
\nonumber \\
&=&  
c_{1} \cdot \left(K^{c_{2} \cdot (1- \gamma)} \cdot 
H_{\gamma}(\mathbf{Q},\mathds{P}^{unif})^{c_{2}} - c_{3} \right)
, 
\qquad c_{1}, c_{2} \in \mathbb{R}\backslash\{0\}, \, 
c_{3} \in \mathbb{R}, 
\label{fo.genentropy1}\\
\mathcal{E}_{c_{4},f}^{R}(\mathbf{Q}) &:=& 
h_{c_{4},f}^{R}\bigg(\sum_{k=1}^{K} q_{k}^{\gamma}\bigg) 
= \frac{c_{4}}{f^{\prime}(0)} \cdot \log\bigg(\sum_{k=1}^{K} q_{k}^{\gamma}\bigg),
\nonumber \\
&=& \frac{c_{4}}{f^{\prime}(0)} \cdot 
\Big( \log\left(
H_{\gamma}(\mathbf{Q},\mathds{P}^{unif})
\right) +
(1- \gamma) \cdot \log(K)
\Big),
\qquad c_{4} \in \mathbb{R}\backslash\{0\}, \, 
\label{fo.genentropy2}
\end{eqnarray}
(which is similar to the entropy-class of Morales et al. \cite{Mor:96}
who use a different, more restrictive parametrization
and probability distributions $\mathds{Q}$),
one gets the following extremum-behaviour:

\begin{itemize}

\item $\mathcal{E}_{\gamma,c_{1},c_{2},c_{3}}(\mathbf{Q})$
is BS-minimziable if $\gamma <0$ and $c_{1} \cdot c_{2} >0$;

\item $\mathcal{E}_{\gamma,c_{1},c_{2},c_{3}}(\mathbf{Q})$
is BS-minimizable if $\gamma \geq 2$ and $c_{1} \cdot c_{2} >0$;

\item $\mathcal{E}_{\gamma,c_{1},c_{2},c_{3}}(\mathbf{Q})$
is BS-minimizable if $\gamma \in \, ]0,1[$ and $c_{1} \cdot c_{2} < 0$;

\item $\mathcal{E}_{\gamma,c_{1},c_{2},c_{3}}(\mathbf{Q})$
is BS-maximizable if $\gamma <0$ and $c_{1} \cdot c_{2} < 0$;

\item $\mathcal{E}_{\gamma,c_{1},c_{2},c_{3}}(\mathbf{Q})$
is BS-maximizable if $\gamma \geq 2$ and $c_{1} \cdot c_{2} < 0$;

\item $\mathcal{E}_{\gamma,c_{1},c_{2},c_{3}}(\mathbf{Q})$
is BS-maximizable if $\gamma \in \, ]0,1[$ and $c_{1} \cdot c_{2} > 0$;

\item $\mathcal{E}_{c_{4},f}^{R}(\mathbf{Q})$
is BS-minimizable if $\gamma <0$ and  $\frac{c_{4}}{f^{\prime}(0)} > 0$;

\item $\mathcal{E}_{c_{4},f}^{R}(\mathbf{Q})$
is BS-minimizable if $\gamma \geq 2$
and  $\frac{c_{4}}{f^{\prime}(0)} > 0$;

\item $\mathcal{E}_{c_{4},f}^{R}(\mathbf{Q})$
is BS-minimizable if $\gamma \in \, ]0,1[$ and  $\frac{c_{4}}{f^{\prime}(0)} < 0$;

\item $\mathcal{E}_{c_{4},f}^{R}(\mathbf{Q})$
is BS-maximizable if $\gamma <0$ and  $\frac{c_{4}}{f^{\prime}(0)} < 0$;

\item $\mathcal{E}_{c_{4},f}^{R}(\mathbf{Q})$
is BS-maximizable if $\gamma \geq 2$
and  $\frac{c_{4}}{f^{\prime}(0)} < 0$;

\item $\mathcal{E}_{c_{4},f}^{R}(\mathbf{Q})$
is BS-maximizable if $\gamma \in \, ]0,1[$ and  $\frac{c_{4}}{f^{\prime}(0)} > 0$.

\end{itemize}

\vspace{0.2cm}
\noindent
From this, one can deduce that our new BS method works for the
constrained minimization/maximization of the 
following well-known, prominently used
measures of entropy respectively measures of diversity, and beyond:

\begin{enumerate}

\item[(E1)] 
$c_{1}=1$, $c_{2} = \frac{1}{\gamma}$, $c_{3}=0$:
the Euclidean \textit{$\gamma-$norm} (also known as \textit{$\gamma-$norm heterogeneity index}, 
see e.g. \cite{VanDerLub:86})
$|| \mathbf{Q} ||_{\gamma} := \Big(\sum_{k=1}^{K} q_{k}^{\gamma}\Big)^{1/\gamma}
= K^{(1- \gamma)/\gamma} \cdot \Big(H_{\gamma}(\mathbf{Q},\mathds{P}^{unif})\Big)^{1/\gamma}$
is BS-minimizable on
$\mathbf{\Omega} = A \cdot \textrm{$\boldsymbol{\Omega}$\hspace{-0.23cm}$\boldsymbol{\Omega}$}$ 
for $\gamma \geq 2$, 
and BS-maximizable on
$\mathbf{\Omega} = A \cdot \textrm{$\boldsymbol{\Omega}$\hspace{-0.23cm}$\boldsymbol{\Omega}$}$
for $\gamma <0$ and $\gamma \in \, ]0,1[$ (note that $|| \mathbf{Q} ||_{1} =A$) ;\\
similarly, the \textit{$\gamma-$mean heterogeneity index} 
(see e.g. \cite{VanDerLub:86}, 
as well as \cite{Jost:06} for its interpretation as 
\textquotedblleft  effective number of species\textquotedblright\ 
respectively as 
\textquotedblleft  numbers equivalent\textquotedblright)
given by 
$\mathcal{E}^{HI}(\mathbf{Q}) := \Big(\sum_{k=1}^{K} q_{k}^{\gamma}\Big)^{1/(\gamma-1)}
= \frac{1}{K} \cdot \Big(H_{\gamma}(\mathbf{Q},\mathds{P}^{unif})\Big)^{1/(\gamma-1)}$
is BS-minimizable on
$\mathbf{\Omega} = A \cdot \textrm{$\boldsymbol{\Omega}$\hspace{-0.23cm}$\boldsymbol{\Omega}$}$ for 
$\gamma \in \, ]0,1[$ and $\gamma \geq 2$, and BS-maximizable on
$\mathbf{\Omega} = A \cdot \textrm{$\boldsymbol{\Omega}$\hspace{-0.23cm}$\boldsymbol{\Omega}$}$
for $\gamma <0$.
Alternatively, $\mathcal{E}^{HI}(\mathbf{Q})$
is also called \textit{($\gamma-$order) Hill diversity index
or Hill number} \cite{Hil:73}, respectively 
\textit{($\gamma-$order) Hannah-Kay index} \cite{Hann:77}, respectively 
\textit{($\gamma-$order) Renyi heterogeneity}
(cf. \cite{Nun:20}), respectively \textit{($\gamma-$order) exponential Renyi entropy
or exponential entropy} (cf. \cite{Camp:66})
since it is equal to $\exp(\mathcal{E}^{gR}(\mathbf{Q}))$ (cf. (E6) below).
The $\gamma-$mean heterogeneity index (under one of the above-mentioned namings)
was recently employed e.g. 
by~\cite{Grei:15}--\cite{Lass:21}. 

\item[(E2)]  
$c_{1}=\frac{1}{2^{1-\gamma}-1}$, $c_{2} = 1$, $c_{3}=1$:
the entropy
\begin{equation}
\mathcal{E}^{gHC}(\mathbf{Q}) := 
\frac{1}{2^{1-\gamma}-1} \cdot \left(\sum_{k=1}^{K} q_{k}^{\gamma} - 1 \right)
= 
\frac{1}{2^{1-\gamma}-1} \cdot \left(K^{1- \gamma} \cdot 
H_{\gamma}(\mathbf{Q},\mathds{P}^{unif}) - 1 \right)
\label{fo.E14}
\end{equation}
is BS-minimizable on
$\mathbf{\Omega} = A \cdot \textrm{$\boldsymbol{\Omega}$\hspace{-0.23cm}$\boldsymbol{\Omega}$}$
for $\gamma <0$, and BS-maximizable on
$\mathbf{\Omega} = A \cdot \textrm{$\boldsymbol{\Omega}$\hspace{-0.23cm}$\boldsymbol{\Omega}$}$
for $\gamma \in \, ]0,1[$ and $\gamma \geq 2$;
the special subcase $A=1$ in \eqref{fo.E14} (and thus, $\mathbf{Q} =\mathds{Q}$ is
a probability vector) corresponds to the  
\textit{$\gamma-$order entropy of Havrda-Charvat} \cite{Hav:67} (also called 
non-additive \textit{$\gamma-$order Tsallis entropy} \cite{Tsa:88} in statistical physics)
where the special case $\gamma=2$ is (a multiple of) 
\textit{Vajda\textquoteright s quadratic entropy}
\cite{Vaj:89} and \textit{Ahlswede\textquoteright s identification entropy} \cite{Ahl:06a}
(see also \cite{Ahl:06b}).
Some exemplary (relatively) recent studies and applications
of $\mathcal{E}^{gHC}(\mathds{Q})$
appear e.g. in~\cite{Sew:20},\cite{Liu4:15},\cite{Rong:20},\cite{Pet:09}--\cite{Rame:21}.
For $\gamma=2$, a directly connected quantity is the  
\textit{measure of concentration} 
(cf. e.g. \cite{DeWet:20})
$\mathcal{E}^{gMC}(\mathds{Q}) : = 1- \frac{1}{K} - \mathcal{E}^{gHC}(\mathds{Q})
= \sum_{k=1}^{K} \left(q_{k} - \frac{1}{K} \right)^{2}$ which (up to a multiple) was introduced by
\cite{Bru:99} as an appropriate measure of information
for quantum experiments.

\item[(E3)] 
$\gamma := \frac{1}{\widetilde{\gamma}}$, $c_{1}=\frac{1}{\widetilde{\gamma}-1}$, 
$c_{2} = \widetilde{\gamma}$, $c_{3}=1$: the entropy
\begin{equation}
\mathcal{E}^{gA}(\mathbf{Q}) :=
\frac{1}{\widetilde{\gamma}-1} \cdot \left(\left(\sum_{k=1}^{K} 
q_{k}^{1/\widetilde{\gamma}}\right)^{\widetilde{\gamma}} - 1 \right) = 
\frac{1}{\widetilde{\gamma}-1} \cdot \left(K^{\widetilde{\gamma} -1} \cdot 
H_{1/\widetilde{\gamma}}(\mathbf{Q},\mathds{P}^{unif})^{\widetilde{\gamma}} - 1 \right)
\label{fo.E15}
\end{equation}
is BS-minimizable on
$\mathbf{\Omega} = A \cdot \textrm{$\boldsymbol{\Omega}$\hspace{-0.23cm}$\boldsymbol{\Omega}$}$
for $\widetilde{\gamma} <0$,
and BS-maximizable on
$\mathbf{\Omega} = A \cdot \textrm{$\boldsymbol{\Omega}$\hspace{-0.23cm}$\boldsymbol{\Omega}$}$ for 
$\widetilde{\gamma} \in \, ]0,\frac{1}{2}]$ and $\widetilde{\gamma} > 1$;
the special subcase $A=1$ in \eqref{fo.E15} (and thus, $\mathbf{Q} =\mathds{Q}$ is
a probability vector) corresponds to the 
\textit{$\widetilde{\gamma}-$order entropy of Arimoto} \cite{Ari:71}.

\item[(E4)] $s \in \mathbb{R}\backslash\{1\}$, $c_{1}=\frac{1}{1-s}$, $c_{2} = \frac{1-s}{1-\gamma}$, $c_{3}=1$:
the entropy
\begin{equation}
\mathcal{E}^{gSM1}(\mathbf{Q})
:= \frac{1}{1-s} \cdot \left(\left(\sum_{k=1}^{K} q_{k}^{\gamma}\right)^{(1-s)/(1-\gamma)} 
- 1 \right) = 
\frac{1}{1-s} \cdot \left(K^{1- s} \cdot 
H_{\gamma}(\mathbf{Q},\mathds{P}^{unif})^{(1-s)/(1-\gamma)} - 1 \right)
\label{fo.E16}
\end{equation}
is BS-minimizable on
$\mathbf{\Omega} = A \cdot \textrm{$\boldsymbol{\Omega}$\hspace{-0.23cm}$\boldsymbol{\Omega}$}$
for $\gamma < 0$ and BS-maximizable 
on $\mathbf{\Omega} = A \cdot \textrm{$\boldsymbol{\Omega}$\hspace{-0.23cm}$\boldsymbol{\Omega}$}$
for $\gamma \in \, ]0,1[$ and $\gamma \geq 2$;
the special subcase $A=1$ in \eqref{fo.E16} (and thus, $\mathbf{Q} =\mathds{Q}$ is
a probability vector) corresponds to the 
\textit{entropy of order $\gamma$ and degree $s$ of Sharma \& Mittal}
\cite{Shar:75} in the scaling of e.g. \cite{Sal:93}.

\item[(E5)] $s \in \mathbb{R}\backslash\{0\}$, 
$\gamma = s+1$, $c_{1}= - \frac{1}{s}$, $c_{2} = 1$, $c_{3}=1$:
the diversity index
\begin{equation}
\mathcal{E}^{gPT}(\mathbf{Q})
:= - \frac{1}{s} \cdot \left(\sum_{k=1}^{K} q_{k}^{s+1} - 1 \right) = 
- \frac{1}{s} \cdot \left(K^{- s} \cdot 
H_{s+1}(\mathbf{Q},\mathds{P}^{unif}) - 1 \right)
\label{fo.E17}
\end{equation}
is BS-minimizable on
$\mathbf{\Omega} = A \cdot \textrm{$\boldsymbol{\Omega}$\hspace{-0.23cm}$\boldsymbol{\Omega}$}$
for $s < -1$ and BS-maximizable on
$\mathbf{\Omega} = A \cdot \textrm{$\boldsymbol{\Omega}$\hspace{-0.23cm}$\boldsymbol{\Omega}$}$
for $s \in \, ]-1,0[$ and $s \geq 1$;
the special subcase $A=1$ in \eqref{fo.E17} (and thus, $\mathbf{Q} =\mathds{Q}$ is
a probability vector) corresponds to the
\textit{diversity index of degree $s$ of Patil \& Taillie} \cite{Pati:82}; 
the case $s=1$ for probability measures $\mathbf{Q}=\mathds{Q}$ gives the 
well-known \textit{Gini-Simpson diversity index}.

\item[(E6)] $c_{4}=\frac{1}{1-\gamma}$, $f(z) =z$: the entropy
\begin{equation}
\mathcal{E}^{gR}(\mathbf{Q})
:= \frac{1}{1-\gamma} \cdot \log\left(\sum_{k=1}^{K} q_{k}^{\gamma}\right)
= \frac{1}{1-\gamma} \cdot 
\Big( \log\left(
H_{\gamma}(\mathbf{Q},\mathds{P}^{unif})
\right) + (1- \gamma) \cdot \log(K) \Big)
= \frac{\log 2}{1-\gamma} \cdot \log_{2}\left(\sum_{k=1}^{K} q_{k}^{\gamma}\right)
\label{fo.E18}
\end{equation}
is BS-minimizable on
$\mathbf{\Omega} = A \cdot \textrm{$\boldsymbol{\Omega}$\hspace{-0.23cm}$\boldsymbol{\Omega}$}$ 
for $\gamma <0$, 
and BS-maximizable on
$\mathbf{\Omega} = A \cdot \textrm{$\boldsymbol{\Omega}$\hspace{-0.23cm}$\boldsymbol{\Omega}$}$
for $\gamma \in \, ]0,1[$ and $\gamma \geq 2$;
the special subcase $A=1$ in \eqref{fo.E18} (and thus, $\mathbf{Q} = \mathds{Q}$ is
a probability vector) corresponds to the prominent 
(additive) \textit{$\gamma-$order Renyi entropy} \cite{Ren:61}.
As well known, there is a vast literature on Renyi
entropies $\mathcal{E}^{gR}(\mathds{Q})$.
Some exemplary (mostly recent) studies and applications
appear e.g. 
in~\cite{Sun:07},\cite{Bun:14},\cite{Kum:19},\cite{Sew:20},\cite{Liu4:15},\cite{Rong:20},\cite{Erg:20},\cite{Nat:75}--\cite{Pan:21}.

\end{enumerate}

\begin{remark}
\label{matrix versions}
(i) For Renyi entropies there are also matrix versions
$ \mathcal{E}^{gR}(X) := 
\frac{1}{1-\gamma} \cdot \log\left(\sum_{i=1}^{K_{1}} 
\sum_{j=1}^{K_{2}} x_{ij}^{\gamma}\right)$
where $X:=(x_{ij})_{i=1,\ldots,K_{1}}^{j=1,\ldots,K_{2}}$ 
is a $K_{1} \times K_{2}-$matrix whose elements $x_{ij}$ are (say) strictly positive
and sum up to $A$. Such a setup with $A=1$ is e.g. used in \textit{time-frequency analyses
of signals} where the $i$\textquoteright s correspond to discrete time points, 
the $j$\textquoteright s to discrete frequencies, and $x_{ij}$
to the probability that $(i,j)$ occurs; see e.g.
\cite{Pop:17}. Another line of application is to use as $X$ the normalized  
communicability matrix of a
directed network (respectively the upper triangular part of $X$ in case of an unweighted 
and undirected network). 
Of course, the matrix version $\mathcal{E}^{gR}(X)$ can be easily 
and equivalently rewritten in our vector version $\mathcal{E}^{gR}(\mathbf{Q})$
by setting $\mathbf{Q} := (q_{1}, \ldots, q_{K_{1} \cdot K_{2}})$
such that $x_{ij} = q_{(i-1)\cdot K_{2} +j}$ ($i=1,\ldots,K_{1}$, $j=1,\ldots,K_{2}$)
and hence $K:= K_{1} \cdot K_{2}$;
accordingly, we can apply our BS method.\\
(ii) The latter conversion works analogously also for 
matrix versions of all the other entropies, divergences, etc. of this paper;
more flexible versions where $i \in \{1,\ldots,K_{1}\}$, $j \in J_{i}$
for some $J_{i} \subseteq \{1,\ldots,K_{2}\}$
as well as multidimensional-array/tensor versions 
can be transformed in a similar book-keeping manner, too.
For instance, within the above-mentioned framework of  
unweighted and undirected networks, \cite{Che3:18} and
\cite{Shi:20} employ communicability matrix versions of 
the Shannon entropy and the Jensen-Shannon divergence (JSD);
see also \cite{Bag:19} for similar network applications of the
JSD. Moreover, \cite{Jena:21} use 
\textquotedblleft  3D versions\textquotedblright\  of Tsallis entropies
for brain magnetic resonance (MR) image segmentation.

\end{remark}

\vspace{0.3cm}

\begin{remark} 
All the above cases which are BS-maximizable can be interpreted as
bare-simulation approach to the solution of
\textit{generalized maximum entropy problems on $\mathbf{\Omega} = A \cdot \textrm{$\boldsymbol{\Omega}$\hspace{-0.23cm}$\boldsymbol{\Omega}$}
$}.\\ 
\end{remark}

\begin{remark}
\label{rem.artificial}
(i) If (all) the above- and below-mentioned entropies
are used for probability vectors 
$\mathds{Q} \in \mathbb{S}^{K}$ --- i.e. 
one employs $\mathcal{E}(\mathds{Q})$ ---
then typically the components $q_{k}$
of $\mathds{Q}$ represent a genuine 
probability mass (frequency) 
$q_{k} = \mathbb{\Pi}[\{ d_{k} \}]$
of some data point (state) $d_{k}$.
However, $\mathds{Q} \in \mathbb{S}^{K}$ may alternatively be \textit{artificially}
generated. For instance, for the purpose of fault detections of mechanical drives,
\cite{Bos:12} use Renyi entropies where the
$q_{k}$\textquoteright s are
normalized squared energy-describing coefficients of the
wavelet packet transform of measured vibration records.
Another exemplary \textquotedblleft  artificial\textquotedblright\ 
operation is concatenation.\\
(ii) An analogous statement holds for the employment of
(all) the above- and below-mentioned divergences $D(\mathds{Q},\mathds{P})$ --- and their transformations ---
between genuine respectively
artificially generated probability vectors 
$\mathds{Q}, \mathds{P}  \in \mathbb{S}^{K}$.

\end{remark}

\vspace{0.2cm}
\noindent
To proceed with our general investigations, let us mention that
the remaining parameter cases $\gamma =0$ and $\gamma =1$ can be treated analogously. For the sake
of brevity, we only deal with the latter. For this,
let $\mathbf{Q} \in A \cdot \mathbb{S}^{K}$
with $A := \sum_{k=1}^{K} q_{k} >0$
and $\mathds{P} := \mathds{P}^{unif}$.
Clearly, $I(\mathbf{Q},\mathds{P}^{unif}) - A \cdot \log K = 
\sum_{k=1}^{K} q_{k} \cdot \log(q_{k})$; 
thus the latter is BS-minimizable on
$\mathbf{\Omega} = A \cdot \textrm{$\boldsymbol{\Omega}$\hspace{-0.23cm}$\boldsymbol{\Omega}$}$. 
More generally, for any continuous strictly increasing (respectively strictly decreasing)
function $h_{1} : \, ] - 1 - A \cdot \log K, \infty[ \  \mapsto \mathbb{R}$,
the quantity $h_{1}\Big(\sum_{k=1}^{K} q_{k} \cdot \log(q_{k})\Big)$ 
is BS-minimizable on
$\mathbf{\Omega} = A \cdot \textrm{$\boldsymbol{\Omega}$\hspace{-0.23cm}$\boldsymbol{\Omega}$}$ 
(respectively BS-maximizable on
$\mathbf{\Omega} = A \cdot \textrm{$\boldsymbol{\Omega}$\hspace{-0.23cm}$\boldsymbol{\Omega}$}$). Important special cases are:

\begin{enumerate}

\item[(E7)] $h_{1}(y) := h_{1}^{-Id}(y) = -y$: the entropy
\begin{equation}
\mathcal{E}^{Sh}(\mathbf{Q}) := 
h_{1}^{-Id}\Big(\sum_{k=1}^{K} q_{k} \cdot \log(q_{k})\Big)
= - \sum_{k=1}^{K} q_{k} \cdot \log(q_{k}) 
\label{fo.E19}
\end{equation}
is BS-maximizable on
$\mathbf{\Omega} = A \cdot \textrm{$\boldsymbol{\Omega}$\hspace{-0.23cm}$\boldsymbol{\Omega}$}$; 
the special subcase $A=1$ in \eqref{fo.E19} (and thus, $\mathbf{Q} = \mathds{Q}$ is
a probability vector) corresponds to the omnipresent \textit{Shannon entropy} \cite{Sha:48};
hence, by our bare-simulation approach we can particularly tackle \textit{maximum entropy problems}
on almost arbitrary sets 
$\textrm{$\boldsymbol{\Omega}$\hspace{-0.23cm}$\boldsymbol{\Omega}$}$
of probability vectors. Analogously, we can treat  
$\frac{1}{\log(K)} \cdot \mathcal{E}^{Sh}(\mathds{Q})$ which is 
called \textit{Pielou\textquoteright s evenness index} \cite{Pie:66},
and $1-\frac{1}{\log(K)} \cdot \mathcal{E}^{Sh}(\mathds{Q}) \in [0,1]$ which is 
sometimes used as
\textit{clonality (clonotype diversity) index} (see e.g. \cite{Gab:19} and 
\cite{Bash:19} (with supplementary private communication)).
As a further example for Remark \ref{rem.artificial},
\cite{Lyu:21} uses $q_{k}$\textquoteright s which are
normalized squared coefficients of an orthogonal wavelet decomposition,
and accordingly, $\frac{1}{\log(K)} \cdot \mathcal{E}^{Sh}(\mathds{Q})$
can be interpreted as the entropy of the distribution of
energy of oscillations at various frequency and time scales.
Some further exemplary 
studies and applications
of the maximization of $\mathcal{E}^{Sh}(\mathds{Q})$
--- aside from the vast physics literature  ---
appear e.g. 
in~\cite{DeS:01}--\cite{Han:21}.

\item[(E8)] $s \in \, ]0,1[ \, \cup \, ]1,\infty[$, 
$h_{1}(y) := h_{s}^{SM2}(y) :=  \frac{e^{(s-1) \cdot y} -1}{1-s}$ (cf. \eqref{fo.hKL1})
with $y \in \mathbb{R}$: the entropy
\begin{equation}
\mathcal{E}^{SM2}(\mathbf{Q}) := 
h_{s}^{SM2}\Big(\sum_{k=1}^{K} q_{k} \cdot \log(q_{k})\Big)
=  \frac{1}{1-s} \cdot \Big( \exp\Big\{
(s-1) \cdot \sum_{k=1}^{K} q_{k} \cdot \log(q_{k})\Big\} -1 \Big)  
\label{fo.E20}
\end{equation}
is BS-maximizable on
$\mathbf{\Omega} = A \cdot \textrm{$\boldsymbol{\Omega}$\hspace{-0.23cm}$\boldsymbol{\Omega}$}$; 
the special subcase $A=1$ in \eqref{fo.E20} (and thus, $\mathbf{Q} = \mathds{Q}$ is
a probability vector) corresponds to the (second type) \textit{entropy
of Sharma \& Mittal} \cite{Shar:75} in the scaling of e.g. \cite{Par:06} (p.20). 

\end{enumerate} 

%
%

\section{Further important deterministic optimization problems 
\label{SectFurther}}

\noindent
In order to support the importance and universality 
of information-theoretic methods for other research fields, 
let us show how our BS method can be used to tackle
--- in a new way --- important classes of (say) deterministic optimization problems,
which are not directly connected to divergences \textit{at first sight}.

\vspace{0.2cm}
\noindent
For instance, as a consequence of the above Subsection \ref{Sub Contruction Princ Power case}
in combination with the solved cases in Section \ref{Sect Cases},
by retransformation we can even apply our BS method to optimizations of nonnegative \textit{linear} objective functions
with constraint sets on Euclidean $\gamma$-norm spheres.
Indeed, for nonnegative $\breve{\mathbf{Q}} := (\breve{q}_{1}, \ldots, \breve{q}_{K})$ 
and $\breve{\mathbf{P}} := (\breve{p}_{1}, \ldots, \breve{p}_{K})$ one can rewrite
their scalar product as $\gamma-$order Hellinger integrals
\begin{eqnarray}
& & \hspace{-1.3cm}
\sum_{k=1}^{K} \breve{q}_{k} \cdot \breve{p}_{k}
\ = \ c_{1} \cdot \sum_{k=1}^{K} q_{k}^{\gamma} \cdot p_{k}^{1-\gamma}  
\ = \ c_{1} \cdot H_{\gamma}(\mathbf{Q},\mathds{P}) \, \hspace{0.5cm} \textrm{where}
\label{scalarprod1}
\\
& & \hspace{-1.4cm}
\textrm{
$\gamma \in \, ]0,1[ \, \cup \, [2,\infty[$ \,  if \, $\breve{\mathbf{Q}} \in [0,\infty[^{K}$, 
$\breve{\mathbf{P}} \in \, ]0,\infty[^{K}$
\hspace{0.4cm} respectively \hspace{0.4cm}
$\gamma \in \, ]-\infty,0[$ \,  if \, $\breve{\mathbf{Q}} \in \, ]0,\infty[^{K}$,
$\breve{\mathbf{P}} \in \, ]0,\infty[^{K}$,
}
\nonumber \\
& & \hspace{-1.3cm} q_{k} \ := \ \breve{q}_{k}^{1/\gamma} , 
\quad p_{k} \ := \ \frac{\breve{p}_{k}^{1/(1-\gamma)}}{\sum_{i=1}^{K} \breve{p}_{i}^{1/(1-\gamma)}} ,
\quad c_{1}:= \Big( \sum_{i=1}^{K} \breve{p}_{i}^{1/(1-\gamma)} \Big)^{1-\gamma} = :
|| \breve{\mathbf{P}} ||_{1/(1-\gamma)} \ .
\label{scalarprod3}
\end{eqnarray}
The required constraint $\sum_{k=1}^{K} q_{k} = A >0$ retransforms to $|| \breve{\mathbf{Q}} ||_{1/\gamma}=A^{\gamma}$
and thus, $\breve{\mathbf{Q}}$ must lie on (the positive/nonnegative part of) the
$\frac{1}{\gamma}-$norm-sphere $\partial B_{1/\gamma}(0,A^{\gamma})$
around the origin with radius $A^{\gamma}$.
Accordingly, for $\gamma \in \, [2,\infty[$ we have
\begin{equation}
\inf_{\breve{\mathbf{Q}} \in \breve{\mathbf{\Omega}}} \, 
\sum_{k=1}^{K} \breve{q}_{k} \cdot \breve{p}_{k}
\ = \ c_{1} \cdot 
\inf_{\mathbf{Q}\in A \cdot \textrm{$\boldsymbol{\Omega}$\hspace{-0.19cm}$\boldsymbol{\Omega}$} }
H_{\gamma}(\mathbf{Q},\mathds{P})
\label{linHell1}
\end{equation}
and we can apply \eqref{brostu3:fo.norweiemp6.new1b} of 
Proposition \ref{brostu3:cor.powdivexpl3}(a)
respectively 
Proposition \ref{brostu3:cor.powdivexplNOR}(a) below
\footnote{
here and analogously henceforth, by this we mean 
the condition \eqref{brostu3:fo.norweiemp6.new1b}
as it appears in the 
Proposition \ref{brostu3:cor.powdivexpl3}(a)
respectively 
Proposition \ref{brostu3:cor.powdivexplNOR}(a)
},
as long as the original constraint set 
$\breve{\mathbf{\Omega}} \in \partial B_{1/\gamma}(0,A^{\gamma}) \, \cap \, [0,\infty[^{K}$
transforms (via $q_{k} = \breve{q}_{k}^{1/\gamma}$) into a constraint
set $A \cdot \textrm{$\boldsymbol{\Omega}$\hspace{-0.23cm}$\boldsymbol{\Omega}$}$
which satisfies the regularity assumption \eqref{regularity} in the relative topology
(as a side remark, notice that $int(\partial B_{1/\gamma}(0,A^{\gamma})) = \emptyset$ 
in the full topology).
For the case $\gamma \in \, ]-\infty,0[$ we also have \eqref{linHell1}
and apply \eqref{brostu3:fo.norweiemp6.new1b} of Proposition \ref{brostu3:cor.powdivexpl2}(a) 
for any original constraint set 
$\breve{\mathbf{\Omega}} \in \partial B_{1/\gamma}(0,A^{\gamma}) \,  \cap \, ]0,\infty[^{K}$
which transforms into $A \cdot \textrm{$\boldsymbol{\Omega}$\hspace{-0.23cm}$\boldsymbol{\Omega}$}$
satisfying \eqref{regularity} in the relative topology.
In contrast, for the case $\gamma \in \, ]0,1[$ we get 
\begin{equation}
\sup_{\breve{\mathbf{Q}} \in \breve{\mathbf{\Omega}}} \, 
\sum_{k=1}^{K} \breve{q}_{k} \cdot \breve{p}_{k}
\ = \ c_{1} \cdot
\sup_{\mathbf{Q}\in A \cdot \textrm{$\boldsymbol{\Omega}$\hspace{-0.19cm}$\boldsymbol{\Omega}$} }
H_{\gamma}(\mathbf{Q},\mathds{P})
\nonumber
\end{equation}
and apply \eqref{brostu3:fo.norweiemp6.new4b} of Proposition \ref{brostu3:cor.powdivexpl1}(a) 
for any original constraint set 
$\breve{\mathbf{\Omega}} \in \partial B_{1/\gamma}(0,A^{\gamma}) \,  \cap \, [0,\infty[^{K}$
which transforms 
into $A \cdot \textrm{$\boldsymbol{\Omega}$\hspace{-0.23cm}$\boldsymbol{\Omega}$}$
satisfying \eqref{regularity} in the relative topology.

\vspace{0.4cm}
\noindent
Analogously to Remark \ref{computingGeneral} in the previous Section \ref{Sect Minimization}
(where we have dealt with constraints sets $\mathbf{\Omega}$ of considerably different 
topological nature than here)
we can also principally tackle all the optimization problems
of Subsection \ref{Sub Contruction Princ Power case},
the Sections \ref{SectRenyi},\ref{SectMaxEnt}
and the upper part of this Section \ref{SectFurther}
by basically \textit{only employing a fast and accurate 
--- pseudo, true, natural, quantum  ---
random number generator},
provided that the constraint set  
$\textrm{$A \cdot \boldsymbol{\Omega}$\hspace{-0.23cm}$\boldsymbol{\Omega}$}$
satisfies the mild assumptions \eqref{regularity} (in the relative topology)
and \eqref{def fi wrt Omega}. Recall that $A>0$ (and for $\varphi_{2}$ even 
$A \in \mathbb{R}\backslash\{0\}$) and that
$\mathbf{Q} \in \textrm{$A \cdot \boldsymbol{\Omega}$\hspace{-0.23cm}$\boldsymbol{\Omega}$}$
implies in particular the constraint $\sum_{k=1}^{K} q_{k} = A$.
The regularity assumption
\eqref{regularity} allows for e.g. high-dimensional 
constraint sets 
$\textrm{$A \cdot \boldsymbol{\Omega}$\hspace{-0.23cm}$\boldsymbol{\Omega}$}$
which are \textit{non-convex} and even \textit{highly disconnected}, and for which
other minimization methods (e.g. pure enumeration, gradient or steepest descent methods, etc.) 
may be problematic or intractable.
For example, \eqref{regularity} covers 
kind of \textquotedblleft  $K-$dimensional 
(not necessarily regular)
polka dot pattern type\textquotedblright\ relaxations
$\textrm{$A \cdot \boldsymbol{\Omega}$\hspace{-0.23cm}$\boldsymbol{\Omega}$} := 
\dot{\bigcup}_{i=1}^{N} \mathcal{U}_{i}(\mathbf{Q}_{i}^{dis})$ of finite 
discrete constraint sets 
$\textrm{$A \cdot \boldsymbol{\Omega}$\hspace{-0.23cm}$\boldsymbol{\Omega}^{dis}$}
:= \{\mathbf{Q}_{1}^{dis}, \ldots, \mathbf{Q}_{N}^{dis}\}$ 
of high cardinality $N$
(e.g. being exponential or factorial in a large $K$), 
where each $K-$dimensional vector $\mathbf{Q}_{i}^{dis}$ has total-sum-of-components equal to $A$ and
is surrounded by some small (\textquotedblleft  flat\textquotedblright , 
i.e. in the relative topology)
neighborhood $\mathcal{U}_{i}(\mathbf{Q}_{i}^{dis})$.
For the sake of brevity, in the following discussion 
we confine ourselves to the deterministic 
setup (e.g. Proposition \ref{brostu3:cor.powdivexplNOR}(a) rather than (b)) 
which particularly
involves $\mathbb{\Pi }[\cdot]$ (rather than $\mathbb{\Pi}_{X_{1}^{n}}[\cdot]$)
and $\boldsymbol{\xi}_{n}^{w\mathbf{W}}$ 
(rather than $\xi _{n,\mathbf{X}}^{w\mathbf{W}}$).
In such a context, all the optimization problems of Subsection 
\ref{Sub Contruction Princ Power case},
the Sections \ref{SectRenyi},\ref{SectMaxEnt}
and the upper part of this Section \ref{SectFurther} ---
subsumed as
(cf. \eqref{brostu3:fo.1} to \eqref{brostu3:fo.2b})
\begin{equation}
\inf_{\mathbf{Q}\in A \cdot 
\textrm{$\boldsymbol{\Omega}$\hspace{-0.19cm}$\boldsymbol{\Omega}$} } \,
\Phi(\mathbf{Q})
\qquad \textrm{respectively} \qquad
\sup_{\mathbf{Q}\in A \cdot 
\textrm{$\boldsymbol{\Omega}$\hspace{-0.19cm}$\boldsymbol{\Omega}$} } \,
\Phi(\mathbf{Q}) \quad \textrm{---}
\nonumber
\end{equation}
can be regarded as a 
\textquotedblleft  BS-tractable\textquotedblright\
\textit{relaxations} of the 
nonlinear discrete (e.g. integer, combinatorial) programming problems
\begin{equation}
\inf_{\mathbf{Q}\in A \cdot 
\textrm{$\boldsymbol{\Omega}$\hspace{-0.19cm}$\boldsymbol{\Omega}^{dis}$} } \,
\Phi(\mathbf{Q})
\qquad \textrm{respectively} \qquad
\sup_{\mathbf{Q}\in A \cdot 
\textrm{$\boldsymbol{\Omega}$\hspace{-0.19cm}$\boldsymbol{\Omega}^{dis}$} } \,
\Phi(\mathbf{Q}) \, ;
\nonumber
\end{equation}
as examples take e.g. $\Phi(\mathbf{Q}) = 
c_{1} \cdot \Big( \Big(\sum_{k=1}^{K} q_{k}^{\gamma}\Big)^{c_{2}}  
- c_{3} \Big)$ (with $\gamma \in \mathbb{R}\backslash \{0,1\}$) \,  or \, 
$\Phi(\mathbf{Q}) = \Phi_{\mathds{P}}(\mathbf{Q}) = 
D_{\widetilde{c}\cdot \varphi _{\gamma}}(\mathbf{Q},\mathds{P})$.
For instance, 
$A \cdot \textrm{$\boldsymbol{\Omega}$\hspace{-0.23cm}$\boldsymbol{\Omega}^{dis}$}$ 
may contain only $K-$dimensional vectors $\mathbf{Q}_{i}^{dis}$ ($i=1,\ldots,N$)
whose components stem from a finite set $\mathcal{B}$ of
nonnegative integers and add up to $A$. If $\mathcal{B} = \{0,1\}$, then we can even deal with
nonnegative \textit{linear} objective functions 
$\Phi(\mathbf{Q}) = \sum_{k=1}^{K} \breve{p}_{k} \cdot q_{k}$ 
where $\mathbf{Q} := (q_{1}, \ldots, q_{K})$
with $q_{k} \in \{0,1\}$ and 
$\breve{\mathbf{P}} := (\breve{p}_{1}, \ldots, \breve{p}_{K})$
has components $\breve{p}_{k} >0$ 
which reflect e.g. the cost associated with the $k-$th state.
Indeed, by noticing that $q_{k}^{\gamma}=q_{k}$ for 
$\gamma \in \, ]0,1[ \, \cup \, [2,\infty[$, 
we can employ \eqref{scalarprod1} and \eqref{scalarprod3} to end up with
\begin{eqnarray}
& & \hspace{-1.1cm}
\inf_{\mathbf{Q}\in A \cdot \textrm{$\boldsymbol{\Omega}$\hspace{-0.19cm}$\boldsymbol{\Omega}^{dis}$} } \, 
\sum_{k=1}^{K} q_{k} \cdot \breve{p}_{k}
 = || \breve{\mathbf{P}} ||_{1/(1-\gamma)} \cdot \negthinspace \negthinspace
\inf_{\mathbf{Q}\in A \cdot \textrm{$\boldsymbol{\Omega}$\hspace{-0.19cm}$\boldsymbol{\Omega}^{dis}$} } \, 
\sum_{k=1}^{K} q_{k}^{\gamma} \cdot \left( 
\frac{\breve{p}_{k}^{1/(1-\gamma)}}{\sum_{i=1}^{K} \breve{p}_{i}^{1/(1-\gamma)}}
\right)^{1-\gamma}
\negthinspace \negthinspace \negthinspace
=  || \breve{\mathbf{P}} ||_{1/(1-\gamma)} \cdot \negthinspace \negthinspace
\inf_{\mathbf{Q}\in A \cdot \textrm{$\boldsymbol{\Omega}$\hspace{-0.19cm}$\boldsymbol{\Omega}^{dis}$} }
H_{\gamma}(\mathbf{Q},\mathds{P}), \  \gamma \in  [2,\infty[, \ \ \ \ \ 
\label{linHell3}
\\
& & \hspace{-1.1cm}
\sup_{\mathbf{Q}\in A \cdot \textrm{$\boldsymbol{\Omega}$\hspace{-0.19cm}$\boldsymbol{\Omega}^{dis}$} } \, 
\sum_{k=1}^{K} q_{k} \cdot \breve{p}_{k}
=  || \breve{\mathbf{P}} ||_{1/(1-\gamma)} \cdot 
\sup_{\mathbf{Q}\in A \cdot \textrm{$\boldsymbol{\Omega}$\hspace{-0.19cm}$\boldsymbol{\Omega}^{dis}$} }
H_{\gamma}(\mathbf{Q},\mathds{P}), \ \ \gamma \in \, ]0,1[ \, .
\label{linHell4}
\end{eqnarray}
The corresponding relaxations are
\begin{eqnarray}
& & \hspace{-1.7cm}
\inf_{\mathbf{Q}\in A \cdot \textrm{$\boldsymbol{\Omega}$\hspace{-0.19cm}$\boldsymbol{\Omega}$} } \, 
\sum_{k=1}^{K} q_{k} \cdot \breve{p}_{k}
\ =  \ || \breve{\mathbf{P}} ||_{1/(1-\gamma)} \cdot 
\inf_{\mathbf{Q}\in A \cdot \textrm{$\boldsymbol{\Omega}$\hspace{-0.19cm}$\boldsymbol{\Omega}$} }
H_{\gamma}(\mathbf{Q},\mathds{P}), \ \  \gamma \in \, [2,\infty[ \, ,
\label{linHell5} \\
& & \hspace{-1.7cm}
\sup_{\mathbf{Q}\in A \cdot \textrm{$\boldsymbol{\Omega}$\hspace{-0.19cm}$\boldsymbol{\Omega}$} } \, 
\sum_{k=1}^{K} q_{k} \cdot \breve{p}_{k}
\ =  \ || \breve{\mathbf{P}} ||_{1/(1-\gamma)} \cdot 
\sup_{\mathbf{Q}\in A \cdot \textrm{$\boldsymbol{\Omega}$\hspace{-0.19cm}$\boldsymbol{\Omega}$} }
H_{\gamma}(\mathbf{Q},\mathds{P}), \ \ \gamma \in \, ]0,1[ \, ;
\label{linHell6}
\end{eqnarray}
for \eqref{linHell5} we can apply \eqref{brostu3:fo.norweiemp6.new1b} of 
Proposition \ref{brostu3:cor.powdivexpl3}(a) respectively 
Proposition \ref{brostu3:cor.powdivexplNOR}(a),
whereas for \eqref{linHell6}
we apply \eqref{brostu3:fo.norweiemp6.new4b} of Proposition \ref{brostu3:cor.powdivexpl1}(a)
---  as long as the relaxation constraint set 
$A \cdot \textrm{$\boldsymbol{\Omega}$\hspace{-0.23cm}$\boldsymbol{\Omega}$}$
satisfies \eqref{regularity} in the relative topology.
For the sake of illustration, let us consider a sum-minimization-type 
\textit{linear assignment problem} 
with side constraints 
(for a comprehensive book on assignment problems see e.g. \cite{Burk:09}).
Suppose that there are $K$ individuals (people, machines, etc.) to carry out $K$ 
tasks (jobs, etc.). Each individual is assigned to carry out exactly one task.
There is cost $c_{ij} >0$ if individual $i$ is assigned to (i.e., carries out) task $j$.
We want to find the minimum total cost amongst all assignments.
There may be side constraints, 
e.g. each assignment has a value $v_{ij} >0$ and the total value of the assignment 
should be above a pregiven threshold.
As usual, the problem can be formulated with the help of binary variables $x_{ij}$ where
$x_{ij} =1$ if individual $i$ is assigned to task $j$, and
$x_{ij} =0$ otherwise. Accordingly, we want to compute
\begin{eqnarray}
&& \inf_{K \times K-\textrm{matrices} \  x=(x_{ij})} \ 
\sum_{i=1}^{K} \sum_{j=1}^{K}
c_{ij} \cdot x_{ij}
\label{Assign1a}
\\
&&
\textrm{subject to}
\nonumber\\
&& \sum_{j=1}^{K} x_{ij} = 1 \quad \textrm{for all } i \in \{1,\ldots,K\},
\qquad \textrm{(i.e. each individual $i$ does one task),}
\label{Assign1b}\\
&& \sum_{i=1}^{K} x_{ij} = 1 \quad \textrm{for all } j\in \{1,\ldots,K\},
\qquad \textrm{(i.e. each task $j$ is done by one individual),}
\label{Assign1c}
\\
&& x_{ij} \in \{0,1\} \quad \textrm{for all } i \in \{1,\ldots,K\}, \, j \in \{1,\ldots,K\},
\label{Assign1d}
\\
&& \textrm{side (i.e. additional) constraints on $x=(x_{ij})_{i,j=1,\ldots,K}$}.
\label{Assign1e}
\end{eqnarray}
Of course, this can be equivalently rewritten in terms of
$K^{2}-$dimensional vectors as follows: let 
$\mathbf{Q} := (q_{1}, \ldots, q_{K^{2}})$ and 
$\breve{\mathbf{P}} := (\breve{p}_{1}, \ldots, \breve{p}_{K^{2}})$
be such that
$c_{ij}= \breve{p}_{(i-1)\cdot K +j}$ and $x_{ij}= q_{(i-1)\cdot K +j}$
for $i,j \in \{1,\ldots,K\}$ and compute
\begin{eqnarray}
&& 
\inf_{\mathbf{Q}\in K \cdot \textrm{$\boldsymbol{\Omega}$\hspace{-0.19cm}$\boldsymbol{\Omega}^{dis}$} } \, 
\sum_{k=1}^{K^{2}} q_{k} \cdot \breve{p}_{k}
\label{Assign2a}
\\
&&
\textrm{where 
$K \cdot \textrm{$\boldsymbol{\Omega}$\hspace{-0.23cm}$\boldsymbol{\Omega}^{dis}$}
 \subset \mathbb{R}^{K^{2}}$ is the set of all
vectors $\mathbf{Q} = (q_{1}, \ldots, q_{K^{2}})$ which satisfy the constraints
}
\nonumber
\\
&& \sum_{j=1}^{K} q_{(i-1)\cdot K +j} \ = \ 1 \quad \textrm{for all } i \in \{1,\ldots,K\},
\label{Assign2b}
\\
&& \sum_{i=1}^{K} q_{(i-1)\cdot K +j} \ = \ 1 \quad \textrm{for all } j \in \{1,\ldots,K\},
\label{Assign2c}
\\
&& q_{k} \in \{0,1\} \quad \textrm{for all } k \in \{1,\ldots,K^{2} \},
\label{Assign2d}
\\
&& \textrm{side constraints on $\mathbf{Q}$}.
\label{Assign2e}
\end{eqnarray}
As seen above, this can be rewritten as $\gamma-$order Hellinger-integral minimization problem \eqref{linHell3},
with $\gamma \geq 2$. 
We can obtain a \textit{highly disconnected} 
\textquotedblleft  non-void-interior-type\textquotedblright\
relaxation of the binary integer programming problem \eqref{Assign2a} to \eqref{Assign2e}
by replacing \eqref{Assign2d} with 
\begin{equation}
q_{k} \in [0,\varepsilon_{1}] \, \cup \, [1-\varepsilon_{2},1]
\quad \textrm{for all } k \in \{1,\ldots,K^{2} \},
\label{Assign2f}
\end{equation}
for some (possibly arbitrarily) small 
$\varepsilon_{1}, \varepsilon_{2} >0$ with $\varepsilon_{1} + \varepsilon_{2} < 1$.
We denote by $K \cdot \textrm{$\boldsymbol{\Omega}$\hspace{-0.23cm}$\boldsymbol{\Omega}$}$
the outcoming set manifested by the constraints \eqref{Assign2b}, \eqref{Assign2c},
\eqref{Assign2e} and \eqref{Assign2f},
and accordingly we end up with a minimization problem 
of type \eqref{linHell5},
which we can tackle by \eqref{brostu3:fo.norweiemp6.new1b} of 
Proposition \ref{brostu3:cor.powdivexpl3}(a)
respectively 
Proposition \ref{brostu3:cor.powdivexplNOR}(a), 
as long as \eqref{regularity} (in the relative topology) is satisfied.
For instance, we can take $\gamma=2$ and basically solve the corresponding
optimization problem by basically simulating $K^{2}-$dimensional Gaussian random variables
(even though the cardinality of 
$K \cdot \textrm{$\boldsymbol{\Omega}$\hspace{-0.23cm}$\boldsymbol{\Omega}^{dis}$}$ may be
high).
As a side remark, let us mention that our relaxation \eqref{Assign2f}
contrasts considerably to the frequently used continuous \textit{linear programming (LP) relaxation}
\begin{equation}
q_{k} \in [0,1]
\quad \textrm{for all } k \in \{1,\ldots,K^{2} \}.
\nonumber
\end{equation}
Let us finally mention that an important special case of a minimization
problem \eqref{Assign1a} to \eqref{Assign1e}
is --- the integer programming formulation of --- 
the omnipresent (asymmetric) \textit{traveling salesman problem (TSP)}
with possible side constraints
\footnote{
see e.g.~\cite{App:06}--\cite{Cook:12} 
for comprehensive books 
on TSP, its variations and its applications to
logistics, machine scheduling, printed circuit board drilling, 
communication-network frequencing, genome sequencing, 
data clustering, and many others.
}.
There, one has $K$ cities and the cost of traveling from city $i$ to city $j\ne i$
is given by $c_{ij}>0$. Moreover, one sets
$x_{ij} =1$ if the traveler goes directly from city $i$ to city $j$
(in that order), and $x_{ij} =0$ otherwise.
For technical reasons, for $i=j$ we attribute a cost $c_{ii}>0$ (e.g. hotel costs),
but we require that always $x_{ii}=0$ which we 
subsume as the first part of the constraints \eqref{Assign1e}.
Then, the constraint \eqref{Assign1b} means that the traveler leaves from city $i$ exactly once,
whereas \eqref{Assign1c} reflects that the traveler arrives at city $j$ exactly once.
The goal is to find a directed tour ---
i.e. a directed cycle/circuit that visits all $K$ cities once --- of minimum cost.
Within this context, the second part of the constraints \eqref{Assign1e} 
should basically exclude solutions which consist of disconnected subtours
(subtour elimination constraints (of e.g. the seminal Dantzig et al. \cite{Dan:54}), 
connectivity constraints, cut-set constraints).
Here, we also allow for 
additional/side constraints 
which we subsume as the third part \eqref{Assign1e} of the constraints.
Hence, our above-mentioned considerations open the gate to \textit{principally tackle} 
such kind of TSP problems with our BS method.

\vspace{0.3cm}
\noindent
For sum-\textit{maximization}-type linear assignment problems with side constraints, 
where e.g. $c_{ij}$ is a profit
(rather than a cost) 
and the ultimate goal is total profit maximization,
we can proceed analogously, by
employing \eqref{linHell4},\eqref{linHell6}
(instead of \eqref{linHell3},\eqref{linHell5}).

\vspace{0.3cm}
\noindent
Another line of applications of our method 
to deterministic optimization problems is  that
the BS minimizability of \eqref{fo.genentropy1}
with $\gamma = 2$ and $c_{1} \cdot c_{2} >0$
(see also \eqref{brostu3:fo.norweiemp6.new3a} 
of Proposition \ref{brostu3:cor.powdivexplNOR}(a) below)
can be employed to solve the following discrete Monge-Kantorovich-type optimal
mass transportation problem (optimal coupling problem)
with \textit{side (i.e. additional) constraints}:\\
given two nonnegative-entries vectors $\boldsymbol{\mu} := 
(\mu_{1}, \ldots  \mu_{K_{1}} ) \in [0,\infty[^{K_{1}}$
and $\boldsymbol{\nu} := (\nu_{1}, \ldots  \nu_{K_{2}} ) \in [0,\infty[^{K_{2}}$ with 
equal total \textquotedblleft  mass\textquotedblright\ $\sum_{k=1}^{K_{1}} \mu_{k} = 
\sum_{k=1}^{K_{2}} \nu_{k} = A >0$, compute

\vspace{-0.4cm}

\begin{eqnarray}
&& \inf_{K_{1} \times K_{2}-\textrm{matrices} \  \pi} 
K_{1} \cdot K_{2} \cdot \sum_{u=1}^{K_{1}} \sum_{v=1}^{K_{2}}
\left(\pi_{u,v} - \frac{1}{K_{1} \cdot K_{2}} \right)^{2}
\label{Monge1a} \qquad \textrm{subject to}
\nonumber
\\[-0.1cm]
&& 
\sum_{v=1}^{K_{2}} \pi_{u,v} = \mu_{u} \quad \textrm{for all } u \in \{1,\ldots,K_{1}\},
\quad
\sum_{u=1}^{K_{1}} \pi_{u,v} = \nu_{v} \quad \textrm{for all } v \in \{1,\ldots,K_{2}\},
\nonumber
\\
&& \pi_{u,v} \in [0,A] \quad \textrm{for all } u \in \{1,\ldots,K_{1}\}, \, v \in \{1,\ldots,K_{2}\},
\quad \textrm{side constraints on $\pi$, $\boldsymbol{\mu}$, $\boldsymbol{\nu}$}.
\nonumber
\end{eqnarray}
Indeed, this problem can be equivalently rewritten in terms 
$K_{1} \cdot K_{2}-$dimensional vectors as follows:
given two nonnegative-entries vectors $\boldsymbol{\mu}$,$\boldsymbol{\nu}$ as above, compute

\vspace{-0.4cm}

\begin{eqnarray}
&& 
\inf_{\mathbf{Q}\in \mathbf{\Omega}} 
K_{1} \cdot K_{2} \cdot\sum_{k=1}^{K_{1} \cdot K_{2}}
\left(q_{k} - \frac{1}{K_{1} \cdot K_{2}} \right)^{2}
\ = \ 
\inf_{\mathbf{Q}\in \mathbf{\Omega}} 
K_{1} \cdot K_{2} \cdot\sum_{k=1}^{K_{1} \cdot K_{2}}
q_{k}^{2} + 1 - 2A
\label{Monge2a}
\\
&&
\textrm{where 
$\mathbf{\Omega} \subset \mathbb{R}^{K_{1}\cdot K_{2}}$ is the set of all
vectors $\mathbf{Q} = (q_{1}, \ldots, q_{K_{1} \cdot K_{2}})$ which satisfy the constraints
}
\nonumber
\\
&& \sum_{j=1}^{K_{2}} q_{(i-1)\cdot K_{2} +j} \ = \ \mu_{i} \quad \textrm{for all } i \in \{1,\ldots,K_{1}\},
\nonumber
\quad
\sum_{i=1}^{K_{1}} q_{(i-1)\cdot K_{2} +j} \ = \ \nu_{j} \quad \textrm{for all } j \in \{1,\ldots,K_{2}\},
\nonumber
\\
&& q_{k} \in [0,A] \quad \textrm{for all } k \in \{1,\ldots,K_{1} \cdot K_{2} \},
\nonumber
\\
&& \textrm{side constraints on $\mathbf{Q}$,  $\boldsymbol{\mu}$, $\boldsymbol{\nu}$}.
\label{Monge2e}
\end{eqnarray}
Clearly, via divisions by $A$, one can equivalently rewrite $\mathbf{\Omega} 
= A \cdot \textrm{$\boldsymbol{\Omega}$\hspace{-0.23cm}$\boldsymbol{\Omega}$}$
for some $\textrm{$\boldsymbol{\Omega}$\hspace{-0.23cm}$\boldsymbol{\Omega}$} \subset 
\mathbb{S}^{K_{1} \cdot K_{2}}$
in the $K_{1}\cdot K_{2}-$dimensional probability simplex.
Hence, we can employ \eqref{fo.genentropy1}
(see also \eqref{brostu3:fo.norweiemp6.new3a})
with $K=c_{1}=K_{1} \cdot K_{2}$, $\gamma=2$,
$c_{2}=1$ and $c_{3} = \frac{2A-1}{K_{1} \cdot K_{2}}$, provided that the side constraints \eqref{Monge2e}
are such that $\mathbf{\Omega}$ satisfies the regularity property \eqref{regularity} and the
finiteness property \eqref{def fi wrt Omega}. 
Notice that \eqref{Monge2a} is equal to   
\, $
\inf_{\mathbf{Q}\in \mathbf{\Omega}} 
D_{2\cdot \varphi_{2}}(\mathbf{Q},\mathds{P}^{unif})
$ \, 
where $\mathds{P}^{unif} := (\frac{1}{K_{1} \cdot K_{2}}, 
\ldots, \frac{1}{K_{1} \cdot K_{2}})$ is the probability vector 
of frequencies of the uniform distribution on $\{1, \ldots, K_{1} \cdot K_{2}\}$,
and $\widetilde{c}=2$.
The special case $A=1$ with side constraint \eqref{Monge2e} of the
form $K_{1} \cdot \min_{i \in \{1,\ldots, K_{1}\}} \mu_{i} + 
K_{2} \cdot \min_{j \in \{1,\ldots, K_{2}\}} \nu_{j} \geq 1$
was explicitly solved by e.g. \cite{Bert:20a} \cite{Bert:22},
who also give applications to cryptographic guessing problems (spy problems),
task partitioning and graph clustering.

\vspace{0.3cm}
\noindent
The importance of the case $\gamma=2$ stems also from the fact
that one can equivalently rewrite \textit{separable quadratic minimization problems}
as minimization problems of Pearson chi-square divergences.
Indeed, one can straightforwardly derive that
\begin{equation}
\inf_{\mathbf{\breve{Q}}\in \mathbf{\breve{\Omega}}} \, 
\sum_{k=1}^{K} ( \, c_{1,k} + c_{2,k} \cdot \breve{q}_{k} + c_{3,k} \cdot \breve{q}_{k}^2 \, ) \, ,
\qquad 
c_{1,k} \in \mathbb{R}, \  
c_{2,k} \in \mathbb{R}\backslash\{0\}, \
c_{3,k} \in \, ]0,\infty[,  
\label{quadopt1}
\end{equation}

\vspace{-0.3cm}
\noindent
is equal to (recall that $\varphi_{2}(t) := \frac{(t - 1)^2}{2}$, cf. \eqref{brostu3:fo.powdivgen})

\begin{equation}
c_{4} + \inf_{\mathbf{Q}\in \mathbf{\Omega}} 
D_{\varphi_{2}}(\mathbf{Q},\mathbf{P}) \, ,
\label{quadopt2}
\end{equation}
where $\mathbf{Q} := (q_{1}, \ldots, q_{K})$ with 
$q_{k} := - c_{2,k}  \cdot \breve{q}_{k}$,
$\mathbf{P} := (p_{1}, \ldots, p_{K})$ with $p_{k}:= \frac{c_{2,k}^{2}}{2 \cdot c_{3,k}} >0$,
$c_{4} := \sum_{k=1}^{K} \big( \, c_{1,k} - \frac{c_{2,k}^{2}}{4 \cdot c_{3,k}} \, \big)$,
and $\mathbf{\Omega}$ is the corresponding reformulation of the constraint set $\mathbf{\breve{\Omega}}$.
To achieve the applicability of our BS method, we further transform \eqref{quadopt2} 
into its equal form (cf. \eqref{min Pb prob1},\eqref{min Pb prob2}) 
\begin{equation}
c_{4} + \inf_{\widetilde{\mathbf{Q}}\in \mathbf{\Omega}/M_{\mathbf{P}}} 
D_{M_{\mathbf{P}} \cdot \varphi_{2}}(\widetilde{\mathbf{Q}},\widetilde{\mathds{P}}) \, 
\label{quadopt3}
\end{equation}
with $M_{\mathbf{P}}:=\sum_{k=1}^{K} p_{k}>0$ and $\widetilde{\mathds{P}}:=\mathbf{P}/M_{\mathbf{P}}$.
If $\mathbf{\Omega}/M_{\mathbf{P}}$ satisfies \eqref{regularity} and \eqref{def fi wrt Omega}
(e.g. it may be highly disconnected),
then we can apply Theorem \ref{brostu3:thm.divW.var}. In contrast,
if $\mathbf{\Omega}/M_{\mathbf{P}} = A \cdot 
\textrm{$\boldsymbol{\Omega}$\hspace{-0.23cm}$\boldsymbol{\Omega}$}$ 
for some $A \in \mathbb{R}\backslash\{0\}$ and some
$\boldsymbol{\Omega}$\hspace{-0.23cm}$\boldsymbol{\Omega}\subset \mathbb{S}_{>0}^{K}$ 
satisfying \eqref{regularity},
then we can apply 
Theorem \ref{brostu3:thm.divnormW.new}, Remark \ref{remark divnormW}(vi),
Lemma \ref{Lemma Indent Rate finite case_new}(a)
(see also Proposition \ref{brostu3:cor.powdivexplNOR}(a) below)
together with Remark \ref{rem.inversions}(c);
for instance, this may appear if $\mathbf{\breve{\Omega}}$ contains 
(amongst others) the original constraint $\sum_{k=1}^{K} \breve{q}_{k} = C$
for some constant $C>0$, and $c_{2,k} =c_{2}$ does not depend on $k$,
which leads to the choice $A = - \frac{c_{2} \cdot C}{M_{\mathbf{P}}}$.
Notice that $A<0$ if $c_{2}>0$. For example, optimization problems
\eqref{quadopt1} with $c_{1,k} >0$, $c_{2,k} >0$, $c_{3,k} >0$ and
constraints $\sum_{k=1}^{K} \breve{q}_{k} = C$, 
$\breve{q}_{k} \in [\underline{\breve{q}}_{k},\overline{\breve{q}}_{k}]$
appear in distributed energy management as \textit{economic dispatch problems} in smart grids
of power generators,
where $\breve{q}_{k}$ is the active power generation of the $k-$th generator, 
$C$ is the total power demand,  
$\underline{\breve{q}}_{k}$ resp. $\overline{\breve{q}}_{k}$ represent
the lower resp. upper bound of the $k-$th generator's output,
and the cost of power generation is $c_{1,k} + c_{2,k} \cdot \breve{q}_{k} + c_{3,k} \cdot \breve{q}_{k}^2$ 
(cf. e.g.~\cite{Yan:13}--\cite{Xu:21}).
Another important special case of \eqref{quadopt1} to \eqref{quadopt3} 
is the omnipresent $L_{2}-$minimization; indeed, with the choices $c_{3,k}=1$,
$c_{2,k}= - 2 v_{k}$, and $c_{1,k}= v_{k}^2$ for some $\mathbf{V} =(v_{1},\ldots,v_{K})$,
the minimization problem \eqref{quadopt1} is nothing but
$\inf_{\mathbf{\breve{Q}}\in \mathbf{\breve{\Omega}}} \, || \mathbf{\breve{Q}} - \mathbf{V} ||_{2}^{2}$;
if $\mathbf{\breve{\Omega}}$ depends on a pregiven $L-$dimensional vector $\mathbf{x}$ (with $L < K$),
this can be regarded as a \textit{non-parametric regression problem} in a wide sense.

%
%

\section{Estimators}
\label{Sect Estimators}

\noindent
We demonstrate how one can \textit{principally} implement our
BS approach; further, deeper analyses are given 
in a follow-up paper.

\vspace{-0.3cm}


\subsection{Estimators for the deterministic minimization problem}
\label{Subsect Estimators determ}

\noindent
We address the minimization problem 
\begin{equation}
D_{\varphi }(\mathbf{\Omega},\mathbf{P})
:= \inf_{\mathbf{Q}\in \mathbf{\Omega} } D_{\varphi }(\mathbf{Q},\mathbf{P}) =
\inf_{\widetilde{\mathbf{Q}}\in \widetilde{\mathbf{\Omega}} }
D_{\widetilde{\varphi} }(\widetilde{\mathbf{Q}},\widetilde{\mathds{P}})
=: D_{\widetilde{\varphi} }(\widetilde{\mathbf{\Omega}},\widetilde{\mathds{P}})
\qquad \textrm{with } \widetilde{\mathbf{\Omega}} :=\mathbf{\Omega} /M_{\mathbf{P}} 
\qquad \textrm{(cf. \eqref{min Pb} and \eqref{min Pb prob2})},
\label{min Pb prob combined}
\end{equation}

\vspace{-0.1cm}
\noindent
whose numerical solution is based on Theorem \ref{brostu3:thm.divW.var}
which basically states that for large integer $n \in \mathbb{N}$ one has 
\vspace{-0.1cm}
\begin{equation}
\inf_{\mathbf{Q}\in \mathbf{\Omega} } D_{\varphi }(\mathbf{Q},\mathbf{P})
\approx
- \frac{1}{n}\log \,\mathbb{\Pi}\negthinspace \left[
\boldsymbol{\xi }_{n}^{\mathbf{\widetilde{W}}}
\in \widetilde{\mathbf{\Omega}}\right]
\label{LDP Minimization approx}
\end{equation}

\vspace{-0.1cm}
\noindent
in terms of $\widetilde{\varphi }:=M_{\mathbf{P}} \cdot \varphi$ and
the random vectors 
$ \boldsymbol{\xi }_{n}^{\mathbf{\widetilde{W}}}
=\Big(\frac{1}{n}\sum_{i\in I_{1}^{(n)}}\widetilde{W}_{i},\ldots ,\frac{1}{n}
\sum_{i\in I_{K}^{(n)}}\widetilde{W}_{i}\Big) \, \text{(cf. \eqref{Xi_n^W vector})} $
with $n_{k}:=\lfloor n \cdot \widetilde{p}_{k}\rfloor $ 
leading to the disjoint index blocks $I_{1}^{(n)}:=\left\{
1,\ldots ,n_{1}\right\} $, $I_{2}^{(n)}:=\left\{ n_{1}+1,\ldots
,n_{1}+n_{2}\right\} $, $\ldots$, 
$I_{K}^{(n)} := \{ \sum_{k=1}^{K-1} n_{k} + 1, \ldots, n \}$. 
Recall that $\mathbf{\widetilde{W}} := (\widetilde{W}_{1}, \ldots, \widetilde{W}_{n})$
is a random vector consisting of components $\widetilde{W}_{i}$
which are i.i.d. copies of the random variable $\widetilde{W}$
whose distribution is 
$\mathbb{\Pi }[\widetilde{W}\in \cdot \,]=\widetilde{\mathbb{\bbzeta}}[\,\cdot \,]$ 
obeying the representation
$\widetilde{\varphi}(t) = 
\sup_{z \in \mathbb{R}} \Big( z\cdot t - \log \int_{\mathbb{R}} e^{zy} d\widetilde{\mathbb{\bbzeta}}(y) \Big), \, 
t \in \mathbb{R},  \, 
\textrm{(cf. \eqref{brostu3:fo.link.var})} .$
Hence, due to \eqref{LDP Minimization approx}, the estimation of 
$D_{\varphi}(\mathbf{\Omega },\mathbf{P})$ 
amounts to the estimation of
$\mathbb{\Pi}\negthinspace \left[
\boldsymbol{\xi }_{n}^{\mathbf{\widetilde{W}}}
\in \widetilde{\mathbf{\Omega}} \right]$.
For the rest of this subsection, we assume that 
$\widetilde{\mathds{P}} \in \mathbb{S}_{> 0}^{K}$,
that $n$ is chosen such that all $n \cdot \widetilde{p}_{k}$ are integers
(and hence, $n = \sum_{k=1}^{K} n_{k}$),
and that $\widetilde{\mathbf{\Omega}} \subset \mathbb{R}^{K}$ satisfies the regularity property
$cl(\widetilde{\mathbf{\Omega}})=cl\big( 
int\big( \widetilde{\mathbf{\Omega}} \big) \big), int\big( \widetilde{\mathbf{\Omega}} \big) \ne \emptyset$
which implies that the same condition holds for $\mathbf{\Omega}$;
moreover, we suppose 
that $D_{\widetilde{\varphi} }(\widetilde{\mathbf{\Omega}},\widetilde{\mathds{P}}) \in \, ]0,\infty[$ 
(and thus $\widetilde{\mathds{P}} \notin cl(\widetilde{\mathbf{\Omega}})$). 
For the ease of the following discussions, we introduce the notations
\[
T\left( \mathbf{x}\right) :=\bigg( \frac{1}{n_{1}}\sum_{i\in I_{1}^{(n)}}x_{i},\ldots,
\frac{1}{n_{K}}\sum_{i\in I_{K}^{(n)}} x_{i}\bigg)
\quad \textrm{for any $\mathbf{x}:=\left( x_{1},..,x_{n}\right) \mathbf{\in }\mathbb{R}^{n}$,}
\]
as well as  $\mathfrak{D}$ for the diagonal matrix with diagonal entries $1/\widetilde{p}_{1},\ldots,1/\widetilde{p}_{K}$
and null entries off the diagonal.
Accordingly, the probability in \eqref{LDP Minimization approx} becomes
$\mathbb{\Pi }\negthinspace \left[
\boldsymbol{\xi }_{n}^{\mathbf{\widetilde{W}}}
\in \widetilde{\mathbf{\Omega}} \right] \ = \
\mathbb{\Pi}\negthinspace \left[ T(\mathbf{\widetilde{W}}) \in \mathbf{\Lambda}
\right] $
where $\mathbf{\Lambda} := \widetilde{\mathbf{\Omega}} \cdot \mathfrak{D}$
is a set of (row) vectors in $\mathbb{R}^{K}$ which is known/derived from the concrete context.
The \textit{naive estimator} $\widehat{\widetilde{\Pi}}_{L}^{naive}$ of 
$\mathbb{\Pi}\negthinspace \left[
\boldsymbol{\xi}_{n}^{\mathbf{\widetilde{W}}}
\in \widetilde{\mathbf{\Omega}} \right]$
is constructed through the following procedure:
simulate independently $L$ copies 
$\mathbf{\widetilde{W}}^{(1)},\ldots,\mathbf{\widetilde{W}}^{(L)}$ of the vector
 $\mathbf{\widetilde{W}}:=\left( \widetilde{W}_{1},\ldots,\widetilde{W}_{n}\right) $, 
with independent entries under $\widetilde{\mathbb{\bbzeta}}$, and define 
(with a slight abuse of notation)
$\widehat{\widetilde{\Pi}}_{L}^{naive} :=\frac{1}{L}\sum_{\ell =1}^{L} \mathbf{1}_{\mathbf{\Lambda}
}\Big( T\Big( \mathbf{\widetilde{W}}^{(\ell)}\Big) \Big) ; $
however this procedure is time costly, since this estimate has a very bad hit rate.
Thus, in the following, a so-called \textquotedblleft efficient Importance Sampling (IS)\textquotedblright\ 
scheme --- in the sense of Sadowsky
\& Bucklew \cite{Sad:90} (denoted [SB] hereunder) --- 
is adapted for the sophisticated (i.e. non-naive) estimation of 
$\mathbb{\Pi}[
\boldsymbol{\xi }_{n}^{\mathbf{\widetilde{W}}}
\in \widetilde{\mathbf{\Omega}}]$. For this,
we employ the following additional Assumption (OM) on
the set $\widetilde{\mathbf{\Omega}}$:

\vspace{0.2cm}
\noindent
(OM) 
For any $\widetilde{\boldsymbol{\omega}} \in cl(\widetilde{\mathbf{\Omega}})$
there exists a vector $\mathbf{x} = \left(x_{1},\ldots,x_{n}\right) \in 
\left] t_{-}^{sc},t_{+}^{sc}\right[^{n}$  such that 
$\widetilde{\boldsymbol{\omega}} = \left( \frac{1}{n}\sum_{i\in I_{1}^{(n)}}x_{i},\ldots,
\frac{1}{n} \sum_{i\in I_{K}^{(n)}}x_{i}\right)$, 
or equivalently, for any $\boldsymbol{\lambda} \in cl(\mathbf{\Lambda})$ there exists a 
vector $\mathbf{x} = \left(x_{1},\ldots,x_{n}\right) \in 
\left] t_{-}^{sc},t_{+}^{sc}\right[^{n}$ 
such that 
$\boldsymbol{\lambda} =T\left(\mathbf{x}\right)$.

\vspace{0.2cm}
\noindent 
In the current setup, Assumption (OM) is for instance feasible if 
for all $\widetilde{\boldsymbol{\omega}} \in cl(\widetilde{\mathbf{\Omega}})$
there holds for all its components $\widetilde{\omega}_{k}
\in \, ]\widetilde{p}_{k} \cdot t_{-}^{sc}, \widetilde{p}_{k} \cdot t_{+}^{sc}[ 
\, = \, ] \frac{n_{k}}{n} \cdot t_{-}^{sc}, \frac{n_{k}}{n} \cdot t_{+}^{sc}[$ ($k=1,\ldots,K$)
\footnote{since then e.g. we can uniformly take $x_{i} :=  \frac{n}{n_{k}} \cdot \widetilde{\omega}_{k}
= \lambda_{k} \in \, \left] \frac{n}{n_{k}} \cdot \frac{n_{k}}{n} \cdot t_{-}^{sc},
\frac{n}{n_{k}} \cdot \frac{n_{k}}{n} \cdot t_{+}^{sc}\right[  \, = \, \left] t_{-}^{sc},t_{+}^{sc}\right[$
for all $i \in I_{k}^{(n)}$ ($k=1,\ldots,K$)
}
--- which e.g. is always satisfied 
in the common case $dom(\widetilde{\varphi}) 
= dom(\varphi) = \, ]a,b[ \, = 
\, \left] t_{-}^{sc},t_{+}^{sc}\right[ \, = \, ]0,\infty[$ (e.g. for
the power-divergence generators $\widetilde{\varphi} 
= \widetilde{c} \cdot \varphi_{\gamma}$, $\gamma \leq 0$, cf. Subsections \ref{Subsect Case1},\ref{Subsect Case5} below)
together with $cl(\widetilde{\mathbf{\Omega}}) \in \mathbb{R}_{>0}^{K}$.

\enlargethispage{0.7cm}

\vspace{0.1cm}
\noindent
To proceed, for any distribution $\widetilde{S}$ on $\mathbb{R}^{n}$ with support included in the
support of the product measure $\widetilde{\mathbb{\bbzeta}}^{\otimes n}$ it holds
(with a slight abuse of notation) 
\vspace{-0.3cm}
\[
\mathbb{\Pi}\negthinspace \left[
\boldsymbol{\xi }_{n}^{\mathbf{\widetilde{W}}}
\in \widetilde{\mathbf{\Omega}}\right]
=E_{\widetilde{\mathbb{\bbzeta}}^{\otimes n}} \negthinspace \negthinspace \left[
\mathbf{1}_{\mathbf{\Lambda} }( T( \mathbf{\widetilde{W}}))\right] =
E_{\widetilde{S}} \bigg[ 
\mathbf{1}_{\mathbf{\Lambda} }\big( T\big( \mathbf{\widetilde{V}}\big) \big) \cdot 
\frac{d\widetilde{\mathbb{\bbzeta}}^{\otimes n}
}{d\widetilde{S}}\left( \mathbf{\widetilde{V}}\right) \bigg]
\]

\vspace{-0.1cm}
\noindent
from where the \textit{improved IS estimator} of 
$\mathbb{\Pi}\negthinspace \left[
\boldsymbol{\xi }_{n}^{\mathbf{\widetilde{W}}}
\in \widetilde{\mathbf{\Omega}} \right]$
is obtained by sampling $L$ i.i.d. replications 
$\mathbf{\widetilde{V}}^{(1)},\ldots,\mathbf{\widetilde{V}}^{(L)}$ of the random vector 
$\mathbf{\widetilde{V}}$ with distribution $\widetilde{S}$ and by defining
\vspace{-0.1cm}
\begin{equation}
\widehat{\widetilde{\Pi}}_{L}^{improved}:=\frac{1}{L}\sum_{\ell=1}^{L}
\mathbf{1}_{\mathbf{\Lambda}}( T( \mathbf{\widetilde{V}}^{(\ell)})) \cdot \frac{d\widetilde{\mathbb{\bbzeta}}^{\otimes n}}{d\widetilde{S}}
\left( \mathbf{\widetilde{V}}^{(\ell)}\right) \ .  
\label{estim improved}
\end{equation}

\vspace{-0.1cm}
\noindent
The precise form of the efficient IS distribution $\widetilde{S}^{opt}$ relies on the definition
of a \textquotedblleft  dominating point\textquotedblright\ 
of $\mathbf{\Lambda}$, which we characterize in the present context.
For $\mathbf{x} := \left( x_{1},\ldots,x_{n}\right)$ in $\mathbb{R}^{n}$ we define
$I_{\mathbf{\widetilde{W}}}(\mathbf{x}):=\sup_{\mathbf{z} \in \mathbb{R}^{n}}
\left(\left\langle 
\mathbf{z}, \mathbf{x} \right\rangle -\log E_{\widetilde{\mathbb{\bbzeta}}^{\otimes n}}
[ \, \exp (\langle \mathbf{z},\mathbf{\widetilde{W}}\rangle) ]
\right), $
and for each 
$\boldsymbol{\lambda} \in cl(\mathbf{\Lambda})$ we let
$ I(\boldsymbol{\lambda}):=\inf \left\{
 I_{\mathbf{\widetilde{W}}}(\mathbf{x}):T(\mathbf{x)=\boldsymbol{\lambda}}
\right\}$ (notice that the set is non-empty because of (OM)).
We call a point $\underline{\boldsymbol{\lambda}} := \left( \underline{\lambda}_{1},\ldots,
\underline{\lambda}_{K}\right)$ 
a \textit{minimal rate point (mrp) of $\mathbf{\Lambda}$} if 
(a) $\underline{\boldsymbol{\lambda}} \in \partial\mathbf{\Lambda}$
and (b)
$I(\underline{\boldsymbol{\lambda}})\leq I(\boldsymbol{\lambda}) \  \text{ for all }
\boldsymbol{\lambda} \in \mathbf{\Lambda}$.
A point $\underline{\boldsymbol{\lambda}}$ is called a 
\textit{dominating point of $\mathbf{\Lambda}$} if (a) $\underline{\boldsymbol{\lambda}}
\in \partial \mathbf{\Lambda}$, and (b) $I\left(\underline{\boldsymbol{\lambda}}\right) 
\leq I(\boldsymbol{\lambda})$ for all 
$\boldsymbol{\lambda} \in \mathbf{\Lambda}$ with attainment, namely there exists 
a vector $\underline{\mathbf{x}} \in \left] t_{-}^{sc},t_{+}^{sc}\right[^{n}$ such
that $I_{\widetilde{W}}\left( \underline{\mathbf{x}}\right) =I\left( 
\underline{\boldsymbol{\lambda}}\right)$ with $\underline{\boldsymbol{\lambda}}=
T\left( \underline{\mathbf{x}}\right)$.
The characterization of a dominating point 
$\underline{\lambda}$ is settled in the following 

\vspace{0.3cm}
\noindent

\begin{lemma}
\label{Lemma minim rate points}
Suppose that Assumption (OM) holds, and let $\underline{\boldsymbol{\lambda}}$ be a (always existent) mrp
of $\mathbf{\Lambda}$. Then, 
$\underline{\boldsymbol{\lambda}}$ is a
dominating point, and
$\inf \left\{ I_{\mathbf{\widetilde{W}}}\left( \mathbf{x}\right): \, T(\mathbf{x})=\underline{\boldsymbol{\lambda}}\right\}$ is reached at some 
vector $\underline{\mathbf{x}}$ in $\left] t_{-}^{sc},t_{+}^{sc}\right[^{n}$ such
that for all $k \in \left\{ 1,\ldots ,K\right\} $ and all $i \in I_{k}^{(n)}$
there holds $\underline{x}_{i}= \underline{\lambda}_{k}$ and  
$I_{\mathbf{\widetilde{W}}}(\underline{\mathbf{x}})=
n \cdot \sum_{k=1}^{K}\widetilde{p}_{k} \cdot 
\widetilde{\varphi}\left(\underline{\lambda}_{k}\right)$.
\end{lemma}

\vspace{0.3cm}
\noindent
The proof of Lemma \ref{Lemma minim rate points} is given in Appendix F.
Notice that (OM) implies the \textit{existence} of a dominating point 
$\underline{\boldsymbol{\lambda}}$, 
but \textit{uniqueness} may not hold. 
In the latter case, one can try to
proceed as in Theorem 2 of [SB] and the discussion thereafter.

\vspace{0.3cm}
\noindent
However, we assume now uniqueness of $\underline{\boldsymbol{\lambda}}$; this allows for
the identification of $\widetilde{S}^{opt}$. By Theorem 1 of [SB]
and Theorem 3.1 of \cite{Csi:75}, the
asymptotically optimal IS distribution $\widetilde{S}^{opt}$ is obtained as the 
Kullback-Leibler-divergence projection of 
$\widetilde{\mathbb{\bbzeta}}^{\otimes n}$ on the set of all probability distributions
on $\mathbb{R}^{n}$ centered at
point $\underline{\mathbf{x}}$, whose coordinates are 
--- according to Lemma \ref{Lemma minim rate points} ---
functions of the coordinates of
$\underline{\mathbf{\widetilde{Q}}} := \underline{\boldsymbol{\lambda}} \cdot \mathfrak{D}^{-1}$ such that 
$T\left( \underline{\mathbf{x}}\right) = \underline{\mathbf{\widetilde{Q}}} \cdot \mathfrak{D}$.

\vspace{0.3cm}
\noindent
The above definition of $\widetilde{S}^{opt}$ presumes the knowledge of 
$\underline{\boldsymbol{\lambda}}$, which cannot be assumed 
(otherwise the minimization problem
is solved in advance). The aim of the following construction is to provide
a proxy $\widetilde{S}$ to $\widetilde{S}^{opt}$, where $\widetilde{S}$
is the Kullback-Leibler-divergence projection of 
$\widetilde{\mathbb{\bbzeta}}^{\otimes n}$ on the set of all probability distributions
on $\mathbb{R}^{n}$ centered at
some point $\mathbf{x}^{\ast}$ which is close to $\underline{\mathbf{x}}$.
For this sake, we need to have at hand a \textit{proxy} of 
$\underline{\boldsymbol{\lambda}}$ or, equivalently, 
a \textit{preliminary guess} $\mathbf{\widetilde{Q}}^{\ast}$ of 
$\underline{\mathbf{\widetilde{Q}}}
:=\arg \inf_{\mathbf{\widetilde{Q}}
\in \widetilde{\mathbf{\Omega}}
}\sum_{k=1}^{K} \widetilde{p}_{k} \cdot \widetilde{\varphi}(\widetilde{q}_{k}/\widetilde{p}_{k})$.
This guess is by no
means produced in order to provide a direct estimate of 
$D_{\widetilde{\varphi} }(\widetilde{\mathbf{\Omega}},\widetilde{\mathds{P}})$
but merely to provide the IS distribution $\widetilde{S}$ which in
turn leads to a sharp estimate of 
$D_{\widetilde{\varphi} }(\widetilde{\mathbf{\Omega}},\widetilde{\mathds{P}})$.

\vspace{0.3cm}
\noindent
\textit{Proxy method 1:} in some cases we might have at hand some particular point 
$\mathbf{\widetilde{Q}}^{\ast} :=\left(
\widetilde{q}_{1}^{\ast},..,\widetilde{q}_{K}^{\ast}\right) $ 
in $\widetilde{\mathbf{\Omega}}$;
the resulting IS distribution $\widetilde{S}$
with \underline{$\mathbf{\widetilde{Q}}$}
substituted by $\mathbf{\widetilde{Q}^{\ast}}$ is not optimal in the sense of 
[SB], but anyhow produces an estimator with good hitting rate, possibly with a loss in
the variance.
One possible way to obtain such a point $\mathbf{\widetilde{Q}^{\ast}}$ in 
$\widetilde{\mathbf{\Omega}}$
is to simulate runs of 
(say) $M-$variate i.i.d. vectors $\mathbf{\widetilde{W}}$
under $\widetilde{\mathbb{\bbzeta}}^{\otimes M}$
until the first time where $\boldsymbol{\xi }_{M}^{\mathbf{\widetilde{W}}}$
belongs to $\widetilde{\mathbf{\Omega}}$; then we set
$\mathbf{\widetilde{Q}^{\ast}} := \boldsymbol{\xi }_{M}^{\mathbf{\widetilde{W}}}$
for the succeeding realization $\mathbf{\widetilde{W}}$.
Before we proceed, it is useful to mention that the need for a drastic fall
in the number of simulation runs pertains for cases when 
$D_{\widetilde{\varphi} }(\widetilde{\mathbf{\Omega}},\widetilde{\mathds{P}})$
is large. The following construction is suited to this
case, which is of relevance in applications both in optimization and in
statistics when choosing between competing models none of which
is assumed to represent the true one, but merely less inadequate ones. 

\vspace{0.3cm}
\noindent
\textit{Proxy method 2:} when 
$D_{\widetilde{\varphi} }(\widetilde{\mathbf{\Omega}},\widetilde{\mathds{P}})$
is presumably large, we make use of
asymptotic approximation to get a proxy of $\underline{\mathbf{\widetilde{Q}}}$.
For this, we define a sampling distribution on $\mathbb{R}^{K}$ fitted to the divergence through
\begin{equation}
f(\mathbf{\widetilde{Q}}):=C \cdot \exp \negthinspace \Big( -\sum_{k=1}^{K} \widetilde{p}_{k} \cdot 
\widetilde{\varphi}(\widetilde{q}_{k}/\widetilde{p}_{k})\Big) \ = \ 
C \cdot \exp \negthinspace \Big( -D_{\widetilde{\varphi}}\left( \mathbf{\widetilde{Q}},
\widetilde{\mathds{P}}\Big) 
\right)
\label{simul distr}
\end{equation}
where $C$ is a normalizing constant. 
Let $\mathbf{T}$ be a $K-$variate random variable with density $f$. The
distribution of $\mathbf{T}$ given $\left( \mathbf{T} \in \widetilde{\mathbf{\Omega}} 
\right)$ concentrates around 
$\arg \inf_{\widetilde{\mathbf{Q}}\in \widetilde{\mathbf{\Omega}} }
D_{\widetilde{\varphi} }(\widetilde{\mathbf{Q}},\widetilde{\mathds{P}})$ when 
$D_{\widetilde{\varphi} }(\widetilde{\mathbf{\Omega}},\widetilde{\mathds{P}})$
is large. Indeed, for any $\widetilde{\mathbf{Q}}\in \widetilde{\mathbf{\Omega}}$
denote by $\mathbf{V}_{\varepsilon}(\widetilde{\mathbf{Q}})$ a small neighborhood
of $\widetilde{\mathbf{Q}}$ in $\mathbb{R}^{K}$ with radius $\varepsilon$;
clearly, the probability of the event 
$\left( \mathbf{T} \in \mathbf{V}_{\varepsilon}(\widetilde{\mathbf{Q}})\right)$ 
when restricted to $\widetilde{\mathbf{Q}}\in \widetilde{\mathbf{\Omega}}$
is maximum when $\widetilde{\mathbf{Q}} = \underline{\widetilde{\mathbf{Q}}}$,
where $\underline{\widetilde{\mathbf{Q}}}$ is the 
\textquotedblleft  
dominating point of $\widetilde{\mathbf{\Omega}}$\textquotedblright\  
\textit{in the sense that} $\underline{\mathbf{\widetilde{Q}}} := \underline{\boldsymbol{\lambda}} 
\cdot \mathfrak{D}^{-1} \in \partial \widetilde{\mathbf{\Omega}}$ 
is the above-defined transform of the dominating point $\underline{\boldsymbol{\lambda}}$
(assuming uniqueness);
a precise argumentation under adequate conditions is
postponed to Appendix F.
Accordingly, we obtain a proxy\textbf{\ }$\mathbf{\widetilde{Q}}^{\ast}$ of 
$\underline{\mathbf{\widetilde{Q}}}$ by simulating a sequence of independent 
$K-$variate random variables $\mathbf{T}_{1},\ldots$ with
distribution \eqref{simul distr} until (say) $\mathbf{T}_{m}$ belongs to $\widetilde{\mathbf{\Omega}}$ and set $\mathbf{\widetilde{Q}}^{\ast}:=\mathbf{T}_{m}$.

\vspace{0.4cm}
\noindent
To proceed with the derivation of the IS sampling distribution $\widetilde{S}$
on $\mathbb{R}^{n}$, we fix 
$\mathbf{\widetilde{Q}}^{\ast} :=\left(
\widetilde{q}_{1}^{\ast},..,\widetilde{q}_{K}^{\ast}\right)$
to be a proxy of $\underline{\mathbf{\widetilde{Q}}}$ or
an initial guess in $\widetilde{\mathbf{\Omega}}$. 
As an intermediate step, we construct
the probability distribution $\widetilde{U}_{k}$ on $\mathbb{R}$ given by 
\begin{equation}
d\widetilde{U}_{k}(v) \ := \ \exp \left( \tau_{k} \cdot v - 
\Lambda_{\widetilde{\mathbb{\bbzeta}}}(\tau_{k})\right) d\widetilde{\mathbb{\bbzeta}}(v)
=\frac{\exp \left(\tau _{k} \cdot v\right)}{MGF_{\widetilde{\mathbb{\bbzeta}}}(\tau_{k})} \, d\widetilde{\mathbb{\bbzeta}}(v)  
\label{Q_k}
\end{equation}
where $\tau_{k} \in int(dom(MGF_{\widetilde{\mathbb{\bbzeta}}}))$ 
is the unique solution of the equation 
$\Lambda_{\widetilde{\mathbb{\bbzeta}}}^{\prime}\left( \tau_{k}\right) =
\frac{\widetilde{q}_{k}^{\ast}}{\widetilde{p}_{k}} \in \, ]t_{-}^{sc},t_{+}^{sc}[$
and thus --- by relation \eqref{proof Ftheorem equ 3} of Appendix E --- we can compute explicitly
$\tau_{k}  = \ \widetilde{\varphi}^{\, \prime} \negthinspace
\left(\frac{\widetilde{q}_{k}^{\ast}}{\widetilde{p}_{k}}\right).$
Therefore, $\widetilde{U}_{k}$ is the Kullback-Leibler-divergence projection of 
$\widetilde{\mathbb{\bbzeta}}$ on the class of all probability distributions
on $\mathbb{R}$ whose expectation is $\widetilde{q}_{k}^{\ast}$.
As a side remark, notice that one possible way of obtaining an explicit form of the probability distribution $\widetilde{U}_{k}$ may be by identification through its moment generating function
\begin{equation}
dom(MGF_{\widetilde{\mathbb{\bbzeta}}})-\tau_{k} \ \ni \ z \ \mapsto
MGF_{\widetilde{U}_{k}}(z) = 
\frac{MGF_{\widetilde{\mathbb{\bbzeta}}}(z+\tau_{k})}{MGF_{\widetilde{\mathbb{\bbzeta}}}(\tau_{k})}
\nonumber
\end{equation}
of which all ingredients are principally available; for instance, this will be used
in the solved-cases Section \ref{Sect Cases} below. From \eqref{Q_k}, we define
$\widetilde{S}_{k}:=\underbrace{\widetilde{U}_{k} \otimes \cdots 
\otimes \widetilde{U}_{k}}_{n_{k}\text{times}} \, 
\textrm{for all $k\in\{1,\ldots,K\}$}, $
whence
\begin{eqnarray}
d\widetilde{S}_{k}\left( v_{k,1},\ldots,v_{k,n_{k}}\right) =
\exp \Big( (\sum_{i\in I_{k}^{(n)}} \tau_{k} \cdot v_{k,i}) - 
n_{k} \cdot \Lambda_{\widetilde{\mathbb{\bbzeta}}}(\tau _{k})\Big) \ 
d\widetilde{\mathbb{\bbzeta}}\left(v_{k,1}\right) 
\cdots
d\widetilde{\mathbb{\bbzeta}}\left(v_{k,n_{k}}\right),
\nonumber
\end{eqnarray}
which manifests $\widetilde{S}_{k}$ as the Kullback-Leibler-divergence projection of
$\underbrace{\widetilde{\mathbb{\bbzeta}} \otimes \cdots \otimes \widetilde{\mathbb{\bbzeta}}}_{n_{k}\text{times}}$
on the class of all probability distributions on $\mathbb{R}^{k}$
whose expectation vector is 
$\mathbf{\widetilde{Q}}^{\ast} = 
\left(\widetilde{q}_{1}^{\ast},\ldots,\widetilde{q}_{K}^{\ast}\right) \in 
\mathbb{R}^{k}$.
Let now 
\begin{equation}
\widetilde{S} := \widetilde{S}_{1} \otimes \cdots \otimes \widetilde{S}_{K} \, , 
\nonumber
\end{equation}
which therefore satisfies (recall that $\sum_{k=1}^{K} n_{k} =n$) 
\begin{equation}
d\widetilde{S}\left( v_{1,1},\ldots v_{1,n_{1}},
\ldots, v_{K,1},\ldots v_{K,n_{K}}\right) =
\exp \Big(\sum_{k=1}^{K} \, (\sum_{i\in
I_{k}^{(n)}}  \tau _{k} \cdot v_{k,i}) - n_{k} \cdot
\Lambda_{\widetilde{\mathbb{\bbzeta}}}(\tau _{k})  \Big) \, 
d\widetilde{\mathbb{\bbzeta}}^{\otimes n}\left( v_{1,1},\ldots v_{1,n_{1}},
\ldots, v_{K,1},\ldots v_{K,n_{K}}\right).
\  
\label{Sdiffform}
\end{equation}
The same procedure with all $\widetilde{q}_{k}^{\ast}$
substituted by the coordinates $\underline{\widetilde{q}}_{k}$
of $\underline{\mathbf{\widetilde{Q}}}$ 
produces $S^{opt}$. Therefore, $\widetilde{S}$ is a substitute
for $S^{opt}$ with the change in the centering from
the unknown vector $\underline{\mathbf{\widetilde{Q}}}$
to its proxy $\mathbf{\widetilde{Q}}^{\ast}$.

\vspace{0.3cm}
\noindent
As a straightforward consequence of \eqref{estim improved} and \eqref{Sdiffform}, 
we obtain the improved IS estimator of 
$\mathbb{\Pi}\negthinspace \left[
\boldsymbol{\xi }_{n}^{\mathbf{\widetilde{W}}}
\in \widetilde{\mathbf{\Omega}}\right]$ as
\begin{equation}
\widehat{\widetilde{\Pi}}_{L}^{improved}\ =\ \frac{1}{L}\sum_{\ell=1}^{L}
\mathbf{1}_{\mathbf{\Lambda}}( T( \mathbf{\widetilde{V}}^{(\ell)})) \cdot 
\prod_{k=1}^{K} IS_{k}(\mathbf{\widetilde{V}}_{k}^{(\ell)})
\label{improved2}
\end{equation}
where $\mathbf{\widetilde{V}}_{k}^{(\ell)} := \left( \widetilde{V}_{i}^{(\ell)} \right)_{i \in I_{k}^{(n)}}$ is the $k-$th block of the $\ell-$th replication 
$\mathbf{\widetilde{V}}^{(\ell)}$ of $\mathbf{\widetilde{V}}$ under $\widetilde{S}$,
and 
the $k-$th importance-sampling factor is \, 
$\widetilde{IS}_{k}(v_{k,1},\ldots,v_{k,n_{k}}) \ := \ 
\frac{d\widetilde{\mathbb{\bbzeta}}^{\otimes n_{k}}}{d\widetilde{S}_{k}}\left( v_{k,1},
\ldots,v_{k,n_{k}}\right)
=  \exp\Big(n_{k} \cdot \Lambda_{\widetilde{\mathbb{\bbzeta}}}(\tau_{k}) \, - \, \tau_{k} 
\cdot \sum_{i=1}^{n_{k}} v_{k,i} \Big)$ \, 
with $n_{k} = card(I_{k}^{(n)})$.

\vspace{0.3cm}
\noindent
Summing up,
we arrive at the following algorithm in case that $\widetilde{\mathbf{\Omega}}$ has a
unique dominating point (in the above-defined sense):

\noindent
\textbf{Step D1.}
Exemplarily, we start with proxy method 2 (the other proxy method 1
works analogously):
get a proxy\textbf{\ }$\mathbf{\widetilde{Q}}^{\ast}$ of 
$\underline{\mathbf{\widetilde{Q}}}$ by simulating a sequence of independent 
$K-$variate random variables $\mathbf{T}_{1},\ldots$ with
distribution \eqref{simul distr} until (say) $\mathbf{T}_{m}$ belongs to $\widetilde{\mathbf{\Omega}}$
and set $\mathbf{\widetilde{Q}}^{\ast}:=\mathbf{T}_{m}$.

\noindent
\textbf{Step D2.}  
For all $k$ in $\{1,\ldots,K\}$ 
compute $\tau _{k}  = \ \widetilde{\varphi}^{\, \prime}
\left(\frac{\widetilde{q}_{k}^{\ast}}{\widetilde{p}_{k}}\right)$.

\noindent
\textbf{Step D3.}
For all $\ell$ in $\{1,\ldots,L\}$ perform a run of $\mathbf{\widetilde{V}}^{(\ell)}$ 
under $\widetilde{S}$ as follows:

\noindent
For all $k$ in $\{1,\ldots,K\}$ simulate $n_{k}$ i.i.d. random variables
$\widetilde{V}_{k_{1}}^{(\ell)},\ldots,\widetilde{V}_{k_{n_{k}}}^{(\ell)}$ with common distribution 
$\widetilde{U}_{k}$ defined in \eqref{Q_k}. Set $\mathbf{\widetilde{V}}_{k}^{(\ell)} := 
(\widetilde{V}_{k_{1}}^{(\ell)},\ldots,\widetilde{V}_{k_{n_{k}}}^{(\ell)})$ to be the 
corresponding row vector.
Construct $\mathbf{\widetilde{V}}^{(\ell)}$ as the row vector obtained by concatenating the 
$\mathbf{\widetilde{V}}_{k}^{(\ell)}$, i.e.
$\mathbf{\widetilde{V}}^{(\ell)}:=\left( \mathbf{\widetilde{V}}_{1}^{(\ell)},\ldots,
\mathbf{\widetilde{V}}_{K}^{(\ell)}\right) \, , $
and make use of $\widehat{\widetilde{\Pi}}_{L}^{improved}$ given in 
\eqref{improved2}
with the $\tau_{k}$'s obtained in Step D2 above to define 
(in the light of \eqref{min Pb prob combined},\eqref{LDP Minimization approx}) 
the \textit{BS minimum-distance estimator}
\begin{equation}
\widehat{D_{\varphi}}(\mathbf{\Omega},\mathbf{P})
\ := \ \widehat{D_{\widetilde{\varphi }}}(\widetilde{\mathbf{\Omega}},\widetilde{\mathds{P}})
\ := \ - \frac{1}{n}\log 
\widehat{\widetilde{\Pi}}_{L}^{improved} \ . 
\label{estimator minimization}
\end{equation}

\vspace{0.3cm}
\noindent
For many cases,
the simulation burden needed for the computation of $\widehat{\widetilde{\Pi}}_{L}^{improved}$ 
--- and thus of $\widehat{D_{\varphi}}(\mathbf{\Omega},\mathbf{P})$ ---
can be drastically reduced, especially for high dimensions $K$
and large sample size $n \cdot L$.
In fact, in terms of the notations
$n_{k}:=card(I_{k}^{(n)})$, 
$\widehat{\mathit{W}}_{k}^{(\ell)}:=\sum_{i\in I_{k}^{(n)}}\widetilde{V}_{i}^{(\ell)}$ and
\begin{equation}
\widetilde{ISF}_{k}(x) \ := \ \frac{d\widetilde{\mathbb{\bbzeta}}^{\ast n_{k}}}{
d\widetilde{U}_{k}^{\ast n_{k}}
}(x) \ = \ 
\exp(n_{k} \cdot \Lambda_{\widetilde{\mathbb{\bbzeta}}}(\tau_{k}) \, - \, x \cdot \tau_{k} ) 
\label{ISK}
\end{equation}
(where $\widetilde{\mathbb{\bbzeta}}^{\ast n_{k}}$ is the $n_{k}-$convolution of the measure 
$\widetilde{\mathbb{\bbzeta}}$),
one can rewrite \eqref{improved2} as
\begin{equation}
\widehat{\widetilde{\Pi}}_{L}^{improved}=\frac{1}{L}\sum_{\ell=1}^{L}
\mathbf{1}_{\mathbf{\Lambda}
}\big( \,
\big(
\frac{1}{n_{1}} \widehat{\mathit{W}}_{1}^{(\ell)},
\ldots, \frac{1}{n_{K}} \widehat{\mathit{W}}_{K}^{(\ell)} \big) \, 
\big) \cdot
\dprod\limits_{k=1}^{K}\widetilde{ISF}_{k}(\widehat{\mathit{W}}_{k}^{(\ell)})  
\label{Improved IS for inf div new}
\end{equation}
with $K-$vector 
$\big(\frac{1}{n_{1}} \widehat{\mathit{W}}_{1}^{(\ell)},
\ldots, \frac{1}{n_{K}} \widehat{\mathit{W}}_{K}^{(\ell)} \big)$.
Clearly, the random variable $\widehat{\mathit{W}}_{k}^{(\ell)}$ ($k=1, \ldots, K$)
has distribution $\widetilde{U}_{k}^{\ast n_{k}}$. Hence, if 
$\widetilde{U}_{k}^{\ast n_{k}}$ can be \textit{explicitly} constructed,
then for the computation of $\widehat{\widetilde{\Pi}}_{L}^{improved}$
it suffices to simulate the $K \cdot L$ random variables $\widehat{\mathit{W}}_{k}^{(\ell)}$ 
rather than the $n \cdot L$ random variables $\widetilde{V}_{i}^{(\ell)}$;
notice that according to the right-hand side of \eqref{ISK}, 
one can explicitly compute $\widetilde{ISF}_{k}\left( \cdot \right)$
which can be interpreted as
\textit{Importance Sampling Factor pertaining to the block $k$}.
In the case that $\widetilde{\mathbb{\bbzeta}}$ is infinitely 
divisible, simulation issues may become especially comfortable.
The tractability 
of this reduction effect will be exemplarily demonstrated 
in the solved-cases Subsections \ref{Subsect Case1} to \ref{Subsect Case6} below, 
for the BS minimization of the important power divergences (for which the 
infinite divisibility holds).

\vspace{0.1cm}
\noindent
Let us finally remark that from the above-mentioned 
Steps D1 to D4  (and analogously S1 to S4 below)  
one can see that for our BS method we 
basically need only a fast and accurate 
--- pseudo, true, natural, quantum  ---
random number generator. The corresponding computations can be principally 
run in parallel, and require relatively moderate computer memory/storage;
a detailed discussion 
is beyond the scope of this paper,
given its current length.

\vspace{-0.1cm}


\subsection{Estimators for the statistical minimization problem}
\label{Subsect Estimators stoch} 

\subsubsection{Estimation of 
$\mathbb{\Pi}_{X_{1}^{n}}[\boldsymbol{\xi}_{n,\mathbf{X}}^{w\mathbf{W}}\in 
\textrm{$\boldsymbol{\Omega}$\hspace{-0.23cm}$\boldsymbol{\Omega}$}]$} \ 

\vspace{0.1cm}
\noindent
In relation with Theorem \ref{brostu3:thm.divnormW.new} --- by making use of the
notations in \eqref{I^(n)_k for stat case},\eqref{cv emp measure X to P vector},\eqref{brostu3:fo.norweiemp.vec} ---
we start by remarking that 
the development of the estimator $\widehat{\Pi}_{L}^{improved}$
works quite analogously to that of $\widehat{\widetilde{\Pi}}_{L}^{improved}$
in the previous Subsection \ref{Subsect Estimators determ}.
To make this even more transparent, we 
label (w.l.o.g.) all random vectors of length $n$ 
in the same way as above: we sort the already given and thus fixed data $X_{i}$'s in such a way 
that the first $n_{1}$ of them share the same value $d_{1}$, and so on, until the last block with length 
$n_{K}$ in which the data have common value $d_{K}$.  
With this, analogously to Subsection \ref{Subsect Estimators determ}
we apply again an \textquotedblleft efficient Importance Sampling (IS)\textquotedblright\ 
scheme in the sense of Sadowsky \& Bucklew \cite{Sad:90}.
This will involve the simulation of $L$
independent $n-$tuples $\mathbf{V}^{(\ell)}\mathbf{:=}\left(
V_{n}^{(\ell)},\ldots,V_{n}^{(\ell)}\right) $ with common distribution $S$ on $\mathbb{R}^{n}$, 
such that $\mathbb{\bbzeta}^{\otimes n}$ is 
(measure-)equivalent with respect to $S$.
In fact, we rewrite 
$\mathbb{\Pi}_{X_{1}^{n}}[\boldsymbol{\xi}_{n,\mathbf{X}}^{w\mathbf{W}}\in 
\textrm{$\boldsymbol{\Omega}$\hspace{-0.23cm}$\boldsymbol{\Omega}$}]$ as

\vspace{-0.25cm}

\begin{equation}
\mathbb{\Pi}_{X_{1}^{n}}[\boldsymbol{\xi}_{n,\mathbf{X}}^{w\mathbf{W}}\in 
\textrm{$\boldsymbol{\Omega}$\hspace{-0.23cm}$\boldsymbol{\Omega}$}] =
E_{S}\Big[\frac{\mathrm{d}\mathbb{\bbzeta} ^{\otimes n}}{\mathrm{d}S}(V_{1},\ldots ,V_{n})\cdot 
\mathbf{1}_{\textrm{$\boldsymbol{\Omega}$\hspace{-0.19cm}$\boldsymbol{\Omega}$}}(
\boldsymbol{\xi}_{n,\mathbf{X}}^{w\mathbf{V}})\Big]
\label{imposa statistical}
\end{equation}
where $S$ designates any IS distribution of the vector 
$\mathbf{V:=}(V_{1},\ldots ,V_{n})$, and 
$E_{S}[ \, \cdot \, ]$ denotes the corresponding expectation operation.
Notice that $S$ is a \textit{random} probability distribution on $\mathbb{R}^{n}$;
in fact, $S$ is a conditional probability distribution given $X_{1}^{n}$,
and thus it would be more precise to write $S | X_{1}^{n}$ instead of $S$;
for the sake of brevity, we omit $| X_{1}^{n}$.
As a consequence of \eqref{imposa statistical}, for adequately chosen $S$, 
an improved estimator of 
$\mathbb{\Pi}_{X_{1}^{n}}[\boldsymbol{\xi}_{n,\mathbf{X}}^{w\mathbf{W}}\in 
\textrm{$\boldsymbol{\Omega}$\hspace{-0.23cm}$\boldsymbol{\Omega}$}]$
is given by 

\vspace{-0.25cm}

\begin{equation}
\widehat{\Pi}_{L}^{improved}
:= \frac{1}{L}\sum_{\ell=1}^{L}\frac{\mathrm{d}
\mathbb{\bbzeta}^{\otimes n}}{\mathrm{d}S}(V_{1}^{(\ell)},\ldots ,V_{n}^{(\ell)})\cdot
\mathbf{1}_{\textrm{$\boldsymbol{\Omega}$\hspace{-0.19cm}$\boldsymbol{\Omega}$}}(\boldsymbol{\xi}_{n,\mathbf{X}}^{w\mathbf{V}^{(\ell)}}) \, , 
\label{Pi-improved power case}
\end{equation}
which by the virtue of \eqref{LDP Normalized Vec} also estimates 
$\inf_{\mathds{Q}\in \textrm{$\boldsymbol{\Omega}$\hspace{-0.19cm}$\boldsymbol{\Omega}$} }
\ \inf_{m\neq 0} 
D_{\varphi}(m\cdot \mathds{Q},\mathds{P})$ (which we here suppose to be in $]0,\infty[$
as well as $\mathds{P}_{n}^{emp} \notin cl(\textrm{$\boldsymbol{\Omega}$\hspace{-0.23cm}$\boldsymbol{\Omega}$})$ 
for large enough $n$).
Let us now deal with the concrete construction of a reasonable $S$.
For this, as above, we need a particular $\mathbf{Q}^{\ast} \in 
int(\textrm{$\boldsymbol{\Omega}$\hspace{-0.23cm}$\boldsymbol{\Omega}$})$.
This (i) may be given in advance or (ii) it may be achieved by simulation;
in the following, we only work with the latter. Indeed,
given some (typically) large integer $M$, we employ a realization 
$\mathbf{W}^{\ast}:=\left( W_{1}^{\ast},\ldots,W_{M}^{\ast}\right)$ 
such that $\mathbf{Q}^{\ast}:=
\boldsymbol{\xi}_{M,\mathbf{X}}^{w\mathbf{W}^{\ast}} \in 
int(\textrm{$\boldsymbol{\Omega}$\hspace{-0.23cm}$\boldsymbol{\Omega}$})$,
by drawing replicates $\mathbf{W} = (W_{1}, \ldots, W_{M})$
under $\mathbb{\bbzeta}^{\otimes M}$ 
until the first time where 
$\boldsymbol{\xi}_{M,\mathbf{X}}^{w\mathbf{W}}$ 
belongs to $int(\textrm{$\boldsymbol{\Omega}$\hspace{-0.23cm}$\boldsymbol{\Omega}$})$;
(only) at this point, for consistency we artificially add $m-1$ copies of each observed data point
and transparently keep the same notations, leading e.g. to
$p_{M,k}^{emp} = p_{n,k}^{emp}$ for $M := m \cdot n$.
Notice that by the nature of $\textrm{$\boldsymbol{\Omega}$\hspace{-0.23cm}$\boldsymbol{\Omega}$}$,
$\mathbf{Q}^{\ast}$ is a probability vector which has the $K$ components

\vspace{-0.25cm}
\begin{equation}
q_{k}^{\ast}:=\sum_{i=1}^{M}\frac{W_{i}^{\ast}}{\sum_{j=1}^{M}W_{j}^{\ast }
}\mathbf{1}_{\{d_{k}\}}(X_{i}),
\hspace{1.5cm} k=1,\ldots,K.
\label{q_k^*}
\end{equation}
Before we proceed, let us give the 
substantial remark that changing 
$\left( V_{1},\ldots,V_{n}\right) $ 
drawn under $S$ to $\left(
c \cdot V_{1}, \ldots ,c \cdot V_{n}\right) $ for any 
$c\neq 0$ 
yields $
\boldsymbol{\xi}_{n,\mathbf{X}}^{w\mathbf{V}}
= \boldsymbol{\xi}_{n,\mathbf{X}}^{w\, c\cdot\mathbf{V}}$ 
so that the distribution $S$ is not uniquely determined. 
Amongst all candidates, we choose the --- uniquely determined --- 
$S$ which is the Kullback-Leibler-divergence projection of $\mathbb{\bbzeta}^{\otimes n}$ 
on the set of all probability distributions on $\mathbb{R}^{n}$
such that the $K$ \textquotedblleft  non-normalized\textquotedblright\ moment constraints

\vspace{-0.2cm}

\begin{equation}
E_{S}[ \boldsymbol{\xi}_{n,\mathbf{X}}^{\mathbf{V}} ] =
\boldsymbol{\xi}_{M,\mathbf{X}}^{\mathbf{W}^{\ast}}
\label{IS moment constraints}
\end{equation}

\vspace{-0.1cm}
\noindent
(rather than the normalized 
$E_{S}[ \boldsymbol{\xi}_{n,\mathbf{X}}^{w\mathbf{V}}] =
\boldsymbol{\xi}_{M,\mathbf{X}}^{w\mathbf{W}^{\ast}}$)
are satisfied, with the non-normalized vectors

\vspace{-0.2cm}

\[
\boldsymbol{\xi}_{M,\mathbf{X}}^{\mathbf{W}^{\ast}} :=
\bigg( \frac{1}{M}\sum_{j=1}^{M}W_{j}^{\ast}\bigg) \cdot
Q^{\ast} =: \overline{W^{\ast}} \cdot Q^{\ast},
\qquad
\boldsymbol{\xi}_{n,\mathbf{X}}^{\mathbf{V}} :=
\bigg( \frac{1}{n}\sum_{j=1}^{n}V_{j}\bigg) \cdot
\boldsymbol{\xi}_{n,\mathbf{X}}^{w\mathbf{V}} \ .
\]

\noindent
As already indicated above,
this projection $S$ is a well-determined unique distribution on $\mathbb{R}^{n}$
and --- as we shall see in Proposition \ref{Proposition S} below --- 
it is such that 
$\boldsymbol{\xi}_{n,\mathbf{X}}^{w\mathbf{V}}$ belongs to 
$\textrm{$\boldsymbol{\Omega}$\hspace{-0.23cm}$\boldsymbol{\Omega}$}$
with probability bounded away from $0$ as $n$
increases, when $\left( V_{1},\ldots,V_{n}\right) $ are drawn under $S$.
Therefore, this IS distribution produces an estimate of
$\mathbb{\Pi}_{X_{1}^{n}}[\boldsymbol{\xi}_{n,\mathbf{X}}^{w\mathbf{W}}\in 
\textrm{$\boldsymbol{\Omega}$\hspace{-0.23cm}$\boldsymbol{\Omega}$}]$.

\vspace{0.3cm}
\noindent
In order to justify the above construction of $S$, we give the following
result, which states that the IS sampling distribution $S$ yields a good
hitting rate. Its proof will be given in Appendix G.

\vspace{0.3cm}

\begin{proposition}
\label{Proposition S}
With the above definition of $S$, $\lim
\inf_{n\rightarrow \infty }$ 
$ S
\left[ \boldsymbol{\xi}_{n,\mathbf{X}}^{w\mathbf{V}} \in
\textrm{$\boldsymbol{\Omega}$\hspace{-0.23cm}$\boldsymbol{\Omega}$}
\right] $
is bounded away from $0$.
\end{proposition}

\vspace{0.3cm}
\noindent
We now come to the detailed construction of $S$. 
The constraints \eqref{IS moment constraints} can be written in explicit form as
\begin{equation}
E_{S}\Big[ \, \frac{1}{n_{k}}\sum_{i\in I_{k}^{(n)}} V_{i} \, \Big] 
\ = \ \overline{W^{\ast}} \cdot q_{k}^{\ast}  \, , \qquad k=1,\ldots,K.
\label{k_th moment constraint}
\end{equation}
The distribution $S$ can be obtained by blocks.
Indeed, let us define $S^{k}$ as the
Kullback-Leibler-divergence (KL) projection of $\mathbb{\bbzeta}^{\otimes n_{k}}$ on the set of all
distributions on $\mathbb{R}^{n_{k}}$ such that \eqref{k_th moment constraint} holds.
We define the resulting $S$ as the product distribution of those $S^{k}$ 's.
To obtain the latter, we start by defining
$U_{k}$ as the KL projection of $\mathbb{\bbzeta} $ on the set of all measures $Q$
on $\mathbb{R}$ under \eqref{k_th moment constraint}. Then, 
\begin{equation}
dU_{k}(v)= \exp (\tau_{k} \cdot v-\Lambda_{\mathbb{\bbzeta} }\left( \tau _{k}\right) ) 
\ d\mathbb{\bbzeta} (v) \, , 
\nonumber 
\end{equation}
where 
$\tau_{k} \in int(dom(MGF_{\mathbb{\bbzeta}}))$ is (under the appropriately adapted Assumption (OM))
the unique solution of the equation
$ \Lambda_{\mathbb{\bbzeta} }^{\prime}\left( \tau_{k}\right) =
\overline{W^{\ast}} \cdot \frac{q_{k}^{\ast}}{ p_{n,k}^{emp} } $
and thus --- by relation \eqref{proof Ftheorem equ 3} of Appendix E --- we can compute explicitly
$\tau_{k} = \varphi^{\prime}\left( \frac{\overline{W^{\ast}} \cdot q_{k}^{\ast}}{
p_{n,k}^{emp}} \right) .
$
The distribution $S^{k}$ is then defined by
$ S^{k}:=\underbrace{U_{k}\otimes \cdots \otimes U_{k}}_{n_{k}\text{times}} $
from which we obtain 
$ S := S^{1}\otimes \cdots \otimes S^{K}. $
With this construction, it holds 
\[
\frac{dS}{d\mathbb{\bbzeta} ^{\otimes n}}(v_{1,1},\ldots v_{1,n_{1}},
\ldots, v_{K,1},\ldots v_{K,n_{K}})=
\exp\bigg(
\sum\limits_{k=1}^{K} \sum_{i\in I_{k}^{(n)}} \Big(
\tau_{k} \cdot v_{k,i}-\Lambda_{\mathbb{\bbzeta}}(
\tau_{k} )\Big) 
\bigg)
\]
which proves that $S$ is indeed the KL projection of $\mathbb{\bbzeta} ^{\otimes
n}$ we aimed at.
Therefore, $\mathbf{V}$ is composed of $K$ independent blocks of length $n_{k}
$ each, and the $k-$th subvector $\mathbf{V}_{k}$ consists of all the 
random variables $V_{i}$ whose index $i$ satisfies $X_{i}=d_{k}.$ Within $\mathbf{V}_{k}$, all
components are i.i.d. with same distribution $U_{k}$ on $\mathbb{R}$ defined
through 
$ \frac{dU_{k}}{d\mathbb{\bbzeta} }(u)=\exp \left\{ \tau _{k}\cdot u-\Lambda_{\mathbb{\bbzeta}
}(\tau _{k})\right\} =\frac{\exp \left\{ \tau _{k}\cdot u\right\} }{
MGF_{\mathbb{\bbzeta} }(\tau _{k})}, $\\
which leads to the moment generating function
$ dom(MGF_{\mathbb{\bbzeta}})-\tau_{k} \ \ni \ z \ \mapsto
 MGF_{U_{k}}(z):=\int_{\mathbb{R}}e^{zy}dU_{k}(y)=\frac{MGF_{\mathbb{\bbzeta}
}(z+\tau _{k})}{MGF_{\mathbb{\bbzeta} }(\tau _{k})}. $
Notice that $U_{k}$ is a distorted
distribution of $\mathbb{\bbzeta}$ with the distortion
parameter $\tau _{k}$
(in some cases, this distortion even becomes a tilting/dampening). 
The estimator $\widehat{\Pi}_{L}^{improved}$ defined in 
\eqref{Pi-improved power case} can be implemented through the following algorithm:

\vspace{0.2cm}
\noindent
\textbf{Step S1.}
\noindent
Choose some (typically large) $M$ and simulate (in the above-described fashion)
repeatedly i.i.d. vectors $\left(
W_{1},\ldots,W_{M}\right) $ --- whose independent components have common
distribution $\mathbb{\bbzeta} $ 
--- until $\boldsymbol{\xi}_{M,\mathbf{X}}^{w\mathbf{W}}$
belongs to $\textrm{$\boldsymbol{\Omega}$\hspace{-0.23cm}$\boldsymbol{\Omega}$}$.
Call  $\left( W_{1}^{\ast },\ldots,W_{M}^{\ast }\right) $ the corresponding
vector and $\overline{W^{\ast }}$ the arithmetic mean of its components.\
Moreover, denote by
$\boldsymbol{\xi}_{M,\mathbf{X}}^{w\mathbf{W}^{\ast }}$
the corresponding normalized weighted empirical measure, identified with the $K-$component vector 
$Q^{\ast} := (q_{1}^{\ast },\ldots,q_{K}^{\ast })$ with $q_{k}^{\ast }$ defined in \eqref{q_k^*}.\\[-0.2cm] 

\noindent
\textbf{Step S2.}
For all $k \in \{1,\ldots,K\}$ 
compute $\tau_{k} = \varphi^{\prime}\left( \frac{\overline{W^{\ast}} \cdot q_{k}^{\ast}}{
p_{n,k}^{emp}
} \right)$.

\noindent
\textbf{Step S3.}
For all $\ell \in \{1,\ldots,L\}$ simulate independently for all 
$k \in \{1,\ldots,K\}$ 
a row vector $\mathbf{V}_{k}^{(\ell)}$ $:=\left(
V_{k_{1}}^{(\ell)},...,V_{k_{n_{k}}}^{(\ell)}\right) $ with independent components
with common distribution $U_{k}$ defined above.
Concatenate these
vectors to the row vector $\mathbf{V}^{(\ell)}.$\\
\noindent
\textbf{Step S4.}
Compute the estimator $\widehat{\Pi }_{L}^{improved}$ by making use of the formula
\eqref{Pi-improved power case} which turns into the explicit form
\begin{eqnarray}
& & \widehat{\Pi }_{L}^{improved} =\frac{1}{L}\sum_{\ell=1}^{L}
\exp\bigg(
\sum\limits_{k=1}^{K}\bigg( 
n_{k} \cdot \Lambda_{\mathbb{\bbzeta} }(\tau_{k}) -
\tau_{k} \cdot \sum_{i\in I_{k}^{(n)}}V_{i}^{(\ell)}
\bigg) \bigg)
\cdot \mathbf{1}_{\textrm{$\boldsymbol{\Omega}$\hspace{-0.19cm}$\boldsymbol{\Omega}$}}
\left( 
\boldsymbol{\xi}_{n,\mathbf{X}}^{w\mathbf{V}^{(\ell)}}
\right).   
\label{improved IS empirical measure stat section}
\end{eqnarray}

\vspace{0.3cm}
\noindent
Analogous to the paragraph right after \eqref{estimator minimization},
in many cases we may improve the simulation burden needed for the computation of 
the estimator $\widehat{\Pi }_{L}^{improved}$. 
In fact, in terms of the notations
$\widehat{\mathit{W}}_{k}^{(\ell)}:=\sum_{i\in I_{k}^{(n)}}V_{i}^{(\ell)}$
we can rewrite \eqref{improved IS empirical measure stat section} as
\begin{eqnarray}
& & \hspace{-3.2cm}
\widehat{\Pi }_{L}^{improved} =\frac{1}{L}\sum_{\ell=1}^{L}
\mathbf{1}_{\textrm{$\boldsymbol{\Omega}$\hspace{-0.19cm}$\boldsymbol{\Omega}$}}
\left( \boldsymbol{\xi}_{n,\mathbf{X}}^{w\mathbf{V}^{(\ell)}}
\right) 
\cdot \prod_{k=1}^{K}
ISF_{k}\left(\widehat{\mathit{W}}_{k}^{(\ell)}
\right)  
\label{improved IS empirical measure stat section 2}
\\
& & \hspace{-3.2cm} \textrm{with} \qquad
ISF_{k}(x) \ := \ 
\exp(n_{k} \cdot \Lambda_{\mathbb{\bbzeta}}(\tau_{k}) \, - \, x \cdot \tau_{k} ) 
\label{ISK2}
\\
\textrm{and} \qquad
\boldsymbol{\xi}_{n,\mathbf{X}}^{w\mathbf{V}^{(\ell)}} &=&
\begin{cases}
\left(\frac{\widehat{\mathit{W}}_{1}^{(\ell)}}{\sum_{k=1}^{K}\widehat{\mathit{W}}_{k}^{(\ell)}},
\ldots, \frac{\widehat{\mathit{W}}_{K}^{(\ell)}}{\sum_{k=1}^{K}\widehat{\mathit{W}}_{k}^{(\ell)}} \right) ,
\qquad \textrm{if } \sum_{k=1}^{K}\widehat{\mathit{W}}_{k}^{(\ell)} \ne 0, \\
\ (\infty, \ldots, \infty) =: \boldsymbol{\infty}, \hspace{2.3cm} \textrm{if }
\sum_{k=1}^{K}\widehat{\mathit{W}}_{k}^{(\ell)} = 0 \, .
\end{cases}
\label{brostu3:fo.norweiemp.vecVell} 
\end{eqnarray}
Clearly, the random variable $\widehat{\mathit{W}}_{k}^{(\ell)}$ 
($k=1, \ldots, K$)
has distribution $U_{k}^{\ast n_{k}}$. Hence, if 
$U_{k}^{\ast n_{k}}$ can be \textit{explicitly} constructed,
then for the computation of $\widehat{\Pi}_{L}^{improved}$
it suffices to independently simulate the $K \cdot L$ random variables 
$\widehat{\mathit{W}}_{k}^{(\ell)}$ 
(rather than the $n \cdot L$ random variables $V_{i}^{(\ell)}$).

\vspace{0.2cm}


\subsubsection{Direct BS-estimability of 
$D_{\varphi}
(\textrm{$\boldsymbol{\Omega}$\hspace{-0.23cm}$\boldsymbol{\Omega}$},\mathds{P})$
} \ 

\vspace{0.1cm}
\noindent
For the cases $\varphi := \widetilde{c}\cdot \varphi_{\gamma}$ (cf. \eqref{brostu3:fo.powdivgen}),
by making use of 
Theorem \ref{brostu3:thm.divnormW.new}, 
Lemma \ref{Lemma Indent Rate finite case_new} (especially
\eqref{Inf in Lemma rate case finite 1},\eqref{brostu3:fo.mmin-KL1},\eqref{brostu3:fo.mmin-RKL1})
and the results of the previous subsection,
we get through inversion the \textit{power-divergence estimators
(BS estimators of power divergences)}
\begin{eqnarray}
&& 
\hspace{-1.9cm}
\widehat{D_{\widetilde{c}\cdot \varphi
_{\gamma}}(\textrm{$\boldsymbol{\Omega}$\hspace{-0.23cm}$\boldsymbol{\Omega}$},\mathds{P})
}
\ := \ -
\frac{\widetilde{c}}{\gamma (\gamma -1)}\left\{ 1-\left( 1+\frac{\gamma }{
\widetilde{c}} \cdot \frac{1}{n} \cdot \log \widehat{\Pi}_{L}^{improved}\right)
^{1-\gamma }\right\} ,
\qquad  
\gamma \in \, ]-\infty,0[ \, \cup \, ]0,1[ \, \cup \, [2,\infty[,
\nonumber
\\
& & 
\hspace{-1.9cm}
\widehat{D_{\widetilde{c}\cdot \varphi
_{0}}(\textrm{$\boldsymbol{\Omega}$\hspace{-0.23cm}$\boldsymbol{\Omega}$},\mathds{P})
} \ := \ 
-\frac{1}{n}\log \widehat{\Pi}_{L}^{improved} ,
\hspace{5.8cm} \gamma =0,
\nonumber
\\
& & 
\hspace{-1.9cm}
\widehat{D_{\widetilde{c}\cdot \varphi
_{1}}(\textrm{$\boldsymbol{\Omega}$\hspace{-0.23cm}$\boldsymbol{\Omega}$},\mathds{P})
} \ : = \ 
- \widetilde{c} \cdot \log \left( 1+\frac{1}{\widetilde{c}} \cdot \frac{1}{n} \cdot
\log \widehat{\Pi}_{L}^{improved}\right) , 
\hspace{3.0cm} \gamma =1.
\nonumber
\end{eqnarray}
For more details including the correspondingly involved simulation distributions $\bbzeta$, see the
solved-cases Section \ref{Sect Cases} below.

\vspace{0.2cm}


\subsubsection{
BS-estimability of bounds of
$D_{\varphi}
(\textrm{$\boldsymbol{\Omega}$\hspace{-0.23cm}$\boldsymbol{\Omega}$},\mathds{P})$
} \ 
\label{subsub-estimator-GC2} 

\vspace{0.1cm}
\noindent
For divergence cases which are not directly BS-estimable, we present now
the algorithm for the BS evaluation of
the bounds
\begin{equation}
\inf_{m\neq 0}
D_{\varphi }\left( m \cdot
\textrm{$\boldsymbol{\Omega}$\hspace{-0.23cm}$\boldsymbol{\Omega}$},\mathds{P} 
\right)
=
\inf_{\mathds{Q}\in \textrm{$\boldsymbol{\Omega}$\hspace{-0.19cm}$\boldsymbol{\Omega}$}}
D_{\varphi }\left( m\left( \mathds{Q}\right) \cdot \mathds{Q},\mathds{P}\right)
\stackrel{(\triangle)}{=} D_{\varphi }\left( m(\mathds{Q}^{\ast})\cdot 
\mathds{Q}^{\ast},\mathds{P}\right) \leq D_{\varphi
}\left( \textrm{$\boldsymbol{\Omega}$\hspace{-0.23cm}$\boldsymbol{\Omega}$},\mathds{P}\right) \leq D_{\varphi }\left( \mathds{Q}^{\ast},\mathds{P}\right) 
\label{estimator-general-divbounds-D(W,P)}
\end{equation}
obtained in Subsection \ref{Sub Bounds General div},
where $\mathds{Q}^{\ast}$
satisfies the above equality $(\triangle)$.
The estimator of the lower bound in \eqref{estimator-general-divbounds-D(W,P)} is\\
$\widehat{D}:=-\frac{1}{n}\log \widehat{\text{ }\Pi }_{L}^{improved}$
defined in \eqref{improved IS empirical measure stat section}.
We now turn to an estimate of the upper bound. 
Consider for any fixed $\mathds{Q}:=\left( q_{1},\ldots,q_{K}\right) $ in 
$\mathbb{S}_{>0}^{K}$ the real number $m_{n}(\mathds{Q})$ which satisfies
$D_{\varphi }\left( m_{n}(\mathds{Q}) \cdot \mathds{Q},
\mathds{P}_{n}^{emp} \right) =
\inf_{m\neq 0}D_{\varphi }\left( m\cdot \textrm{$\boldsymbol{\Omega}$\hspace{-0.23cm}$\boldsymbol{\Omega}$}, \mathds{P}_{n}^{emp}
\right) $
where $\mathds{P}_{n}^{emp}$ was defined in the course of \eqref{I^(n)_k for stat case}.
Such $m_{n}(\mathds{Q})$ is well defined for all $\mathds{Q}$ 
since it satisfies the equation (in $m$)
\begin{equation}
\frac{d}{dm}D_{\varphi }\left( m \cdot \mathds{Q},
\mathds{P}_{n}^{emp}
\right) =\sum_{k=1}^{K} q_{k} \cdot \varphi^{\prime }\left( \frac{m \cdot q_{k}}{
p_{n,k}^{emp}
}\right) =0 .
 \label{equ m(Q)}
\end{equation}
Since the mapping $m\rightarrow D_{\varphi }\left( m \cdot \mathds{Q},\mathds{P}\right) $ is convex
and differentiable, existence and uniqueness of $m_{n}(\mathds{Q})$ hold; furthermore, 
$m_{n}(\mathds{Q})\in \left] \min_{k}
p_{n,k}^{emp}
/q_{k},\max_{k}
p_{n,k}^{emp}
/q_{k}\right[ $
since $\frac{d}{dm}D_{\varphi }\left( m \cdot \mathds{Q},
\mathds{P}_{n}^{emp}
\right) $ is negative when 
$m=\min_{k}
p_{n,k}^{emp}
/q_{k}$ and positive when $m=\max_{k}
p_{n,k}^{emp}
/q_{k}$.
An estimate of the distribution $\mathds{Q}^{\ast}$ is required. This can be
achieved as follows:
\begin{itemize}
\item Estimate $\inf_{m\neq 0}D_{\varphi }\left( m \cdot
\textrm{$\boldsymbol{\Omega}$\hspace{-0.23cm}$\boldsymbol{\Omega}$}, \mathds{P}\right)$ 
through 
$\widehat{D} := -\frac{1}{n}\log \widehat{\Pi}_{L}^{improved}$
defined in \eqref{improved IS empirical measure stat section}.

\item Set $i=0.$

\item Get some $\mathds{Q}_{i} := (q_{i,1}, \ldots, q_{i,K})$ in 
$\textrm{$\boldsymbol{\Omega}$\hspace{-0.23cm}$\boldsymbol{\Omega}$}$;
 this can be obtained by simulating runs of
vectors $\left( W_{1},\ldots\right) $ through  i.i.d. sampling under $\mathbb{\bbzeta}$.
Evaluate $m_{n}(\mathds{Q}_{i})$ by solving \eqref{equ m(Q)}
(with $q_{i,k}$ instead of $q_{k}$) for $m$, which is a fast calculation by the
bisection method.

\item If $D_{\varphi }\left( m_{n}(\mathds{Q}_{i}) \cdot 
\mathds{Q}_{i},
\mathds{P}_{n}^{emp}
\right) <\widehat{D}+\eta $ 
for some small $\eta >0$, then the proxy of $\mathds{Q}^{\ast}$ is $\mathds{Q}_{i},$
denoted by $\widehat{\mathds{Q}^{\ast}}$.

\item Else set $i\leftarrow i+1$ and get $\mathds{Q}_{i}$ in 
$\textrm{$\boldsymbol{\Omega}$\hspace{-0.23cm}$\boldsymbol{\Omega}$} 
\cap \left\{
\mathds{Q} \, : \, D_{\varphi }\left( \mathds{Q},
\mathds{P}_{n}^{emp}
\right) <D_{\varphi }\left( \mathds{Q}_{i-1},
\mathds{P}_{n}^{emp}
\right)
\right\} $ and iterate.
\end{itemize}
That this algorithm converges in the sense that it produces some 
$\widehat{\mathds{Q}^{\ast }}$ is clear.
Since by \eqref{estimator-general-divbounds-D(W,P)} 
\[
D_{\varphi }\left( m(\mathds{Q}^{\ast}) \cdot \mathds{Q}^{\ast },\mathds{P}\right) 
\leq D_{\varphi }\left(
\textrm{$\boldsymbol{\Omega}$\hspace{-0.23cm}$\boldsymbol{\Omega}$},\mathds{P}\right) 
\leq D_{\varphi }\left( \mathds{Q}^{\ast},\mathds{P}\right) ,
\]
we have obtained both estimated lower and upper bounds for 
$D_{\varphi}\left( \textrm{$\boldsymbol{\Omega}$\hspace{-0.23cm}$\boldsymbol{\Omega}$} ,
\mathds{P}\right)$.

That the upper bound is somehow optimal can be seen from the power-divergence Cases 1 to 6
developed in the solved-cases Section \ref{Sect Cases} below.
Indeed, in this context
the solution of equation \eqref{equ m(Q)} is explicit and produces $m(\mathds{Q})$ as
a function of $D_{\varphi }\left( \mathds{Q},\mathds{P}\right)$ through a Hellinger integral,
and the mapping $\mathds{Q}\rightarrow D_{\varphi }\left( m(\mathds{Q})
\cdot \mathds{Q},\mathds{P}\right)$ is
increasing with respect to $D(\mathds{Q},\mathds{P})$. 
Hence, $\mathds{Q} \rightarrow \inf_{m\neq 0}  
D_{\varphi }\left( m \cdot \mathds{Q},\mathds{P}\right)$ is minimal
when $D_{\varphi }\left( \mathds{Q},\mathds{P}\right) $ is minimal as 
$\mathds{Q} \in \textrm{$\boldsymbol{\Omega}$\hspace{-0.23cm}$\boldsymbol{\Omega}$}$.
Therefore, 
$\mathds{Q}^{\ast}\in \arg \inf_{\mathds{Q}\in
\textrm{$\boldsymbol{\Omega}$\hspace{-0.19cm}$\boldsymbol{\Omega}$}}
D_{\varphi }\left( m(\mathds{Q}) \cdot \mathds{Q},\mathds{P}\right) $ also satisfies 
$\mathds{Q}^{\ast}\in
\arg \inf_{\mathds{Q}\in \textrm{$\boldsymbol{\Omega}$\hspace{-0.19cm}$\boldsymbol{\Omega}$}}
D_{\varphi }\left( \mathds{Q},\mathds{P}\right)$.

%
%

\section{Finding/Constructing the distribution of the weights \label{SectFind}}

\noindent
Recall first that in Theorem \ref{brostu3:thm.divnormW.new},
one crucial component is the sequence $(W_{i})_{i\in \mathbb{N}}$ of weights
being i.i.d. copies of a random variable $W$ whose probability distribution 
is $\mathbb{\bbzeta}$ (i.e. $\mathbb{\Pi}[W \in \cdot \, ] = \mathbb{\bbzeta}[ \, \cdot \,]$),
where the latter has to be connected with the divergence generator 
$\varphi \in \Upsilon (]a,b[)$ through the representation 
\begin{equation*}
\varphi (t)=\sup_{z\in \mathbb{R}}\bigg( z\cdot t-\log \int_{\mathbb{R}
}e^{zy}d\mathbb{\bbzeta} (y)\bigg), \qquad t\in \mathbb{R},\ \ 
\hspace{1.0cm} (cf. \ \eqref{Phi Legendre of mgf(W)})
\end{equation*}
under the additional requirement that the
function $z\mapsto MGF_{\mathbb{\bbzeta} }(z):=\int_{\mathbb{R}}e^{zy}d\mathbb{\bbzeta} (y)$ is
finite on some open interval containing zero (\textquotedblleft light-tailedness\textquotedblright );
for Theorem \ref{brostu3:thm.divW.var},
we need the corresponding variant \eqref{brostu3:fo.link.var} for 
$M_{\mathbf{P}} \cdot \varphi \in \Upsilon (]a,b[)$ (rather than $\varphi$). 

Hence, finding such \textquotedblleft BS-associated pairs $(\varphi,\mathbb{\bbzeta})$\textquotedblright\ is an 
important --- but highly nontrivial --- issue. 
Indeed, one approach is to start from a given divergence generator $\varphi \in \widetilde{\Upsilon}(]a,b[)$ having some additional properties, switch to its Fenchel-Legendre transform 
$\varphi^{*}$ (and some exponentially-linear
transforms thereof), and verify some sufficient conditions for the outcome 
to be a moment-generating function $MGF_{\mathbb{\bbzeta}}$ of a unique probability distribution $\mathbb{\bbzeta}$ which has light tails.
For finding the concrete $\mathbb{\bbzeta}$, one typically should know the explicit form of $\varphi^{*}$.
However, it is well known that it can sometimes be hard 
to determine the explicit form of the  
Fenchel-Legendre transform of a convex function.
This hardness issue also applies for the reverse direction of starting from a concrete
probability distribution $\mathbb{\bbzeta}$ with light tails, computing its log-moment-generating function
(called cumulant-generating function) 
$z \mapsto \Lambda_{\mathbb{\bbzeta}}(z) := \log MGF_{\mathbb{\bbzeta}}(z)$
and the corresponding Fenchel-Legendre transform $\Lambda_{\mathbb{\bbzeta}}^{*}$ 
which is nothing but the associated divergence generator $\varphi$ 
(cf. \eqref{Phi Legendre of mgf(W)}). 

\vspace{0.2cm}
\noindent
As will be demonstrated via numerous solved cases in the following Section \ref{Sect Cases}, 
the  --- \textquotedblleft kind of intermediate\textquotedblright\ --- new construction method
given in the below-mentioned Theorem \ref{brostu3:thm.Wfind.new} can help to ease 
these two tasks. 
To formulate this, we employ the class $\textgoth{F}$ of functions $F: \, ]-\infty,\infty[ \, \mapsto [-\infty,\infty]$
with the following properties:

\begin{enumerate}

\item[(F1)] $int(dom(F)) = \, ]a_{F},b_{F}[$ for some $-\infty \leq a_{F} < 1 < b_{F} \leq \infty$;

\item[(F2)] $F$ is smooth (infinitely continuously differentiable) on $]a_{F},b_{F}[$;

\item[(F3)] $F$ is strictly increasing on $]a_{F},b_{F}[$.

\end{enumerate}

\noindent
Clearly, for any $F \in \textgoth{F}$ one gets the existence of
$F(a_{F}) := \lim_{t \downarrow a_{F}} F(t) \in [-\infty, \infty[$ and
$F(b_{F}) := \lim_{t \uparrow b_{F}} F(t) \in \, ]-\infty, \infty]$;
moreover, its inverse $F^{-1}: \mathcal{R}(F) \mapsto [a_{F},b_{F}]$
exists, where $\mathcal{R}(F) := \{F(t): t \in dom(F) \}$. Furthermore, $F^{-1}$
is strictly increasing and smooth (infinitely continuously differentiable)
on the open interval $int(\mathcal{R}(F)) = \{F(t): t \in \, ]a_{F},b_{F}[ \}
=]F(a_{F}),F(b_{F})[$,
and $F^{-1}(int(\mathcal{R}(F))) = \, ]a_{F},b_{F}[$. Within such a context,
we obtain

\vspace{0.2cm}

\begin{theorem} 
\label{brostu3:thm.Wfind.new}
Let $F \in \textgoth{F}$  and fix an arbitrary point $c \in int(\mathcal{R}(F))$. 
Moreover, introduce the notations\footnote{for the sake of brevity, we avoid
here the more complete notation $\lambda_{-}^{F,c}$, $\lambda_{+}^{F,c}$,
$t_{-}^{sc,F,c}$, $t_{+}^{sc,F,c}$ indicating the dependence on $F$ and $c$.
}
 $]\lambda_{-},\lambda_{+}[ \, := \, int(\mathcal{R}(F)) -c$ \, 
and \, $]t_{-}^{sc},t_{+}^{sc}[ \, := \, ]1+a_{F}-F^{-1}(c),1+b_{F}-F^{-1}(c)[$ \, 
(which implies $\lambda_{-} < 0 < \lambda_{+}$ and $t_{-}^{sc} < 1 < t_{+}^{sc}$).\\
Furthermore, define the 
functions $\Lambda : \ ]-\infty,\infty[ \ \mapsto \, [-\infty, \infty]$
and $\varphi : \ ]-\infty,\infty[ \  \mapsto \, [0, \infty]$ by
\begin{eqnarray}
\hspace{-0.7cm} \Lambda(z) := \Lambda^{(c)}(z) \hspace{-0.2cm} &:=& \hspace{-0.2cm}
\begin{cases}
\int\displaylimits_{0}^{z} F^{-1}(u+c) \, du
+ z \cdot (1-F^{-1}(c)) \ \in \, ]-\infty, \infty[,
\hspace{2.85cm} \textrm{if }  z \in \, ]\lambda_{-},\lambda_{+}[, 
\\
\int\displaylimits_{0}^{\lambda_{-}} F^{-1}(u+c) \, du
+ \lambda_{-} \cdot (1-F^{-1}(c))  
\  \in [-\infty,\infty],
\hspace{2.5cm} \textrm{if }  z = \lambda_{-} > -\infty,
\\
\int\displaylimits_{0}^{\lambda_{+}} F^{-1}(u+c) \, du
+ \lambda_{+} \cdot (1-F^{-1}(c))  
\  \in \ [-\infty,\infty],
\hspace{2.35cm} \textrm{if }  z = \lambda_{+} < \infty, 
\\
\infty, \hspace{9.90cm} \textrm{else},
\end{cases}
\label{brostu3:fo.Wfind1.new} 
\end{eqnarray}
where the second respectively third line are meant as \,   
$\lim_{z \downarrow \lambda_{-}} \big( \int\displaylimits_{0}^{z} F^{-1}(u+c) \, du
+ z \cdot (1-F^{-1}(c)) \big)$ \
respectively \\
$\lim_{z \uparrow \lambda_{+}} \big( \int\displaylimits_{0}^{z} F^{-1}(u+c) \, du
+ z \cdot (1-F^{-1}(c)) \big)$, and 

\begin{eqnarray}
\hspace{-0.3cm} 
\varphi(t) := \varphi^{(c)}(t) \hspace{-0.2cm} &:=& \hspace{-0.2cm}
\begin{cases}
(t+F^{-1}(c)-1) \cdot
[ F\left(t+F^{-1}(c)-1 \right) - c \, ]
-\int\displaylimits_{0}^{F\left(t+F^{-1}(c)-1 \right) - c} F^{-1}(u+c) \, du
\ \in \, [0,\infty[,\\
\hspace{11.5cm}
\quad \textrm{if } 
t \in \, ]t_{-}^{sc},t_{+}^{sc}[,
\\
(t_{-}^{sc}+F^{-1}(c)-1) \cdot
[ F\left(t_{-}^{sc}+F^{-1}(c)-1 \right) - c \, ]
-\int\displaylimits_{0}^{F\left(t_{-}^{sc}+F^{-1}(c)-1 \right) - c} F^{-1}(u+c) \, du
\ \in \ ]0,\infty],\\
\hspace{11.5cm}
\quad \textrm{if } 
t = t_{-}^{sc} > -\infty,
\\
(t_{+}^{sc}+F^{-1}(c)-1) \cdot
[ F\left(t_{+}^{sc}+F^{-1}(c)-1 \right) - c \, ]
-\int\displaylimits_{0}^{F\left(t_{+}^{sc}+F^{-1}(c)-1 \right) - c} F^{-1}(u+c) \, du
\ \in \ ]0,\infty],\\
\hspace{11.5cm}
\quad \textrm{if } 
t = t_{+}^{sc} < \infty,
\\ 
\varphi(t_{-}^{sc}) + 
\lambda_{-} 
\cdot (t- t_{-}^{sc}) \ \in \ ]0,\infty],
\hspace{4.5cm} \textrm{if } 
t_{-}^{sc} > - \infty \ \textrm{and} \  t \in \, ]-\infty, t_{-}^{sc}[, 
\\ 
\varphi(t_{+}^{sc}) + 
\lambda_{+} 
\cdot (t - t_{+}^{sc}) \ \in \ ]0,\infty],
\hspace{4.5cm} \textrm{if } 
t_{+}^{sc} < \infty \ \textrm{and} \  t \in \, ]t_{+}^{sc}, \infty[, 
\\
\infty, \hspace{11.3cm} \textrm{else},
\end{cases}
\label{brostu3:fo.Wfind2.new}
\end{eqnarray}
where the second respectively third line are again meant as lower respectively upper limit.\\
Then, $\Lambda$ and $\varphi$
have the following properties:\\
\noindent
(i) \, On $]\lambda_{-},\lambda_{+}[$, the function $\Lambda$ is smooth and strictly convex
and consequently, $\exp(\Lambda)$ is smooth and strictly log-convex;
moreover, there holds $\Lambda(0) =0$, $\Lambda^{\prime}(0) =1$.
\\  
\noindent
(ii) \, $\varphi \in \widetilde{\Upsilon}(]a,b[)$,
where $a := t_{-}^{sc} \cdot \textfrak{1}_{ \{ -\infty \} }(\lambda_{-}) - \infty \cdot \textfrak{1}_{]-\infty,0[}(\lambda_{-})$, 
\ $b := t_{+}^{sc} \cdot \textfrak{1}_{ \{ \infty \} }(\lambda_{+}) + \infty \cdot \textfrak{1}_{]0,\infty[}(\lambda_{+})$. 
\\
\noindent
(iii) \, 
$
\varphi (t) = \Lambda^{*}(t) = 
\sup_{z \in ]-\infty,\infty[} \left( z\cdot t - 
\Lambda(z)\right) = 
\sup_{z \in ]\lambda_{-},\lambda_{+}[
} \left( z\cdot t - 
\Lambda(z)\right)$ for all $t \in \mathbb{R}$.\\
(iv) \  
$\Lambda(z) = \varphi^{*}(z) = 
\sup_{t \in ]-\infty,\infty[} \left( t\cdot z - 
\varphi(t)\right) = 
\sup_{t \in ]a,b[
} \left( t\cdot z - 
\varphi(t) \right)$ for all $z \in \mathbb{R}$.\\

\end{theorem}

\vspace{0.2cm}
\noindent
The proof of Theorem \ref{brostu3:thm.Wfind.new} will be given Appendix E.

\vspace{0.3cm}
\noindent

\begin{remark}
(a) Notice that the newly constructed $\Lambda$ and $\varphi$
(cf. \eqref{brostu3:fo.Wfind1.new},\eqref{brostu3:fo.Wfind2.new}) depend on the
choice of the \textit{anchor point} $c$; this is
e.g. illustrated in Subsection \ref{Subsect Case6} below.
Hence, as a side effect, by using whole families $(F_{\vartheta})_{\vartheta}$ 
together with different anchor points $c$, via Theorem \ref{brostu3:thm.Wfind.new}
one can generate new classes
(and new classifications)
of $\varphi-$divergence generators --- and thus of corresponding $\varphi-$divergences ---
which can be of great use, even in other contexts beyond our BS optimization framework.\\
(b) If $F$ satisfies $F(1)=0$ and thus $F^{-1}(0)=1$, then the natural choice $c:=0$ 
induces $]\lambda_{-},\lambda_{+}[ \, = int(\mathcal{R}(F))$ and 
$]t_{-}^{sc},t_{+}^{sc}[ \, = \, ]a_{F},b_{F}[$,
and consequently (due to $F^{-1}(c)-1 =0$) leads to the simplification of 
\textquotedblleft  the first lines of\textquotedblright\
\eqref{brostu3:fo.Wfind1.new},\eqref{brostu3:fo.Wfind2.new} to
\begin{eqnarray}
& & \hspace{-0.7cm} \Lambda(z) := \Lambda^{(0)}(z) :=
\int\displaylimits_{0}^{z} F^{-1}(u) \, du,
\hspace{2.3cm} z \in int(\mathcal{R}(F)),
\label{brostu3:fo.Wfind1b}
\\[-0.2cm]
& & \hspace{-0.7cm} \varphi(t) := \varphi^{(0)}(t) := 
t \cdot F\left(t\right)
-\int\displaylimits_{0}^{F\left(t\right)} F^{-1}(u) \, du,
\qquad 
t \in \, ]a_{F},b_{F}[ ;
\label{brostu3:fo.Wfind2b}
\end{eqnarray}
the simplifications of the respective other lines of 
\eqref{brostu3:fo.Wfind1.new} and \eqref{brostu3:fo.Wfind2.new} 
are straightforward.
\end{remark}

\vspace{0.2cm}
\noindent
One can even prove many more properties of $\varphi$ given by 
\eqref{brostu3:fo.Wfind2.new}
which strongly indicate that 
the $F-$constructed function $z \mapsto \exp(\Lambda(z)) = \exp(\varphi^{*}(z))$ is a 
\textit{good candidate}
for a moment generating function of a probability distribution
$\mathbb{\bbzeta}$, and hence for the representability \eqref{Phi Legendre of mgf(W)}
(i.e. $\varphi \in \Upsilon (]a,b[)$). However, one still needs to 
verify one of the conditions (a) to (c) of the following Proposition \ref{sufficiency}.
This may go wrong, as the case of power divergences $\varphi_{\gamma}$ with $\gamma \in \, ]1,2[$
indicates (cf. the conjecture of 
Subsection \ref{Subsubsect Case6c}
below). 

\vspace{0.3cm}

\begin{proposition}
\label{sufficiency} 
Let $\varphi$ be given by 
\eqref{brostu3:fo.Wfind2.new} of Theorem \ref{brostu3:thm.Wfind.new}.
Then, $\varphi \in \Upsilon (]a,b[)$ if one of the following three conditions
holds: \\
\noindent 
(a) $a >-\infty$, $\lambda_{-} = - \infty$, and the function $z \mapsto 
M(z) := e^{-a\cdot z + \varphi^{*}(z)}$ 
is absolutely monotone on $]-\infty,0[$\\ 
(i.e. all derivatives exist
and satisfy $\frac{\partial^{k}}{\partial z^k} M(z) \geq 0$ for all $k\in \mathbb{N}_{0}$, $z \in \, ]-\infty,0[$ ),
\\
\noindent 
(b) $b < \infty$, $\lambda_{+} = \infty$, and the function $z \mapsto 
M(z) := e^{b\cdot z + \varphi^{*}(- z)}$ is absolutely
monotone on $]-\infty,0[$, 
\\
\noindent 
(c) $a = -\infty$, $b = -\infty$, and
the function $z \mapsto M(z) := e^{\varphi^{*}(z)}$ is exponentially convex
on $]\lambda_{-}, \lambda_{+}[$\\
(i.e. $M(\cdot)$ is continuous and satisfies
$\sum_{i=1}^{n} \sum_{j=1}^{n} c_{i} \cdot c_{j} \cdot M\Big(\frac{z_{i} + z_{j}}{2}\Big) 
\geq 0$ for all $n \in \mathbb{N}$, $c_{i}, c_{j} \in \mathbb{R}$ and  $z_{i}, z_{j} \in 
\, ]\lambda_{-}, \lambda_{+}[$ ).
\\
If one of the three conditions (a) to (c) holds, then
the associated probability distribution $\mathbb{\bbzeta}$ 
(cf. \eqref{Phi Legendre of mgf(W)})
has expectation  $\int_{\mathbb{R}} y \, d\mathbb{\bbzeta}(y) =1$ and 
finite moments of all orders, i.e. 
$\int_{\mathbb{R}} y^{j} \, d\mathbb{\bbzeta}(y) < \infty$ for all $j \in \mathbb{N}_{0}$;
in terms of $\mathbb{\bbzeta}[ \cdot \, ] := \mathbb{\Pi}[W \in \cdot \, ]$
this means that $E_{\mathbb{\Pi}}[W] =1$ and $E_{\mathbb{\Pi}}[W^{j}] < \infty$.

\end{proposition}

\vspace{0.3cm}
\noindent
Proposition \ref{sufficiency}(a),(b) follow from 
Theorem \ref{brostu3:thm.Wfind.new} and the well-known
(probability-version of) \textit{Bernstein\textquoteright s theorem} \cite{Bern:29} 
(see e.g. also 
\cite{Schi:12}) --- which needs to be appropriately shifted ---
whereas (c) follows from 
Theorem \ref{brostu3:thm.Wfind.new} and the well-known
(probability-version of)
\textit{Widder\textquoteright s theorem}
\cite{Wid:34} 
\footnote{
for the relevant conversion between the involved Riemann-Stieltjes
integral with nondecreasing (but not necessarily right-continuous)
integrator into a measure integral, one can apply e.g. the
general theory in Chapter 6 of \cite{Chow:97}.
}
(see e.g. also~\cite{Wid:41}--\cite{Kote:20}).
As far as applicability is concerned, it is well known that
verifying absolute monotonicity is typically 
more comfortable than verifying exponential convexity.
Fortunately, one can often use the former, 
since for many known divergence generators there holds $a > -\infty$
(often $a=0$) or/and $b < \infty$.

\vspace{0.3cm}
\noindent
For the identification of light-tailed \textit{semi-/half-lattice} distributions, we obtain the following two sets of conditions, which even allow for the desired
explicit determination of $\mathbb{\bbzeta}$:

\vspace{0.2cm}

\begin{proposition}
\label{lattice1} 
Let $\varphi$ be given by 
\eqref{brostu3:fo.Wfind2.new} of Theorem \ref{brostu3:thm.Wfind.new},
with some $a > -\infty$.
Furthermore, assume that there exists some constant $\breve{c} >0$ as well as some 
function $H: [0,\infty[ \, \mapsto [0,\infty[$ which is continuous on $[0,1]$ 
with $H(1)=1$ and absolutely monotone on $]0,1[$,
such that
$e^{\varphi^{*}(\frac{z}{\breve{c}}) - a\cdot \frac{z}{\breve{c}} } = H(e^{z})$
($z \in \, ]-\infty, \breve{c} \cdot \lambda_{+}[$ ).
Then one has $\varphi \in \Upsilon (]a,b[)$ and 
$\mathbb{\bbzeta} = \sum_{n=0}^{\infty} p_{n} \cdot \delta_{a + \breve{c} \cdot n}$ 
with $p_{n} := \frac{1}{n !} \cdot \frac{\mathrm{d}^{n}H}{\mathrm{d}t}(0) 
$,
i.e. $\mathbb{\Pi}[W = a + \breve{c} \cdot n \, ] = p_{n}$ \ ($n \in \mathbb{N}_{0}$).

\end{proposition}

\vspace{0.4cm}

\begin{proposition}
\label{lattice2}  
Let $\varphi$ be given by 
\eqref{brostu3:fo.Wfind2.new} of Theorem \ref{brostu3:thm.Wfind.new},
with some $b < \infty$.
Furthermore, assume that there exists some constant $\breve{c} >0$ as well as some 
function $H: [0,\infty[ \, \mapsto [0,\infty[$ which is continuous on $[0,1]$ 
with $H(1)=1$ and absolutely monotone on $]0,1[$, such that
$e^{\varphi^{*}(- \frac{z}{\breve{c}}) + b \cdot \frac{z}{\breve{c}} } = H(e^{z})$  
($z \in \, ]-\infty, - \breve{c} \cdot \lambda_{-}[ \, )$. 
Then one has $\varphi \in \Upsilon (]a,b[)$ and 
$\mathbb{\bbzeta} = \sum_{n=0}^{\infty} p_{n} \cdot \delta_{b - \breve{c} \cdot n}$ 
with $p_{n} := \frac{1}{n !} \cdot \frac{\mathrm{d}^{n}H}{\mathrm{d}t}(0)$, 
i.e. $\mathbb{\Pi}[W = b - \breve{c} \cdot n \, ] = p_{n}$ \ ($n \in \mathbb{N}_{0}$).

\end{proposition}

\vspace{0.4cm}
\noindent
The Propositions \ref{lattice1} and \ref{lattice2} follow 
from some straightforward transformations
and a well-known characterization of probability generating functions $H$ (see
e.g. in Theorem 1.2.10 of 
\cite{Stro:11}). 

\vspace{0.3cm}
\noindent
As an incentive for the following investigations, let us recall
the discussion in the surroundings of Condition \ref{Condition  Fi Tilda in Minimization}
pertaining to the minimization problem \eqref{min Pb prob2},
where we have addressed possible connections between 
the two representabilities \eqref{Phi Legendre of mgf(W)} 
(needed e.g. for Theorem \ref{brostu3:thm.divnormW.new})
and \eqref{brostu3:fo.link.var} (needed e.g. for Theorem \ref{brostu3:thm.divW.var});
this strongly relates to the question, for which constants $\widetilde{c} >0$ the validity $\varphi \in \Upsilon (]a,b[)$ 
triggers the validity of $\widetilde{c} \cdot \varphi \in \Upsilon (]a,b[)$. 
To begin with, it is straightforward to see that $\varphi \in \Upsilon (]a,b[)$
\textit{always implies} $\widetilde{c} \cdot \varphi \in \Upsilon (]a,b[)$
for all \textit{integers} $\widetilde{c} \in \mathbb{N}$; indeed, 
if $\varphi$ satisfies \eqref{Phi Legendre of mgf(W)}
for some $\mathbb{\bbzeta} = \mathbb{\Pi}[ W \in \cdot \, ]$,
then for each integer $\widetilde{c} \in \mathbb{N}$
one gets that $\widetilde{c} \cdot \varphi$ satisfies 
\eqref{Phi Legendre of mgf(W)} for 
$\widetilde{\mathbb{\bbzeta}} = \mathbb{\Pi}[ \sum_{j=1}^{\widetilde{c}} \frac{W_{j}}{\widetilde{c}} \in \cdot \, ]$;
in the latter, the $W_{j}$\textquoteright s are i.i.d. copies of $W$.
Clearly, $MGF_{\widetilde{\mathbb{\bbzeta}}}$ is then finite on some open interval containing zero
(differing from the one for $MGF_{\mathbb{\bbzeta}}$
only by a scaling with $1/\widetilde{c}$).

\vspace{0.2cm}
\noindent
For the following family
of distributions, one can even trigger $\widetilde{c} \cdot \varphi \in \Upsilon (]a,b[)$
for \textit{all} $\widetilde{c} >0$: for the sake of a corresponding precise formulation,
recall first the common knowledge that, generally speaking, a probability distribution $\mathbb{\bbzeta}$ on $\mathbb{R}$
with light tails --- in the sense that its moment generating function
$z\mapsto MGF_{\mathbb{\bbzeta} }(z):=\int_{\mathbb{R}}e^{z \cdot y}d\mathbb{\bbzeta} (y)$ is
finite on some open interval $]\lambda_{-},\lambda_{+}[$ containing zero ---
is (said to be) \textit{infinitely divisible} if there holds
\begin{equation}
\textrm{for each $n \in \mathbb{N}$ there exists
a probability distribution $\mathbb{\bbzeta}_{n}$ on $\mathbb{R}$ such that}
\int_{\mathbb{R}} e^{z \cdot y} d\mathbb{\bbzeta} (y) = 
\Big(\int_{\mathbb{R}} e^{z \cdot y} d\mathbb{\bbzeta}_{n} (y) \Big)^{n}, \quad z \in \, ]\lambda_{-},\lambda_{+}[ ;
\label{IDmeas}
\end{equation}
in fact, \eqref{IDmeas} means that the (light-tailed) moment generating function $MGF_{\mathbb{\bbzeta} }$
is \textit{infinitely divisible} in the sense that each $n-$th root $(MGF_{\mathbb{\bbzeta} })^{1/n}$ must be the moment
generating function of some (light-tailed) probability distribution (denoted here by $\mathbb{\bbzeta}_{n}$).
In particular, \eqref{IDmeas} implies that $\mathbb{\bbzeta}_{n}$ is unique, and
that $\mathbb{\bbzeta}$ must necessarily have (one-sided or two-sided) unbounded support 
$supp(\mathbb{\bbzeta})$. The latter may differ from $supp(\mathbb{\bbzeta}_{n})$.
In our BS context \eqref{Phi Legendre of mgf(W)}, \eqref{IDmeas} equivalently means that
the associated random variable $W$ is \textit{infinitely divisible}
(with light-tailed distribution), in the sense that 
\begin{equation}
\textrm{for each $n \in \mathbb{N}$ there exists a sequence of i.i.d. random variables $Y_{n,1}, \cdots, Y_{n,n}$ such that }
W \stackrel{\text{\tiny d}}{=} Y_{n,1} + \cdots + Y_{n,n} ,
\nonumber
\end{equation}
where $\stackrel{\text{\tiny d}}{=}$ means \textquotedblleft  have equal probability distributions\textquotedblright\
and $\mathbb{\Pi}[W \in \cdot \, ] = \mathbb{\bbzeta}[ \, \cdot \,]$, 
$\mathbb{\Pi}[Y_{n,1} \in \cdot \, ] = \mathbb{\bbzeta}_{n}[ \, \cdot \,]$.

\vspace{0.2cm}
\noindent
For the above-mentioned context, we obtain the useful assertion
(which will be proved in Appendix D):

\vspace{0.2cm}

\begin{proposition}
\label{ID1} 
Suppose that $\varphi \in \Upsilon (]a,b[)$ with connected
probability distribution $\mathbb{\bbzeta}$ from \eqref{Phi Legendre of mgf(W)}.
Then there holds:
\begin{equation*}
\textrm{$\widetilde{c} \cdot \varphi \in \Upsilon (]a,b[)$
for all $\widetilde{c} > 0$} \ \ 
\Longleftrightarrow \ \ 
\textrm{$\mathbb{\bbzeta}$ is infinitely divisible.} 
\end{equation*}

\end{proposition}

\noindent
Notice that Proposition \ref{ID1} covers especially the important
prominent \textit{power divergences} 
(cf. the solved-cases Sections \ref{Subsect Case1} to \ref{Subsect Case6} below)
for which we provide the corresponding infinitely divisible distributions
explicitly. More generally, 
for the identification of light-tailed infinitely divisible distributions, we obtain the following 
three sets of sufficient conditions:

\vspace{0.2cm}

\begin{proposition}
\label{ID2} 
Let $\varphi$ be given by 
\eqref{brostu3:fo.Wfind2.new} of Theorem \ref{brostu3:thm.Wfind.new}. 
Then, 
$\varphi \in \Upsilon (]a,b[)$ and 
the associated probability distribution $\mathbb{\bbzeta}$ 
is infinitely divisible,  if one of the following three conditions
holds: \\
\noindent 
(a) $a >-\infty$, $\lambda_{-} = - \infty$, and the function 
$z \mapsto \varphi^{* \prime}(z) - a = (\varphi^{\prime})^{-1}(z) - a $ 
is absolutely monotone on $]-\infty,0[$,
\\
\noindent 
(b) $b < \infty$, $\lambda_{+} = \infty$, and the function 
$z \mapsto - \varphi^{* \prime}(- z) + b
= - (\varphi^{\prime})^{-1}(-z) +b $ 
is absolutely monotone on $]-\infty,0[$, 
\\
\noindent 
(c) $a = -\infty$, $b = \infty$, and
the function 
$z \mapsto \frac{\varphi^{* \prime \prime}(z)}{\varphi^{* \prime \prime}(0)}
= \frac{\varphi^{\prime \prime}(1)}{\varphi^{\prime \prime}((\varphi^{\prime})^{-1}(z))} $ \ 
is exponentially convex
on $]\lambda_{-}, \lambda_{+}[$.  \\
In the first case (a) there automatically follows $b=\infty$,
whereas in the second case (b) one automatically gets $a =- \infty$.

\end{proposition}

\vspace{0.3cm}
\noindent
The proof of Proposition \ref{ID2} is given in Appendix D.

\vspace{0.2cm}
\noindent
Let us end this section by giving some further comments on
the task of finding concretely the 
probability distribution (if existent)  
$\mathbb{\bbzeta}[ \, \cdot \,] = \mathbb{\Pi}[W \in \cdot \, ]$
from the Fenchel-Legendre transform  $\Lambda = \varphi^{*}$
of a pregiven divergence generator $\varphi$,
which should satisfy
$ MGF(z) :=
\exp(\Lambda(z)) = \int_{\mathbb{R}}e^{z \cdot y} \, d\mathbb{\bbzeta} (y) 
=  E_{\mathbb{\Pi}}[\exp(z \cdot W)]$    
($z \in \mathbb{R}$).
Recall that this is used for the
simulation of the weights $(W_{i})_{i\in \mathbb{N}}$ which are 
i.i.d. copies of $W$ and which are the crucial building ingredients 
of $\boldsymbol{\xi }_{n}^{\mathbf{W}}$ in Theorem \ref{brostu3:thm.divW.var},
respectively, of $\xi _{n,\mathbf{X}}^{w\mathbf{W}}$ 
in Theorem \ref{brostu3:thm.divnormW.new}.
The search for $\mathbb{\bbzeta}$ can be done 
e.g. by inversion of a candidate moment generating function MGF, or by 
search in tables or computer software which list distributions and their MGF. 
Also notice that $\mathbb{\bbzeta}$ needs not necessarily be explicitly known in full detail
(e.g. in terms of a computationally tractable density or frequency); for instance,
as well known from insurance applications, 
for --- comfortably straightforwardly simulable --- doubly-random sums $W := \sum_{i=1}^{N} A_{i}$ 
of nonnegative i.i.d. random variables $(A_{i})_{i\in \mathbb{N}}$ with known law $\Pi_A[\, \cdot \, ]
:= \Pi[A \in \cdot \, ]$
being independent of a counting-type random variable $N$ with known law $\Pi_N$, one can mostly compute
explicitly $MGF_{\mathbb{\bbzeta}}(z) = PGF_{\Pi_N}(MGF_{\Pi_A}(z))$ 
with the help of the probability generating function
$PGF_{\Pi_N}$ of $\Pi_N$, but 
the corresponding density/frequency of $\mathbb{\bbzeta}$ may not be known explicitly in a tractable form. The below-mentioned solved Case 2 in Subsection \ref{Subsect Case2} 
of power divergences with generator $\varphi_{\gamma}$ ($\gamma \in \, ]0,1[$)
manifests such a situation. 
In the end, if no explicit distribution $\mathbb{\bbzeta}$ and
no comfortably simulable $W-$construction are available, one can still try
to simulate an i.i.d. sequence
$(W_{i})_{i\in \mathbb{N}}$ from the pregiven moment generating function 
(which is $\exp(\Lambda(z))$ here);
see e.g. \cite{McLei:14} and references therein which
also contains saddle point methods approximation techniques. 

\vspace{0.2cm}
\noindent
Let us finally mention that a more complete picture on finding the distribution $\bbzeta$ 
of the weights $W$ (including necessary and sufficient conditions, boundary behaviours, etc.)
can be found in our paper's full arXiv-version \cite{Bro:21}.

\newpage

%
%

\section{Explicitly solved cases \label{Sect Cases}}

\noindent
With the help of Theorem \ref{brostu3:thm.Wfind.new} we can comfortably generate 
various important solved cases, which we demonstrate now.
Notice especially in the Cases \ref{Subsect Case1} to \ref{Subsect Case6}
the astonishing effect that 
the \textquotedblleft  homogeneous/continuous\textquotedblright\ class of
power-divergence generators $(\varphi_{\gamma})_{\gamma \in \mathbb{R}}$ 
(cf. \eqref{brostu3:fo.powdivgen}) are connected
to a \textquotedblleft  very inhomogeneous\textquotedblright\  
family of underlying ``cornerstone simulation laws'' 
$(\mathbb{\bbzeta}_{\gamma})_{\gamma \in \mathbb{R}}$ 
of $W-$distributions:
discrete, continuous, mixture of discrete and continuous,
as the parameter $\gamma$ varies.


\subsection{Case 1
\label{Subsect Case1}}

\noindent
For $\gamma < 0$, $\widetilde{c} \in \, ]0,\infty[$  and
$]a_{F_{\gamma,\widetilde{c}}},b_{F_{\gamma,\widetilde{c}}}[ \,  := \, ]0,\infty[$ 
we define
\begin{eqnarray}
F_{\gamma,\widetilde{c}}(t) &:=&
\begin{cases}
\frac{\widetilde{c}}{\gamma-1} \cdot (t^{\gamma-1}-1), \qquad \textrm{if } \  
t \in \, ]0,\infty[, \\
- \infty, \hspace{2.5cm}  \textrm{if } \  t \in \, ]-\infty,0]. 
\end{cases}
\label{brostu3:Fcase1and2} 
\end{eqnarray}
Clearly, $\mathcal{R}(F_{\gamma,\widetilde{c}})= \big]-\infty, \frac{\widetilde{c}}{1-\gamma}\big[$.
Furthermore, $F_{\gamma,\widetilde{c}}(\cdot)$ is strictly increasing and smooth on 
$]a_{F_{\gamma,\widetilde{c}}},b_{F_{\gamma,\widetilde{c}}}[$,
and thus, $F_{\gamma,\widetilde{c}} \in \textgoth{F}$. 
Since $F_{\gamma,\widetilde{c}}(1)=0$, let us  
choose the natural anchor point $c:=0 \in int(\mathcal{R}(F_{\gamma,\widetilde{c}}))$,
which leads to $]\lambda_{-},\lambda_{+}[ \, := int(\mathcal{R}(F_{\gamma,\widetilde{c}}))
-c = \big]-\infty, \frac{\widetilde{c}}{1-\gamma}\big[$,
$]t_{-}^{sc},t_{+}^{sc}[ \, := \, ]1+a_{F_{\gamma,\widetilde{c}}}-F_{\gamma,\widetilde{c}}^{-1}(c),1+b_{F_{\gamma,\widetilde{c}}}-F_{\gamma,\widetilde{c}}^{-1}(c)[ \, = \, ]0,\infty[$ and
$]a,b[ \, =  \, ]0,\infty[$ since
$a := t_{-}^{sc} \cdot \textfrak{1}_{ \{ -\infty \} }(\lambda_{-}) - \infty \cdot \textfrak{1}_{]-\infty,0[}(\lambda_{-}) = 0$, 
\ $b := t_{+}^{sc} \cdot \textfrak{1}_{ \{ \infty \} }(\lambda_{+}) + \infty \cdot \textfrak{1}_{]0,\infty[}(\lambda_{+}) = \infty$.
By using $F_{\gamma,\widetilde{c}}^{-1}(x) = (1+\frac{(\gamma -1) \cdot x}{\widetilde{c}})^{\frac{1}{\gamma-1}}$ for 
$x \in int(\mathcal{R}(F_{\gamma,\widetilde{c}}))$, 
by straightforward calculations we can deduce
from formula \eqref{brostu3:fo.Wfind2.new}  
(see also \eqref{brostu3:fo.Wfind2b}) 
\begin{eqnarray}
\varphi_{\gamma,\widetilde{c}}(t) := \varphi_{\gamma,\widetilde{c}}^{(0)}(t)
\hspace{-0.2cm} &=& \hspace{-0.2cm}
\begin{cases}
\widetilde{c} \cdot \frac{t^\gamma-\gamma \cdot t+ \gamma - 1}{\gamma \cdot (\gamma-1)}
\ \in \ [0,\infty[,
\qquad
\quad \textrm{if } 
t \in \, ]0,\infty[,
\\
\infty, \hspace{4.35cm} \textrm{if }  t \in \, ]-\infty,0], 
\end{cases}
\nonumber
\end{eqnarray}
which coincides with $\widetilde{c} \cdot \varphi_{\gamma}(t)$
for $\varphi_{\gamma}(t)$ from \eqref{brostu3:fo.powdivgen} and 
which generates the $\gamma-$corresponding power divergences given in \eqref{brostu3:fo.powdiv.new}. 
Moreover, we can derive from formula \eqref{brostu3:fo.Wfind1.new} 
(see also \eqref{brostu3:fo.Wfind1b}) 
\begin{eqnarray}
\Lambda_{\gamma,\widetilde{c}}(z) := \Lambda_{\gamma,\widetilde{c}}^{(0)}(z) &=&
\begin{cases}
\frac{\widetilde{c}}{\gamma} \cdot
\left\{ \left( \frac{\gamma -1}{\widetilde{c}} \cdot z + 
1 \right)^{\frac{\gamma}{\gamma-1}} -1  \right\}, 
\qquad \textrm{if } z \in \big]-\infty, \frac{\widetilde{c}}{1-\gamma}\big[, \\
- \frac{\widetilde{c}}{\gamma} >0, \hspace{3.7cm} 
\textrm{if }  z = \frac{\widetilde{c}}{1-\gamma}, \\
\infty, \hspace{4.5cm}  \textrm{if } z \in \big]\frac{\widetilde{c}}{1-\gamma}, \infty \big[. 
\end{cases}
\nonumber
\end{eqnarray}
The latter is the cumulant generating function of a 
\textquotedblleft tilted (i.e. negatively distorted) stable distribution\textquotedblright\ 
$\mathbb{\bbzeta}[ \, \cdot \,] = \mathbb{\Pi}[W \in \cdot \, ]$
of a random variable $W$,
which can be constructed as follows: let $Z$ be an auxiliary random variable  
(having density $f_{Z}$ and support $supp(Z) = [0,\infty[$) of a stable law
with parameter-quadruple
$(\frac{-\gamma }{1-\gamma },1,0,-\frac{\widetilde{c}^{1/(1-\gamma )} \cdot
(1-\gamma )^{-\gamma /(1-\gamma )}}{\gamma })$ 
in terms of the \textquotedblleft form-B notation\textquotedblright\
on p.12 in~\cite{Zol:86}; 
by applying a general Laplace-transform result on p.112 of the same text we can deduce
\begin{eqnarray}
M_{Z}(z) := E_{\mathbb{\Pi}}[\exp(z \cdot Z)] 
= \int_{0}^{\infty} \exp(z \cdot y) \cdot f_{Z}(y) \, dy  \hspace{-0.2cm} &=& \hspace{-0.2cm}
\begin{cases}
\exp\Big(
\frac{\widetilde{c}^{1/(1-\gamma )} \cdot (1-\gamma )^{-\gamma /(1-\gamma )}}{\gamma } 
\cdot (-z)^{\alpha} \Big),  
\quad \textrm{if } \  z \in ]-\infty,0] , \\
\infty, \hspace{5.3cm} \textrm{if } \  z \in \, ]0,\infty[, \\
\end{cases}
\label{brostu3:tiltstab1}
\end{eqnarray}
where $\alpha := - \frac{\gamma}{1-\gamma} \in \, ]0,1[$. Since $0 \notin int(dom(M_{Z}))$
(and thus, $Z$ does not have light-tails) we have to tilt (dampen) the density
in order to extend the effective domain. Accordingly, let $W$ be a random variable having 
density\footnote{
in the classical sense, with respect to Lebesgue measure} 
\begin{equation}
f_{W}(y)\ :=\ \frac{\exp \{-\frac{y \cdot \widetilde{c}}{1-\gamma }\}}{\exp \{\widetilde{c}/\gamma \}}
\cdot f_{Z}(y)\cdot 
\textfrak{1}_{]0,\infty[}(y),
\qquad y \in \mathbb{R}.
\label{brostu3:fo.norweiemp7a}
\end{equation}
Then one can straightforwardly deduce from \eqref{brostu3:tiltstab1} that  
$\int_{0}^{\infty} f_{W}(y) \, dy =1$ and that
\begin{eqnarray}
M_{W}(z) := E_{\mathbb{\Pi}}[\exp(z \cdot W)] 
= \int_{0}^{\infty} \exp(z \cdot y) \cdot f_{W}(y) \, dy  \hspace{-0.2cm} &=& \hspace{-0.2cm}
\begin{cases}
\exp\left(\frac{\widetilde{c}}{\gamma} \cdot
\left\{ \left( \frac{\gamma -1}{\widetilde{c}} \cdot z + 
1 \right)^{\frac{\gamma}{\gamma-1}} -1  \right\} \right),  
\qquad \textrm{if } \  z \in ]-\infty,\frac{\widetilde{c}}{1-\gamma}] , \\
\infty, \hspace{5.5cm} \textrm{if } \  z \in \, ]\frac{\widetilde{c}}{1-\gamma},\infty[ . \\
\end{cases}
\nonumber
\end{eqnarray}
Notice that $\mathbb{\bbzeta}$ is an infinitely divisible (cf. Proposition \ref{ID1}) continuous distribution with density $f_{W}$, and that
$\mathbb{\bbzeta}[ \, ] 0,\infty[ \, ] = \mathbb{\Pi}[W > 0]=1$.
Concerning the important Remark \ref{dist of components}(i), 
for i.i.d. copies $(W_{i})_{i \in \mathbb{N}}$ of $W$ we obtain the 
probability distribution 
$\mathbb{\bbzeta}^{\ast n_{k}}[\cdot] :=  \mathbb{\Pi}[\breve{W} \in \cdot \, ]$ 
of $\breve{W} := \sum_{i\in I_{k}^{(n)}} W_{i}$ 
having the density
\vspace{-0.2cm}
\begin{equation}
f_{\breve{W}}(y)\ :=\ \frac{\exp \{-\frac{y \cdot \widetilde{c}}{1-\gamma }\}}{\exp \{\widetilde{c} 
\cdot card(I_{k}^{(n)})/\gamma \}}
\cdot f_{\breve{Z}}(y)\cdot 
\textfrak{1}_{]0,\infty[}(y),
\qquad y \in \mathbb{R},
\label{fbreveW}
\end{equation}
where $\breve{Z}$ is a random variable with density $f_{\breve{Z}}$ of a stable law
with parameters
$(\frac{-\gamma }{1-\gamma },1,0,-card(I_{k}^{(n)}) \cdot \frac{\widetilde{c}^{1/(1-\gamma )} \cdot
(1-\gamma )^{-\gamma /(1-\gamma )}}{\gamma })$.
We are now in the position to state explicitly the 
bare-simulation-minimizations (respectively maximizations)
of the corresponding ($\gamma-$order) power divergences, 
Renyi divergences, Hellinger integrals
and measures of entropy (diversity), which can be deduced from 
Theorem \ref{brostu3:thm.divnormW.new}, 
Remark \ref{remark divnormW}(vi),
Lemma \ref{Lemma Indent Rate finite case_new}(a), \eqref{fo.hgamma1},
\eqref{def Renyi}, \eqref{fo.genentropy1}, \eqref{fo.genentropy2}:

\vspace{0.2cm}

\begin{proposition}
\label{brostu3:cor.powdivexpl2} 
(a) Consider $\varphi := \widetilde{c} \cdot \varphi_{\gamma}$ with $\gamma <0$, 
and let $\mathds{P} \in \mathbb{S}_{>0}^{K}$ as well as $\widetilde{c}>0$ be arbitrary but fixed. 
Furthermore, let $W:=(W_{i})_{i\in \mathbb{N}}$
be an i.i.d. sequence of non-negative real-valued random variables  
having density \eqref{brostu3:fo.norweiemp7a}. Then for all $A>0$ and all
$\boldsymbol{\Omega}$\hspace{-0.23cm}$\boldsymbol{\Omega}\subset \mathbb{S}_{>0}^{K}$ 
with \eqref{regularity} there holds
\begin{equation}
\hspace{-0.7cm}-\lim_{n\rightarrow \infty }\frac{1}{n} \log \,
\mathbb{\Pi}\negthinspace \left[\boldsymbol{\xi}_{n}^{w\mathbf{W}}
\in \textrm{$\boldsymbol{\Omega}$\hspace{-0.23cm}$\boldsymbol{\Omega}$} \right] =
\inf_{\mathbf{Q}\in A \cdot \textrm{$\boldsymbol{\Omega}$\hspace{-0.19cm}$\boldsymbol{\Omega}$}}
\frac{\widetilde{c}}{\gamma }\cdot \left[ 1-
A^{\gamma/(\gamma-1)} \cdot \left[ 1+ \gamma \cdot (A-1) + 
\frac{\gamma \cdot \left( \gamma -1\right) }{\widetilde{c}}\cdot
D_{\widetilde{c}\cdot \varphi _{\gamma }}(\mathbf{Q},\mathds{P})\right]
^{-1/\left( \gamma -1\right) }\right]
\label{brostu3:fo.norweiemp5.new}
\end{equation}
as well as the BS minimizabilities/maximizabilites (cf. Definition \ref{brostu3:def.1}) 
\begin{eqnarray}
& & \hspace{-0.9cm}
\inf_{\mathbf{Q}\in A \cdot \textrm{$\boldsymbol{\Omega}$\hspace{-0.19cm}$\boldsymbol{\Omega}$}}
D_{\widetilde{c}\cdot \varphi _{\gamma}}(\mathbf{Q},\mathds{P})
=\lim_{n\rightarrow \infty }\frac{\widetilde{c}}{\gamma \cdot \left(
\gamma -1\right) }\cdot \left\{ A^{\gamma} \cdot \left( 1+\frac{\gamma }{\widetilde{c}}
\cdot \frac{1}{n}\cdot 
\log \, \mathbb{\Pi}\negthinspace \left[\boldsymbol{\xi}_{n}^{w\mathbf{W}}
\in \textrm{$\boldsymbol{\Omega}$\hspace{-0.23cm}$\boldsymbol{\Omega}$} \right]
\right) ^{1-\gamma } + \gamma \cdot (1-A) -1 \right\} ,\qquad \ 
\label{brostu3:fo.norweiemp6.new}
\\
& & \hspace{-0.9cm}
\inf_{\mathbf{Q}\in A \cdot \textrm{$\boldsymbol{\Omega}$\hspace{-0.19cm}$\boldsymbol{\Omega}$}}
H_{\gamma}(\mathbf{Q},\mathds{P})
=\lim_{n\rightarrow \infty }  A^{\gamma} \cdot \left( 1+ \gamma
\cdot \frac{1}{n}\cdot 
\log \, \mathbb{\Pi}\negthinspace \left[\breve{\boldsymbol{\xi}}_{n}^{w\mathbf{W}}
\in \textrm{$\boldsymbol{\Omega}$\hspace{-0.23cm}$\boldsymbol{\Omega}$} \right]
\right) ^{1-\gamma }  ,\qquad \ 
\label{brostu3:fo.norweiemp6.new1b}
\\
& & \hspace{-0.9cm}
\inf_{\mathbf{Q}\in A \cdot \textrm{$\boldsymbol{\Omega}$\hspace{-0.19cm}$\boldsymbol{\Omega}$}}
c_{1} \negthinspace \cdot \negthinspace \Big(H_{\gamma}(\mathbf{Q},\mathds{P})^{c_{2}} - c_{3} \Big)
=\lim_{n\rightarrow \infty }  c_{1} \negthinspace \cdot \negthinspace \left\{ A^{c_{2} \cdot \gamma} \cdot
\negthinspace \left( 1+ 
\frac{\gamma}{n}\cdot 
\log \, \mathbb{\Pi}\negthinspace \left[\breve{\boldsymbol{\xi}}_{n}^{w\mathbf{W}}
\in \textrm{$\boldsymbol{\Omega}$\hspace{-0.23cm}$\boldsymbol{\Omega}$} \right]
\right) ^{c_{2} \cdot (1-\gamma)} \negthinspace \negthinspace - c_{3} \right\} ,\ \ 
\textrm{if $c_{1} \cdot c_{2} >0$, $c_{3} \in \mathbb{R}$}, \quad \ 
\label{brostu3:fo.norweiemp6.new1c}
\\
& & \hspace{-0.9cm}
\sup_{\mathbf{Q}\in A \cdot \textrm{$\boldsymbol{\Omega}$\hspace{-0.19cm}$\boldsymbol{\Omega}$}}
c_{1} \negthinspace \cdot \negthinspace \Big(H_{\gamma}(\mathbf{Q},\mathds{P})^{c_{2}} - c_{3} \Big)
=\lim_{n\rightarrow \infty }  c_{1} \negthinspace \cdot \negthinspace \left\{ A^{c_{2} \cdot \gamma} \cdot
\negthinspace \left( 1+ 
\frac{\gamma}{n}\cdot 
\log \, \mathbb{\Pi}\negthinspace \left[\breve{\boldsymbol{\xi}}_{n}^{w\mathbf{W}}
\in \textrm{$\boldsymbol{\Omega}$\hspace{-0.23cm}$\boldsymbol{\Omega}$} \right]
\right) ^{c_{2} \cdot (1-\gamma)} \negthinspace \negthinspace - c_{3} \right\} ,\ \ 
\textrm{if $c_{1} \cdot c_{2} < 0$, $c_{3} \in \mathbb{R}$}, \quad \ 
\label{brostu3:fo.norweiemp6.new1d}
\\
& & \hspace{-0.9cm}
\inf_{\mathbf{Q}\in A \cdot \textrm{$\boldsymbol{\Omega}$\hspace{-0.19cm}$\boldsymbol{\Omega}$}}
R_{\gamma}(\mathbf{Q},\mathds{P})
= \lim_{n\rightarrow \infty } \frac{1}{\gamma \cdot (\gamma-1)} \cdot 
\log\Big( A^{\gamma} \cdot \Big( 1+ \gamma
\cdot \frac{1}{n}\cdot 
\log \, \mathbb{\Pi}\negthinspace \left[\breve{\boldsymbol{\xi}}_{n}^{w\mathbf{W}}
\in \textrm{$\boldsymbol{\Omega}$\hspace{-0.23cm}$\boldsymbol{\Omega}$} \right]
\Big) ^{1-\gamma } \Big), 
\label{brostu3:fo.norweiemp6.new2}
\\
& & \hspace{-0.9cm}
\inf_{\mathbf{Q}\in A \cdot \textrm{$\boldsymbol{\Omega}$\hspace{-0.19cm}$\boldsymbol{\Omega}$}}
c_{1} \negthinspace \cdot \negthinspace \Big(
\Big(\sum_{k=1}^{K} q_{k}^{\gamma}\Big)^{c_{2}} \negthinspace
-  c_{3} \Big)
=\lim_{n\rightarrow \infty }  c_{1} \negthinspace \cdot \negthinspace \left\{ 
K^{c_{2}\cdot(1-\gamma)} \cdot A^{c_{2} \cdot \gamma} \cdot
\negthinspace \left( 1+ 
\frac{\gamma}{n}\cdot 
\log \, \mathbb{\Pi}\negthinspace \left[\breve{\breve{\boldsymbol{\xi}}}_{n}^{w\mathbf{W}}
\in \textrm{$\boldsymbol{\Omega}$\hspace{-0.23cm}$\boldsymbol{\Omega}$} \right]
\right) ^{c_{2} \cdot (1-\gamma)} \negthinspace \negthinspace - c_{3} \right\} ,\ \
\nonumber \\
& & \hspace{13.0cm} 
\textrm{if $c_{1} \cdot c_{2} >0$, $c_{3} \in \mathbb{R}$}, \quad \ 
\label{brostu3:fo.norweiemp6.new3a}
\end{eqnarray}
\begin{eqnarray}
& & \hspace{-0.9cm}
\sup_{\mathbf{Q}\in A \cdot \textrm{$\boldsymbol{\Omega}$\hspace{-0.19cm}$\boldsymbol{\Omega}$}}
c_{1} \negthinspace \cdot \negthinspace \Big(
\Big(\sum_{k=1}^{K} q_{k}^{\gamma}\Big)^{c_{2}}  \negthinspace
-   c_{3} \Big)
=\lim_{n\rightarrow \infty }  c_{1} \negthinspace \cdot \negthinspace \left\{ 
K^{c_{2}\cdot(1-\gamma)} \cdot A^{c_{2} \cdot \gamma} \cdot
\negthinspace \left( 1+ 
\frac{\gamma}{n}\cdot 
\log \, \mathbb{\Pi}\negthinspace \left[\breve{\breve{\boldsymbol{\xi}}}_{n}^{w\mathbf{W}}
\in 
\textrm{$\boldsymbol{\Omega}$\hspace{-0.23cm}$\boldsymbol{\Omega}$} \right]
\right) ^{c_{2} \cdot (1-\gamma)} \negthinspace \negthinspace - c_{3} \right\} ,\ \
\nonumber \\
& & \hspace{13.0cm} 
\textrm{if $c_{1} \cdot c_{2} < 0$, $c_{3} \in \mathbb{R}$}, \quad \ 
\label{brostu3:fo.norweiemp6.new3b}
\\
& & \hspace{-0.9cm}
\inf_{\mathbf{Q}\in A \cdot \textrm{$\boldsymbol{\Omega}$\hspace{-0.19cm}$\boldsymbol{\Omega}$}}
\frac{1}{1-\gamma} \negthinspace \cdot \negthinspace \log\Big(\sum_{k=1}^{K} q_{k}^{\gamma}\Big)
= \lim_{n\rightarrow \infty } \frac{1}{1-\gamma} \negthinspace \cdot \negthinspace \left[
\log\Big( A^{\gamma} \negthinspace \cdot \negthinspace \Big( 1+ \gamma
\cdot \frac{1}{n}\cdot 
\log \, \mathbb{\Pi}\negthinspace \left[\breve{\breve{\boldsymbol{\xi}}}_{n}^{w\mathbf{W}}
\in \textrm{$\boldsymbol{\Omega}$\hspace{-0.23cm}$\boldsymbol{\Omega}$} \right]
\Big) ^{1-\gamma } \Big) + (1-\gamma) \cdot \log(K) \right],
\label{brostu3:fo.norweiemp6.new3c}
\end{eqnarray}
where $\boldsymbol{\xi} _{n}^{w\mathbf{W}}$ is the normalized randomly
weighted empirical measure given in \eqref{brostu3:fo.norweiemp.vec det},
$\breve{\boldsymbol{\xi}}_{n}^{w\mathbf{W}}$ 
is its special case for $\widetilde{c}=1$, and 
$\breve{\breve{\boldsymbol{\xi}}}_{n}^{w\mathbf{W}}$
is its special case for $\widetilde{c}=1$ together with $\mathds{P} = \mathds{P}^{unif}$
\footnote{the latter two notations 
will be also used in the following Propositions 
\ref{brostu3:cor.powdivexpl1} to \ref{brostu3:cor.powdivexpl5} 
}.  
From this, the BS-minimizability/maximizability of the important norms/entropies/diversity indices 
(E1) to (E6) follow immediately as special cases.\\
(b) The special case $\varphi := \widetilde{c}\cdot \varphi_{\gamma}$   
($\gamma <0$)
of Theorem \ref{brostu3:thm.divnormW.new} works analogously to (a), 
with the differences that we employ (i) additionally
a sequence $(X_{i})_{i\in \mathbb{N}}$  of random variables
being independent of $(W_{i})_{i\in \mathbb{N}}$ 
and satisfying condition \eqref{cv emp measure X to P}
(resp. \eqref{cv emp measure X to P vector}), 
(ii) $A=1$ (instead of arbitrary $A>0$),
(iii) $\mathbb{\Pi}_{X_{1}^{n}}[\cdot]$ (instead of $\mathbb{\Pi }[\cdot]$),
(iv) $\boldsymbol{\xi}_{n,\mathbf{X}}^{w\mathbf{W}}$
(instead of $\boldsymbol{\xi}_{n}^{w\mathbf{W}}$),
(v) $\breve{\boldsymbol{\xi}}_{n,\mathbf{X}}^{w\mathbf{W}}$
(instead of $\breve{\boldsymbol{\xi}}_{n}^{w\mathbf{W}}$),
and 
(vi) $\breve{\breve{\boldsymbol{\xi}}}_{n,\mathbf{X}}^{w\mathbf{W}}$
(instead of $\breve{\breve{\boldsymbol{\xi}}}_{n}^{w\mathbf{W}}$).

\end{proposition}

\vspace{0.2cm}
\noindent
Within the context of Subsection \ref{Subsect Estimators determ}, 
for the concrete simulative estimation 
$\widehat{D_{\widetilde{c} \cdot \varphi_{\gamma}}}(\mathbf{\Omega},\mathbf{P})$
via \eqref{estimator minimization} and \eqref{Improved IS for inf div new}, we
obtain --- in terms of 
$M_{\mathbf{P}}:=\sum_{i=1}^{K}p_{i}>0$,
$n_{k} = n \cdot \widetilde{p}_{k} \in \mathbb{N}$ and
$\widetilde{q}_{k}^{\ast}$ from proxy method 1 or 2
--- that $\widetilde{U}_{k}^{\ast n_{k}}$
has the (Lebesgue-)density
\[
f_{\widetilde{U}_{k}^{\ast n_{k}}
}(x) \ :=  \ \frac{\exp((\tau_{k} - \frac{\widetilde{c} \cdot M_{\mathbf{P}}}{1-\gamma})\cdot x)}{
\exp\left(n_{k} \cdot \frac{\widetilde{c} \cdot M_{\mathbf{P}}}{\gamma}\cdot
(1+\frac{\gamma-1}{\widetilde{c} \cdot M_{\mathbf{P}}} \cdot \tau_{k})^{\gamma/(\gamma-1)}\right)} 
\cdot f_{\breve{\breve{Z}}}(x) \cdot
\textfrak{1}_{]0,\infty[}(x),
\qquad x \in \mathbb{R}, 
\]
where $\tau_{k} = \widetilde{c} \cdot M_{\mathbf{P}} \cdot
\frac{1-\big(\frac{\widetilde{q}_{k}^{\ast}}{\widetilde{p}_{k}}\big)^{\gamma-1}}{1-\gamma}$
for $\widetilde{q}_{k}^{\ast} >0$, and  
$\breve{\breve{Z}}$ is a random variable with density $f_{\breve{\breve{Z}}}$ of a stable law
with parameter-quadruple
$(\frac{-\gamma }{1-\gamma },1,0,- n_{k} \cdot \frac{(\widetilde{c}\cdot M_{\mathbf{P}})^{1/(1-\gamma )} \cdot (1-\gamma )^{-\gamma /(1-\gamma )}}{\gamma })$
(analogously to $\breve{Z}$ of \eqref{fbreveW}
but with $\widetilde{c}$ replaced by $\widetilde{c}\cdot M_{\mathbf{P}}$). Also,
\[
\widetilde{ISF}_{k}(x)=e^{-\tau _{k} \cdot x} 
\cdot \exp \left( \frac{n_{k}\cdot \widetilde{c} \cdot M_{\mathbf{P}}}{\gamma }
\cdot \left(\left(1+ \frac{\gamma -1}{\widetilde{c} \cdot M_{\mathbf{P}}} \cdot \tau _{k}\right) ^{\frac{
\gamma }{\gamma -1}}-1 \right)\right), \qquad x >0. 
\]
For the above random variables,
algorithms for simulation can be obtained by adapting e.g.
the works of \cite{Dev:09} and \cite{Dev:14}.

\noindent
Within the different context of Subsection \ref{Subsect Estimators stoch},
the corresponding estimators $\widehat{\Pi }_{L}^{improved}$ can be obtained 
as follows:

\begin{enumerate}

\item[(i)] proceed as above but set $M_{\mathbf{P}}=1$,
replace $\widetilde{q}_{k}^{\ast}$ by $\overline{W^{\ast}} \cdot q_{k}^{\ast}$
as well as $\widetilde{p}_{k}$ by 
$ p_{n,k}^{emp} $;
accordingly, $\widetilde{U}_{k}^{\ast n_{k}}$ turns into 
$U_{k}^{\ast n_{k}}$ and $\widetilde{ISF}_{k}$ into $ISF_{k}$;

\item[(ii)] simulate independently the random variables $\widehat{\mathit{W}}_{k}^{(\ell)}$
from $U_{k}^{\ast n_{k}}$ \, ($k \in \{1, \ldots, K\}$, $\ell\in \{1,\ldots,L\}$);

\item[(iii)] plug in the results of (i),(ii)
into \eqref{improved IS empirical measure stat section 2},
\eqref{ISK2}, and \eqref{brostu3:fo.norweiemp.vecVell} 
in order to concretely compute $\widehat{\Pi }_{L}^{improved}$.

\end{enumerate}

\noindent
From this, we can easily generate improved estimators of 
the power divergences
$\inf_{\mathbf{Q}\in \textrm{$\boldsymbol{\Omega}$\hspace{-0.19cm}$\boldsymbol{\Omega}$}}
D_{\widetilde{c}\cdot \varphi _{\gamma}}(\mathbf{Q},\mathds{P})$
--- and more generally, improved estimators of all the infimum-quantities 
(e.g. Renyi divergences) respectively supremum-quantities in the parts (b) of the 
Proposition \ref{brostu3:cor.powdivexpl2} 
with $A=1$ --- by simply replacing 
$\mathbb{\Pi}_{X_{1}^{n}}[\boldsymbol{\xi}_{n}^{w\mathbf{W}}
\in \textrm{$\boldsymbol{\Omega}$\hspace{-0.23cm}$\boldsymbol{\Omega}$} ]$
(respectively, its variants) by the corresponding 
estimator $\widehat{\Pi }_{L}^{improved}$. 
If --- in the light of Remark \ref{remark divnormW}(vi) --- 
the $\mathds{P} = (p_{1}, \ldots, p_{K})$ is a pregiven known probability vector 
\footnote{
e.g. the uniform distribution $\mathds{P}^{unif}$ on $\{1,\ldots,K\}$
}
(rather than the 
limit of the vector of empirical frequencies/masses
of a sequence of random variables $X_{i}$, cf. \eqref{cv emp measure X to P vector}),
then we proceed analogously as above by replacing 
$ p_{n,k}^{emp}$ with $p_{k}$;
correspondingly, we obtain 
improved estimators of all the infimum-quantities respectively supremum-quantities 
(e.g. Renyi entropies, diversity indices) in the parts 
(a) of Proposition \ref{brostu3:cor.powdivexpl2} 
with $A=1$. For the sake of brevity, 
here we only present explicitly the outcoming improved 
estimator for the power divergences 
(in the \textquotedblleft  $X_{i}-$context\textquotedblright\ ). 
Indeed, we simply replace the $\mathbb{\Pi}_{X_{1}^{n}}[\boldsymbol{\xi}_{n}^{w\mathbf{W}}
\in \textrm{$\boldsymbol{\Omega}$\hspace{-0.23cm}$\boldsymbol{\Omega}$} ]$
in the ``part-(b)-version of'' 
formula \eqref{brostu3:fo.norweiemp6.new}
(with $A=1$) by the improved estimator 
$\widehat{\Pi }_{L}^{improved}$
obtained through (i) to (iii); for arbitrarily fixed $\widetilde{c} >0$, 
this leads to the \textit{improved power-divergence estimators
(BS estimators of power divergences)}
\begin{eqnarray}
&& 
\hspace{-1.9cm}
\widehat{D_{\widetilde{c}\cdot \varphi
_{\gamma}}(\textrm{$\boldsymbol{\Omega}$\hspace{-0.23cm}$\boldsymbol{\Omega}$},\mathds{P})
}
\ := \ -
\frac{\widetilde{c}}{\gamma (\gamma -1)}\left\{ 1-\left( 1+\frac{\gamma }{
\widetilde{c}} \cdot \frac{1}{n} \cdot \log \widehat{\Pi}_{L}^{improved}\right)
^{1-\gamma }\right\} .
\label{fo.Est1.Case1} 
\end{eqnarray}


\subsection{Case 2
\label{Subsect Case2}}

\noindent
For $\gamma \in \, ]0,1[$, $\widetilde{c} \in \, ]0,\infty[$  and
$]a_{F_{\gamma,\widetilde{c}}},b_{F_{\gamma,\widetilde{c}}}[ \,  := \, ]0,\infty[$ 
we obtain the same $F_{\gamma,\widetilde{c}}(t)$ of \eqref{brostu3:Fcase1and2},
$\mathcal{R}(F_{\gamma,\widetilde{c}})= \big]-\infty, \frac{\widetilde{c}}{1-\gamma}\big[$,
$]\lambda_{-},\lambda_{+}[ \, = \big]-\infty, \frac{\widetilde{c}}{1-\gamma}\big[$
(with $c:=0$),
$]t_{-}^{sc},t_{+}^{sc}[ \, = \, ]0,\infty[$ and
$]a,b[ \, =  \, ]0,\infty[$. Accordingly, we deduce
from \eqref{brostu3:fo.Wfind2.new},\eqref{brostu3:fo.Wfind2b} 
\begin{eqnarray}
\varphi_{\gamma,\widetilde{c}}(t) := \varphi_{\gamma,\widetilde{c}}^{(0)}(t)
\hspace{-0.2cm} &=& \hspace{-0.2cm}
\begin{cases}
\widetilde{c} \cdot \frac{t^\gamma-\gamma \cdot t+ \gamma - 1}{\gamma \cdot (\gamma-1)}
\ \in \ [0,\infty[,
\qquad
\quad \textrm{if } t \in \, ]0,\infty[,\\
\frac{\widetilde{c}}{\gamma} > 0, \hspace{3.8cm} \textrm{if }  t = 0, \\
\infty, \hspace{4.35cm} \textrm{if }  t \in \, ]-\infty,0[, 
\end{cases}
\nonumber
\end{eqnarray}
which coincides with $\widetilde{c} \cdot \varphi_{\gamma}(t)$
for $\varphi_{\gamma}(t)$ from \eqref{brostu3:fo.powdivgen} and 
which generates the $\gamma-$corresponding power divergences given in \eqref{brostu3:fo.powdiv.new}. 
Moreover, we can derive from \eqref{brostu3:fo.Wfind1.new}, \eqref{brostu3:fo.Wfind1b} 
\begin{eqnarray}
\Lambda_{\gamma,\widetilde{c}}(z) := \Lambda_{\gamma,\widetilde{c}}^{(0)}(z) &=&
\begin{cases}
\frac{\widetilde{c}}{\gamma} \cdot
\left\{ \left( \frac{\gamma -1}{\widetilde{c}} \cdot z + 
1 \right)^{\frac{\gamma}{\gamma-1}} -1  \right\}, 
\qquad \textrm{if } z \in \big]-\infty, \frac{\widetilde{c}}{1-\gamma}\big[, \\
\infty, \hspace{4.5cm}  \textrm{if } z \in \big[\frac{\widetilde{c}}{1-\gamma}, \infty \big[, 
\end{cases}
\nonumber
\end{eqnarray}
which is the cumulant generating function of the Compound-Poisson-Gamma 
distribution $\mathbb{\bbzeta} = C(POI(\theta),GAM(\alpha,\beta))$
with $\theta = \frac{\widetilde{c}}{\gamma} > 0$, 
rate parameter (inverse scale parameter) $\alpha = \frac{\widetilde{c}}{1-\gamma} >0$, 
and shape parameter $\beta = \frac{\gamma}{1-\gamma} >0$. 
In other words, $W$ has the comfortably simulable
form $W = \sum_{i=1}^{N} Z_{i}$ 
\footnote{with the usual convention $\sum_{i=1}^{0} Z_{i} := 0$}
for some i.i.d. sequence $( Z_{i})_{i\in \mathbb{N}}$ of 
Gamma $GAM(\alpha,\beta)$ distributed random variables
(with parameter-pair $(\alpha,\beta)$)
\footnote{
here and henceforth, we use the notation that a Gamma distribution $GAM(\alpha,\beta)$
with rate parameter (inverse scale parameter) $\alpha >0$ and 
shape parameter $\beta>0$ has (Lebesgue-)density
$f(y) := \frac{\alpha^{\beta} \cdot 
y^{\beta-1} \cdot e^{-\alpha \cdot y} }{\Gamma(\beta)} 
\cdot \textfrak{1}_{]0,\infty[}(y)$, 
\  $y \in \mathbb{R}$; \ its cumulant generating function is
$\Lambda(z) = \beta \cdot \log(\frac{\alpha}{\alpha-z})$ for $z \in \, ]-\infty,\alpha[$
(and $\Lambda(z) = \infty$ for $z \geq \alpha$).
}
 and some independent 
$POI(\theta)-$distributed random variable $N$.
Notice that 
$\mathbb{\bbzeta}$ is an infinitely divisible distribution 
(cf. Proposition \ref{ID1}) which is a mixture of a one-point distribution at zero and a
continuous distribution on $[0,\infty[$, with
$\mathbb{\bbzeta}[ \, [0,\infty[ \, ] = \mathbb{\Pi}[W \geq 0]=1$,
$\mathbb{\bbzeta}[\{0\}] = \mathbb{\Pi}[W = 0]= e^{-\theta}$ and 
$\mathbb{\bbzeta}[B] = \mathbb{\Pi}[W \in B] = \int_{B} f_{\widetilde{c},\gamma}(u) \, du$ 
for every (measurable) subset of $]0,\infty[$ having density 
\begin{eqnarray}
& & \hspace{-0.8cm} 
f_{C(POI(\theta),GAM(\alpha,\beta))}(y)
: = \frac{\exp\left(- \alpha \cdot y - \theta \right)}{y}  
\cdot \sum_{k=1}^{\infty} \frac{\theta^{k} \cdot (\alpha y)^{k\beta}}{k! \cdot \Gamma(k\beta)} 
\cdot  \textfrak{1}_{]0,\infty[}(y)
\nonumber
\\
& & \hspace{-0.8cm}
= \frac{1}{y} \cdot \exp\left(- \, \widetilde{c} \cdot 
\left(\frac{y}{1-\gamma} + \frac{1}{\gamma} \right) \right)
\cdot \sum_{k=1}^{\infty} \frac{a_{k}}{k!} \cdot \widetilde{c}^{k/(1-\gamma)} \cdot \gamma^{-k} 
\cdot(1-\gamma)^{-k\gamma/(1-\gamma)} \cdot y^{k\gamma/(1-\gamma)} 
\cdot \textfrak{1}_{[0,\infty[}(y)
=: f_{\widetilde{c},\gamma}(y),
\quad y \in \mathbb{R},
\nonumber
\end{eqnarray}
where $a_{k} := 1/\Gamma(\frac{k \cdot \gamma}{1-\gamma} )$
(see e.g. \cite{Aal:92} with a different parametrization).
Concerning the important Remark \ref{dist of components}(i), 
for i.i.d. copies $(W_{i})_{i \in \mathbb{N}}$ of $W$ we obtain the 
probability distribution 
$\mathbb{\bbzeta}^{\ast n_{k}}[\cdot] :=  \mathbb{\Pi}[\breve{W} \in \cdot \, ]$ 
of $\breve{W} := \sum_{i\in I_{k}^{(n)}} W_{i}$ 
to be $C(POI(\breve{\theta}),GAM(\alpha,\beta))$
with $\breve{\theta} = \frac{\widetilde{c} \cdot card(I_{k}^{(n)})}{\gamma} > 0$, 
$\alpha = \frac{\widetilde{c}}{1-\gamma} >0$, $\beta = \frac{\gamma}{1-\gamma} >0$.
For the bare-simulation-minimizations (respectively maximizations)\\

\noindent
of the corresponding ($\gamma-$order) power divergences, 
Renyi divergences,
Hellinger integrals
and measures of entropy (diversity),
we obtain from
Theorem \ref{brostu3:thm.divnormW.new},
Remark \ref{remark divnormW}(vi),
Lemma \ref{Lemma Indent Rate finite case_new}(a), \eqref{fo.hgamma1},
\eqref{def Renyi}, \eqref{fo.genentropy1}, \eqref{fo.genentropy2} 
the following:

\vspace{0.2cm}

\begin{proposition}
\label{brostu3:cor.powdivexpl1} 
(a) Consider $\varphi := \widetilde{c}\cdot \varphi_{\gamma}$ with $\gamma \in \, ]0,1[$, 
and let $\mathds{P} \in \mathbb{S}_{>0}^{K}$ as well as $\widetilde{c}>0$ be arbitrary but fixed. 
Furthermore, let $W:=(W_{i})_{i\in \mathbb{N}}$
be an i.i.d. sequence of random variables  
with Compound-Poisson-Gamma distribution $\mathbb{\bbzeta} =C(POI(\theta ),GAM(\alpha ,\beta
))$ having parameters $\theta =\frac{\widetilde{c}}{\gamma }>0$, 
$\alpha =\frac{\widetilde{c}}{1-\gamma }>0$, $\beta =\frac{\gamma }{1-\gamma }>0$.
Then for all $A >0$ and all 
$\boldsymbol{\Omega}$\hspace{-0.23cm}$\boldsymbol{\Omega}\subset \mathbb{S}^{K}$ 
with \eqref{regularity}
there hold \eqref{brostu3:fo.norweiemp5.new}, \eqref{brostu3:fo.norweiemp6.new},
\eqref{brostu3:fo.norweiemp6.new2}
as well as
\begin{eqnarray}
& & \hspace{-0.9cm}
\sup_{\mathbf{Q}\in A \cdot \textrm{$\boldsymbol{\Omega}$\hspace{-0.19cm}$\boldsymbol{\Omega}$}}
H_{\gamma}(\mathbf{Q},\mathds{P})
=\lim_{n\rightarrow \infty }  A^{\gamma} \cdot \left( 1+ \gamma
\cdot \frac{1}{n}\cdot \log \,
\mathbb{\Pi}\negthinspace \left[\breve{\boldsymbol{\xi}}_{n}^{w\mathbf{W}}
\in \textrm{$\boldsymbol{\Omega}$\hspace{-0.23cm}$\boldsymbol{\Omega}$} \right]
\right) ^{1-\gamma }  ,\qquad \ 
\label{brostu3:fo.norweiemp6.new4b}
\\
& & \hspace{-0.9cm}
\sup_{\mathbf{Q}\in A \cdot \textrm{$\boldsymbol{\Omega}$\hspace{-0.19cm}$\boldsymbol{\Omega}$}}
c_{1} \negthinspace \cdot \negthinspace \Big(H_{\gamma}(\mathbf{Q},\mathds{P})^{c_{2}} - c_{3} \Big)
=\lim_{n\rightarrow \infty }  c_{1} \negthinspace \cdot \negthinspace \left\{ A^{c_{2} \cdot \gamma} \cdot
\negthinspace \left( 1+ 
\frac{\gamma}{n}\cdot \log \,
\mathbb{\Pi}\negthinspace \left[\breve{\boldsymbol{\xi}}_{n}^{w\mathbf{W}}
\in \textrm{$\boldsymbol{\Omega}$\hspace{-0.23cm}$\boldsymbol{\Omega}$} \right]
\right) ^{c_{2} \cdot (1-\gamma)} \negthinspace \negthinspace - c_{3} \right\} ,\ \ 
\textrm{if $c_{1} \cdot c_{2} >0$, $c_{3} \in \mathbb{R}$}, \quad \ 
\nonumber
\\
& & \hspace{-0.9cm}
\inf_{\mathbf{Q}\in A \cdot \textrm{$\boldsymbol{\Omega}$\hspace{-0.19cm}$\boldsymbol{\Omega}$}}
c_{1} \negthinspace \cdot \negthinspace \Big(H_{\gamma}(\mathbf{Q},\mathds{P})^{c_{2}} - c_{3} \Big)
=\lim_{n\rightarrow \infty }  c_{1} \negthinspace \cdot \negthinspace \left\{ A^{c_{2} \cdot \gamma} \cdot
\negthinspace \left( 1+ 
\frac{\gamma}{n}\cdot \log \,
\mathbb{\Pi}\negthinspace \left[\breve{\boldsymbol{\xi}}_{n}^{w\mathbf{W}}
\in \textrm{$\boldsymbol{\Omega}$\hspace{-0.23cm}$\boldsymbol{\Omega}$} \right]
\right) ^{c_{2} \cdot (1-\gamma)} \negthinspace \negthinspace - c_{3} \right\} ,\ \ 
\textrm{if $c_{1} \cdot c_{2} < 0$, $c_{3} \in \mathbb{R}$}, \quad \ 
\nonumber 
\\
& & \hspace{-0.9cm}
\sup_{\mathbf{Q}\in A \cdot \textrm{$\boldsymbol{\Omega}$\hspace{-0.19cm}$\boldsymbol{\Omega}$}}
c_{1} \negthinspace \cdot \negthinspace \Big(
\Big(\sum_{k=1}^{K} q_{k}^{\gamma}\Big)^{c_{2}} \negthinspace
-  c_{3} \Big)
=\lim_{n\rightarrow \infty }  c_{1} \negthinspace \cdot \negthinspace \left\{ 
K^{c_{2}\cdot(1-\gamma)} \cdot A^{c_{2} \cdot \gamma} \cdot
\negthinspace \left( 1+ 
\frac{\gamma}{n}\cdot \log \,
\mathbb{\Pi}\negthinspace \left[\breve{\breve{\boldsymbol{\xi}}}_{n}^{w\mathbf{W}}
\in \textrm{$\boldsymbol{\Omega}$\hspace{-0.23cm}$\boldsymbol{\Omega}$} \right]
\right) ^{c_{2} \cdot (1-\gamma)} \negthinspace \negthinspace - c_{3} \right\} ,\ \
\nonumber \\
& & \hspace{13.2cm} 
\textrm{if $c_{1} \cdot c_{2} >0$, $c_{3} \in \mathbb{R}$}, \quad \ 
\nonumber
\end{eqnarray}
\begin{eqnarray}
& & \hspace{-0.9cm}
\inf_{\mathbf{Q}\in A \cdot \textrm{$\boldsymbol{\Omega}$\hspace{-0.19cm}$\boldsymbol{\Omega}$}}
c_{1} \negthinspace \cdot \negthinspace \Big(
\Big(\sum_{k=1}^{K} q_{k}^{\gamma}\Big)^{c_{2}}  \negthinspace
-   c_{3} \Big)
=\lim_{n\rightarrow \infty }  c_{1} \negthinspace \cdot \negthinspace \left\{ 
K^{c_{2}\cdot(1-\gamma)} \cdot A^{c_{2} \cdot \gamma} \cdot
\negthinspace \left( 1+ 
\frac{\gamma}{n}\cdot \log \,
\mathbb{\Pi}\negthinspace \left[\breve{\breve{\boldsymbol{\xi}}}_{n}^{w\mathbf{W}}
\in \textrm{$\boldsymbol{\Omega}$\hspace{-0.23cm}$\boldsymbol{\Omega}$} \right]
\right) ^{c_{2} \cdot (1-\gamma)} \negthinspace \negthinspace - c_{3} \right\} ,\ \
\nonumber \\
& & \hspace{13.2cm} 
\textrm{if $c_{1} \cdot c_{2} < 0$, $c_{3} \in \mathbb{R}$} . \quad \ 
\nonumber \\
& & \hspace{-0.9cm}
\sup_{\mathbf{Q}\in A \cdot \textrm{$\boldsymbol{\Omega}$\hspace{-0.19cm}$\boldsymbol{\Omega}$}}
\frac{1}{1-\gamma} \negthinspace \cdot \negthinspace \log\Big(\sum_{k=1}^{K} q_{k}^{\gamma}\Big)
= \lim_{n\rightarrow \infty } \frac{1}{1-\gamma} \negthinspace \cdot \negthinspace \left[
\log\Big( A^{\gamma} \negthinspace \cdot \negthinspace \Big( 1+ \gamma
\cdot \frac{1}{n}\cdot 
\log \, \mathbb{\Pi}\negthinspace \left[\breve{\breve{\boldsymbol{\xi}}}_{n}^{w\mathbf{W}}
\in \textrm{$\boldsymbol{\Omega}$\hspace{-0.23cm}$\boldsymbol{\Omega}$} \right]
\Big) ^{1-\gamma } \Big) + (1-\gamma) \cdot \log(K) \right] \negthinspace \negthinspace .
\label{brostu3:fo.norweiemp6.new4a}
\end{eqnarray}
From this, the BS-minimizability/maximizability of the important 
norms/entropies/diversity indices 
(E1) to (E6) follows immediately as special cases.\\
(b) The special case $\varphi := \widetilde{c}\cdot \varphi_{\gamma}$  
($\gamma \in ]0,1[$ )
of Theorem \ref{brostu3:thm.divnormW.new} works analogously to (a), 
with the differences that we employ (i) additionally
a sequence $(X_{i})_{i\in \mathbb{N}}$  of random variables
being independent of $(W_{i})_{i\in \mathbb{N}}$ 
and satisfying condition \eqref{cv emp measure X to P}
(resp. \eqref{cv emp measure X to P vector}), 
(ii) $A=1$ (instead of arbitrary $A>0$),
(iii) $\mathbb{\Pi}_{X_{1}^{n}}[\cdot]$ (instead of $\mathbb{\Pi }[\cdot]$),
(iv) $\boldsymbol{\xi}_{n,\mathbf{X}}^{w\mathbf{W}}$
(instead of $\boldsymbol{\xi}_{n}^{w\mathbf{W}}$),
(v) $\breve{\boldsymbol{\xi}}_{n,\mathbf{X}}^{w\mathbf{W}}$
(instead of $\breve{\boldsymbol{\xi}}_{n}^{w\mathbf{W}}$),
and 
(vi) $\breve{\breve{\boldsymbol{\xi}}}_{n,\mathbf{X}}^{w\mathbf{W}}$
(instead of $\breve{\breve{\boldsymbol{\xi}}}_{n}^{w\mathbf{W}}$).

\end{proposition}

\vspace{0.3cm}
\noindent
Within the context of Subsection \ref{Subsect Estimators determ}, 
for the concrete simulative estimation 
$\widehat{D_{\widetilde{c} \cdot \varphi_{\gamma}}}(\mathbf{\Omega},\mathbf{P})$
via \eqref{estimator minimization} and \eqref{Improved IS for inf div new}, we
derive --- in terms of $M_{\mathbf{P}}:=\sum_{i=1}^{K}p_{i}>0$,
$n_{k} = n \cdot \widetilde{p}_{k} \in \mathbb{N}$ and
$\widetilde{q}_{k}^{\ast}$ from proxy method 1 or 2
--- that \\
$ \widetilde{U}_{k}^{\ast n_{k}}
=C\big( POI( n_{k}\cdot \breve{\theta}),
GAM\big(\frac{\widetilde{c} \cdot M_{\mathbf{P}}}{1-\gamma} - \tau_{k},
\frac{\gamma}{1-\gamma} \big) \big)$ with 
$\breve{\theta}:= \frac{\widetilde{c} \cdot M_{\mathbf{P}}}{\gamma}
\cdot \big(\frac{(\gamma -1) \cdot \tau_{k}}{\widetilde{c} \cdot M_{\mathbf{P}}}
+1\big)^{\gamma/(\gamma -1)}$ and 
$\tau_{k} = \widetilde{c} \cdot M_{\mathbf{P}} \cdot
\frac{
1-\big(\frac{\widetilde{q}_{k}^{\ast}}{\widetilde{p}_{k}}\big)^{\gamma-1}}{1-\gamma}$
for $\widetilde{q}_{k}^{\ast} >0$. Furthermore, 
\[
\widetilde{ISF}_{k}(x)=e^{-\tau _{k}x} \cdot \exp \left( \frac{n_{k}\cdot \widetilde{c} \cdot M_{\mathbf{P}}}{\gamma }
\cdot \left(\left(1+ \frac{\gamma -1}{\widetilde{c} \cdot M_{\mathbf{P}}} \cdot \tau _{k}\right) ^{\frac{
\gamma }{\gamma -1}}-1 \right)\right), \qquad x \geq 0 .
\]
Within the different context of Subsection \ref{Subsect Estimators stoch},
the corresponding estimators $\widehat{\Pi }_{L}^{improved}$ can be obtained 
analogously to the last paragraph of Subsection \ref{Subsect Case1},
with Proposition \ref{brostu3:cor.powdivexpl1} instead of 
Proposition \ref{brostu3:cor.powdivexpl2} (and \eqref{fo.Est1.Case1} remains the same).


\subsection{Case 3
\label{Subsect Case3}}

\noindent
For $\gamma >2$, $\widetilde{c} \in \, ]0,\infty[$  and
$]a_{F_{\gamma,\widetilde{c}}},b_{F_{\gamma,\widetilde{c}}}[ \,  := \, ]0,\infty[$  
we obtain 
\begin{eqnarray}
F_{\gamma,\widetilde{c}}(t) &:=&
\begin{cases}
\frac{\widetilde{c}}{\gamma-1} \cdot (t^{\gamma-1}-1), \qquad \textrm{if } \  
t \in \, ]0,\infty[, \\
-\frac{\widetilde{c}}{\gamma-1}, \hspace{2.2cm} 
\textrm{if } \  t=0, \\
- \infty, \hspace{2.5cm}  \textrm{if } \  t \in \, ]-\infty,0[, 
\end{cases}
\label{brostu3:Fcase3} 
\end{eqnarray}
$\mathcal{R}(F_{\gamma,\widetilde{c}})= \big[-\frac{\widetilde{c}}{\gamma-1},\infty \big[$,
$]\lambda_{-},\lambda_{+}[ \, = \big]-\frac{\widetilde{c}}{\gamma-1},\infty \big[$
(with $c:=0$),
$]t_{-}^{sc},t_{+}^{sc}[ \, = \, ]0,\infty[$ and
$]a,b[ \, =  \, ]-\infty,\infty[$.
Correspondingly, we deduce
from \eqref{brostu3:fo.Wfind2.new},\eqref{brostu3:fo.Wfind2b}
\begin{eqnarray}
\varphi_{\gamma,\widetilde{c}}(t) := \varphi_{\gamma,\widetilde{c}}^{(0)}(t)
\hspace{-0.2cm} &=& \hspace{-0.2cm}
\begin{cases}
\widetilde{c} \cdot \frac{t^\gamma-\gamma \cdot t+ \gamma - 1}{\gamma \cdot (\gamma-1)}
\ \in \ [0,\infty[,
\qquad
\quad \textrm{if } \ 
t \in \, ]0,\infty[,
\\
\frac{\widetilde{c}}{\gamma} > 0, \hspace{3.8cm} \textrm{if } \ t = 0, \\
\frac{\widetilde{c}}{\gamma} -\frac{\widetilde{c}}{\gamma-1} \cdot t \ \in \ ]0,\infty[,
\hspace{1.6cm} \textrm{if } \  t \in \, ]-\infty,0[, 
\end{cases}
\label{brostu3:fo.expowlink2.Case3}
\end{eqnarray}
which coincides with $\widetilde{c} \cdot \varphi_{\gamma}(t)$
for $\varphi_{\gamma}(t)$ from \eqref{brostu3:fo.powdivgen} and 
which generates the $\gamma-$corresponding power divergences given in \eqref{brostu3:fo.powdiv.new}. 
Moreover, we can derive from formula \eqref{brostu3:fo.Wfind1.new} 
(see also \eqref{brostu3:fo.Wfind1b}) 
\begin{eqnarray}
\Lambda_{\gamma,\widetilde{c}}(z) := \Lambda_{\gamma,\widetilde{c}}^{(0)}(z) &=&
\begin{cases}
\frac{\widetilde{c}}{\gamma} \cdot
\left\{ \left( \frac{\gamma -1}{\widetilde{c}} \cdot z + 
1 \right)^{\frac{\gamma}{\gamma-1}} -1  \right\}, \qquad \textrm{if } \ z \in \big]-\frac{\widetilde{c}}{\gamma-1},\infty \big[, \\
- \frac{\widetilde{c}}{\gamma} < 0, \hspace{3.7cm} \textrm{if } \  
  z = -\frac{\widetilde{c}}{\gamma-1}, \\
\infty, \hspace{4.5cm}  \textrm{if } \  
  z \in \, ]-\infty, -\frac{\widetilde{c}}{\gamma-1}[ . 
\end{cases}
\label{brostu3:fo.expowlink1.Case3} 
\end{eqnarray}
The latter
is the cumulant generating function of a 
\textquotedblleft distorted stable distribution\textquotedblright\ 
$\mathbb{\bbzeta}[ \, \cdot \,] = \mathbb{\Pi}[W \in \cdot \, ]$
of a random variable $W$,
which can be constructed as follows: let $Z$ be an auxiliary random variable 
(having density $f_{Z}$ and support $supp(Z) = ]-\infty,\infty[$) of a stable law
with parameter-quadruple
$(\frac{\gamma }{\gamma -1},1,0,\frac{\widetilde{c}^{1/(1-\gamma )} \cdot (\gamma -1)^{\gamma /(\gamma -1)}}{\gamma })$
 in terms of the above-mentioned \textquotedblleft form-B notation\textquotedblright ;
by applying a general Laplace-transform result on p.~112 of~\cite{Zol:86} 
we can derive
\begin{eqnarray}
M_{Z}(z) := E_{\mathbb{\Pi}}[\exp(z \cdot Z)] 
= \int_{-\infty}^{\infty} \exp(z \cdot y) \cdot f_{Z}(y) \, dy  \hspace{-0.2cm} &=& \hspace{-0.2cm}
\begin{cases}
\exp\Big(
\frac{\widetilde{c}^{1/(1-\gamma )} \cdot (\gamma -1 )^{\gamma /(\gamma -1)}}{\gamma } 
\cdot (-z)^{\alpha} \Big),  
\quad \textrm{if } \  z \in ]-\infty,0] , \\
\infty, \hspace{5.1cm} \textrm{if } \  z \in \, ]0,\infty[, \\
\end{cases}
\label{brostu3:diststab1}
\end{eqnarray}
where $\alpha := \frac{\gamma}{\gamma -1} \in \, ]1,2[$. Since $0 \notin int(dom(M_{Z}))$
(and thus, $Z$ does not have light-tails) we have to distort the density
in order to extend the effective domain. Accordingly, let $W$ be a random variable having density
\begin{equation}
f_{W}(y)\ :=\ \frac{\exp \{\frac{y \cdot \widetilde{c}}{\gamma-1}\}}{\exp \{\widetilde{c}/\gamma \}}
\cdot f_{Z}(- y), \qquad y \in \mathbb{R}.
\label{brostu3:fo.norweiemp7atwo}
\end{equation}
Then one can straightforwardly deduce from \eqref{brostu3:diststab1} that  
$\int_{-\infty}^{\infty} f_{W}(y) \, dy =1$ and that
\begin{eqnarray}
M_{W}(z) := E_{\mathbb{\Pi}}[\exp(z \cdot W)] 
= \int_{-\infty}^{\infty} \exp(z \cdot y) \cdot f_{W}(y) \, dy  \hspace{-0.2cm} &=& \hspace{-0.2cm}
\begin{cases}
\exp\left(\frac{\widetilde{c}}{\gamma} \cdot
\left\{ \left( \frac{\gamma -1}{\widetilde{c}} \cdot z + 
1 \right)^{\frac{\gamma}{\gamma-1}} -1  \right\} \right),  
\qquad \textrm{if } \  z \in [- \frac{\widetilde{c}}{\gamma -1},\infty[ , \\
\infty, \hspace{5.5cm} \textrm{if } \  z \in \, ]-\infty, -\frac{\widetilde{c}}{\gamma-1}[ . \\
\end{cases}
\nonumber
\end{eqnarray}
Notice that $\mathbb{\bbzeta}$ is an infinitely divisible (cf. Proposition \ref{ID1}) continuous distribution with density $f_{W}$,
and that
$\mathbb{\bbzeta}[ \, ]0,\infty[ \, ] = \mathbb{\Pi}[W > 0]=  
\int_{0}^{\infty} f_{W}(u) \, du \in \, ]0,1[$,
$\mathbb{\bbzeta}[ \, \{0\} \, ] = \mathbb{\Pi}[W = 0]= 0$. 
Concerning the important Remark \ref{dist of components}(i), 
for i.i.d. copies $(W_{i})_{i \in \mathbb{N}}$ of $W$ we obtain the 
probability distribution 
$\mathbb{\bbzeta}^{\ast n_{k}}[\cdot] :=  \mathbb{\Pi}[\breve{W} \in \cdot \, ]$ 
of $\breve{W} := \sum_{i\in I_{k}^{(n)}} W_{i}$ 
having the density
\begin{equation}
f_{\breve{W}}(y)\ :=\ \frac{\exp \{\frac{y \cdot \widetilde{c}}{\gamma -1}\}}{\exp \{\widetilde{c} \cdot card(I_{k}^{(n)})/\gamma \}}
\cdot f_{\breve{Z}}(- y), \qquad y \in \mathbb{R},
\label{brostu3:fo.norweiemp7a.sum}
\end{equation}
where $\breve{Z}$ is a random variable with density $f_{\breve{Z}}$ of a stable law
with parameters
$(\frac{\gamma }{\gamma -1},1,0,card(I_{k}^{(n)}) 
\cdot \frac{\widetilde{c}^{1/(1-\gamma )} \cdot (\gamma -1)^{\gamma /(\gamma -1)}}{\gamma })$.
For the bare-simulation-minimizations (respectively maximizations)
of the corresponding ($\gamma-$order) power divergences, 
Renyi divergences,
Hellinger integrals
and measures of entropy (diversity),
we obtain from Theorem 
\ref{brostu3:thm.divnormW.new} respectively Remark \ref{remark divnormW}(vi),
Lemma \ref{Lemma Indent Rate finite case_new}(a), \eqref{fo.hgamma1},
\eqref{def Renyi}, \eqref{fo.genentropy1}, \eqref{fo.genentropy2} the following  

\vspace{0.2cm}

\begin{proposition}
\label{brostu3:cor.powdivexpl3} 
(a) Consider $\varphi := \widetilde{c}\cdot \varphi_{\gamma}$ with $\gamma >2$, 
and let $\mathds{P} \in \mathbb{S}_{>0}^{K}$ as well as $\widetilde{c}>0$ be arbitrary but fixed. 
Furthermore, let $W:=(W_{i})_{i\in \mathbb{N}}$
be an i.i.d. sequence of real-valued random variables having density
\eqref{brostu3:fo.norweiemp7atwo}.
Then for all $A >0$ and 
$\boldsymbol{\Omega}$\hspace{-0.23cm}$\boldsymbol{\Omega}\subset \mathbb{S}^{K}$ with \eqref{regularity}
there hold all the BS-extremizabilites
\eqref{brostu3:fo.norweiemp5.new} to \eqref{brostu3:fo.norweiemp6.new3b}
as well as \eqref{brostu3:fo.norweiemp6.new4a}.
From this, the BS-minimizability/maximizability of the important 
norms/entropies/diversity indices 
(E1) to (E6) follow immediately as special cases.\\
(b) The special case $\varphi := \widetilde{c}\cdot \varphi_{\gamma}$   
($\gamma >2$)
of Theorem \ref{brostu3:thm.divnormW.new} works analogously to (a), 
with the differences that we employ (i) additionally
a sequence $(X_{i})_{i\in \mathbb{N}}$  of random variables
being independent of $(W_{i})_{i\in \mathbb{N}}$ 
and satisfying condition \eqref{cv emp measure X to P}
(resp. \eqref{cv emp measure X to P vector}), 
(ii) $A=1$ (instead of arbitrary $A>0$),
(iii) $\mathbb{\Pi}_{X_{1}^{n}}[\cdot]$ (instead of $\mathbb{\Pi }[\cdot]$),
(iv) $\boldsymbol{\xi}_{n,\mathbf{X}}^{w\mathbf{W}}$
(instead of $\boldsymbol{\xi}_{n}^{w\mathbf{W}}$),
(v) $\breve{\boldsymbol{\xi}}_{n,\mathbf{X}}^{w\mathbf{W}}$
(instead of $\breve{\boldsymbol{\xi}}_{n}^{w\mathbf{W}}$),
and 
(vi) $\breve{\breve{\boldsymbol{\xi}}}_{n,\mathbf{X}}^{w\mathbf{W}}$
(instead of $\breve{\breve{\boldsymbol{\xi}}}_{n}^{w\mathbf{W}}$).

\end{proposition}

\vspace{0.3cm}
\noindent
Within the context of Subsection \ref{Subsect Estimators determ}, 
for the concrete simulative estimation 
$\widehat{D_{\widetilde{c} \cdot \varphi_{\gamma}}}(\mathbf{\Omega},\mathbf{P})$
via \eqref{estimator minimization} and \eqref{Improved IS for inf div new}, we
derive --- in terms of 
$M_{\mathbf{P}}:=\sum_{i=1}^{K}p_{i}>0$,
$n_{k} = n \cdot \widetilde{p}_{k} \in \mathbb{N}$ and
$\widetilde{q}_{k}^{\ast}$ from proxy method 1 or 2
--- that $\widetilde{U}_{k}^{\ast n_{k}}$
has the (Lebesgue-)density
\[
f_{\widetilde{U}_{k}^{\ast n_{k}}
}(x) \ :=  \ \frac{\exp((\tau_{k} + \frac{\widetilde{c} \cdot M_{\mathbf{P}}}{\gamma-1})\cdot x)}{
\exp\left(n_{k} \cdot \frac{\widetilde{c} \cdot M_{\mathbf{P}}}{\gamma}\cdot
(1+\frac{\gamma-1}{\widetilde{c} \cdot M_{\mathbf{P}}} \cdot \tau_{k})^{\gamma/(\gamma-1)}\right)} 
\cdot f_{\breve{\breve{Z}}}(-x) , 
\qquad x \in \mathbb{R}, 
\]

\newpage

\noindent
where $\tau_{k} = - \frac{\widetilde{c} \cdot M_{\mathbf{P}}}{\gamma-1} \cdot
\big(1- \big(\frac{\widetilde{q}_{k}^{\ast}}{\widetilde{p}_{k}}\big)^{\gamma-1}
\cdot \textfrak{1}_{]0,\infty[}(\widetilde{q}_{k}^{\ast}) \big)$
for $\widetilde{q}_{k}^{\ast} \in \mathbb{R}$, and  
$\breve{\breve{Z}}$ is a random variable with density $f_{\breve{\breve{Z}}}$ of a stable law
with parameter-quadruple
$(\frac{\gamma }{\gamma-1},1,0,n_{k} \cdot \frac{(\widetilde{c}\cdot M_{\mathbf{P}})^{1/(1-\gamma )} \cdot (\gamma-1)^{\gamma /(\gamma-1)}}{\gamma })$
(analogously to $\breve{Z}$ of 
\eqref{brostu3:fo.norweiemp7a.sum}
but with $\widetilde{c}$ replaced by $\widetilde{c}\cdot M_{\mathbf{P}}$). Furthermore,
$\widetilde{ISF}_{k}(x)=e^{-\tau _{k} \cdot x} \cdot \exp \Big( \frac{n_{k}\cdot \widetilde{c} \cdot M_{\mathbf{P}}}{\gamma }
\cdot \Big(\left(1+ \frac{\gamma -1}{\widetilde{c} \cdot M_{\mathbf{P}}} \cdot \tau _{k}\right) ^{\frac{
\gamma }{\gamma -1}}-1 \Big)\Big), \ x \in \mathbb{R} $.
For the above random variables,
algorithms for simulation can be obtained by adapting e.g.
the works of \cite{Dev:09} and \cite{Dev:14}.
Within the different context of Subsection \ref{Subsect Estimators stoch},
the corresponding estimators $\widehat{\Pi }_{L}^{improved}$ can be obtained 
analogously to the last paragraph of Subsection \ref{Subsect Case1},
with Proposition \ref{brostu3:cor.powdivexpl3} instead of 
Proposition \ref{brostu3:cor.powdivexpl2} (and \eqref{fo.Est1.Case1} remains the same).


\subsection{Case 4
\label{Subsect Case4}}

\noindent
For $\gamma =2$, $\widetilde{c} \in \, ]0,\infty[$  and
$]a_{F_{\gamma,\widetilde{c}}},b_{F_{\gamma,\widetilde{c}}}[ \,  = \, ]-\infty,\infty[$ 
we define $F_{2,\widetilde{c}}(t) :=
\widetilde{c} \cdot (t-1)$ ($t \in \, ]-\infty,\infty[)$.
Clearly, $\mathcal{R}(F_{2,\widetilde{c}})= \, ]-\infty, \infty [$ 
and $F_{2,\widetilde{c}} \in \textgoth{F}$. 
Since $F_{2,\widetilde{c}}(1)=0$, let us  
choose the natural anchor point $c:=0$,
which leads to $]\lambda_{-},\lambda_{+}[\, 
= \, ]-\infty, \infty [$, 
$]t_{-}^{sc},t_{+}^{sc}[ = \, ]-\infty, \infty [$ and 
$]a,b[ \, = \, ]-\infty,\infty[$.
By using $F_{2,\widetilde{c}}^{-1}(x) = 1+\frac{x}{\widetilde{c}}$ for 
$x \in int(\mathcal{R}(F_{2,\widetilde{c}}))$, 
from \eqref{brostu3:fo.Wfind2.new},\eqref{brostu3:fo.Wfind2b} 
we deduce 
\begin{eqnarray}
\varphi_{2,\widetilde{c}}(t) := \varphi_{2,\widetilde{c}}^{(0)}(t)
&=& 
\widetilde{c} \cdot \frac{(t-1)^{2}}{2} 
\ \in \ [0,\infty[,
\qquad
t \in \, ]-\infty,\infty[,
\nonumber
\end{eqnarray}
which coincides with $\widetilde{c} \cdot \varphi_{2}(t)$
for $\varphi_{2}(t)$ from \eqref{brostu3:fo.powdivgen}
which generates the $\widetilde{c}-$fold of  
the half Pearson-chisquare divergence
given in 
the sixth line of \eqref{brostu3:fo.powdiv.new}. 
Moreover, from 
\eqref{brostu3:fo.Wfind1.new},\eqref{brostu3:fo.Wfind1b}
we derive
\begin{eqnarray}
\Lambda_{2,\widetilde{c}}(z) := \Lambda_{2,\widetilde{c}}^{(0)}(z) &=&
\frac{z^2}{2 \widetilde{c}} + z
\ = \ \frac{\widetilde{c}}{2} \cdot
\Big\{ \Big( \frac{1}{\widetilde{c}} \cdot z + 
1 \Big)^{2} -1  \Big\},   
\qquad  z \in \, ]-\infty,\infty [, 
\nonumber
\end{eqnarray}
which is the well-known 
cumulant generating function of the Normal distribution
(Gaussian distribution) $\mathbb{\bbzeta} = NOR(1,\frac{1}{\widetilde{c}})$
with mean $1$ and variance $\frac{1}{\widetilde{c}}$.
Notice that $\mathbb{\bbzeta}$ is an infinitely divisible 
(cf. Proposition \ref{ID1}) continuous distribution with density  
$f_{NOR(1,\frac{1}{\widetilde{c}})}(y) := \sqrt{\frac{\widetilde{c}}{2 \pi}} \cdot \exp(- \frac{\widetilde{c} \cdot (y-1)^2}{2} )$
($y \in \mathbb{R}$) and that
$\mathbb{\bbzeta}[ \, ]0,\infty[ \, ] = \mathbb{\Pi}[W > 0]=  
\int_{0}^{\infty} f_{NOR(1,\frac{1}{\widetilde{c}})}(u) \, du \in \, ]0,1[$, \ 
$\mathbb{\bbzeta}[ \, \{0\} \, ] = \mathbb{\Pi}[W = 0]= 0$. 
Concerning the important Remark \ref{dist of components}(i), 
for i.i.d. copies $(W_{i})_{i \in \mathbb{N}}$ of $W$ the 
probability distribution 
$\mathbb{\bbzeta}^{\ast n_{k}}[\cdot] :=  \mathbb{\Pi}[\breve{W} \in \cdot \, ]$ 
of $\breve{W} := \sum_{i\in I_{k}^{(n)}} W_{i}$  
is $NOR(card(I_{k}^{(n)}),\frac{card(I_{k}^{(n)})}{\widetilde{c}})$.
For the desired bare-simulation-optimizations
we obtain from
Theorem \ref{brostu3:thm.divnormW.new} respectively Remark \ref{remark divnormW}(vi),
Lemma \ref{Lemma Indent Rate finite case_new}(a), \eqref{fo.hgamma1},
\eqref{def Renyi}, \eqref{fo.genentropy1}, \eqref{fo.genentropy2} the following  

\vspace{0.2cm}

\begin{proposition}
\label{brostu3:cor.powdivexplNOR}
(a) Consider $\varphi := \widetilde{c}\cdot \varphi_{\gamma}$ with $\gamma =2$, 
and let $\mathds{P} \in \mathbb{S}_{>0}^{K}$ as well as $\widetilde{c}>0$ be arbitrary but fixed. 
Furthermore, let $W:=(W_{i})_{i\in \mathbb{N}}$
be an i.i.d. sequence of real-valued random variables  
with probability distribution $\mathbb{\bbzeta} =NOR(1,\frac{1}{\widetilde{c}})$.
Then for all $A >0$ and 
$\boldsymbol{\Omega}$\hspace{-0.23cm}$\boldsymbol{\Omega}\subset \mathbb{S}^{K}$ with \eqref{regularity} there hold all the BS-extremizabilites
\eqref{brostu3:fo.norweiemp5.new} to \eqref{brostu3:fo.norweiemp6.new3b}
as well as \eqref{brostu3:fo.norweiemp6.new4a} 
with plugging-in $\gamma=2$.
From this, the BS-minimizability/maximizability of the important norms/entropies/diversity indices 
(E1) to (E6) follow as special cases.
By Remark \ref{rem.inversions}(c), one can even take $A<0$ in
\eqref{brostu3:fo.norweiemp5.new} to \eqref{brostu3:fo.norweiemp6.new3b}
and \eqref{brostu3:fo.norweiemp6.new4a}
as well as in (E1),(E2),(E4),(E6).
\\
(b) The special case $\varphi := \widetilde{c}\cdot \varphi_{\gamma}$ \,  
($\gamma =2$)
of Theorem \ref{brostu3:thm.divnormW.new} works analogously to (a), 
with the differences that we employ (i) additionally
a sequence $(X_{i})_{i\in \mathbb{N}}$  of random variables
being independent of $(W_{i})_{i\in \mathbb{N}}$ 
and satisfying condition \eqref{cv emp measure X to P}
(resp. \eqref{cv emp measure X to P vector}), 
(ii) $A=1$ (instead of arbitrary $A>0$),
(iii) $\mathbb{\Pi}_{X_{1}^{n}}[\cdot]$ (instead of $\mathbb{\Pi }[\cdot]$),
(iv) $\boldsymbol{\xi}_{n,\mathbf{X}}^{w\mathbf{W}}$
(instead of $\boldsymbol{\xi}_{n}^{w\mathbf{W}}$),
(v) $\breve{\boldsymbol{\xi}}_{n,\mathbf{X}}^{w\mathbf{W}}$
(instead of $\breve{\boldsymbol{\xi}}_{n}^{w\mathbf{W}}$),
and 
(vi) $\breve{\breve{\boldsymbol{\xi}}}_{n,\mathbf{X}}^{w\mathbf{W}}$
(instead of $\breve{\breve{\boldsymbol{\xi}}}_{n}^{w\mathbf{W}}$).

\end{proposition}

\vspace{0.3cm}
\noindent
Within the context of Subsection \ref{Subsect Estimators determ}, 
for the concrete simulative estimation 
$\widehat{D_{\widetilde{c} \cdot \varphi_{2}}}(\mathbf{\Omega},\mathbf{P})$
via \eqref{estimator minimization} and \eqref{Improved IS for inf div new}, we
derive --- in terms of $M_{\mathbf{P}}:=\sum_{i=1}^{K}p_{i}>0$,
$n_{k} = n \cdot \widetilde{p}_{k} \in \mathbb{N}$ and
$\widetilde{q}_{k}^{\ast}$ from proxy method 1 or 2
--- that 
$\widetilde{U}_{k}^{\ast n_{k}}
=NOR(n_{k}\cdot(1+\frac{\tau_{k}}{\widetilde{c} \cdot M_{\mathbf{P}}}),
\frac{n_{k}}{\widetilde{c} \cdot M_{\mathbf{P}}})$ 
with $\tau_{k}=\widetilde{c} \cdot M_{\mathbf{P}} \cdot 
( \frac{\widetilde{q}_{k}^{\ast}}{\widetilde{p}_{k}} - 1 ) $ 
for $\widetilde{q}_{k}^{\ast} \in \mathbb{R}$.
Moreover, for all $x \in \mathbb{R}$ one obtains
$\widetilde{ISF}_{k}(x)= \exp\big(\frac{n_{k} \cdot \tau_{k}^2}{2\widetilde{c} 
\cdot M_{\mathbf{P}}}    
- (x- n_{k}) \cdot \tau_{k} \big)$.
Within the different context of Subsection \ref{Subsect Estimators stoch},
the corresponding estimators $\widehat{\Pi }_{L}^{improved}$ can be obtained 
analogously to the last paragraph of Subsection \ref{Subsect Case1},
with Proposition \ref{brostu3:cor.powdivexplNOR} instead of 
Proposition \ref{brostu3:cor.powdivexpl2} (and \eqref{fo.Est1.Case1} remains the same).


\subsection{Case 5
\label{Subsect Case5}}

\noindent
For $\gamma =0$, $\widetilde{c} \in \, ]0,\infty[$  and
$]a_{F_{\gamma,\widetilde{c}}},b_{F_{\gamma,\widetilde{c}}}[ \,  := \, ]0,\infty[$ 
we obtain the same $F_{\gamma,\widetilde{c}}(t)$ of \eqref{brostu3:Fcase1and2},
$\mathcal{R}(F_{\gamma,\widetilde{c}})= \big]-\infty, \frac{\widetilde{c}}{1-\gamma}\big[$,
$]\lambda_{-},\lambda_{+}[ \, = \big]-\infty, \frac{\widetilde{c}}{1-\gamma}\big[$
(with $c:=0$),
$]t_{-}^{sc},t_{+}^{sc}[ \, = \, ]0,\infty[$ and
$]a,b[ \, =  \, ]0,\infty[$. 
By using $F_{0,\widetilde{c}}^{-1}(x) = 
\frac{1}{1- \frac{x}{\widetilde{c}}} $ for 
$x \in int(\mathcal{R}(F_{0,\widetilde{c}})) = ]-\infty, \widetilde{c}[$, 
we can deduce from \eqref{brostu3:fo.Wfind2.new},\eqref{brostu3:fo.Wfind2b} 
\begin{eqnarray}
\varphi_{0,\widetilde{c}}(t) := \varphi_{0,\widetilde{c}}^{(0)}(t)
\hspace{-0.2cm} &=& \hspace{-0.2cm}
\begin{cases}
\widetilde{c} \cdot \left(- \log t + t -1  \right)
\ \in \ [0,\infty[,
\qquad
\quad \textrm{if } 
t \in \, ]0,\infty[,
\\
\infty, \hspace{5.1cm} \textrm{if } t \in \, ]-\infty, 0] ,
\end{cases}
\nonumber
\end{eqnarray}
which coincides with $\widetilde{c} \cdot \varphi_{0}(t)$
for the generator $\varphi_{0}(t)$ from \eqref{brostu3:fo.powdivgen}
which generates the reverse Kullback-Leibler divergence (reverse relative entropy) 
given in the second line of \eqref{brostu3:fo.powdiv.new} with $\widetilde{c}=1$. 
Furthermore, we can derive from \eqref{brostu3:fo.Wfind1.new},\eqref{brostu3:fo.Wfind1b} 
\begin{eqnarray}
\Lambda_{0,\widetilde{c}}(z) := \Lambda_{0,\widetilde{c}}^{(0)}(z) &=&
\begin{cases}
- \widetilde{c} \cdot \log\left( 1 - \frac{z}{\widetilde{c}} \right), \qquad \textrm{if } \  
 z \in \big]-\infty,\widetilde{c} \big[, \\
\infty, \hspace{2.75cm}  \textrm{if } \  z \in \big[\widetilde{c}, \infty \big[ , 
\end{cases}
\nonumber
\end{eqnarray}
which is the cumulant generating function of
the Gamma distribution
$\mathbb{\bbzeta} = GAM(\widetilde{c},\widetilde{c})$
with rate parameter (inverse scale parameter) $\widetilde{c}$ and 
shape parameter $\widetilde{c}$; 
the special case $\widetilde{c} =1$   
leads to $\mathbb{\bbzeta} = GAM(1,1) = EXP(1)$ being the exponential distribution with mean 1.
Notice that $\mathbb{\bbzeta}$ is an infinitely divisible (cf. Proposition \ref{ID1}) continuous distribution with density 
$f(y) := \frac{\widetilde{c}^{\widetilde{c}} \cdot 
y^{\widetilde{c}-1} \cdot e^{-\widetilde{c} \cdot y} }{\Gamma(\widetilde{c})} \cdot \textfrak{1}_{]0,\infty[}(y)$
($y \in \mathbb{R}$),
and that $\mathbb{\bbzeta}[ \, ]0,\infty[ \, ] = \mathbb{\Pi}[W>0]=1$.
Concerning the important Remark \ref{dist of components}(i), 
for i.i.d. copies $(W_{i})_{i \in \mathbb{N}}$ of $W$ the 
probability distribution 
$\mathbb{\bbzeta}^{\ast n_{k}}[\cdot] :=  \mathbb{\Pi}[\breve{W} \in \cdot \, ]$ 
of $\breve{W} := \sum_{i\in I_{k}^{(n)}} W_{i}$ 
is $GAM(\widetilde{c},\widetilde{c} \cdot card(I_{k}^{(n)}))$.
For the desired bare-simulation-optimizations we obtain from
Theorem \ref{brostu3:thm.divnormW.new}, 
Remark \ref{remark divnormW}(vi) and
Lemma \ref{Lemma Indent Rate finite case_new}(c)
the following

\vspace{0.2cm}

\begin{proposition}
\label{brostu3:cor.powdivexpl4} 
(a) Consider $\varphi := \widetilde{c}\cdot \varphi_{\gamma}$ with $\gamma =0$, 
and let $\mathds{P} \in \mathbb{S}_{>0}^{K}$ as well as $\widetilde{c}>0$ be arbitrary but fixed. 
Furthermore, let $W:=(W_{i})_{i\in \mathbb{N}}$
be an i.i.d. sequence of non-negative real-valued random variables  
with Gamma distribution $\mathbb{\bbzeta} =GAM(\widetilde{c},\widetilde{c})$.  
Then for all $A >0$ and all 
$\boldsymbol{\Omega}$\hspace{-0.23cm}$\boldsymbol{\Omega}\subset \mathbb{S}_{>0}^{K}$ with \eqref{regularity} 
there hold the BS minimizabilites (cf. \eqref{brostu3:fo.2})
\begin{eqnarray}
& & \inf_{\mathbf{Q}\in A \cdot \textrm{$\boldsymbol{\Omega}$\hspace{-0.19cm}$\boldsymbol{\Omega}$}}
D_{\widetilde{c}\cdot \varphi_{0}}(\mathbf{Q},\mathds{P}) = 
-\lim_{n\rightarrow \infty }\frac{1}{n} \log \,
\mathbb{\Pi}\negthinspace \left[\boldsymbol{\xi}_{n}^{w\mathbf{W}}
\in \textrm{$\boldsymbol{\Omega}$\hspace{-0.23cm}$\boldsymbol{\Omega}$} \right] + \widetilde{c} \cdot (A-1- \log A),
\nonumber
\\
& & \inf_{\mathbf{Q}\in A \cdot \textrm{$\boldsymbol{\Omega}$\hspace{-0.19cm}$\boldsymbol{\Omega}$}}
\widetilde{I}(\mathbf{Q},\mathds{P}) = 
\inf_{\mathbf{Q}\in A \cdot \textrm{$\boldsymbol{\Omega}$\hspace{-0.19cm}$\boldsymbol{\Omega}$}}
\sum\displaylimits_{k=1}^{K} p_{k} \cdot \log\left( \frac{p_{k}}{q_{k}} \right)
= -\lim_{n\rightarrow \infty }\frac{1}{n} \log \,
\mathbb{\Pi}\negthinspace \left[\boldsymbol{\xi}_{n}^{w\mathbf{W}}
\in \textrm{$\boldsymbol{\Omega}$\hspace{-0.23cm}$\boldsymbol{\Omega}$} \right] - \log A .
\nonumber
\end{eqnarray}
(b) The special case $\varphi := \widetilde{c}\cdot \varphi_{\gamma}$ \,  
($\gamma =0$)
of Theorem \ref{brostu3:thm.divnormW.new} works analogously to (a), 
with the differences that we employ (i) additionally
a sequence $(X_{i})_{i\in \mathbb{N}}$  of random variables
being independent of $(W_{i})_{i\in \mathbb{N}}$ 
and satisfying condition \eqref{cv emp measure X to P}
(resp. \eqref{cv emp measure X to P vector}), 
(ii) $A=1$ (instead of arbitrary $A>0$),
(iii) $\mathbb{\Pi}_{X_{1}^{n}}[\cdot]$ (instead of $\mathbb{\Pi }[\cdot]$),
and (iv) $\boldsymbol{\xi}_{n,\mathbf{X}}^{w\mathbf{W}}$
(instead of $\boldsymbol{\xi}_{n}^{w\mathbf{W}}$).

\end{proposition}

\vspace{0.3cm}
\noindent
Within the context of Subsection \ref{Subsect Estimators determ}, 
for the concrete simulative estimation 
$\widehat{D_{\widetilde{c} \cdot \varphi_{0}}}(\mathbf{\Omega},\mathbf{P})$
via \eqref{estimator minimization} and \eqref{Improved IS for inf div new}, we
derive --- in terms of $M_{\mathbf{P}}:=\sum_{i=1}^{K}p_{i}>0$,
$n_{k} = n \cdot \widetilde{p}_{k} \in \mathbb{N}$ and
$\widetilde{q}_{k}^{\ast}$ from proxy method 1 or 2
--- that
$\widetilde{U}_{k}^{\ast n_{k}}
=GAM\left( \widetilde{c} \cdot M_{\mathbf{P}} - \tau_{k},
n_{k}\cdot \widetilde{c} \cdot M_{\mathbf{P}}\right)$, with 
$\tau_{k}=\widetilde{c} \cdot M_{\mathbf{P}} \cdot ( 1- 
\frac{\widetilde{p}_{k}}{\widetilde{q}_{k}^{\ast}})$
for $\widetilde{q}_{k}^{\ast} >0$ (the latter is equivalent to $\tau_{k} < \widetilde{c} \cdot M_{\mathbf{P}}$).
Moreover, for all $x >0$ one gets $\widetilde{ISF}_{k}(x)=
\left( \frac{\widetilde{c} \cdot M_{\mathbf{P}}}{
\widetilde{c}\cdot M_{\mathbf{P}} - \tau _{k}}\right)^{n_{k}\cdot \widetilde{c}\cdot M_{\mathbf{P}}}
\cdot e^{-\tau_{k} \cdot x}$.
Within the different context of Subsection \ref{Subsect Estimators stoch},
the corresponding estimators $\widehat{\Pi }_{L}^{improved}$ can be obtained 
analogously to the last paragraph of Subsection \ref{Subsect Case1},
with Proposition \ref{brostu3:cor.powdivexpl4} instead of 
Proposition \ref{brostu3:cor.powdivexpl2} 
and with 
$\widehat{D_{\widetilde{c}\cdot \varphi
_{0}}(\textrm{$\boldsymbol{\Omega}$\hspace{-0.23cm}$\boldsymbol{\Omega}$},\mathds{P})
}
\ := \ 
-\frac{1}{n}\log \widehat{\Pi}_{L}^{improved} 
$
instead of \eqref{fo.Est1.Case1}.


\subsection{Case 6
\label{Subsect Case6}}


\subsubsection{Case 6a: anchor point $c=0$}

\noindent
for $\gamma =1$, $\widetilde{c} \in \, ]0,\infty[$  and
$]a_{F_{1,\widetilde{c}}},b_{F_{1,\widetilde{c}}}[ \, = \, ]0,\infty[$ 
we define
\begin{eqnarray}
F_{1,\widetilde{c}}(t)  &:=&
\begin{cases}
\widetilde{c} \cdot \log t = 
\lim_{\gamma \rightarrow 1} F_{\gamma,\widetilde{c}}(t), \qquad \textrm{if } t \in \, ]0,\infty[, \\
-\infty, \hspace{3.8cm} \textrm{if } t \in \, ]-\infty,0]. \\
\end{cases}
\nonumber
\end{eqnarray}
Clearly, $\mathcal{R}(F_{1,\widetilde{c}})= ]-\infty,\infty [$ 
and $F_{1,\widetilde{c}} \in \textgoth{F}$. 
Since $F_{1,\widetilde{c}}(1)=0$, let us \textit{first} 
choose the natural anchor point $c:=0$,
which leads to $]\lambda_{-},\lambda_{+}[ \, = int(\mathcal{R}(F_{1,\widetilde{c}})) = 
\, ]-\infty,\infty [$, $]t_{-}^{sc},t_{+}^{sc}[ \, = \, ]0,\infty[$
and $]a,b[ \, = \, ]0,\infty[$.
By using $F_{1,\widetilde{c}}^{-1}(x) = \exp(\frac{x}{\widetilde{c}})$ for 
$x \in \mathcal{R}(F_{1,\widetilde{c}})$, 
we deduce from \eqref{brostu3:fo.Wfind2.new},\eqref{brostu3:fo.Wfind2b}
\begin{eqnarray}
\varphi_{1,\widetilde{c}}(t) := \varphi_{1,\widetilde{c}}^{(0)}(t)
\hspace{-0.2cm} &:=& \hspace{-0.2cm}
\begin{cases}
\widetilde{c} \cdot \left( t \cdot \log t + 1 - t  \right)
\ \in \ [0,\infty[,
\qquad
\quad \textrm{if } t \in \, ]0,\infty[,
\\
\widetilde{c}, \hspace{5.3cm} \textrm{if }  t = 0, \\
\infty, \hspace{5.1cm} \textrm{if } t \in \, ]-\infty,0[ ,
\end{cases}
\nonumber
\end{eqnarray}
which coincides with $\widetilde{c} \cdot \varphi_{1}(t)$
for the generator $\varphi_{1}(t)$ from \eqref{brostu3:fo.powdivgen}
which generates the Kullback-Leibler divergence (relative entropy) 
given in the fourth line of \eqref{brostu3:fo.powdiv.new} with $\widetilde{c}=1$.
Moreover, we derive from \eqref{brostu3:fo.Wfind1.new},\eqref{brostu3:fo.Wfind1b} 
\begin{eqnarray}
& & \hspace{-0.7cm} \Lambda_{1,\widetilde{c}}(z) := \Lambda_{1,\widetilde{c}}^{(0)}(z) :=
\int\displaylimits_{0}^{z} F_{1,\widetilde{c}}^{-1}(u) \, du
=  \widetilde{c} \cdot \left( \exp\Big(\frac{z}{\widetilde{c}}\Big) - 1 \right),
\ \  z \in \, 
]-\infty,\infty [ ,
\nonumber
\end{eqnarray}
which is the cumulant generating function of 
$\mathbb{\bbzeta} = \frac{1}{\widetilde{c}} \cdot POI(\widetilde{c})$ 
being the \textquotedblleft $\frac{1}{\widetilde{c}}-$fold
Poisson distribution with mean $\widetilde{c}$\textquotedblright\,
which means that 
$W = \frac{1}{\widetilde{c}} \cdot Z$ for a $POI(\widetilde{c})-$distributed
random variable $Z$;
for the special case $\widetilde{c} =1$ one particularly gets  
$\mathbb{\bbzeta} = POI(1)$ to be the Poisson distribution with mean 1.
Notice that $\mathbb{\bbzeta}$ is an infinitely divisible (cf. Proposition \ref{ID1}) discrete distribution with the
frequencies 
$\mathbb{\Pi}[W=\ell \cdot \frac{1}{\widetilde{c}}]= \exp(-\widetilde{c}) \cdot 
\frac{ \widetilde{c}^{\ell}
}{\ell!}$ for all nonnegative integers $\ell \in \mathbb{N}_{0}$
(and zero elsewhere).
Hence, $\mathbb{\Pi}[W\geq 0]=1$, $\mathbb{\Pi}[W=0]= \exp(-\widetilde{c})$.
Concerning the important Remark \ref{dist of components}(i),
for i.i.d. copies $(W_{i})_{i \in \mathbb{N}}$ of $W$ the 
probability distribution of $\breve{W} := \sum_{i\in I_{k}^{(n)}} W_{i}$ 
is $\frac{1}{\widetilde{c}} \cdot POI(\widetilde{c} \cdot card(I_{k}^{(n)}))$.

\vspace{0.4cm}

\noindent
For the desired bare-simulation-optimizations we obtain from
Theorem \ref{brostu3:thm.divnormW.new}, 
Remark \ref{remark divnormW}(vi) and
Lemma \ref{Lemma Indent Rate finite case_new}(b)
the following

\vspace{0.2cm}

\begin{proposition}
\label{brostu3:cor.powdivexpl5} 
(a) Consider $\varphi := \widetilde{c}\cdot \varphi_{\gamma}$ with $\gamma =1$, 
and let $\mathds{P} \in \mathbb{S}_{>0}^{K}$ as well as $\widetilde{c}>0$ be arbitrary but fixed. 
Furthermore, let $W:=(W_{i})_{i\in \mathbb{N}}$
be an i.i.d. sequence of non-negative real-valued random variables  
with distribution $\mathbb{\bbzeta} =\frac{1}{\widetilde{c}
}\cdot POI(\widetilde{c})$. 
Then for all $A >0$ and all 
$\boldsymbol{\Omega}$\hspace{-0.23cm}$\boldsymbol{\Omega}\subset \mathbb{S}^{K}$ with \eqref{regularity}
there holds 
\begin{eqnarray}
\hspace{-0.7cm}-\lim_{n\rightarrow \infty }\frac{1}{n}
\log \, \mathbb{\Pi}\negthinspace \left[\boldsymbol{\xi}_{n}^{w\mathbf{W}}
\in \textrm{$\boldsymbol{\Omega}$\hspace{-0.23cm}$\boldsymbol{\Omega}$} \right]
&=& \inf_{\mathds{Q}\in A \cdot
\textrm{$\boldsymbol{\Omega}$\hspace{-0.19cm}$\boldsymbol{\Omega}$} }\ 
\widetilde{c}\cdot \left[ 1-
A \cdot \exp \left( -\frac{1}{A \cdot \widetilde{c}}\cdot 
D_{\widetilde{c}\cdot \varphi
_{1}}(\mathbf{Q},\mathds{P}) + \frac{1}{A} -1 \right) \right]
\nonumber
\end{eqnarray}
and the BS minimizabilities/maximizabilites (cf. Definition \ref{brostu3:def.1})
\begin{eqnarray}
& & \hspace{-1.2cm}
\inf_{\mathbf{Q}\in A \cdot 
\textrm{$\boldsymbol{\Omega}$\hspace{-0.19cm}$\boldsymbol{\Omega}$} }D_{\widetilde{c}\cdot \varphi_{1}}(\mathbf{Q},\mathds{P})=
\lim_{n\rightarrow \infty } \widetilde{c}\cdot \left\{ 
1 - A \cdot \left[ 1 + \log \left( \frac{1}{A} \cdot \Big(1+
\frac{1}{\widetilde{c}}\cdot \frac{1}{n}\cdot \log \,
\mathbb{\Pi}\negthinspace \left[\boldsymbol{\xi}_{n}^{w\mathbf{W}}
\in \textrm{$\boldsymbol{\Omega}$\hspace{-0.23cm}$\boldsymbol{\Omega}$} \right] \Big)
\right) \right] \right\}, 
\nonumber \\
& & \hspace{-1.2cm}
\inf_{\mathbf{Q}\in A \cdot 
\textrm{$\boldsymbol{\Omega}$\hspace{-0.19cm}$\boldsymbol{\Omega}$} } \,
I(\mathbf{Q},\mathds{P})=
\inf_{\mathbf{Q}\in A \cdot 
\textrm{$\boldsymbol{\Omega}$\hspace{-0.19cm}$\boldsymbol{\Omega}$} } \, 
\sum\displaylimits_{k=1}^{K} q_{k} \cdot \log\left( \frac{q_{k}}{p_{k}} \right) 
= - \lim_{n\rightarrow \infty } A \cdot \log \left( \frac{1}{A} \cdot \Big(1+
\frac{1}{n}\cdot \log \,
\mathbb{\Pi}\negthinspace \left[\breve{\boldsymbol{\xi}}_{n}^{w\mathbf{W}}
\in \textrm{$\boldsymbol{\Omega}$\hspace{-0.23cm}$\boldsymbol{\Omega}$} \right]
\Big) \right),
\nonumber\\
& & \hspace{-1.2cm}
\max_{\mathbf{Q}\in A \cdot 
\textrm{$\boldsymbol{\Omega}$\hspace{-0.19cm}$\boldsymbol{\Omega}$} } \,
\mathcal{E}^{Sh}(\mathbf{Q})
= \max_{\mathbf{Q}\in A \cdot 
\textrm{$\boldsymbol{\Omega}$\hspace{-0.19cm}$\boldsymbol{\Omega}$} } \,
(-1) \cdot \sum_{k=1}^{K} q_{k} \cdot \log(q_{k})
= \lim_{n\rightarrow \infty} 
A \cdot \log K + A \cdot \log \left( \frac{1}{A} \cdot \Big(1+
\frac{1}{n}\cdot \log \,
\mathbb{\Pi}\negthinspace \left[\breve{\breve{\boldsymbol{\xi}}}_{n}^{w\mathbf{W}}
\in \textrm{$\boldsymbol{\Omega}$\hspace{-0.23cm}$\boldsymbol{\Omega}$} \right]
\Big)
\right),  
\label{fo.BS.Shannon} \\
& & \hspace{-1.2cm}
\max_{\mathbf{Q}\in A \cdot 
\textrm{$\boldsymbol{\Omega}$\hspace{-0.19cm}$\boldsymbol{\Omega}$} } \,
\mathcal{E}^{gSM2}(\mathbf{Q})
= \max_{\mathbf{Q}\in A \cdot 
\textrm{$\boldsymbol{\Omega}$\hspace{-0.19cm}$\boldsymbol{\Omega}$} } \,
\frac{1}{1-s} \cdot \exp\Big\{
(s-1) \cdot \sum_{k=1}^{K} q_{k} \cdot \log(q_{k}) -1\Big\}
\nonumber\\
& & \hspace{-1.2cm}
= \lim_{n\rightarrow \infty}
\frac{1}{1-s} \cdot \exp\left\{
(1-s) \cdot 
\left[
A \cdot \log K + A \cdot \log \left( \frac{1}{A} \cdot \Big(1+
\frac{1}{n}\cdot \log \,
\mathbb{\Pi}\negthinspace \left[\breve{\breve{\boldsymbol{\xi}}}_{n}^{w\mathbf{W}}
\in \textrm{$\boldsymbol{\Omega}$\hspace{-0.23cm}$\boldsymbol{\Omega}$} \right]
\Big)
\right) 
\right]
-1 \right\},
\ \  s \in \, ]0,1[ \, \cup \, ]1,\infty[ .
\nonumber
\end{eqnarray}
The special subcase $A=1$ in \eqref{fo.BS.Shannon} (and thus, $\mathbf{Q}$ is
a probability vector) corresponds to the
\textit{maximum entropy problem} for the Shannon entropy $\mathcal{E}^{Sh}(\cdot)$.
This can hence be tackled by our BS approach for
almost arbitrary sets 
$\textrm{$\boldsymbol{\Omega}$\hspace{-0.23cm}$\boldsymbol{\Omega}$}$
of probability vectors.\\
(b) The special case $\varphi := \widetilde{c}\cdot \varphi_{\gamma}$ \,  
($\gamma =1$)
of Theorem \ref{brostu3:thm.divnormW.new} works analogously to (a), 
with the differences that we employ (i) additionally
a sequence $(X_{i})_{i\in \mathbb{N}}$  of random variables
being independent of $(W_{i})_{i\in \mathbb{N}}$ 
and satisfying condition \eqref{cv emp measure X to P}
(resp. \eqref{cv emp measure X to P vector}), 
(ii) $A=1$ (instead of arbitrary $A>0$),
(iii) $\mathbb{\Pi}_{X_{1}^{n}}[\cdot]$ (instead of $\mathbb{\Pi }[\cdot]$),
(iv) $\boldsymbol{\xi}_{n,\mathbf{X}}^{w\mathbf{W}}$
(instead of $\boldsymbol{\xi}_{n}^{w\mathbf{W}}$),
(v) $\breve{\boldsymbol{\xi}}_{n,\mathbf{X}}^{w\mathbf{W}}$
(instead of $\breve{\boldsymbol{\xi}}_{n}^{w\mathbf{W}}$),
and 
(vi) $\breve{\breve{\boldsymbol{\xi}}}_{n,\mathbf{X}}^{w\mathbf{W}}$
(instead of $\breve{\breve{\boldsymbol{\xi}}}_{n}^{w\mathbf{W}}$).

\end{proposition}

\vspace{0.3cm}
\noindent
Within the context of Subsection \ref{Subsect Estimators determ}, 
for the concrete simulative estimation 
$\widehat{D_{\widetilde{c} \cdot \varphi_{1}}}(\mathbf{\Omega},\mathbf{P})$
via \eqref{estimator minimization} and \eqref{Improved IS for inf div new}, we
derive --- in terms of 
$M_{\mathbf{P}}:=\sum_{i=1}^{K}p_{i}>0$,
$n_{k} = n \cdot \widetilde{p}_{k} \in \mathbb{N}$ and
$\widetilde{q}_{k}^{\ast}$ from proxy method 1 or 2
--- that 
$\widetilde{U}_{k}^{\ast n_{k}}$
is the probability distribution \\
$\frac{1}{\widetilde{c} \cdot M_{\mathbf{P}}} \cdot 
POI\left(n_{k} \cdot \widetilde{c} \cdot M_{\mathbf{P}} \cdot 
\exp(\frac{\tau _{k}}{\widetilde{c} \cdot M_{\mathbf{P}}})\right) $ 
with support on the lattice 
$\left\{ \frac{j}{\widetilde{c} \cdot M_{\mathbf{P}}}, \, j\in \mathbb{N}_{0}\right\}$,
where $\tau_{k}=\widetilde{c} \cdot \log \left( \frac{\widetilde{q}_{k}^{\ast}}{\widetilde{p}_{k}}\right) $ 
for $\widetilde{q}_{k}^{\ast} >0$.
Moreover,  for all $j \in \mathbb{N}_{0}$ we obtain 
(by setting $x:= \frac{j}{\widetilde{c} \cdot M_{\mathbf{P}}}$)
$$
\widetilde{ISF}_{k}\left(\frac{j}{\widetilde{c} \cdot M_{\mathbf{P}}}\right)
= \exp\left(
n_{k} \cdot \widetilde{c} \cdot M_{\mathbf{P}} \cdot
\left( \exp \left( \frac{\tau_{k}}{\widetilde{c} \cdot M_{\mathbf{P}}}\right) -1 \right)
- j \cdot  \frac{\tau_{k}}{\widetilde{c} \cdot M_{\mathbf{P}}} 
\right) . $$
Within the different context of Subsection \ref{Subsect Estimators stoch},
the corresponding estimators $\widehat{\Pi }_{L}^{improved}$ can be obtained 
analogously to the last paragraph of Subsection \ref{Subsect Case1},
with Proposition \ref{brostu3:cor.powdivexpl5} instead of 
Proposition \ref{brostu3:cor.powdivexpl2} 
and with\\
$\widehat{D_{\widetilde{c}\cdot \varphi
_{1}}(\textrm{$\boldsymbol{\Omega}$\hspace{-0.23cm}$\boldsymbol{\Omega}$},\mathds{P})
} \ : = \ 
- \widetilde{c} \cdot \log \left( 1+\frac{1}{\widetilde{c}} \cdot \frac{1}{n} \cdot
\log \widehat{\Pi}_{L}^{improved}\right) $
instead of \eqref{fo.Est1.Case1}.


\vspace{0.2cm}

\subsubsection{Case 6b: Different anchor point
\label{Subsubsect Case6b}} \ 

\vspace{0.2cm}
\noindent
For several fields of applications 
it is important 
to have $\varphi-$divergences
which also allow for vectors with negative components. In order
to construct corresponding generators $\varphi$
which are e.g. (i) close to classical ones for nonnegative entries, and (ii) which
can be used for our BS method, one can appropriately vary the anchor point $c$ in
Theorem \ref{brostu3:thm.Wfind.new}. This is exemplarily shown in the following.
Indeed, let $\gamma =1$,
$]a_{F_{1,\widetilde{c}}},b_{F_{1,\widetilde{c}}}[ \, = \, ]0,\infty[$ 
and $F_{1,\widetilde{c}}(t)$ as in Case 6a above; for brevity let us fix
(say) $\widetilde{c}:=1$.
By choosing a \textit{general} anchor point 
$c \in \mathcal{R}(F_{1,1})= \, ]-\infty,\infty [$ (instead of $c=0$),
we obtain $]\lambda_{-},\lambda_{+}[ \, = int(\mathcal{R}(F_{1,1}))- c = \, ]-\infty,\infty [$,
$]t_{-}^{sc},t_{+}^{sc}[ \, = \, ]1+a_{F_{1,1}}-F_{1,1}^{-1}(c),1+b_{F_{1,1}}-F_{1,1}^{-1}(c)[ \, 
= \, ]1-e^{c},\infty[$
and $]a,b[ \, = \, ]1-e^{c},\infty[$. 
From \eqref{brostu3:fo.Wfind2.new},\eqref{brostu3:fo.Wfind2b} we deduce
\begin{eqnarray}
\varphi_{1,1}(t) := \varphi_{1,1}^{(c)}(t)
\hspace{-0.2cm} &:=& \hspace{-0.2cm}
\begin{cases}
(t + e^{c} -1) \cdot  [\log(t + e^{c} -1) -c\, ] + 1 - t
\ \in \ [0,\infty[,
\qquad
\quad \textrm{if } 
t \in \, ]1-e^{c},\infty[,
\\
e^{c}, \hspace{8.45cm} \textrm{if } t = 1-e^{c}, \\
\infty, \hspace{8.4cm} \textrm{if } t \in \, ]-\infty,1-e^{c}[ .
\end{cases}
\nonumber
\end{eqnarray}
The corresponding divergence is
\begin{eqnarray}
& & D_{\varphi_{1,1}^{(c)}}(\mathbf{Q},\mathbf{P}) := 
 \sum\limits_{k=1}^{K} \Big(q_{k} + p_{k} \cdot (e^{c}-1) \Big)
 \cdot  \Big\{ \log \Big(\frac{q_{k}}{p_{k}} + e^{c}-1 \Big) - c \Big\}
- \sum\limits_{k=1}^{K} q_{k} + \sum\limits_{k=1}^{K} p_{k} ,
\nonumber \\
& & \hspace{4.0cm} \textrm{if }   \ 
\mathbf{P} \in \mathbb{R}_{> 0}^{K} \ \textrm{and } \mathbf{Q} \in \mathbb{R}^{K}
\textrm{with }  \mathbf{Q} \in [ (1-e^{c}) \cdot \mathbf{P}, \boldsymbol{\infty} [
\ \textrm{component-wise}, 
\nonumber
\end{eqnarray}
which for the special anchor-point choice $c=0$ coincides with the
Kullback-Leibler divergence (relative entropy)
given in the fourth line of \eqref{brostu3:fo.powdiv.new}.
Notice that $D_{\varphi_{1,1}^{(c)}}(\mathbf{Q},\mathbb{P})$
has been recently used in \cite{Bro:19a}
for the important task of testing mixtures of probability distributions;
in fact, in order to get considerable comfort in testing mixture-type hypotheses
against corresponding marginal-type alternatives, 
they employ choices $c>0$ since then $\varphi_{1,1}^{(c)}(t)$ is finite especially for some
range of negative values $t<0$. 
Returning to our general considerations, 
we can employ \eqref{brostu3:fo.Wfind1.new},\eqref{brostu3:fo.Wfind1b}
to derive  
\begin{eqnarray}
& & \hspace{-0.7cm} \Lambda_{1,1}(z) := \Lambda_{1,1}^{(c)}(z) :=
\int\displaylimits_{0}^{z} F_{1,1}^{-1}(u+c) \, du
+ z \cdot (1-F_{1,1}^{-1}(c))
=  e^{c} \cdot \left( e^{z} - 1 \right) + z \cdot (1- e^{c}),
\qquad  z \in \, ]-\infty,\infty[, \qquad \ 
\nonumber
\end{eqnarray}
which is the cumulant generating function of
the \textquotedblleft shifted Poisson distribution\textquotedblright\ 
$\mathbb{\bbzeta} = POI(e^{c}) + 1-e^{c}$, \, 
i.e. $W := Z + 1-e^{c}$ with a $POI(e^{c})-$distributed
random variable $Z$. 
Notice that $\mathbb{\bbzeta}$ is a discrete distribution with 
frequencies $\mathbb{\Pi}[W=\ell + 1- e^{c}]= \exp(-e^{c}) \cdot 
\frac{ e^{c \cdot \ell}
}{\ell!}$ for all $\ell \in \mathbb{N}_{0}$
(and zero elsewhere). Moreover,
$\mathbb{\Pi}[W > 0]=1$ iff $c <0$, 
$\mathbb{\Pi}[W < 0] >0$ iff $c >0$,
$\mathbb{\Pi}[W=0] \ne 0$ iff \textquotedblleft $c =\log(1+k)$
for some $k \in \mathbb{N}_{0}$\textquotedblright .
Concerning the important Remark \ref{dist of components}(i), 
for i.i.d. copies $(W_{i})_{i \in \mathbb{N}}$ of $W$ the 
probability distribution 
$\mathbb{\bbzeta}^{\ast n_{k}}[\cdot] :=  \mathbb{\Pi}[\breve{W} \in \cdot \, ]$ 
of $\breve{W} := \sum_{i\in I_{k}^{(n)}} W_{i}$ 
is $POI(card(I_{k}^{(n)}) \cdot e^{c}) + (1-e^{c}) \cdot card(I_{k}^{(n)})$.

\vspace{0.2cm}
\noindent
Within the context of Subsection \ref{Subsect Estimators determ},
for our concrete simulations  
we derive --- in terms of (say) 
$M_{\mathbf{P}}:=\sum_{i=1}^{K}p_{i}=1$,
$n_{k} = n \cdot \widetilde{p}_{k} \in \mathbb{N}$ and
$\widetilde{q}_{k}^{\ast}$ from proxy method 1 or 2
--- that $\widetilde{U}_{k}^{\ast n_{k}}$
is the shifted Poisson distribution\\ 
$ POI\left(n_{k} \cdot 
e^{c +\tau_{k}}\right) + n_{k} \cdot (1-e^{c})$ 
with support on the lattice 
$\left\{ j +  n_{k} \cdot (1-e^{c}), \, j\in \mathbb{N}_{0}\right\}$,
where $\tau_{k}= \log\big(\frac{\widetilde{q}_{k}^{\ast}}{\widetilde{p}_{k}} + e^{c}-1\big)-c$ 
for $\widetilde{q}_{k}^{\ast} > \widetilde{p}_{k} \cdot (1-e^{c})$.
Furthermore,  for all $j \in \mathbb{N}_{0}$ we obtain
(by setting $x:= j +  n_{k} \cdot (1-e^{c})$)
$$
\widetilde{ISF}_{k}\left(j +  n_{k} \cdot (1-e^{c})\right)
= \exp\left(n_{k} \cdot e^{c} \cdot (e^{\tau_{k}} -1) - j \cdot \tau_{k}  \right).
$$
Notice that the mass of $\widetilde{U}_{k}^{\ast n_{k}}$ 
at zero depends crucially on the
value of the anchor point $c$, since 
$\widetilde{U}_{k}^{\ast n_{k}}
[\{0\}] >0$
if and only if $c=\log (1+\frac{\ell}{n_{k}})$ for some $\ell\in \mathbb{N}_{0}$;
moreover, 
$\widetilde{U}_{k}^{\ast n_{k}}
\big[ \ ]0,\infty[ \ \big] =1$ if $c <0$
and 
$\widetilde{U}_{k}^{\ast n_{k}}
\big[ \ ]-\infty,0[ \ \big] >0$ if $c >0$.

\vspace{0.4cm}


\subsubsection{Case 6c: The remaining $\gamma-$constellations
\label{Subsubsect Case6c}} \ 

\vspace{0.2cm}
\noindent
For $\gamma \in \, ]1,2[$, $\widetilde{c} \in \, ]0,\infty[$, 
$]a_{F_{\gamma,\widetilde{c}}},b_{F_{\gamma,\widetilde{c}}}[ \,  := \, ]0,\infty[$ 
and anchor point $c:=0$,
one can proceed as in 
Subsection \ref{Subsect Case3}
with $F_{\gamma,\widetilde{c}}$ from  
\eqref{brostu3:Fcase3} 
and deduce
from \eqref{brostu3:fo.Wfind2.new},\eqref{brostu3:fo.Wfind2b}
the same $\varphi_{\gamma,\widetilde{c}}$ of \eqref{brostu3:fo.expowlink2.Case3}
which coincides with $\widetilde{c} \cdot \varphi_{\gamma}(t)$
for $\varphi_{\gamma}(t)$ from \eqref{brostu3:fo.powdivgen} and 
which generates the $\gamma-$corresponding power divergences given in \eqref{brostu3:fo.powdiv.new}. 
Moreover, we derive from formula \eqref{brostu3:fo.Wfind1.new},\eqref{brostu3:fo.Wfind1b} 
the same $\Lambda_{\gamma,\widetilde{c}}$ of \eqref{brostu3:fo.expowlink1.Case3};
however, in contrast to
Subsection \ref{Subsect Case3}, 
one gets for the therein involved crucial exponent $\frac{\gamma}{\gamma -1} > 2$.
From this, we conjecture that there is no \textit{probability} measure 
$\mathbb{\bbzeta}$ such that $\Lambda_{\gamma,\widetilde{c}}$
is the cumulant generating function of $\mathbb{\bbzeta}$.
Indeed, we conjecture that $\mathbb{\bbzeta}$
becomes a \textit{signed} 
finite measure with total mass 1, 
i.e. it has a density (with respect to some dominating measure)
with positive \textit{and negative} values 
which \textquotedblleft  integrates to 1\textquotedblright ; 
accordingly, our BS method can not be applied to this situation.

\vspace{0.3cm}

\begin{remark}
The characterization of the probability distribution $\bbzeta$
in \eqref{Phi Legendre of mgf(W)} which may result from
Theorem \ref{brostu3:thm.Wfind.new} ---
as seen through the above Cases 1 to 6b --- 
considerably improves
other approaches which make use of their identification through the
concept of power variance functions of Natural Exponential Families,
as developed by~\cite{Twe:47}--\cite{Let:90}
and others. The latter approach has been used in
\cite{Bro:17} in a similar perspective as developed here,
but can not be extended outside the range of 
power divergences, in contrast with the 
following Cases 7 to 10 
which can only be handled as a consequence of Theorem \ref{brostu3:thm.Wfind.new}.

\end{remark}


\subsection{Case 7}

\noindent
Consider the interesting ``\textit{generalization}'' of the Kullback-Leibler 
divergence: for $\widetilde{c} >0$ 
and $\alpha \in \, ]-1,0[ \ \cup \ ]0,\infty[$ define
\begin{eqnarray}
F_{gKL,\alpha,\widetilde{c}}(t) &:=&
\begin{cases}
\widetilde{c} \cdot \log\left( \frac{(1+\alpha) \cdot t}{1+ \alpha \cdot t} \right), \qquad 
\textrm{if \{ $\alpha \in \, ]0,\infty[$ and $t \in \, ]0,\infty[$ \}
or \{ $\alpha \in \, ]-1,0[$ and $t \in \, ]0,-\frac{1}{\alpha}[$ \},} \\
- \infty, \hspace{2.5cm} \textrm{if $\alpha \in \, ]-1,0[ \ \cup \ ]0,\infty[$ and
 $t \in \, ]-\infty,0]$,} \\
\infty, \hspace{2.76cm} \textrm{if $\alpha \in \, ]-1,0[$ and $t \in \, [-\frac{1}{\alpha},\infty[$,}
\end{cases}
\nonumber
\end{eqnarray}
(notice that $\lim_{\alpha \rightarrow 0_{+}} F_{gKL,\alpha,\widetilde{c}}(t)
= F_{1,\widetilde{c}}(t)$, cf. 
Case 6a). Clearly, 
$]a_{F_{gKL,\alpha,\widetilde{c}}},b_{F_{gKL,\alpha,\widetilde{c}}}[ \, := \, ]0,\infty[$
for $\alpha \in \, ]0,\infty[$ and\\
$]a_{F_{gKL,\alpha,\widetilde{c}}},b_{F_{gKL,\alpha,\widetilde{c}}}[ \,  := \, ]0,-\frac{1}{\alpha}[$
for $\alpha \in \, ]-1,0[$. 
Moreover, $\mathcal{R}(F_{gKL,\alpha,\widetilde{c}})= 
\, ]-\infty, \widetilde{c} \cdot \log(1+ \frac{1}{\alpha})[$ for $\alpha \in \, ]0,\infty[$
and $\mathcal{R}(F_{gKL,\alpha,\widetilde{c}})= \, ]-\infty,\infty[$
for $\alpha \in \, ]-1,0[$. 
Furthermore, $F_{gKL,\alpha,\widetilde{c}}(\cdot)$ is strictly increasing and smooth on 
the respective $]a_{F_{gKL,\alpha,\widetilde{c}}},b_{F_{gKL,\alpha,\widetilde{c}}}[$,
and thus, $F_{gKL,\alpha,\widetilde{c}} \in \textgoth{F}$. 
Since $F_{gKL,\alpha,\widetilde{c}}(1)=0$, let us 
choose the natural anchor point $c:=0$,
which leads to $]\lambda_{-},\lambda_{+}[ \, = int(\mathcal{R}(F_{gKL,\alpha,\widetilde{c}})) = \, 
]-\infty, \widetilde{c} \cdot \log(1+ \frac{1}{\alpha})[$
and $]t_{-}^{sc},t_{+}^{sc}[ \, = \, ]0,\infty[ \, = \, ]a,b[$
for the case $\alpha \in \, ]0,\infty[$, respectively,
to $]\lambda_{-},\lambda_{+}[ \, = int(\mathcal{R}(F_{gKL,\alpha,\widetilde{c}})) = \, ]-\infty,\infty[$ and 
$]t_{-}^{sc},t_{+}^{sc}[ \, = \, ]0,-\frac{1}{\alpha}[ \, = \, ]a,b[$
for the case $\alpha \in \, ]-1,0[$.
By employing $F_{gKL,\alpha,\widetilde{c}}^{-1}(x) = 
\frac{1}{(1+\alpha) \cdot e^{-x/\widetilde{c}} - \alpha}$ for  
$x \in \, ]\lambda_{-},\lambda_{+}[$,
we derive from formula \eqref{brostu3:fo.Wfind2.new}  
(see also \eqref{brostu3:fo.Wfind2b}) 
\begin{eqnarray}
& & \hspace{-0.7cm} \varphi_{gKL,\alpha,\widetilde{c}}(t) := \varphi_{gKL,\alpha,\widetilde{c}}^{(0)}(t)
\nonumber \\
& & \hspace{-0.7cm}  := 
\begin{cases}
\widetilde{c} \cdot \left[ \, t \cdot \log t + (t+\frac{1}{\alpha}) \cdot 
\log\Big( \frac{1+\alpha}{1+\alpha \cdot t}  \Big) \, \right ]
\ \in \ [0,\infty[,
\quad \textrm{if \{ $\alpha \in \, ]0,\infty[$ and $t \in \, ]0,\infty[$ \}
or \{ $\alpha \in \, ]-1,0[$ and $t \in \, ]0,-\frac{1}{\alpha}[$ \},} 
\\
\frac{\widetilde{c}}{\alpha} \cdot \log(1+\alpha)
\ \in \ ]0,\infty[, \hspace{3.75cm} 
\textrm{if $\alpha \in \, ]-1,0[ \ \cup \ ]0,\infty[$ and
 $t =0$,} \\
\infty, \hspace{6.95cm} 
\textrm{if $\alpha \in \, ]-1,0[ \ \cup \ ]0,\infty[$ and
 $t \in \, ]-\infty,0[$,} \\
 \infty, \hspace{6.95cm} 
\textrm{if $\alpha \in \, ]-1,0[$ and
 $t \in \, [-\frac{1}{\alpha},\infty[$;} 
\end{cases}
\nonumber
\end{eqnarray}
notice the \textit{new effect} that this divergence generator is finite only
on a \textit{finite} interval, in case of $\alpha \in \, ]-1,0[$.
In any $\alpha-$case, we build from $\varphi_{gKL,\alpha,\widetilde{c}}$ the corresponding divergence (cf. \eqref{brostu3:fo.div})
\begin{eqnarray}
& & 
\hspace{-1.5cm}
D_{\varphi_{gKL,\alpha,\widetilde{c}}}(\mathbf{Q},\mathbf{P})
= \widetilde{c} \cdot \Big\{ \sum\limits_{k=1}^{K} 
q_{k} \cdot \log \Big(\frac{q_{k}}{(1-\frac{1}{1+\alpha}) \cdot q_{k} + \frac{1}{1+\alpha} \cdot p_{k}} \Big)
+ \frac{1}{\alpha} \cdot \sum\limits_{k=1}^{K} p_{k} \cdot \log 
\Big(\frac{p_{k}}{(1-\frac{1}{1+\alpha}) \cdot q_{k} + \frac{1}{1+\alpha} \cdot p_{k}} \Big) \Big\} 
\nonumber
\\ 
& &
\textrm{if \{ $\alpha \in \, ]0,\infty[$, 
$\mathbf{P} \in \mathbb{R}_{\gneqq 0}^{K}$ and 
$\mathbf{Q} \in \mathbb{R}_{\geq 0}^{K}$ \}
or \{ $\alpha \in \, ]-1,0[$,
$\mathbf{P} \in \mathbb{R}_{> 0}^{K}$ and 
$\mathbf{Q} \in \mathbb{R}_{\geq 0}^{K}$  
with $\mathbf{Q} \leq - \frac{1}{\alpha} \cdot \mathbf{P}$ \}
} .
\nonumber
\end{eqnarray}
Notice that the symmetry 
$D_{\varphi_{gKL,\alpha,\widetilde{c}}}(\mathbf{Q},\mathbf{P})
= D_{\varphi_{gKL,\alpha,\widetilde{c}}}(\mathbf{P},\mathbf{Q})$
generally holds only if $\alpha=1$; indeed, this special case leads to
\begin{eqnarray}
\varphi_{snKL,\widetilde{c}}(t) 
:= \varphi_{gKL,1,\widetilde{c}}(t) 
\hspace{-0.2cm} &:=& \hspace{-0.2cm}
\begin{cases}
\widetilde{c} \cdot \left[ \, t \cdot \log t + (t+1) \cdot \log\Big( \frac{2}{t+1}  \Big) \, \right ]
\ \in \ [0,\infty[,
\qquad
\quad \textrm{if } 
t \in \, ]0,\infty[,
\\
\widetilde{c} \cdot \log 2, \hspace{6.6cm} \textrm{if }  t = 0, \\
\infty, \hspace{7.35cm} \textrm{if }  t \in \, ]-\infty,0[ ,
\end{cases}
\nonumber
\end{eqnarray}
and
\begin{equation}
D_{\varphi_{snKL,\widetilde{c}}}(\mathbf{Q},\mathbf{P})
:= D_{\varphi_{gKL,1,\widetilde{c}}}(\mathbf{Q},\mathbf{P}) 
= \widetilde{c} \cdot \Big\{ \sum\limits_{k=1}^{K} q_{k} \cdot \log \Big(\frac{2 q_{k}}{q_{k} + p_{k}} \Big)
+ \sum\limits_{k=1}^{K} p_{k} \cdot \log \Big(\frac{2 p_{k}}{q_{k} + p_{k}} \Big) \Big\}
\ \ \ \textrm{if} \ \mathbf{P}, \mathbf{Q} \in \mathbb{R}_{\gneqq 0}^{K}
\ \textrm{with} \ \mathbf{P} + \mathbf{Q} \in \mathbb{R}_{> 0}^{K}. 
\label{brostu3:fo.snKLdiv}
\end{equation}
For the special subcase that $\widetilde{c} =1$ and 
that $\mathbf{P} = \mathds{P}$, $\mathbf{Q} = \mathds{Q}$ 
are probability vectors, the divergence \eqref{brostu3:fo.snKLdiv} can be
rewritten as a sum of two Kullback-Leibler divergences (cf. \eqref{brostu3:fo.powdiv.new}) 
\begin{equation}
D_{\varphi_{snKL,1}}(\mathds{Q},\mathds{P})
= D_{\varphi_{1}}(\mathds{Q},(\mathds{Q}+\mathds{P})/2)
+ D_{\varphi_{1}}(\mathds{P},(\mathds{Q}+\mathds{P})/2), 
\qquad \textrm{if} \ \mathds{P}, \mathds{Q} \in \mathbb{S}^{K}
\ \textrm{with} \ \frac{\mathds{P} + \mathds{Q}}{2} \in \mathbb{S}_{> 0}^{K},
\nonumber
\end{equation}
which is the well-known
(cf.~\cite{Bur:82},\cite{Lin2:91}--\cite{Sas:15})
\textit{Jensen-Shannon divergence} 
(being also called symmetrized and normalized Kullback-Leibler divergence, 
symmetrized and normalized relative entropy, capacitory discrimination);
this is equal to the $(2\log 2)-$fold of a special (namely, equally-weighted two-population) 
case of the Sibson information radius of order $1$ (cf. \cite{Sib:69})
which has also been addressed e.g. by
\cite{Rao:77} for genetic cluster analysis.
By the way, for $\alpha >0$ the divergence 
$D_{\varphi_{gKL,\alpha,\widetilde{c}}}(\mathds{Q},\mathds{P})$
can also be interpreted as a multiple of a special non-equally-weighted
Sibson information radius of order $1$.
In a context of comparison of --- not necessarily connected --- networks where  $\mathds{Q}$, $\mathds{P}$ are probability vectors
derived from matrices (cf. Remark \ref{matrix versions}) which are
transforms of corresponding graph invariants (e.g. network portraits),
the (matrix-equivalent of the) Jensen-Shannon divergence $D_{\varphi_{snKL,1}}(\mathds{Q},\mathds{P})$
is also called the \textit{network portrait divergence}, 
cf. \cite{Bag:19}.
There is a vast literature on recent applications of the Jensen-Shannon divergence in 
different research fields,
for instance it appears exemplarily
in~\cite{Che2:21},\cite{Ghol:20},\cite{Bare:21},\cite{Kvi:13}--\cite{ZhangZeh:21}. 

\vspace{0.3cm}

\begin{remark}
Let us transform
$\varphi_{gSH,\alpha}(t) := \frac{1-t}{\alpha} \cdot \log(1+\alpha) 
- \varphi_{gKL,\alpha,1}(t)
= -  t \cdot \log t  
+ \frac{1}{\alpha} \cdot (1 + \alpha \cdot t) \cdot \log(1 + \alpha \cdot t)
- \frac{1}{\alpha} \cdot (1+\alpha) \cdot t \cdot
\log(1+\alpha)$ \, (for $t \in [0,1]$).
The function $\varphi_{gSH,\alpha}(\cdot)$ is strictly concave on $[0,1]$ 
with $\varphi_{gSH,\alpha}(0)= \varphi_{gSH,\alpha}(1)=0$.
Hence, for probability vectors $\mathds{Q} =(q_{k})_{k=1,\ldots,K}$, 
the $\varphi-$entropy $\sum_{k=1}^{K}
\varphi_{gSH,\alpha}(q_{k})$ is Kapur\textquoteright s \cite{Kap:86}
generalization of the Shannon entropy
(which corresponds to $\alpha=0$ in the limit)
whose maximization has been connected with generalizations of the Bose-Einstein statistics  
and the Fermi-Dirac statistics e.g. in \cite{Kap:92}.
\end{remark}

\vspace{0.2cm}
\noindent
To proceed with our general considerations,
one can deduce from formula \eqref{brostu3:fo.Wfind1.new} 
(see also \eqref{brostu3:fo.Wfind1b}) 
\begin{eqnarray}
& & \Lambda_{gKL,\alpha,\widetilde{c}}(z) := \Lambda_{gKL,\alpha,\widetilde{c}}^{(0)}(z) 
\nonumber \\
& & :=
\begin{cases}
\int\displaylimits_{0}^{z} F_{gKL,\alpha,\widetilde{c}}^{-1}(u) \, du
=  - \frac{\widetilde{c}}{\alpha} \cdot \log((1+\alpha) - \alpha \cdot e^{z/\widetilde{c}}), 
\qquad  \textrm{
if $\alpha \in \, ]0,\infty[$ and $z \in \, ]-\infty, \widetilde{c} \cdot 
\log(1+ \frac{1}{\alpha})[$}, \\
\int\displaylimits_{0}^{z} F_{gKL,\alpha,\widetilde{c}}^{-1}(u) \, du
=  - \frac{\widetilde{c}}{\alpha} \cdot \log((1+\alpha) - \alpha \cdot e^{z/\widetilde{c}}), 
\qquad  \textrm{
if $\alpha \in \, ]-1,0[$ and $z \in \, ]-\infty, \infty[$}, \\
\infty, \hspace{7.65cm} 
\textrm{
if $\alpha \in \, ]0,\infty[$ and $z \in \, [\widetilde{c} \cdot 
\log(1+ \frac{1}{\alpha}), \infty[$}. \\
\end{cases}
\label{brostu3:fo.expowlink10new} 
\end{eqnarray}
(a) Subcase $\alpha \in \, ]0,\infty[$: \, 
the $\Lambda_{gKL,\alpha,\widetilde{c}}$ of \eqref{brostu3:fo.expowlink10new} is
the cumulant generating function of $\mathbb{\bbzeta} = \frac{1}{\widetilde{c}} \cdot NB(\frac{\widetilde{c}}{\alpha},\frac{1}{1+\alpha})$ 
being the 
\textquotedblleft $\frac{1}{\widetilde{c}}-$fold
Negative-Binomial distribution with parameters $\frac{\widetilde{c}}{\alpha}$ and 
$\frac{1}{1+\alpha}$\textquotedblright\,
which means that 
$W = \frac{1}{\widetilde{c}} \cdot Z$ for a $NB(\frac{\widetilde{c}}{\alpha},\frac{1}{1+\alpha})-$distributed
random variable $Z$. For the special case $\widetilde{c}=1$, $\alpha=1$ 
this reduces to  
$\mathbb{\bbzeta} = NB(1,\frac{1}{2})$ to be 
the Negative-Binomial distribution with 
parameters $1$ and $\frac{1}{2}$.
Notice that $\mathbb{\bbzeta}$ is an infinitely divisible 
(cf. Proposition \ref{ID1}) discrete distribution with frequencies 
$\mathbb{\Pi}[W=\ell \cdot \frac{1}{\widetilde{c}}]= 
(-1)^{\ell} \cdot \binom{- \frac{\widetilde{c}}{\alpha}}{\ell} 
\cdot {\alpha}^{\ell} \cdot (1+\alpha)^{-\ell - \widetilde{c}/\alpha}$ 
for all nonnegative integers $\ell \in \mathbb{N}_{0}$
(and zero elsewhere).
Moreover, one has $\mathbb{\Pi}[W\geq 0]=1$, 
$\mathbb{\Pi}[W=0]= \frac{1}{(1+\alpha)^{\widetilde{c}/\alpha}}$.
Concerning the important Remark \ref{dist of components}(i), 
for i.i.d. copies $(W_{i})_{i \in \mathbb{N}}$ of $W$ the 
probability distribution 
$\mathbb{\bbzeta}^{\ast n_{k}}[\cdot] :=  \mathbb{\Pi}[\breve{W} \in \cdot \, ]$ 
of $\breve{W} := \sum_{i\in I_{k}^{(n)}} W_{i}$ 
is $\frac{1}{\widetilde{c}} \cdot NB(\frac{\widetilde{c}}{\alpha} 
\cdot card(I_{k}^{(n)}),\frac{1}{1+\alpha})$.
Within the context of Subsection \ref{Subsect Estimators determ}, 
for the concrete simulative estimation 
$\widehat{D_{gKL,\alpha,\widetilde{c}}}(\mathbf{\Omega},\mathbf{P})$
via \eqref{estimator minimization} and \eqref{Improved IS for inf div new}, we
obtain --- in terms of 
$M_{\mathbf{P}}:=\sum_{i=1}^{K}p_{i}>0$,
$n_{k} = n \cdot \widetilde{p}_{k} \in \mathbb{N}$ and
$\widetilde{q}_{k}^{\ast}$ from proxy method 1 or 2
--- that $\widetilde{U}_{k}^{\ast n_{k}}
= \frac{1}{\widetilde{c} \cdot M_{\mathbf{P}}} \cdot 
NB\Big(\frac{\widetilde{c} \cdot M_{\mathbf{P}}}{\alpha} \cdot card(I_{k}^{(n)}),
1-\frac{\alpha}{1+\alpha}\cdot \exp(\frac{\tau_{k}}{\widetilde{c} \cdot M_{\mathbf{P}}})\Big)$
where $\tau_{k} = F_{gKL,\alpha,\widetilde{c}}\big(\frac{\widetilde{q}_{k}^{\ast}}{\widetilde{p}_{k}}\big)$
for $\widetilde{q}_{k}^{\ast} >0$; moreover,
$\widetilde{ISF}_{k}$ can be straightforwardly computed by \eqref{ISK}.

\vspace{0.2cm}
\noindent
(b) Subcase $\alpha \in \, ]-1,0[$: \, 
for any integer $m \in \mathbb{N}$ being strictly larger than $\widetilde{c}$ 
and the choice $\alpha= - \frac{\widetilde{c}}{m}$, we obtain
$\Lambda_{gKL,-\widetilde{c}/m,\widetilde{c}}(z) =
m \cdot \log((1 - \frac{\widetilde{c}}{m}) + \frac{\widetilde{c}}{m} \cdot e^{z/\widetilde{c}})$
(cf. \eqref{brostu3:fo.expowlink10new}) which
is the cumulant generating function of $\mathbb{\bbzeta} = \frac{1}{\widetilde{c}} \cdot BIN(m,\frac{\widetilde{c}}{m})$ 
being the 
\textquotedblleft $\frac{1}{\widetilde{c}}-$fold
Binomial distribution with parameters $m$ and 
$\frac{\widetilde{c}}{m}$\textquotedblright\,
which means that 
$W = \frac{1}{\widetilde{c}} \cdot Z$ for a $BIN(m,\frac{\widetilde{c}}{m})-$distributed
random variable $Z$. 
Notice that $\mathbb{\bbzeta}$ is a \textit{non-}infinitely divisible 
discrete distribution with frequencies 
$\mathbb{\Pi}[W=\ell \cdot \frac{1}{\widetilde{c}}]= 
\binom{m}{\ell} 
\cdot (\frac{\widetilde{c}}{m})^{\ell} \cdot 
(1-\frac{\widetilde{c}}{m})^{m-\ell}$ 
for $\ell \in \{0,1, \ldots, m\}$
(and zero elsewhere). Furthermore, 
there holds $\mathbb{\Pi}[W\geq 0]=1$, 
$\mathbb{\Pi}[W=0]= (1-\frac{\widetilde{c}}{m})^{m}$.
Concerning the important Remark \ref{dist of components}(i), 
for i.i.d. copies $(W_{i})_{i \in \mathbb{N}}$ of $W$ the 
probability distribution 
$\mathbb{\bbzeta}^{\ast n_{k}}[\cdot] :=  \mathbb{\Pi}[\breve{W} \in \cdot \, ]$ 
of $\breve{W} := \sum_{i\in I_{k}^{(n)}} W_{i}$ 
is $\frac{1}{\widetilde{c}} \cdot 
BIN(m \cdot card(I_{k}^{(n)}),\frac{\widetilde{c}}{m})$.
Analogously to (a), for
$\widehat{D_{gKL,\alpha,\widetilde{c}}}(\mathbf{\Omega},\mathbf{P})$
we employ that $\widetilde{U}_{k}^{\ast n_{k}}
= \frac{1}{\widetilde{c} \cdot M_{\mathbf{P}}} \cdot 
BIN(m \cdot card(I_{k}^{(n)}),\breve{p})$
where $m \in \mathbb{N}$ is strictly larger than $\widetilde{c} \cdot M_{\mathbf{P}}$,
$\breve{p} := \frac{
\widetilde{c} \cdot M_{\mathbf{P}} \cdot \exp(\frac{\tau_{k}}{\widetilde{c} \cdot M_{\mathbf{P}}})
}{m - \widetilde{c} \cdot M_{\mathbf{P}}
+ \widetilde{c} \cdot M_{\mathbf{P}} \cdot \exp(\frac{\tau_{k}}{\widetilde{c} \cdot M_{\mathbf{P}}})}$
and
$\tau_{k} = F_{gKL,\alpha,\widetilde{c}}\big(\frac{\widetilde{q}_{k}^{\ast}}{\widetilde{p}_{k}}\big)$
for $\widetilde{q}_{k}^{\ast} \in \, ]0, - \frac{\widetilde{p}_{k}}{\alpha}[$; furthermore, for
$\widetilde{ISF}_{k}$ we use \eqref{ISK}.

\vspace{0.3cm}
\noindent
As far as the construction of bounds of 
$D_{\varphi_{gKL,\alpha,\widetilde{c}}}\left(
\textrm{$\boldsymbol{\Omega}$\hspace{-0.23cm}$\boldsymbol{\Omega}$} ,\mathds{P}\right)$ 
in the light of \eqref{estimator-general-divbounds-two}
in Subsection \ref{Sub Bounds General div} 
is concerned, for the sake of brevity we confine ourselves to the
above-mentioned important Jensen-Shannon divergence 
special case $\alpha=1$ and $\widetilde{c}=1$,
and thus $\varphi_{gKL,1,1}=\varphi_{snKL,1}$.
Correspondingly, we abbreviate \eqref{brostu3:fo.snKLdiv}
as 
\[
J(\mathbf{Q},\mathbf{P}) :=
D_{\varphi_{gKL,1,1}}\left(\mathbf{Q},\mathbf{P}\right) =
I(\mathbf{Q},(\mathbf{Q}+\mathbf{P})/2)+
I(\mathbf{P},(\mathbf{Q}+\mathbf{P})/2)
\qquad \textrm{if} \ \mathbf{P}, \mathbf{Q} \in \mathbb{R}_{\gneqq 0}^{K}
\ \textrm{with} \ \mathbf{P} + \mathbf{Q} \in \mathbb{R}_{> 0}^{K},
\]
where we use (as an extension of \eqref{brostu3:fo.divpow.Kull1})
$I(\breve{\mathbf{Q}},\breve{\mathbf{P}}) := 
\sum\displaylimits_{k=1}^{K} \breve{q}_{k} \cdot \log( \frac{\breve{q}_{k}}{\breve{p}_{k}})$
for  $\breve{\mathbf{P}} \in \mathbb{R}_{> 0}^{K}$
and $\breve{\mathbf{Q}} \in \mathbb{R}_{\geq 0}^{K}$.
We explore the sharpness of the bounds for
 $J(\textrm{$\boldsymbol{\Omega}$\hspace{-0.23cm}$\boldsymbol{\Omega}$} ,\mathds{P})$ 
as defined in \eqref{estimator-general-divbounds-two}.
For this, we consider a given probability distribution $\mathds{P}$ on $\mathcal{Y}$ with strictly positive entries; the set $\textrm{$\boldsymbol{\Omega}$\hspace{-0.23cm}$\boldsymbol{\Omega}$}$ 
consists of all probability distributions $\mathds{Q}$ on $\mathcal{Y}$ whose
total variation distance 
$V(\mathds{Q},\mathds{P}):=\sum_{k=1}^{K}\left\vert q_{k}-p_{k}\right\vert 
$ \footnote{
notice that $V(\mathds{Q},\mathds{P})$ always takes values in the interval $[0,2[$
}
to $\mathds{P}$ 
lies between $v$ and $v+h$ for $v>0$ and small $h$ 
and which also satisfies 
\, $
\sup \left( \sup_{k=1,\ldots,K}\frac{p_{k}}{q_{k}},\sup_{k=1,\ldots,K}\frac{q_{k}}{p_{k}}\right)
\leq L
$ \, 
for some strictly positive finite $L$. This set $\textrm{$\boldsymbol{\Omega}$\hspace{-0.23cm}$\boldsymbol{\Omega}$}$
defines a class of distributions 
$\mathds{Q}$ away from $\mathds{P}$ still keeping some regularity w.r.t. 
$\mathds{P}$. Also, $\textrm{$\boldsymbol{\Omega}$\hspace{-0.23cm}$\boldsymbol{\Omega}$}$
satisfies \eqref{regularity}. We will prove that the bounds in \eqref{estimator-general-divbounds-two}
are sharp in this
case. Notice that $J(m \cdot \mathds{Q},\mathds{P}) = \infty$ for $m<0$
and hence \, $
\inf_{\mathds{Q}\in \textrm{$\boldsymbol{\Omega}$\hspace{-0.19cm}$\boldsymbol{\Omega}$}
}\inf_{m \ne 0}J(m \cdot \mathds{Q},\mathds{P}) =
\inf_{\mathds{Q}\in \textrm{$\boldsymbol{\Omega}$\hspace{-0.19cm}$\boldsymbol{\Omega}$}
}\inf_{m>0}J(m \cdot \mathds{Q},\mathds{P})$. 
We first  provide a lower bound for 
the latter. It holds for all $m>0$ and $\mathds{Q}$ in 
$\textrm{$\boldsymbol{\Omega}$\hspace{-0.23cm}$\boldsymbol{\Omega}$}$
that
\begin{eqnarray}
& & 
J(m \cdot \mathds{Q},\mathds{P}) = 
(m+1) \cdot \log \left( 2 \right)
- (m+1) \cdot 
\log \left( m+1\right)
+m \cdot \log m+
I^{\alpha }(\mathds{P},\mathds{Q})+m \cdot I^{1-\alpha }(\mathds{Q},\mathds{P})
\nonumber 
\end{eqnarray}
where $\alpha :=1/\left( m+1\right) $ and $I^{\alpha }(\mathds{P},\mathds{Q})$ is the 
$\alpha -$skewed Kullback-Leibler divergence between $\mathds{P}$ and $\mathds{Q}$ 
defined through
$
I^{\alpha }(\mathds{P},\mathds{Q}):=I(\mathds{P},\alpha \mathds{P}+(1-\alpha )\mathds{Q})
$.
By inequality (27) in \cite{Yam:19} one gets
$
I^{\alpha }(\mathds{P},\mathds{Q})\geq -\log \left( 1-\frac{\alpha ^{2}}{4} \cdot V(\mathds{Q},\mathds{P})^{2}\right). 
$
Since $(m+1) \cdot \log \left(2\right) -(m+1) \cdot \log (m+1)+ m \cdot \log m$ is non-negative for
all $m>0$ and takes its minimal value $0$ for $m=1$, we obtain
$\inf_{m>0} J(m \mathds{Q},\mathds{P})\geq \inf_{m>0}K(m)$
where 
$K(m):=-\log \left( 1-\frac{1}{4\left( m+1\right) ^{2}} \cdot 
V(\mathds{Q},\mathds{P})^{2}\right) -m \cdot 
\log\left( 1- \frac{m^{2}}{4\left( m+1\right)^{2}}
\cdot V(\mathds{Q},\mathds{P})^{2}\right) . $
Since $-\log (1-x)\geq x$ for all $x<1$ and both 
$\frac{1}{4\left(m+1\right) ^{2}} \cdot V(\mathds{Q},\mathds{P})^{2}$ and 
$\frac{m^{2}}{4\left( m+1\right)^{2}}
\cdot V(\mathds{Q},\mathds{P})^{2}$ are less than $1$, it follows that
$ K(m)\geq \frac{V(\mathds{Q},\mathds{P})^{2}}{4} \cdot
\frac{m^{3} +1}{\left(m+1\right)^{2}} $
where the right-hand side attains its minimal value 
on $]0,\infty[$ at $m^{*}= \sqrt{3}-1 \approx 0.73$. Hence, we obtain 
$\inf_{m>0}J(m\cdot \mathds{Q},\mathds{P})\geq 
\frac{V(\mathds{Q},\mathds{P})^{2}}{4}
\cdot (2\sqrt{3}-3)
> 0.116 \, v^2$.
Now by (19) in \cite{Yam:19}, for any $\mathds{Q}$ one gets
$J(\mathds{Q},\mathds{P})\leq \frac{1}{4}\underline{J}(\mathds{Q},\mathds{P})$
where $\underline{J}(\mathds{Q},\mathds{P}):=I(\mathds{Q},\mathds{P})+I(\mathds{P},\mathds{Q})$ 
is the Jensen divergence (also called symmetrized
Kullback-Leibler divergence) between $\mathds{Q}$ and $\mathds{P}$. 
Since (see \cite{Dra:00}) 
$I(\mathds{P},\mathds{Q})\leq \sum_{k=1}^{K}\sqrt{\frac{p_{k}}{q_{k}}}\cdot \left\vert
q_{k}-q_{k}\right\vert ,$
it follows that 
$J(\mathds{Q}^{\ast },\mathds{P})\leq \frac{1}{2}\sqrt{L} \cdot V(\mathds{Q}^{\ast},\mathds{P})$
which provides
$0.116 \cdot v^2 \leq  \inf_{m >0}
J(m \cdot \mathds{Q}^{\ast},\mathds{P}) 
=J(m(\mathds{Q}^{\ast })\cdot \mathds{Q}^{\ast},\mathds{P}) \leq  
 J(\textrm{$\boldsymbol{\Omega}$\hspace{-0.23cm}$\boldsymbol{\Omega}$},\mathds{P})
 \leq  J(\mathds{Q}^{\ast },\mathds{P})  \leq  \frac{1}{2}\sqrt{L} \cdot 
(v+h). $
For small $v$, the
difference between the RHS and the LHS in the above display is  $cst\cdot v + o(v) +
\frac{1}{2} \sqrt{L} \cdot h$
which proves that the bounds are sharp locally, with non-trivial lower bound.
Other upper bounds can be adapted to sets $\textrm{$\boldsymbol{\Omega}$\hspace{-0.23cm}$\boldsymbol{\Omega}$}$ defined through
tighter conditions on $\sup_{\mathds{Q}\in \textrm{$\boldsymbol{\Omega}$\hspace{-0.19cm}$\boldsymbol{\Omega}$}}
\sup_{k=1,\ldots,K}\frac{p_{k}}{q_{k}}$
and $\sup_{\mathds{Q}\in \textrm{$\boldsymbol{\Omega}$\hspace{-0.19cm}$\boldsymbol{\Omega}$}}
\sup_{k=1,\ldots,K}\frac{q_{k}}{p_{k}}$ (of e.g. \cite{Dra:00}).


\subsection{Case 8}

\noindent
As a continuation of the discussion
at the beginning of Case 6b (cf. Subsection \ref{Subsubsect Case6b}),
we give another $\varphi-$divergence
which also allows for vectors with negative components
(i.e. $\varphi$ is finite especially for some
range of negative values $t<0$). Indeed,
for $\beta \in \, ]0,1]$, $\widetilde{c} \in \, ]0,\infty[$  and
$]a_{F_{bw,\beta,\widetilde{c}}},b_{F_{bw,\beta,\widetilde{c}}}[ \,  
= \, ]1-\frac{1}{\beta},\infty[$ 
let us define
\begin{eqnarray}
F_{bw,\beta,\widetilde{c}}(t) &:=&
\begin{cases}
\frac{\widetilde{c}}{2\beta} \cdot \Big(
1-\frac{1}{(\beta \cdot t + 1 - \beta)^2}
\Big), \qquad \textrm{if } \  
t \in \, ]1-\frac{1}{\beta},\infty[, \\
- \infty, \hspace{3.3cm}  \textrm{if } \  t \in \, ]-\infty,1-\frac{1}{\beta}]. 
\end{cases}
\nonumber
\end{eqnarray}
Clearly, $\mathcal{R}(F_{bw,\beta,\widetilde{c}})= \big]-\infty,\frac{\widetilde{c}}{2\beta}\big[$ 
and $F_{bw,\beta,\widetilde{c}} \in \textgoth{F}$. 
Since $F_{bw,\beta,\widetilde{c}}(1)=0$, let us choose the natural anchor point $c:=0$,
which leads to $]\lambda_{-},\lambda_{+}[ \, = int(\mathcal{R}(F_{bw,\beta,\widetilde{c}}))
= \big]-\infty,\frac{\widetilde{c}}{2\beta}\big[$,
$]t_{-}^{sc},t_{+}^{sc}[ \, =  \, 
]a_{F_{bw,\beta,\widetilde{c}}},b_{F_{bw,\beta,\widetilde{c}}}[ 
\, = \, ]1-\frac{1}{\beta},\infty[$
and $]a,b[ \, = \, ]1-\frac{1}{\beta},\infty[$.
By using $F_{bw,\beta,\widetilde{c}}^{-1}(x) = 
\frac{1}{\beta} \cdot \Big\{ \frac{1}{\sqrt{1-2\beta\cdot x/\widetilde{c}}} + \beta -1 \Big\}
$ for 
$x \in int(\mathcal{R}(F_{bw,\beta,\widetilde{c}}))$, 
from formula \eqref{brostu3:fo.Wfind2.new},\eqref{brostu3:fo.Wfind2b}
we can deduce 
\begin{eqnarray}
\varphi_{bw,\beta,\widetilde{c}}(t) := \varphi_{bw,\beta,\widetilde{c}}^{(0)}(t)
\hspace{-0.2cm} &:=& \hspace{-0.2cm}
\begin{cases}
\widetilde{c} \cdot \frac{(t-1)^{2}}{2(\beta \cdot t +1 -\beta)}
\ \in \ [0,\infty[,
\qquad
\quad \textrm{if } 
t \in \, ]1-\frac{1}{\beta},\infty[,
\\
\infty, \hspace{4.25cm} \textrm{if } \ t \in \, ]-\infty,1-\frac{1}{\beta}] .
\end{cases}
\label{brostu3:fo.expowlinkbw2}
\end{eqnarray}

\noindent
Note that $1-\frac{1}{\beta} <0$ so that negative $t$ are allowed here. For $t\geq0$, $\varphi_{bw,\beta,\widetilde{c}}(t)$
is known as \textit{Rukhin's generator} (cf. \cite{Ruk:94}, see e.g.\ also 
\cite{Marh:05}, \cite{Par:06}). 
From the generator $\varphi_{bw,\beta,\widetilde{c}}$ given in \eqref{brostu3:fo.expowlinkbw2},
we build the corresponding divergence (cf. \eqref{brostu3:fo.div})
\begin{eqnarray}
& &
D_{\varphi_{bw,\beta,\widetilde{c}}}(\mathbf{Q},\mathbf{P})
\ = \ \widetilde{c} \cdot 
\sum_{k=1}^{K} p_{k} \cdot
\frac{(\frac{q_{k}}{p_{k}}-1)^{2}}{2(\beta \cdot \frac{q_{k}}{p_{k}} +1 -\beta)} 
\ = \ \frac{\widetilde{c}}{2} \cdot \sum\limits_{k=1}^{K} 
\frac{(q_{k}-p_{k})^{2}}{\beta \cdot q_{k} + (1 -\beta)\cdot p_{k}},
\nonumber\\
& & 
\hspace{4.5cm} 
\textrm{if $\mathbf{P} \in \mathbb{R}_{\gneqq 0}^{K}$ and $\mathbf{Q} \in \mathbb{R}^{K}$
with $\mathbf{Q} \in \, ] \, \mathbf{P} \cdot (1-\frac{1}{\beta}),\boldsymbol{\infty}[$ component-wise;}
\nonumber
\end{eqnarray}
for the special subcase $\widetilde{c}=1$ and $\mathbf{Q} \in \mathbb{R}_{> 0}^{K}$,
$D_{\varphi_{bw,\beta,1}}(\mathbf{Q},\mathbf{P})$ can be interpreted as
--- \textquotedblleft non-probability version\textquotedblright\ of  
--- the well-known \textit{blended weight chi-square divergence of Lindsay} \cite{Lind:94}
(see e.g. 
also~\cite{Bas:94}--\cite{Bas:11}).
The special case $\widetilde{c}=1$ and $\beta =\frac{1}{2}$ for probability vectors, i.e.  
$D_{\varphi_{bw,1/2,1}}(\mathds{Q},\mathds{P})$, is
equal to (a multiple of the matrix-vector-converted (cf. Remark \ref{matrix versions})) \textit{Sanghvi\textquoteright s genetic difference measure}
\cite{San:53} and equal to
the double of the so-called \textit{(squared) Vincze-Le Cam distance}
(cf.~\cite{Vin:81},\cite{LeCam:86},
see also e.g. \cite{Top:00} -- who used the alternative naming 
\textit{triangular discrimination} -- and \cite{Vaj:09});
$D_{\varphi_{bw,1/2,1}}(\mathds{Q},\mathds{P})$
has been used e.g. 
in \cite{Liu2:11} for
a machine learning context 
where $\mathds{Q}$ and $\mathds{P}$ are appropriate histograms of RGB color.  

\vspace{0.3cm}

\begin{remark}
(a) By straightforward calculations, one can show that 
$\varphi_{bw,1,\widetilde{c}}$ is equal to the power-divergence generator 
$\varphi_{\gamma,\widetilde{c}} = \widetilde{c} \cdot \varphi_{\gamma}$ 
(cf. \eqref{brostu3:fo.powdivgen})
with $\gamma=-1$; the corresponding divergence 
$D_{\varphi_{bw,1,\widetilde{c}}}(\mathbf{Q},\mathbf{P})$
is thus equal to the power divergence 
$D_{\widetilde{c} \cdot \varphi_{-1}}(\mathbf{Q},\mathbf{P})$
(cf. \eqref{brostu3:fo.powdiv.new})
which is nothing but the $\widetilde{c}/2-fold$
--- \textquotedblleft non-probability version\textquotedblright\ ---
of \textit{Neyman's chi-square divergence}.\\
(b) For the case $\beta =0$ --- which has been excluded 
above for technical brevity ---
the divergence generator $\varphi_{bw,0,\widetilde{c}}$
corresponds to the power-divergence generator $\varphi_{\gamma,\widetilde{c}}$ 
with $\gamma=2$; the corresponding divergence 
$D_{\varphi_{bw,0,\widetilde{c}}}(\mathbf{Q},\mathbf{P})$
is thus equal to the power divergence 
$D_{\widetilde{c} \cdot \varphi_{2}}(\mathbf{Q},\mathbf{P})$
(cf. \eqref{brostu3:fo.powdiv.new})
which is nothing but the $\widetilde{c}/2-fold$
--- \textquotedblleft non-probability version\textquotedblright\ ---
of Pearson's (i.e. the \textit{classical}) chi-square divergence.
\end{remark}

\vspace{0.2cm}
\noindent
To continue with our general considerations,
we can derive from formula \eqref{brostu3:fo.Wfind1.new} 
(see also \eqref{brostu3:fo.Wfind1b}) 
for all $\beta \in \, ]0,1]$ 
\begin{eqnarray}
\Lambda_{bw,\beta,\widetilde{c}}(z) := 
\Lambda_{bw,\beta,\widetilde{c}}^{(0)}(z) &=&
\begin{cases}
-(\frac{1}{\beta}-1) \cdot z + \frac{\widetilde{c}}{\beta^{2}} \cdot
\Big\{ 1 - \sqrt{1-\frac{2\beta}{\widetilde{c}} \cdot z}  \ \Big\} , 
\qquad 
\textrm{if } z \in \big]-\infty,\frac{\widetilde{c}}{2\beta} \big],  \\
\infty, \hspace{6.28cm}  
\textrm{if } z \in \big]\frac{\widetilde{c}}{2\beta}, \infty \big[, 
\end{cases}
\nonumber
\end{eqnarray}
which
is the cumulant generating function of a probability distribution
$\mathbb{\bbzeta}[ \, \cdot \,] = \mathbb{\Pi}[\check{W} \in \cdot \, ]$
of a random variable $\check{W}$,
which can be constructed as follows: 
$\check{W} := \frac{W}{\beta} - (\frac{1}{\beta} - 1)$, where
$W$ is the random variable constructed in 
Case 1 (cf. Subsection \ref{Subsect Case1})
with $\gamma=-1$ and with $\widetilde{c}$ replaced by $\frac{\widetilde{c}}{\beta^{2}}$
(recall that $W$ has a tilted stable distribution).
In other words, $\mathbb{\bbzeta}$ is a special kind of 
\textit{modified tilted stable distribution}.
Notice that $\mathbb{\bbzeta}$ is an infinitely divisible (cf. Proposition \ref{ID1}) continuous distribution with density $f_{\check{W}}(u) := \beta \cdot f_{W}(\beta \cdot u + 1 -\beta) 
\cdot \mathbf{1}_{]- (\frac{1}{\beta} -1),\infty[}(u)$ ($u \in \mathbb{R}$),
where $f_{W}(\cdot)$ is given in \eqref{brostu3:fo.norweiemp7a}
with $\gamma=-1$ and with $\widetilde{c}$ replaced by $\frac{\widetilde{c}}{\beta^{2}}$.
Moreover, $\mathbb{\bbzeta}[ \, ] 0,\infty[ \, ] = 
\mathbb{\Pi}[\check{W} > 0] >0$.
Concerning Remark \ref{dist of components}(i), for 
i.i.d. copies $(\check{W}_{i})_{i \in \mathbb{N}}$ of $\check{W}$, the 
probability distribution 
$\mathbb{\bbzeta}^{\ast n_{k}}[\cdot] :=  \mathbb{\Pi}[\breve{\check{W}} \in \cdot \, ]$
of $\breve{\check{W}} := \sum_{i\in I_{k}^{(n)}} \check{W}_{i}
= \frac{1}{\beta} \cdot \sum_{i\in I_{k}^{(n)}} W_{i} - n_{k} 
\cdot (\frac{1}{\beta} -1)$ 
has density
$f_{\breve{\check{W}}}(u) := \beta \cdot f_{\breve{W}}(\beta \cdot u + (1 -\beta)\cdot n_{k}) 
\cdot \mathbf{1}_{]- n_{k} \cdot (\frac{1}{\beta} -1),\infty[}(u)$ ($u \in \mathbb{R}$),
where $f_{\breve{W}}(\cdot)$ is given in 
\eqref{fbreveW} 
with $\gamma=-1$ and with $\widetilde{c}$ replaced by $\frac{\widetilde{c}}{\beta^{2}}$;
accordingly, the determination of the corresponding $\widetilde{U}_{k}^{\ast n_{k}}$
and $\widetilde{ISF}_{k}$ works analogously.


\subsection{Case 9
\label{Subsect Case9}}

\noindent
Let us fix any $z_{1},z_{2} \in \mathbb{R}$, $p \in \, ]0,1[$ 
which satisfy $z_{1} < 1 < z_{2}$ and $z_{1} \cdot p + z_{2} \cdot (1-p) =1$
(and thus $p= \frac{z_{2} -1}{z_{2} - z_{1}}$).
On $]a_{F_{twop}},b_{F_{twop}}[ \, := \, ]z_{1},z_{2}[$  we define
\begin{eqnarray} 
F_{twop}(t) 
\ := \ 
\frac{1}{z_{2}-z_{1}} \cdot  \log\left( \frac{(t-z_{1})\cdot p}{(z_{2}-t)\cdot (1- p)} \right) 
\ = \  
\frac{1}{z_{2}-z_{1}} \cdot  \log\left( \frac{(t-z_{1})\cdot (z_{2} -1)}{(z_{2}-t)\cdot (1-z_{1})} \right) \, ,
\qquad t \in \,  
]z_{1},z_{2}[ ,
\nonumber
\end{eqnarray}
where for the last equality we have used the above constraint
(in order to obtain a two-parameter representation).
Straightforwardly, we have $\mathcal{R}(F_{twop})= ]-\infty,\infty[$
and $F_{twop} \in \textgoth{F}$. 
Since $F_{twop}(1)=0$, let us 
choose the natural anchor point $c:=0$,
which leads to $]\lambda_{-},\lambda_{+}[ \, = int(\mathcal{R}(F_{twop})) = 
\, ]-\infty, \infty[$,
$]t_{-}^{sc},t_{+}^{sc}[ \, = \, ]z_{1},z_{2}[ \, = \, ]a,b[$.
By using 
$$F_{twop}^{-1}(x) = \frac{
p \cdot z_{1} + (1- p) \cdot z_{2} \cdot e^{(z_{2}-z_{1}) \cdot x}
}{
p + (1- p) \cdot e^{(z_{2}-z_{1}) \cdot x}
} \, , \qquad
x \in \, ]-\infty, \infty[,
$$  
from formula \eqref{brostu3:fo.Wfind2.new}  
(see also \eqref{brostu3:fo.Wfind2b}) we deduce
\begin{eqnarray}
\varphi_{twop}(t) := \varphi_{twop}^{(0)}(t)
\hspace{-0.2cm} &:=& \hspace{-0.2cm}
\begin{cases}
 \frac{t-z_{1}}{z_{2}-z_{1}} \cdot \log\left( \frac{(t-z_{1})\cdot (z_{2}-1)}{(z_{2}-t)\cdot (1-z_{1})} \right)
- \log\left( \frac{z_{2}-1}{z_{2}-t} \right) 
\ \in \ [0,\infty[,
\qquad
\quad \textrm{if } 
t \in \, ]z_{1}, z_{2}[,
\\
\log\left( \frac{z_{2} - z_{1}}{z_{2} -1} \right)
\in \ ]0,\infty[
, \hspace{5.45cm} \textrm{if } \  t = z_{1}, \\
\log\left( \frac{z_{2} - z_{1}}{1 - z_{1}} \right)
\in \ ]0,\infty[
, \hspace{5.45cm} \textrm{if } \  t = z_{2}, \\
\infty, \hspace{8.2cm} \textrm{if } \ t \in \, ]-\infty,z_{1}[ \, \cup \, ]z_{2}, \infty[ ;
\end{cases}
\nonumber
\end{eqnarray}
notice that this generator 
is finite only on a \textit{finite} interval.
From it, we build the new divergence (cf. \eqref{brostu3:fo.div})
\begin{eqnarray}
& & D_{\varphi_{twop}}(\mathbf{Q},\mathbf{P})
= 
\sum\limits_{k=1}^{K} \frac{q_{k} - z_{1} \cdot p_{k}}{z_{2} - z_{1}}  \cdot 
\log \Big(\frac{(z_{2}-1) \cdot (q_{k} - z_{1} \cdot p_{k})}{(1-z_{1}) \cdot (z_{2} \cdot p_{k} - q_{k})} \Big)
- \sum\limits_{k=1}^{K} p_{k} \cdot  
\log \Big(\frac{(z_{2}-1) \cdot  p_{k}}{z_{2} \cdot p_{k} - q_{k}}  \Big) 
\nonumber
\\
& & \hspace{7.2cm}
\textrm{for }
\mathbf{P} \in \mathbb{R}_{> 0}^{K}, \,  
\mathbf{Q} \in  [z_{1} \mathbf{P}, z_{2} \mathbf{P}] \textrm{ component-wise.}
\nonumber
\end{eqnarray}
Furthermore,
we derive from formula \eqref{brostu3:fo.Wfind1.new} 
(see also \eqref{brostu3:fo.Wfind1b}) 
\vspace{-0.2cm}
\begin{eqnarray}
& & \hspace{-0.7cm} \Lambda_{twop}(z) := \Lambda_{twop}^{(0)}(z) :=
\int\displaylimits_{0}^{z} F_{twop}^{-1}(u) \, du
=  \log\Big( p \cdot e^{z_{1}  \cdot z} + (1- p) \cdot e^{z_{2}  \cdot z} \Big),
\qquad z \in \, ]-\infty, \infty[ , 
\label{brostu3:fo.expowlink12}
\nonumber
\end{eqnarray}

\vspace{-0.2cm}
\noindent
which is the well-known cumulant generating function of
the two-point probability distribution 
$\mathbb{\bbzeta} = p \cdot \delta_{z_{1}} +  (1-p) \cdot \delta_{z_{2}}$,
where $z_{1} < 1 < z_{2}$ and $p= \frac{z_{2} -1}{z_{2} - z_{1}}$.
Of course, $\mathbb{\bbzeta}$ is a discrete distribution with frequencies 
$\mathbb{\Pi}[W=z_{1}] = p$, $\mathbb{\Pi}[W=z_{2}] = 1-p$ (and zero elsewhere).
Moreover, $\mathbb{\Pi}[W > 0]=1$ iff $z_{1} >0$, $\mathbb{\Pi}[W=0] > 0$
iff $z_{1}=0$.
Concerning the important Remark \ref{dist of components}(i),
for i.i.d. copies $(W_{i})_{i \in \mathbb{N}}$ of $W$ the 
probability distribution $\mathbb{\bbzeta}^{\ast n_{k}}[\cdot] :=  \mathbb{\Pi}[\breve{W} \in \cdot \, ]$
of $\breve{W} := \sum_{i\in I_{k}^{(n)}} W_{i}$ 
is the distribution of the $card(I_{k}^{(n)})-$th step of a generalized random walk
starting at zero;
this has a nice explicit (\textquotedblleft  binomial-type\textquotedblright ) 
expression in the special case $z_{1} = - z_{2}$,
namely $\sum_{\ell=0}^{card(I_{k}^{(n)})} \binom{card(I_{k}^{(n)})}{\ell} 
\cdot p^{card(I_{k}^{(n)}) - \ell} \cdot (1-p)^{\ell} \cdot \delta_{z_{2} 
\cdot (2 \ell- card(I_{k}^{(n)}))}$.
Within the context of Subsection \ref{Subsect Estimators determ}, 
for the concrete simulative estimation 
$\widehat{D_{\frac{1}{M_{\mathbf{P}}} \cdot \varphi_{twop}}}(\mathbf{\Omega},\mathbf{P})$
via \eqref{estimator minimization} and \eqref{Improved IS for inf div new}, we
obtain --- in terms of 
$M_{\mathbf{P}}:=\sum_{i=1}^{K}p_{i}>0$,
$n_{k} = n \cdot \widetilde{p}_{k} \in \mathbb{N}$ and
$\widetilde{q}_{k}^{\ast}$ from proxy method 1 or 2
--- that $\widetilde{U}_{k}^{\ast n_{k}}$
is the distribution of the $card(I_{k}^{(n)})-$th step of a generalized random walk
starting at zero, but with $p$ replaced by
$\breve{p} := \frac{p \cdot \exp(z_{1} \cdot \tau_{k})}{p \cdot \exp(z_{1} \cdot \tau_{k}) + 
(1-p) \cdot \exp(z_{2} \cdot \tau_{k})}$
where $\tau_{k} = \frac{1}{M_{\mathbf{P}}} \cdot F_{twop}\big(\frac{\widetilde{q}_{k}^{\ast}}{\widetilde{p}_{k}}\big)$
for $\widetilde{q}_{k}^{\ast} \in ]z_{1} \cdot \widetilde{p}_{k}, z_{2} \cdot \widetilde{p}_{k}]$; moreover,
$\widetilde{ISF}_{k}$ can be straightforwardly computed by \eqref{ISK}.


\subsection{Case 10}

\noindent
It is known that some types of robustness properties of minimum-divergence
estimators are connected with the \textit{boundedness}
of the derivative $\varphi^{\prime}$ of the divergence
generator $\varphi$; this property is
satisfied for the following cases,
which lead to the new classes of divergences \eqref{brostu3:fo.genLap4b}
and \eqref{brostu3:fo.genLap4fa}.

\vspace{0.2cm}


\subsubsection{Case 10a} \ 

\noindent
For any parameter-quadrupel
$\alpha,\beta_{1},\beta_{2},\widetilde{c} \in \, ]0,\infty[$ we choose 
\vspace{-0.1cm}
\begin{equation}
]a_{F},b_{F}[ 
\ \, := \ ]a_{F_{\alpha,\beta_{1},\beta_{2},\widetilde{c}}},
b_{F_{\alpha,\beta_{1},\beta_{2},\widetilde{c}}}[ \ \,  :=  \  
]-\infty, \,
\infty [ 
\nonumber
\end{equation}

\vspace{-0.1cm}
\noindent
and define with $\breve{\theta} := 1 + \alpha \cdot \big(\frac{1}{\beta_{2}} 
- \frac{1}{\beta_{1}} \big)$ (which is in $]-\infty,1[$ for $\beta_{1} < \beta_{2}$,
respectively, in $]1,1+\frac{\alpha}{\beta_{2}}[$ for $\beta_{1} > \beta_{2}$)
\vspace{-0.2cm} 
\begin{eqnarray}
\hspace{-0.6cm}
F_{\alpha,\beta_{1},\beta_{2},\widetilde{c}}(t) 
\negthinspace \negthinspace &:=& \negthinspace \negthinspace
\begin{cases}
\widetilde{c} \cdot \frac{\beta_{1}-\beta_{2}}{2}
+ \frac{\widetilde{c}}{\frac{1-t}{\alpha} + \frac{1}{\beta_{2}} - \frac{1}{\beta_{1}}}
\cdot \Big(
1 - \frac{1}{2} \cdot \sqrt{4 + \big( 
\frac{1-t}{\alpha} + \frac{1}{\beta_{2}} - \frac{1}{\beta_{1}}
\big)^{2} \cdot (\beta_{1}+\beta_{2})^{2}}
\, \Big),
\quad \textrm{if } \  
t \in \, ]a_{F},b_{F}[ \backslash\{\breve{\theta}\}, \\
\widetilde{c} \cdot \frac{\beta_{1}-\beta_{2}}{2}, \hspace{9.75cm} 
\textrm{if } \  t= \breve{\theta}. 
\end{cases}
\nonumber
\end{eqnarray}

\vspace{-0.1cm}
\noindent
One has the continuity 
$\lim_{t \rightarrow \breve{\theta}} 
F_{\alpha,\beta_{1},\beta_{2},\widetilde{c}}(t) = \widetilde{c} \cdot \frac{\beta_{1}-\beta_{2}}{2}$.
Moreover, one can see in a straightforward way that
$F_{\alpha,\beta_{1},\beta_{2},\widetilde{c}}(\cdot)$ is strictly increasing
and that
$\mathcal{R}(F_{\alpha,\beta_{1},\beta_{2},\widetilde{c}}) = \, 
]-\widetilde{c}\cdot \beta_{2},\widetilde{c}\cdot \beta_{1} [$.
Furthermore, $F_{\alpha,\beta_{1},\beta_{2},\widetilde{c}}(\cdot)$ is smooth on $]a_{F},b_{F}[$,
and thus $F_{\alpha,\beta_{1},\beta_{2},\widetilde{c}} \in \textgoth{F}$. 
Since $F_{\alpha,\beta_{1},\beta_{2},\widetilde{c}}(1)=0$, let us  
choose the natural anchor point $c:=0$,
which leads to $]\lambda_{-},\lambda_{+}[ \, 
= int(\mathcal{R}(F_{\alpha,\beta_{1},\beta_{2},\widetilde{c}})
= \, ]-\widetilde{c}\cdot \beta_{2},\widetilde{c}\cdot \beta_{1} [$ and
$]t_{-}^{sc},t_{+}^{sc}[ \, = \, ]a_{F},b_{F}[ \, = \, ]-\infty,\infty[ \, =  \, ]a,b[$.
Also, the corresponding inverse is
\begin{equation}
F_{\alpha,\beta_{1},\beta_{2},\widetilde{c}}^{-1}(x) = 
1 + \alpha \cdot \Big(\frac{1}{\beta_{2}} 
- \frac{1}{\beta_{1}} \Big) - \alpha  \cdot 
\frac{
\frac{1}{\beta_{2}} - \frac{1}{\beta_{1}} - 
\frac{2 x}{\widetilde{c} \cdot \beta_{1} \cdot \beta_{2} }
}{
1+ \frac{x}{\widetilde{c}} \cdot  
\Big(\frac{1}{\beta_{2}} - \frac{1}{\beta_{1}} \Big)
- \frac{x^{2}}{\widetilde{c}^{2} \cdot \beta_{1} \cdot \beta_{2} }
}, 
\qquad x \in int(\mathcal{R}(F_{\alpha,\beta_{1},\beta_{2},\widetilde{c}})) ;
\label{brostu3:fo.genLap1b}
\end{equation}
from this and 
\eqref{brostu3:fo.Wfind2.new}  
(see also \eqref{brostu3:fo.Wfind2b}) we can deduce
\begin{eqnarray}
\hspace{-1.0cm}
\varphi_{\alpha,\beta_{1},\beta_{2},\widetilde{c}}(t) 
\negthinspace := \negthinspace
\varphi_{\alpha,\beta_{1},\beta_{2},\widetilde{c}}^{(0)}(t)
& = &  
\widetilde{c} \cdot \alpha \cdot \Big\{
\frac{\sqrt{4 + (\beta_{1} + \beta_{2})^{2} 
\cdot (\frac{1-t}{\alpha} + \frac{1}{\beta_{2}} - \frac{1}{\beta_{1}})^2}
\,  - \, (\frac{1-t}{\alpha} + \frac{1}{\beta_{2}} - \frac{1}{\beta_{1}}) \cdot
(\beta_{1} -  \beta_{2}) \, - \, 2}{2} 
\nonumber\\ 
\hspace{-1.0cm} & &
+ \log\frac{\sqrt{4 + (\beta_{1} + \beta_{2})^{2} 
\cdot (\frac{1-t}{\alpha} + \frac{1}{\beta_{2}} \, - \, \frac{1}{\beta_{1}})^2} \, - \, 2
}{
\beta_{1} \beta_{2} \cdot (\frac{1-t}{\alpha} + \frac{1}{\beta_{2}} \, - \, \frac{1}{\beta_{1}})^{2}
}  \Big\}
\ \in [0,\infty[,
\qquad  
t \in \, ]-\infty,\infty[.
\label{brostu3:fo.genLap3}
\end{eqnarray}
Notice that  
$\varphi_{\alpha,\beta_{1},\beta_{2},\widetilde{c}}(1) = 0$, 
$\varphi_{\alpha,\beta_{1},\beta_{2},\widetilde{c}}^{\prime}(1) = 0$,
$\varphi_{\alpha,\beta_{1},\beta_{2},\widetilde{c}}(-\infty) = \infty$
and $\varphi_{\alpha,\beta_{1},\beta_{2},\widetilde{c}}(\infty) = \infty$. 
Moreover,  
$\varphi_{\alpha,\beta_{1},\beta_{2},\widetilde{c}}^{\prime}(-\infty) = 
\varphi_{\alpha,\beta_{1},\beta_{2},\widetilde{c}}^{\prime}(a_{F}) =
-\widetilde{c} \cdot \beta_{2}$ and 
$\varphi_{\alpha,\beta_{1},\beta_{2},\widetilde{c}}^{\prime}(\infty) = 
\widetilde{c} \cdot \beta_{1}$. 
From \eqref{brostu3:fo.genLap3},
we construct the corresponding divergence (cf. \eqref{brostu3:fo.div})
\begin{eqnarray}
& &
D_{\varphi_{\alpha,\beta_{1},\beta_{2},\widetilde{c}}}(\mathbf{Q},\mathbf{P})
= \sum\limits_{k=1}^{K} p_{k} \cdot 
\varphi_{\alpha,\beta_{1},\beta_{2},\widetilde{c}}\Big(\frac{q_{k}}{p_{k}}\Big)
\nonumber \\
& &
= \sum\limits_{k=1}^{K} 
p_{k} \cdot \bigg[
\widetilde{c} \cdot \alpha \cdot \Big\{
\frac{\sqrt{4 + (\beta_{1} + \beta_{2})^{2} 
\cdot (\frac{1-\frac{q_{k}}{p_{k}}}{\alpha} + \frac{1}{\beta_{2}} - \frac{1}{\beta_{1}})^2}
\,  - \, (\frac{1-\frac{q_{k}}{p_{k}}}{\alpha} + \frac{1}{\beta_{2}} - \frac{1}{\beta_{1}}) \cdot
(\beta_{1} -  \beta_{2}) \, - \, 2}{2} 
\nonumber \\ 
& &
+ \log\frac{\sqrt{4 + (\beta_{1} + \beta_{2})^{2} 
\cdot (\frac{1-\frac{q_{k}}{p_{k}}}{\alpha} + \frac{1}{\beta_{2}} \, - \, \frac{1}{\beta_{1}})^2} \, - \, 2
}{
\beta_{1} \beta_{2} \cdot (\frac{1-\frac{q_{k}}{p_{k}}}{\alpha} + \frac{1}{\beta_{2}} \, - \, \frac{1}{\beta_{1}})^{2}
}  \Big\} \bigg],  
\qquad \textrm{if } \mathbf{P} \in \mathbb{R}_{\gneqq 0}^{K}, \mathbf{Q} \in \mathbb{R}^{K} .
\label{brostu3:fo.genLap4b}
\end{eqnarray} 
Notice that we can particularly include the case where $p_{k}=0$, since for $q_{k} = 0$ 
we have $0 \cdot \varphi_{\alpha,\beta_{1},\beta_{2},\widetilde{c}}(\frac{0}{0}) = 0$ 
by the convention right after \eqref{brostu3:fo.div}, and for
$q_{k} \ne 0$ we have
$\lim_{t \rightarrow 0_{+}} t \cdot 
\varphi_{\alpha,\beta_{1},\beta_{2},\widetilde{c}}(\frac{1}{t}) 
= \widetilde{c} \cdot \beta_{1}$ and 
$\lim_{t \rightarrow 0_{-}} t \cdot 
\varphi_{\alpha,\beta_{1},\beta_{2},\widetilde{c}}(\frac{1}{t}) 
= - \widetilde{c} \cdot \beta_{2}$
which are both finite, and hence 
$p_{k} \cdot \varphi_{\alpha,\beta_{1},\beta_{2},\widetilde{c}}(\frac{q_{k}}{p_{k}}) 
= q_{k} \cdot \frac{p_{k}}{q_{k}} \cdot 
\varphi_{\alpha,\beta_{1},\beta_{2},\widetilde{c}}(\frac{q_{k}}{p_{k}}) $
stays finite as $p_{k}$ tends to zero.
To proceed with our general investigations, with the help of \eqref{brostu3:fo.genLap1b}
we can derive from \eqref{brostu3:fo.Wfind1.new},\eqref{brostu3:fo.Wfind1b}
\begin{eqnarray}
\Lambda_{\alpha,\beta_{1},\beta_{2},\widetilde{c}}(z) := 
\Lambda_{\alpha,\beta_{1},\beta_{2},\widetilde{c}}^{(0)}(z) 
\hspace{-0.2cm} &=& \hspace{-0.2cm}
\begin{cases}
\breve{\theta} \cdot z - \widetilde{c} \cdot \alpha \cdot \log\Big(
1+ \frac{z}{\widetilde{c}} \cdot  
\Big(\frac{1}{\beta_{2}} - \frac{1}{\beta_{1}} \Big)
- \frac{z^{2}}{\widetilde{c}^{2} \cdot \beta_{1} \cdot \beta_{2} }
\Big), 
\quad \textrm{if } \  
z \in \, ]-\widetilde{c}\cdot \beta_{2},\widetilde{c}\cdot \beta_{1} [, 
\\
\infty, \hspace{7.1cm}  \textrm{if $z \in \, ]-\infty,-\widetilde{c}\cdot \beta_{2}]
\, \cup \, [\widetilde{c}\cdot \beta_{1},\infty[$}. 
\end{cases}
\nonumber
\end{eqnarray}
Notice that $\Lambda_{\alpha,\beta_{1},\beta_{2},\widetilde{c}}(0) = 0$,
$\lim_{z \rightarrow - \widetilde{c}\cdot \beta_{2}} 
\Lambda_{\alpha,\beta_{1},\beta_{2},\widetilde{c}}(z) = \infty$ 
\, and \, $\lim_{z \rightarrow \widetilde{c}\cdot \beta_{1}} 
\Lambda_{\alpha,\beta_{1},\beta_{2},\widetilde{c}}(z) = \infty$. 
Moreover, \\
$\Lambda_{\alpha,\beta_{1},\beta_{2},\widetilde{c}}^{\prime}( -\widetilde{c}\cdot \beta_{2}) 
= - \infty = a_{F}$ and 
$\Lambda_{\alpha,\beta_{1},\beta_{2},\widetilde{c}}^{\prime}(\widetilde{c}\cdot \beta_{1}) = \infty
= b_{F}$ (which have to be interpreted as limits, as usual).

\vspace{0.2cm}


\subsubsection{Case 10b} \ 

\noindent
The analysis for the case $\beta_{1} = \beta_{2} =: \beta$ can be obtained
by taking $\lim_{\beta_1 \rightarrow \beta_{2}}$ in Case 10a. 
Alternatively, one can start afresh. Due to its importance and its particularities, 
we nevertheless state the corresponding results explicitly.
To begin with, for any parameter-triple
$\alpha,\beta,\widetilde{c} \in \, ]0,\infty[$
we choose 
$]a_{F},b_{F}[ 
\ \, := \ ]a_{F_{\alpha,\beta,\widetilde{c}}},
b_{F_{\alpha,\beta,\widetilde{c}}}[ \ \,  :=  \  
]-\infty, \infty \, [ $
and define with $\breve{\theta} := 1$ 
\begin{eqnarray}
\hspace{-0.6cm}
F_{\alpha,\beta,\widetilde{c}}(t) 
\negthinspace \negthinspace &:=& \negthinspace \negthinspace
\begin{cases}
\frac{\widetilde{c} \cdot \alpha}{1-t}
\cdot \Big(1 - \sqrt{1 + \big(\frac{1-t}{\alpha} \big)^{2} \cdot 
\beta^{2}}
\, \Big),
\qquad \textrm{if } \  
t \in \, ]a_{F},b_{F}[ \backslash\{\breve{\theta}\}, \\
0, \hspace{5.1cm} 
\textrm{if } \  t= \breve{\theta}. 
\end{cases}
\nonumber
\end{eqnarray}
Clearly, one has the continuity 
$\lim_{t \rightarrow \breve{\theta}} 
F_{\alpha,\beta,\widetilde{c}}(t) = 0$.
Moreover, one can see in a straightforward way that
$F_{\alpha,\beta,\widetilde{c}}(\cdot)$ is strictly increasing and that
$\mathcal{R}(F_{\alpha,\beta,\widetilde{c}})= \, 
]-\widetilde{c}\cdot \beta,\widetilde{c}\cdot \beta [$.
Furthermore, $F_{\alpha,\beta,\widetilde{c}}(\cdot)$ is smooth on $]a_{F},b_{F}[$,
and thus $F_{\alpha,\beta,\widetilde{c}} \in \textgoth{F}$. 
Since $F_{\alpha,\beta,\widetilde{c}}(1)=0$, let us  
choose the natural anchor point $c:=0$,
which leads to the choice $]\lambda_{-},\lambda_{+}[ \, 
= int(\mathcal{R}(F_{\alpha,\beta,\widetilde{c}})
= \, ]-\widetilde{c}\cdot \beta,\widetilde{c}\cdot \beta [$
and $]t_{-}^{sc},t_{+}^{sc}[ \, = \, ]a_{F},b_{F}[ \, = \, ]-\infty,\infty[
\, = \, ]a,b[$.
The inverse in \eqref{brostu3:fo.genLap1b} collapses to
\begin{equation}
F_{\alpha,\beta,\widetilde{c}}^{-1}(x) = 
1 + \alpha  \cdot 
\frac{ 
\frac{2 x}{\widetilde{c} \cdot \beta^{2} }
}{
1 - \frac{x^{2}}{\widetilde{c}^{2} \cdot \beta^{2} }
}, 
\qquad x \in int(\mathcal{R}(F_{\alpha,\beta,\widetilde{c}})) ;
\nonumber
\end{equation}
from this,  
we can derive from formula \eqref{brostu3:fo.Wfind1.new} 
(see also \eqref{brostu3:fo.Wfind1b}) 
\begin{eqnarray}
\hspace{-0.7cm}
\Lambda_{\alpha,\beta,\widetilde{c}}(z) := 
\Lambda_{\alpha,\beta,\widetilde{c}}^{(0)}(z) &=&
\begin{cases}
\breve{\theta} \cdot z - \widetilde{c} \cdot \alpha \cdot \log\Big(
1 - \frac{z^{2}}{\widetilde{c}^{2} \cdot \beta^{2} }
\Big), 
\qquad \textrm{if } \  
z \in \, ]-\widetilde{c}\cdot \beta,\widetilde{c}\cdot \beta [, 
\\
\infty, \hspace{4.6cm}  
\textrm{if } \  
z \in \, ]-\infty,-\widetilde{c}\cdot \beta] \, \cup
\, [\widetilde{c}\cdot \beta, \infty [. 
\end{cases}
\nonumber
\end{eqnarray}
Notice that $\Lambda_{\alpha,\beta,\widetilde{c}}(0) = 0$, 
$\lim_{z \rightarrow - \widetilde{c}\cdot \beta} 
\Lambda_{\alpha,\beta,\widetilde{c}}(z) = \infty$
\, and \, 
$\lim_{z \rightarrow \widetilde{c}\cdot \beta} 
\Lambda_{\alpha,\beta,\widetilde{c}}(z) = \infty$.
Furthermore, 
$\lim_{z \rightarrow - \widetilde{c}\cdot \beta} 
\Lambda_{\alpha,\beta,\widetilde{c}}^{\prime}(z) = -\infty$
\, and \, 
$\lim_{z \rightarrow \widetilde{c}\cdot \beta} 
\Lambda_{\alpha,\beta,\widetilde{c}}^{\prime}(z) = \infty$. 
To proceed, 
the formula \eqref{brostu3:fo.genLap3}
collapses to
\begin{equation}
\varphi_{\alpha,\beta,\widetilde{c}}(t) 
:=\varphi_{\alpha,\beta,\widetilde{c}}^{(0)}(t)
= \widetilde{c} \cdot \alpha \cdot \Big\{
\sqrt{1 + \beta^{2} 
\cdot \Big(\frac{1-t}{\alpha}\Big)^2} \, - \, 1 \, 
+ \, \log\frac{2 \cdot \Big(
\sqrt{1 + \beta^{2} 
\cdot \Big(\frac{1-t}{\alpha} \Big)^2} \, - \, 1
\Big)
}{
\beta^{2} \cdot \Big(\frac{1-t}{\alpha}\Big)^{2}
}  \Big\}
\ \in [0,\infty[,
\ \  
t \in \, ]-\infty, \infty[ .
\label{brostu3:fo.genLap3ba}
\end{equation}
Notice that  
$\varphi_{\alpha,\beta,\widetilde{c}}(1) = 0$, 
$\varphi_{\alpha,\beta,\widetilde{c}}^{\prime}(1) = 0$,
$\varphi_{\alpha,\beta,\widetilde{c}}(-\infty) = \infty$
and $\varphi_{\alpha,\beta,\widetilde{c}}(\infty) = \infty$. 
Moreover,  
$\varphi_{\alpha,\beta,\widetilde{c}}^{\prime}(-\infty) = 
-\widetilde{c} \cdot \beta$ and 
$\varphi_{\alpha,\beta,\widetilde{c}}^{\prime}(\infty) = 
\widetilde{c} \cdot \beta$.  
From \eqref{brostu3:fo.genLap3ba},
we construct the corresponding divergence (cf. \eqref{brostu3:fo.div})

\vspace{-0.3cm}
\begin{eqnarray}
& & \hspace{-1.4cm}
D_{\varphi_{\alpha,\beta,\widetilde{c}}}(\mathbf{Q},\mathbf{P})
= \sum\limits_{k=1}^{K} p_{k} \cdot 
\varphi_{\alpha,\beta,\widetilde{c}}\Big(\frac{q_{k}}{p_{k}}\Big)
\nonumber \\
& & \hspace{-1.4cm}
= \widetilde{c} \cdot \alpha \cdot  \sum\limits_{k=1}^{K} 
p_{k} \cdot
\Big\{
\sqrt{1 + \beta^{2} 
\cdot \Big(\frac{1-\frac{q_{k}}{p_{k}}}{\alpha}\Big)^2} \, - \, 1 
+ \log\frac{2 \cdot \Big(
\sqrt{1 + \beta^{2} 
\cdot \Big(\frac{1-\frac{q_{k}}{p_{k}}}{\alpha} \Big)^2} \, - \, 1
\Big)
}{
\beta^{2} \cdot \Big(\frac{1-\frac{q_{k}}{p_{k}}}{\alpha}\Big)^{2}
}  \Big\} ,  
\qquad \textrm{if } \mathbf{P} \in \mathbb{R}_{\gneqq 0}^{K}, \,  \mathbf{Q} \in \mathbb{R}^{K}.
\label{brostu3:fo.genLap4fa}
\end{eqnarray}

\vspace{0.2cm}


\subsubsection{Simulation distributions for the Cases 10a,b} \ 

\noindent
As far as the identification of the corresponding simulation laws $\bbzeta$ 
(cf. \eqref{Phi Legendre of mgf(W)}) is concerned,
let us first notice that for
$\alpha,\beta_{1},\beta_{2},\widetilde{c} \in \, ]0,\infty[$, and
anchor point $c=0$ one can see that
--- in terms of $\breve{\theta} := 1 + \alpha \cdot \Big(\frac{1}{\beta_{2}} 
- \frac{1}{\beta_{1}} \Big)$ ---
the derived quantity
$$\Lambda_{\alpha,\beta_{1},\beta_{2},\widetilde{c}}(z) =  
\breve{\theta} \cdot z - \widetilde{c} \cdot \alpha \cdot \log\Big(
1+ \frac{z}{\widetilde{c}} \cdot  
\Big(\frac{1}{\beta_{2}} - \frac{1}{\beta_{1}} \Big)
- \frac{z^{2}}{\widetilde{c}^{2} \cdot \beta_{1} \cdot \beta_{2} }
\Big) \, ,  \qquad 
z \in \, ]- \widetilde{c} \cdot \beta_{2},
\widetilde{c} \cdot  \beta_{1}[,
$$
is the cumulant generating function of 
a \textit{generalized asymmetric Laplace distribution} 
$\mathbb{\bbzeta}[ \, \cdot \,] = \mathbb{\Pi}[W \in \cdot \, ]$
of a random variable $W:= \breve{\theta} + Z_{1} - Z_{2}$, where
$Z_{1}$ and $Z_{2}$ are auxiliary random variables which are 
independent and
$GAM(\widetilde{c} \cdot \beta_{1},\widetilde{c} \cdot \alpha)-$distributed  
respectively $GAM(\widetilde{c} \cdot \beta_{2},\widetilde{c} \cdot \alpha)-$distributed; 
for the special case $\widetilde{c} =1$, $\alpha=1$, $\beta_{1}=\beta_{2} =: \beta$
(and hence, $\breve{\theta}=1$) one gets that    
$\mathbb{\bbzeta}$ is a \textit{classical Laplace distribution} 
(two-tailed exponential distribution,
bilateral exponential law)
with location parameter $1$ and scale parameter $\frac{1}{\beta}$.
Returning to the general constellation,
notice that $\mathbb{\bbzeta}$ is an infinitely divisible 
(cf. Proposition \ref{ID1}) continuous distribution with 
density 
\begin{equation}
f(u) :=  \frac{
\sqrt{2} \cdot \exp\{ \frac{1}{\sigma \cdot 
\sqrt{2}} \cdot (\frac{1}{\kappa} - \kappa) \cdot (u-\theta) \} 
}{\sqrt{\pi} \cdot \sigma^{\tau + 1/2} \cdot \Gamma(\tau)
} \cdot
\left( \frac{\sqrt{2} \cdot |u - \theta |}{\kappa + \frac{1}{\kappa}}  \right)^{\tau - 1/2} 
\negthinspace \negthinspace \cdot K_{\tau - 1/2}\left( \frac{1}{\sigma \cdot 
\sqrt{2}} \cdot  \Big(\kappa + \frac{1}{\kappa}\Big) \cdot |u - \theta | \right),
\quad u \in \mathbb{R}\backslash \{ \theta \},
\label{fo.genLapdens}
\end{equation}
where $(\theta,\kappa,\sigma,\tau)$ is given in Remark \ref{rem.genLap}
below and $K_{\lambda}$ is the modified Bessel function of the third kind 
with index $\lambda$; for the above-mentioned special case of the classical Laplace distribution,
this considerably simplifies to
$f(u):= \frac{\beta}{2} \exp\{- \beta \cdot |u -1 | \} $.
Moreover, note that in the general case one has
$\mathbb{\bbzeta}[ \, ]0,\infty[ \, ] = \mathbb{\Pi}[W > 0]=  
\int_{0}^{\infty} f(u) \, du \in \, ]0,1[$, \ \ 
$\mathbb{\bbzeta}[ \, \{0\} \, ] = \mathbb{\Pi}[W = 0]= 0$.
Concerning the important Remark \ref{dist of components}(i),
for i.i.d. copies $(W_{i})_{i \in \mathbb{N}}$ of $W$, the 
probability distribution 
$\mathbb{\bbzeta}^{\ast n_{k}}[\cdot] :=  \mathbb{\Pi}[\breve{W} \in \cdot \, ]$
of $\breve{W} := \sum_{i\in I_{k}^{(n)}} W_{i}$ 
is the same as that of a random variable 
$\breve{\widetilde{W}} := \breve{\theta} \cdot card(I_{k}^{(n)}) + 
\breve{Z}_{1} - \breve{Z}_{2}$, where
$\breve{Z}_{1}$ and  $\breve{Z}_{2}$ are auxiliary random variables which are 
independent and 
$GAM(\widetilde{c} \cdot \beta_{1},\widetilde{c} \cdot \alpha  
\cdot card(I_{k}^{(n)}))-$distributed  
respectively $GAM(\widetilde{c} \cdot \beta_{2},\widetilde{c} 
\cdot \alpha \cdot card(I_{k}^{(n)}))-$distributed. 
Within the context of Subsection \ref{Subsect Estimators determ}, 
for the concrete simulative estimation 
$\widehat{D_{\alpha,\beta_{1},\beta_{2},\widetilde{c}}}(\mathbf{\Omega},\mathbf{P})$
via \eqref{estimator minimization} and \eqref{Improved IS for inf div new}, we
obtain --- in terms of 
$M_{\mathbf{P}}:=\sum_{i=1}^{K}p_{i}>0$,
$n_{k} = n \cdot \widetilde{p}_{k} \in \mathbb{N}$ and
$\widetilde{q}_{k}^{\ast}$ from proxy method 1 or 2
--- that the distribution $\widetilde{U}_{k}^{\ast n_{k}}$
is the same as that of a random variable 
$\breve{\breve{\widetilde{W}}} := \breve{\theta} \cdot card(I_{k}^{(n)}) + 
\breve{\breve{Z}}_{1} - \breve{\breve{Z}}_{2}$, where
$\breve{\breve{Z}}_{1}$ and  $\breve{\breve{Z}}_{2}$ 
are auxiliary random variables which are 
independent and 
$GAM(\widetilde{c} \cdot M_{\mathbf{P}} \cdot \beta_{1} - \tau_{k},
\widetilde{c} \cdot M_{\mathbf{P}} \cdot \alpha  
\cdot card(I_{k}^{(n)}))-$distributed  
respectively $GAM(\widetilde{c} \cdot M_{\mathbf{P}} \cdot \beta_{2} + \tau_{k},
\widetilde{c} \cdot M_{\mathbf{P}}
\cdot \alpha \cdot card(I_{k}^{(n)}))-$distributed; 
here, $\tau_{k} = F_{\alpha,\beta_{1},\beta_{2},\widetilde{c}}\big(
\frac{\widetilde{q}_{k}^{\ast}}{\widetilde{p}_{k}}\big)$
for $\widetilde{q}_{k}^{\ast} \in \mathbb{R}$. Moreover,
$\widetilde{ISF}_{k}$ can be straightforwardly computed by \eqref{ISK}.

\vspace{0.3cm}

\begin{remark}
\label{rem.genLap}
In the book \cite{Kot:01} one can find 
a very comprehensive study on
generalized asymmetric Laplace distributions 
(also known as Bessel function distributions, McKay distributions),
their close relatives (such as e.g. the financial-econometric 
\textit{variance gamma model}
of \cite{Mad:90}) as well as their applications;
see also e.g. \cite{Kla:15} for connections with some
other Gamma difference distributions.
\cite{Kot:01} uses a different parametrization  
$(\theta,\kappa,\sigma,\tau)$
which is one-to-one with our parametrization
$(\breve{\theta},\alpha,\beta_{1},\beta_{2},\widetilde{c}=1)$, as follows: 
$\theta = \breve{\theta}$, $\tau = \widetilde{c} \cdot \alpha$,
$\sigma = \frac{1}{\widetilde{c}} \cdot \sqrt{\frac{2}{\beta_{1} \cdot \beta_{2}}}$,
$\kappa = \sqrt{\frac{\beta_{1}}{\beta_{2}}}$. 
In particular, this implies that we cover \textit{all}
generalized asymmetric Laplace distributions with mean $1$.
For better comparability, we have used the parametrization $(\theta,\kappa,\sigma,\tau)$
in the above-mentioned representation \eqref{fo.genLapdens} of the density
(due to \cite{Kot:01}).

\end{remark}

%
%

\section{Concluding Remarks}

\noindent
This paper presents a new approach for the optimization of various different 
non-linear --- deterministic respectively statistical --- functionals on 
$\mathbb{R}^{K}$ under fairly general (e.g. non-convex, highly disconnected,
large-dimensional) constraints, 
in terms of an appropriately constructed dimension-free bare simulation method
which is straightforward to implement and which converges.
Algorithms and numerous detailed cases --- e.g. for $\varphi-$divergences,
Renyi divergences, generalized entropies, integer-programming relaxations etc. ---
are presented with explicit solutions
pertaining to the simulations to be performed.
As argued, our newly developed optimization
procedure does \textit{not} rely on the search for minimizers as a first stage, 
in contrast with existing methods for similar problems which thus require
some involved regularity on the constraint set and also suffer
from the so-called ``curse of dimensionality''.

Extensions of our new method can be pursuited e.g. in two directions.
Firstly, the search for optimizers can be handled, either by making use of
the simulated values which have been used in order to get the approximation
of the optimal value of the objective function, or through dichotomous
search. Secondly, the case where the objective function is
defined on an infinite-dimensional space (rather than $\mathbb{R}^K$)
is of interest; for instance, in
the statistical context this amounts to consider adequacy between a
probability distribution $P$ and a model $\Omega$ which consists of \textit{continuous}
distributions. The basic asymptotics which are used in this paper can be
extended to these situations, both in the deterministic case and in the
statistical context. This will be part of a follow-up paper.

%
%

\appendices



%
%

\section{Proofs --- Part 1}

\noindent
\textbf{Proof of Theorem \ref{brostu3:thm.divW.var}.} 
This is a straightforward application of the classical Cramer-type Large Deviation
Theorem in the vector case (see Theorem 2.2.30 and Corollary 6.1.6
in \cite{Dem:09}). 
Recall that above we have transformed the original problem  
into a context where the second argument in $D_{\varphi }(\cdot,\cdot)$ is a probability vector, 
as follows: in terms of $M_{\mathbf{P}}:=\sum_{i=1}^{K}p_{i}>0$ we normalized
$\widetilde{\mathds{P}}:=\mathbf{P}/M_{\mathbf{P}},$ and
 $\widetilde{\mathbf{Q}}:=\mathbf{Q}/M_{\mathbf{P}}$ for $\mathbf{Q}$ in $\mathbf{\Omega}$.
With  
$\widetilde{\varphi} \in \Upsilon (]a,b[)$ defined through $\widetilde{\varphi }:=M_{\mathbf{P}} \cdot \varphi $, 
we have obtained
\begin{equation*}
D_{\varphi }(\mathbf{Q},\mathbf{P})=\sum_{k=1}^{K}p_{k}\cdot \varphi \left( \frac{q_{k}}{p_{k}}
\right) =\sum_{k=1}^{K}M_{\mathbf{P}}\cdot \widetilde{p_{k}}\cdot \varphi
\left( \frac{M_{\mathbf{P}}\cdot \widetilde{q_{k}}}{M_{\mathbf{P}}\cdot \widetilde{p_{k}}}
\right) =D_{\widetilde{\varphi }}(\widetilde{\mathbf{Q}},\widetilde{\mathds{P}})
\qquad \textrm{(cf. \eqref{min Pb prob1})}.
\end{equation*}
It has followed that the solution of \eqref{min Pb} coincides with the one
of the problem of finding
\begin{equation*}
\widetilde{\Phi}_{\widetilde{\mathds{P}}}(\widetilde{\mathbf{\Omega}}) := \inf_{\widetilde{\mathbf{Q}}\in \widetilde{\mathbf{\Omega}} }
D_{\widetilde{\varphi} }(\widetilde{\mathbf{Q}},\widetilde{\mathds{P}}),  
\qquad \textrm{with } \widetilde{\mathbf{\Omega}}:=\mathbf{\Omega} /M_{\mathbf{P}}
\qquad \textrm{(cf. \eqref{min Pb prob2})}.
\end{equation*}
So let us continue by tackling \eqref{min Pb prob2}. 
From the assumptions on $\widetilde{\varphi}$ and the requirement \eqref{brostu3:fo.link.var} 
one can see that
\begin{equation}
\text{$\widetilde{W}_{1}$ has moment generating function
$MGF_{\widetilde{\mathbb{\bbzeta}}}(z) = E_{\mathbb{\Pi}}[e^{z \cdot \widetilde{W}_{1}}]$
which is finite on a non-void neighborhood of $0$}, 
\label{Finite mgf_new}
\end{equation}
\begin{equation}
E_{\mathbb{\Pi}}[\widetilde{W}_1] = 1 ,   
\nonumber
\end{equation} 
since $\widetilde{\varphi}(1)=0=\widetilde{\varphi}^{\prime}(1)$.
With the help of these, we obtain the following

\vspace{0.2cm}

\begin{proposition}
\label{PropLDPWEM-finitease}
Under the assumptions of Theorem \ref{brostu3:thm.divW.var},
for any set $\widetilde{\mathbf{\Omega}} 
\subset \mathcal{M} := \mathbb{R}^{K}$
with \eqref{regularity} one has
\begin{eqnarray}
-\inf_{\widetilde{\mathbf{Q}} \in int(\widetilde{\mathbf{\Omega}}) } 
D_{\varphi}( \widetilde{\mathbf{Q}},\widetilde{\mathds{P}} ) &\leq &\lim
\inf_{n\rightarrow \infty }\frac{1}{n}\log \mathbb{\Pi}\negthinspace \left[ 
\boldsymbol{\xi}_{n}^{\widetilde{\textbf{W}}}\in
\widetilde{\mathbf{\Omega}} \right] \nonumber \\
&\leq &\lim \sup_{n\rightarrow \infty }\frac{1}{n}\log \mathbb{\Pi}\negthinspace \left[
\boldsymbol{\xi}_{n}^{\widetilde{\textbf{W}}}\in \widetilde{\mathbf{\Omega}} \right] 
\leq -\inf_{\widetilde{\mathbf{Q}}\in cl(\widetilde{\mathbf{\Omega}}) }
 D_{\varphi}( \widetilde{\mathbf{Q}},\widetilde{\mathds{P}}) .
\label{LDPsigned:new}
\end{eqnarray}

\end{proposition}

\vspace{0.4cm}
\noindent
\textbf{Proof of Proposition \ref{PropLDPWEM-finitease}.}\ 
Recall that $n_{k} := card(I_{k}^{(n)})$ denotes the number of the elements
of the block $I_{k}^{(n)}$ defined right after \eqref{fo.freqlim}
($k=1,\ldots,K$).  
We follow the line of proof of Theorem 2.2.30 in \cite{Dem:09},
which states the large deviation principle (LDP) for the vector of partial sums of random vectors in $
\mathbb{R}^{K}$, where we also use Corollary 6.1.6 in \cite{Dem:09} in relation
with condition \eqref{Finite mgf_new}.
Indeed, since by definition the $k-$th component of the vector $\boldsymbol{\xi}_{n}^{\mathbf{\widetilde{W}}}$
is equal to $\frac{1}{n} \sum_{i\in I_{k}^{(n)}} \widetilde{W}_{i}$, 
the current proof will follow from a
similar treatment as for the standard Cramer LDP in $\mathbb{R}^{K}.$ The
only difference lies in two facts: firstly, the number of the summands for the
$k-th$ coordinate is $n_{k}$ instead
of $n$ in the standard case; secondly, we will need to substitute $n_{k}$
by its equivalent $n \cdot \widetilde{p}_{k}$, which adds an approximation step. For the
upper bound, the proof is based on the corresponding result for 
$\mathbf{B} := B_{1}\times \cdots \times B_{K}$ where the $B_{k}$\textquoteright s are open bounded intervals
on $\mathbb{R}^{+}$. Since $\lim_{n\rightarrow \infty}\frac{n_{k}}{n}= \widetilde{p}_{k}$
(cf. \eqref{fo.freqlim}),
there holds 
\begin{eqnarray}
&& \hspace{-1.0cm}
\frac{1}{n} \cdot \log 
\mathbb{\Pi} \negthinspace \left[\boldsymbol{\xi}_{n}^{\mathbf{\widetilde{W}}} \in \mathbf{B} \right]   
\ = \ \frac{1}{n}\log \mathbb{\Pi} \bigg[ \dbigcap\limits_{k=1}^{K}\bigg( 
\frac{1}{n} \sum_{i\in I_{k}^{(n)}} \widetilde{W}_{i}
\in B_{k}\bigg) \bigg]  \ = \ 
\frac{1}{n}\dsum\limits_{k=1}^{K}\log \mathbb{\Pi} \bigg[ 
\frac{1+o(1)}{
n_{k}} \sum_{i\in I_{k}^{(n)}} \widetilde{W}_{i}
\in \frac{1}{\widetilde{p}_{k}} B_{k}\bigg],  
\label{expEquiv.new}
\\
& &\hspace{-1.0cm} 
\textrm{and hence } \ \ 
\lim \sup_{n\rightarrow \infty }\frac{1}{n}\log \mathbb{\Pi} \negthinspace \left[ 
\boldsymbol{\xi}_{n}^{\mathbf{\widetilde{W}}}
\in \mathbf{B} \right]
\ \leq \ \dsum\limits_{k=1}^{K} \widetilde{p}_{k} \cdot \lim \sup_{n_{k}\rightarrow \infty }\frac{1
}{n_{k}}\log \mathbb{\Pi} \bigg[ \frac{1}{n_{k}}
\sum_{i\in I_{k}^{(n)}} \widetilde{W}_{i}
\in \frac{1}{p_{k}}B_{k}\bigg]   
\nonumber \\
& & \hspace{5.0cm}
\leq \ -\dsum\limits_{k=1}^{K}\inf_{x_{k}\in cl(B_{k})} \widetilde{p}_{k} \cdot \varphi \left( 
\frac{x_{k}}{\widetilde{p}_{k}}\right) .  
\label{fin proof.new}
\end{eqnarray}
To deduce \eqref{fin proof.new} from \eqref{expEquiv.new}, we have used (i) the fact 
that for all $k$
the random variables $\frac{1}{n_{k}}\left( 1+o(1))\right) \cdot
\sum_{i\in I_{k}^{(n)}} \widetilde{W}_{i}
$ and $\frac{1}{n_{k}}
\sum_{i\in I_{k}^{(n)}} \widetilde{W}_{i}
$ are exponentially equivalent in the
sense that their difference $\Delta _{n_{k}}$ satisfies 
$
\lim \sup_{n_{k}\rightarrow \infty }\frac{1}{n_{k}}\log \mathbb{\Pi}\negthinspace \left[ \, 
\left\vert \Delta _{n_{k}}\right\vert >\eta \, \right] =-\infty , 
$
making use of the Chernoff inequality for all positive $\eta $,
as well as (ii) Theorem 4.2.13 in \cite{Dem:09}. 
Now the summation and the inf-operations can be
permuted in \eqref{fin proof.new} which proves the claim for the hyper-rectangle $\mathbf{B}$.
As in \cite{Dem:09}, for a compact set $\widetilde{\mathbf{\Omega}}$ we
consider its finite covering by such open hyper-rectangles $\mathbf{B}$ and conclude; for 
$\widetilde{\mathbf{\Omega}}$ being a closed set, a tightness argument holds, following 
\cite{Dem:09} Theorem 2.2.30 verbatim. For the lower bound consider the same
hyper-rectangle $\mathbf{B}$. The argument which locates the tilted distribution at the
center of $\mathbf{B}$, together with the use of the LLN for the corresponding r.v.\textquoteright s
as in \cite{Dem:09}, in combination with the same approximations as
above to handle the approximation of $n_{k}$ by $n \cdot \widetilde{p}_{k}$, complete the proof.
We omit the details. 
\hspace{0.5cm}  $\blacksquare$  

\vspace{0.3cm}
\noindent
Let us continue with the proof of Theorem \ref{brostu3:thm.divW.var},
by giving the following two helpful lemmas
for 
\begin{equation}
\Phi_{\mathds{P}}(\mathbf{A}) := \inf_{\mathbf{Q} \in \mathbf{A}} D_{\varphi}\left(\mathbf{Q},\mathds{P}\right),
\qquad
\mathbf{A} \subset \mathcal{M} := \mathbb{R}^{K}.
\label{phishort}
\end{equation}

\begin{lemma}
\label{LDP-lemma4}
For any open set $\mathbf{A} \subset \mathcal{M} := \mathbb{R}^{K}$
one has $\Phi_{\mathds{P}}(\mathbf{A}) = \Phi_{\mathds{P}}(cl(\mathbf{A}))$. 
\end{lemma}

\vspace{0.1cm}
\noindent
This is clear from the continuity of $\Phi_{\mathds{P}}$.

\vspace{0.2cm}

\begin{lemma}
\label{LDP-lemma5}
For any $\mathbf{A} \subset \mathcal{M} := \mathbb{R}^{K}$ satisfying 
\eqref{regularity}
one has $\Phi_{\mathds{P}}(cl(\mathbf{A})) =  \Phi_{\mathds{P}}(\mathbf{A})
 = \Phi_{\mathds{P}}(int(\mathbf{A}))$.
\end{lemma}

\vspace{0.2cm}
\noindent
\textbf{Proof of Lemma \ref{LDP-lemma5}.} \
Assume first that $\Phi_{\mathds{P}}(\mathbf{A})$ is finite. Then 
suppose that $\mathbf{A}$ satisfies 
\eqref{regularity} and $\Phi_{\mathds{P}}(cl(\mathbf{A})) < \Phi_{\mathds{P}}(int(\mathbf{A}))$.
The latter implies the existence of a point $\mathbf{a} \in cl(\mathbf{A})$ such that 
$\mathbf{a} \notin int(\mathbf{A})$ and $D_{\varphi}(\mathbf{a},\mathds{P}) = \Phi_{\mathds{P}}(cl(\mathbf{A}))$. 
But then, by Lemma \ref{LDP-lemma4} and 
\eqref{regularity}
one gets $\Phi_{\mathds{P}}(int(\mathbf{A})) = \Phi_{\mathds{P}}(cl(int(\mathbf{A}))) = \Phi_{\mathds{P}}(cl(\mathbf{A})) = 
D_{\varphi}(\mathbf{a},\mathds{P})$
which leads to a contradiction.
When $\Phi_{\mathds{P}}(\mathbf{A}) = \infty$ then $\Phi_{\mathds{P}} (cl(\mathbf{A}))=\Phi_{\mathds{P}} (int(\mathbf{A}
))=\Phi_{\mathds{P}} (\mathbf{A})=\infty $. \hspace{0.5cm} $\blacksquare$

\vspace{0.2cm}
\noindent
Finally, the asymptotic assertion \eqref{LDP Minimization}
follows from \eqref{LDPsigned:new}, 
\eqref{regularity} and Lemma \ref{LDP-lemma5}.
This completes the proof of Theorem \ref{brostu3:thm.divW.var}. 
\hspace{0.2cm} $\blacksquare$  

%
%

\section{Proofs --- Part 2}

\noindent
Before we tackle the proof of Theorem \ref{brostu3:thm.divnormW.new},
let us introduce the following 

\vspace{0.1cm}

\begin{lemma}
If  $\boldsymbol{\Omega}$\hspace{-0.23cm}$\boldsymbol{\Omega}\subset \mathbb{S}^{K}$
satisfies condition \eqref{regularity},
then $\widetilde{\widetilde{\textrm{$\boldsymbol{\Omega}$\hspace{-0.23cm}$\boldsymbol{\Omega}$}}} 
:= \bigcup\limits_{m\neq 0} 
cl(m \cdot \textrm{$\boldsymbol{\Omega}$\hspace{-0.23cm}$\boldsymbol{\Omega}$})$
has the property \eqref{regularity}.
\end{lemma}

\vspace{0.1cm}
\noindent
This can be deduced in a straightforward way: the assumption implies that 
$cl(\textrm{$\boldsymbol{\Omega}$\hspace{-0.23cm}$\boldsymbol{\Omega}$})$
satisfies \eqref{regularity}, and thus also
$m \cdot cl(\textrm{$\boldsymbol{\Omega}$\hspace{-0.23cm}$\boldsymbol{\Omega}$})$ 
satisfies \eqref{regularity}.
But this implies the validity of 
\eqref{regularity} 
for the
\textquotedblleft cone\textquotedblright\ $\bigcup\limits_{m\neq 0} m 
\cdot cl(\textrm{$\boldsymbol{\Omega}$\hspace{-0.23cm}$\boldsymbol{\Omega}$})$
which is nothing but $\bigcup\limits_{m\neq 0} cl(m \cdot 
\textrm{$\boldsymbol{\Omega}$\hspace{-0.23cm}$\boldsymbol{\Omega}$})$.

\vspace{0.2cm}
\noindent
\textbf{Proof of Theorem \ref{brostu3:thm.divnormW.new}.}\ 
Recall the interpretations of the two vectors
$\boldsymbol{\xi}_{n,\mathbf{X}}^{\mathbf{W}}$
respectively
$\boldsymbol{\xi}_{n,\mathbf{X}}^{w\mathbf{W}}$
given in \eqref{brostu3:fo.weiemp.var2} respectively \eqref{brostu3:fo.norweiemp.vec},
and that the sum 
of their $k$ components are
$\sum_{k=1}^{K} \frac{1}{n} \sum_{i\in I_{k}^{(n)}} W_{i} =\frac{1}{n}\sum_{i=1}^{n} W_{i}$ respectively 
$\sum_{k=1}^{K}
\frac{\sum_{i \in I_{k}^{(n)}}W_{i}}{\sum_{k=1}^{K}\sum_{i \in I_{k}^{(n)}}W_{i}}=
1$ (in case of $\sum_{i=1}^{n}W_{i} \ne 0$).
In the light of these, for 
$\textrm{$\boldsymbol{\Omega}$\hspace{-0.23cm}$\boldsymbol{\Omega}$}  \subset \mathbb{S}^{K}$ 
one gets the set identification
\[
\left\{ \boldsymbol{\xi}_{n,\mathbf{X}}^{w\mathbf{W}} \in \textrm{$\boldsymbol{\Omega}$\hspace{-0.23cm}$\boldsymbol{\Omega}$} \right\}
 =\bigcup\limits_{m\neq 0}\bigg\{
\boldsymbol{\xi}_{n,\mathbf{X}}^{\mathbf{W}}
\in m\cdot \textrm{$\boldsymbol{\Omega}$\hspace{-0.23cm}$\boldsymbol{\Omega}$} ,
\frac{1}{n}\sum_{i=1}^{n} W_{i} = m 
\bigg\}
\]
since $\left\{ \sum_{i=1}^{n}W_{i}=0\right\} $ amounts to $m=0$, which cannot
hold when $\left\{ \boldsymbol{\xi}_{n,\mathbf{X}}^{w\mathbf{W}} \in 
\textrm{$\boldsymbol{\Omega}$\hspace{-0.23cm}$\boldsymbol{\Omega}$} \right\}$. Now 
\begin{eqnarray*}
\mathbb{\Pi}_{X_{1}^{n}}\negthinspace \left[ 
\boldsymbol{\xi}_{n,\mathbf{X}}^{w\mathbf{W}}
\in \textrm{$\boldsymbol{\Omega}$\hspace{-0.23cm}$\boldsymbol{\Omega}$} \right] 
&=&\mathbb{\Pi}_{X_{1}^{n}} \negthinspace\bigg[ \bigcup\limits_{m\neq 0}\bigg\{ \boldsymbol{\xi}_{n,\mathbf{X}}^{\mathbf{W}} \in 
m \cdot \textrm{$\boldsymbol{\Omega}$\hspace{-0.23cm}$\boldsymbol{\Omega}$},
\frac{1}{n}\sum_{i=1}^{n} W_{i} =m \bigg\}
\bigg]  \\
&=&\mathbb{\Pi}_{X_{1}^{n}}\negthinspace \bigg[ \bigcup\limits_{m\neq 0}\Big\{ \boldsymbol{\xi}_{n,\mathbf{X}}^{\mathbf{W}} \in 
m\cdot \textrm{$\boldsymbol{\Omega}$\hspace{-0.23cm}$\boldsymbol{\Omega}$}
\Big\} \bigg]
= \mathbb{\Pi}_{X_{1}^{n}}\negthinspace \bigg[ \boldsymbol{\xi}_{n,\mathbf{X}}^{\mathbf{W}} 
\in \bigcup\limits_{m\neq0} m\cdot 
\textrm{$\boldsymbol{\Omega}$\hspace{-0.23cm}$\boldsymbol{\Omega}$} \bigg] 
\end{eqnarray*}
since $\left\{ \boldsymbol{\xi}_{n,\mathbf{X}}^{\mathbf{W}} \in 
m\cdot \textrm{$\boldsymbol{\Omega}$\hspace{-0.23cm}$\boldsymbol{\Omega}$}\right\} \subset 
\left\{ \frac{1}{n}\sum_{i=1}^{n} W_{i} =m
\right\}$. Therefore 
\begin{equation}
\frac{1}{n}\log \mathbb{\Pi}_{X_{1}^{n}}\negthinspace \left[\boldsymbol{\xi}_{n,\mathbf{X}}^{w\mathbf{W}}
\in \textrm{$\boldsymbol{\Omega}$\hspace{-0.23cm}$\boldsymbol{\Omega}$} \right] =
\frac{1}{n}\log \mathbb{\Pi}_{X_{1}^{n}}\negthinspace \bigg[ \boldsymbol{\xi}_{n,\mathbf{X}}^{\mathbf{W}} \in \bigcup\limits_{m\neq
0}m\cdot \textrm{$\boldsymbol{\Omega}$\hspace{-0.23cm}$\boldsymbol{\Omega}$} \bigg] .
\label{equ.00}
\end{equation}
Analogously to the proof of Proposition \ref{PropLDPWEM-finitease} --- applied to 
$\widetilde{\textrm{$\boldsymbol{\Omega}$\hspace{-0.23cm}$\boldsymbol{\Omega}$}} := 
\bigcup\limits_{m\neq 0}m \cdot \textrm{$\boldsymbol{\Omega}$\hspace{-0.23cm}$\boldsymbol{\Omega}$}$ --- one gets
in terms of (\ref{phishort})
\begin{eqnarray}
-\Phi_{\mathds{P}}\Big(int\Big(\bigcup\limits_{m\neq 0}m\cdot \textrm{$\boldsymbol{\Omega}$\hspace{-0.23cm}$\boldsymbol{\Omega}$}\Big)\Big) 
&\leq &
\lim\inf_{n\rightarrow \infty }\frac{1}{n}\log \mathbb{\Pi}_{X_{1}^{n}}\negthinspace \bigg[ \boldsymbol{\xi}_{n,\mathbf{X}}^{\mathbf{W}} \in
\bigcup\limits_{m\neq 0}m\cdot \textrm{$\boldsymbol{\Omega}$\hspace{-0.23cm}$\boldsymbol{\Omega}$} \bigg] \nonumber \\
&\leq &\lim \sup_{n\rightarrow \infty }\frac{1}{n}\log \mathbb{\Pi}_{X_{1}^{n}}\negthinspace \bigg[
\boldsymbol{\xi}_{n,\mathbf{X}}^{\mathbf{W}} 
\in \bigcup\limits_{m\neq 0}m\cdot \textrm{$\boldsymbol{\Omega}$\hspace{-0.23cm}$\boldsymbol{\Omega}$} \bigg] \leq 
-\Phi_{\mathds{P}}\Big(cl\Big(\bigcup\limits_{m\neq 0}m\cdot \textrm{$\boldsymbol{\Omega}$\hspace{-0.23cm}$\boldsymbol{\Omega}$}\Big)\Big) .
\label{equ.0}
\end{eqnarray}
\begin{eqnarray}
\hspace{-6.2cm}
\text{But}
& & \Phi_{\mathds{P}}\Big(int\Big(\bigcup\limits_{m\neq 0}m\cdot \textrm{$\boldsymbol{\Omega}$\hspace{-0.23cm}$\boldsymbol{\Omega}$}\Big)\Big) \leq
\Phi_{\mathds{P}}\Big(\bigcup\limits_{m\neq 0}int\Big(m\cdot \textrm{$\boldsymbol{\Omega}$\hspace{-0.23cm}$\boldsymbol{\Omega}$}\Big)\Big) =
\inf_{m\neq 0} \Phi_{\mathds{P}}(int(m\cdot \textrm{$\boldsymbol{\Omega}$\hspace{-0.23cm}$\boldsymbol{\Omega}$}))
\label{equ.1}\\
\hspace{-6.2cm}
\text{and}
& & \Phi_{\mathds{P}}\Big(cl\Big(\bigcup\limits_{m\neq 0}m\cdot \textrm{$\boldsymbol{\Omega}$\hspace{-0.23cm}$\boldsymbol{\Omega}$}\Big)\Big) \geq
\Phi_{\mathds{P}}\Big(\bigcup\limits_{m\neq 0}cl\Big(m\cdot \textrm{$\boldsymbol{\Omega}$\hspace{-0.23cm}$\boldsymbol{\Omega}$}\Big)\Big) =
\inf_{m\neq 0} \Phi_{\mathds{P}}(cl(m\cdot \textrm{$\boldsymbol{\Omega}$\hspace{-0.23cm}$\boldsymbol{\Omega}$})) .
\label{equ.2}
\end{eqnarray}
In fact, the inequality in (\ref{equ.1}) is straightforward because of
$\bigcup\limits_{m\neq 0}int(m\cdot \textrm{$\boldsymbol{\Omega}$\hspace{-0.23cm}$\boldsymbol{\Omega}$})
 \subset int(\bigcup\limits_{m\neq 0}m\cdot \textrm{$\boldsymbol{\Omega}$\hspace{-0.23cm}$\boldsymbol{\Omega}$})$
(since the latter is the largest open set contained in 
$\bigcup\limits_{m\neq 0}m\cdot \textrm{$\boldsymbol{\Omega}$\hspace{-0.23cm}$\boldsymbol{\Omega}$}$);
the inequality in (\ref{equ.2}) follows from
\begin{eqnarray}
& & \Phi_{\mathds{P}}\Big(cl\Big(\bigcup\limits_{m\neq 0}m\cdot \textrm{$\boldsymbol{\Omega}$\hspace{-0.23cm}$\boldsymbol{\Omega}$}\Big)\Big) \geq
\Phi_{\mathds{P}}\Big(cl\Big(\bigcup\limits_{m\neq 0}cl\Big(m\cdot \textrm{$\boldsymbol{\Omega}$\hspace{-0.23cm}$\boldsymbol{\Omega}$}\Big)\Big)\Big) =
\Phi_{\mathds{P}}\Big(\bigcup\limits_{m\neq 0}cl\Big(m\cdot \textrm{$\boldsymbol{\Omega}$\hspace{-0.23cm}$\boldsymbol{\Omega}$}\Big)\Big).
\nonumber
\end{eqnarray}
An application of Lemma \ref{LDP-lemma5} yields
$\Phi_{\mathds{P}}(int(m\cdot \textrm{$\boldsymbol{\Omega}$\hspace{-0.23cm}$\boldsymbol{\Omega}$})) = 
\Phi_{\mathds{P}}(m\cdot \textrm{$\boldsymbol{\Omega}$\hspace{-0.23cm}$\boldsymbol{\Omega}$}) = 
\Phi_{\mathds{P}}(cl(m\cdot \textrm{$\boldsymbol{\Omega}$\hspace{-0.23cm}$\boldsymbol{\Omega}$}))$ for all $m \neq 0$, and hence
\begin{equation}
\inf_{m\neq 0} \Phi_{\mathds{P}}(int(m\cdot \textrm{$\boldsymbol{\Omega}$\hspace{-0.23cm}$\boldsymbol{\Omega}$})) = 
\inf_{m\neq 0} \Phi_{\mathds{P}}(m\cdot \textrm{$\boldsymbol{\Omega}$\hspace{-0.23cm}$\boldsymbol{\Omega}$})
= \inf_{m\neq 0} \Phi_{\mathds{P}}(cl(m\cdot \textrm{$\boldsymbol{\Omega}$\hspace{-0.23cm}$\boldsymbol{\Omega}$})) .
\label{equ.4}
\end{equation}
By combining (\ref{equ.00}) to (\ref{equ.4}),
one arrives at
\begin{eqnarray}
&&\lim_{n \rightarrow \infty} \frac{1}{n}\log \mathbb{\Pi}_{X_{1}^{n}} \negthinspace \left[ 
\boldsymbol{\xi}_{n,\mathbf{X}}^{w\mathbf{W}}
\in 
\textrm{$\boldsymbol{\Omega}$\hspace{-0.23cm}$\boldsymbol{\Omega}$} \right]
=  \lim_{n \rightarrow \infty} \frac{1}{n}\log \mathbb{\Pi}_{X_{1}^{n}} \negthinspace \bigg[ \boldsymbol{\xi}_{n,\mathbf{X}}^{\mathbf{W}}
\in \bigcup\limits_{m\neq0}
m\cdot \textrm{$\boldsymbol{\Omega}$\hspace{-0.23cm}$\boldsymbol{\Omega}$} \bigg]
 \nonumber \\
&& = - \inf_{m\neq 0} \Phi_{\mathds{P}}(m\cdot \textrm{$\boldsymbol{\Omega}$\hspace{-0.23cm}$\boldsymbol{\Omega}$}) 
= - \inf_{m\neq 0} \inf_{\mathbf{Q} \in m\cdot \textrm{$\boldsymbol{\Omega}$\hspace{-0.19cm}$\boldsymbol{\Omega}$}} 
D_{\varphi}\left(\mathbf{Q},\mathds{P}\right)
= - \inf_{m\neq 0} \inf_{\mathds{Q} \in \textrm{$\boldsymbol{\Omega}$\hspace{-0.19cm}$\boldsymbol{\Omega}$}} 
D_{\varphi}\left(m\cdot \mathds{Q}, \mathds{P}\right) , 
\nonumber
\end{eqnarray}
where in the second last equality we have \textquotedblleft reverted\textquotedblright\ the notation (\ref{phishort}). Note that we did not assume \eqref{regularity} 
for $\dbigcup\limits_{m\neq 0} m \cdot \textrm{$\boldsymbol{\Omega}$\hspace{-0.23cm}$\boldsymbol{\Omega}$}$.
\hspace{0.5cm} $\blacksquare$  \\

\vspace{0.3cm}

%
%

\begin{center}
{\large
The remaining APPENDICES \ref{Sect AppProofs3} ff. are placed in the
Supplementary Material of this paper.}

\end{center}

\vspace{0.3cm}

%
%

\section*{Acknowledgment}

\noindent 
The authors would like to thank the reviewers and the communicating Associate Editor 
for their great patience to carefully
read through the significantly longer first-submission version 
(which can be found 
on arXiv:2107.01693 under a minorly different title), and for their very helpful suggestions
on reordering respectively outsourcing some parts thereof; this lead to a more comfortable
readability, indeed. 
W. Stummer is grateful to the Sorbonne Universit\'{e} 
Paris for its multiple partial financial support and especially to the LPSM 
for its multiple great hospitality.
M. Broniatowski thanks very much the FAU Erlangen-N{\"u}rnberg
for its partial financial support and hospitality.
Moreover, W. Stummer would like to thank Rene Schilling
for an interesting discussion on complex-valued foundations
of the Bernstein-Widder theorem.

%
%

\ifCLASSOPTIONcaptionsoff
  \newpage
\fi



%

\newpage

%
%

\begin{center}
{\huge
A precise bare simulation approach to 
the minimization of some
distances. I. Foundations\\[0.5cm]
--- Supplementary Material ---}\\[0.7cm]
{\Large 
Michel Broniatowski\footnote{ 
LPSM, Sorbonne Universit\'{e}, 4 place Jussieu, 75252 Paris, France.
ORCID 0000-0001-6301-5531.}
and Wolfgang Stummer\footnote{Department of Mathematics, 
University of Erlangen--N\"{u}rnberg (FAU),
Cauerstrasse $11$, 91058 Erlangen, Germany; e-mail: stummer@math.fau.de.
ORCID 0000-0002-7831-4558. Corresponding author.}
}

\end{center}

%
%

\section{Proofs --- Part 3}
\label{Sect AppProofs3}

\noindent
\textbf{Proof of Lemma \ref{Lemma Indent Rate finite case_new}.}\\
From \eqref{brostu3:fo.powdiv.new} one gets straightforwardly
for arbitrary $\widetilde{c} >0$ 
\begin{eqnarray}
D_{\widetilde{c} \cdot \varphi_{\gamma}}(m \cdot \mathbf{Q},\mathds{P}) \hspace{-0.2cm} &:=& \hspace{-0.2cm}
\begin{cases}
\frac{\widetilde{c} \cdot \left(m^{\gamma} \cdot H_{\gamma} - m \cdot A \cdot \gamma + \gamma -1 \right)}{\gamma \cdot (\gamma-1)}, 
\hspace{2.65cm} \textrm{if }  \gamma \in \, ]-\infty,0[, \  
\mathds{P} \in \mathbb{S}_{\geq 0}^{K}, \ \mathbf{Q} \in A \cdot \mathbb{S}_{> 0}^{K} \ \ \textrm{and }  m >0,  \\
\widetilde{c} \cdot (- \log m + \widetilde{I} -1 + m \cdot A),
\hspace{1.50cm} \textrm{if }  \gamma = 0, \  
\mathds{P} \in \mathbb{S}_{\geq 0}^{K}, \ A \cdot \mathbf{Q} \in \mathbb{S}_{> 0}^{K} \ \ \textrm{and }  m >0,
\\ 
\frac{\widetilde{c} \cdot \left(m^{\gamma} \cdot H_{\gamma} - m \cdot A \cdot \gamma + \gamma -1 \right)}{\gamma \cdot (\gamma-1)},
\hspace{2.65cm} \textrm{if }  \gamma \in \, ]0,1[, \  
\mathds{P} \in \mathbb{S}_{\geq 0}^{K}, \ \mathbf{Q} \in A \cdot \mathbb{S}_{\geq 0}^{K} \ \ \textrm{and } m \geq 0,  \\
\widetilde{c} \cdot (A \cdot m \cdot \log m + m \cdot (I-A) +1), 
\hspace{0.5cm} \textrm{if }  \gamma = 1, \  
\mathds{P} \in \mathbb{S}_{> 0}^{K}, \ \mathbf{Q} \in A \cdot \mathbb{S}_{\geq 0}^{K} \ \ \textrm{and } m \geq 0, \\
\frac{\widetilde{c} \cdot \left(m^{\gamma} \cdot H_{\gamma} \cdot \textfrak{1}_{[0,\infty[}(m)
- m \cdot A \cdot \gamma + \gamma -1 \right)}{\gamma \cdot (\gamma-1)},
\hspace{1.4cm} \textrm{if }  \gamma \in \, ]1,2[, \  
\mathds{P} \in \mathbb{S}_{>0}^{K}, \ \mathbf{Q} \in A \cdot \mathbb{S}^{K} \ \ \textrm{and } m \in \, ]-\infty,\infty[, 
\\
\frac{\widetilde{c} \cdot \left(m^{2} \cdot H_{2} - m \cdot A \cdot 2 + 2 -1 \right)}{2 \cdot (2-1)}, 
\hspace{2.7cm}  \textrm{if }  \gamma = 2, \  
\mathds{P} \in \mathbb{S}_{>0}^{K}, \ \mathbf{Q} \in A \cdot \mathbb{S}^{K} \ \ \textrm{and } m \in \, ]-\infty,\infty[, \\
\frac{\widetilde{c} \cdot \left(m^{\gamma} \cdot H_{\gamma} \cdot \textfrak{1}_{[0,\infty[}(m)
- m \cdot A \cdot \gamma + \gamma -1 \right)}{\gamma \cdot (\gamma-1)},
\hspace{1.4cm} \textrm{if }  \gamma \in \, ]2,\infty[, \  
\mathds{P} \in \mathbb{S}_{>0}^{K}, \ \mathbf{Q} \in A \cdot \mathbb{S}^{K} \ \ \textrm{and } m \in \, ]-\infty,\infty[,
\\
\infty, \hspace{5.5cm} \textrm{else},
\end{cases}
 {\ \ }
\label{proof Lemma 11} 
\end{eqnarray}
where we have used the three $m-$independent abbreviations 
\begin{eqnarray}
H_{\gamma} := \sum\displaylimits_{k=1}^{K} (q_{k})^{\gamma} \cdot (p_{k})^{1-\gamma}
\ = \ 1 + \gamma \cdot (A-1) +
\frac{\gamma \cdot (\gamma-1)}{\widetilde{c}} \cdot D_{\widetilde{c} \cdot \varphi_{\gamma}}(\mathbf{Q}, \mathds{P}),
\qquad \text{(cf. \eqref{brostu3:fo.divpow.hellinger1})}
\nonumber
\end{eqnarray}
\begin{eqnarray}
I:= \sum\displaylimits_{k=1}^{K} q_{k} \cdot \log\left( \frac{q_{k}}{p_{k}} \right)
\ = \ \frac{1}{\widetilde{c}} \cdot D_{\widetilde{c} \cdot \varphi_{1}}(\mathbf{Q}, \mathds{P}) + A - 1,
\qquad \text{(cf. \eqref{brostu3:fo.divpow.Kull1})}
\nonumber
\end{eqnarray}
\begin{eqnarray}
\widetilde{I} := \sum\displaylimits_{k=1}^{K} p_{k} \cdot \log\left( \frac{p_{k}}{q_{k}} \right)
\ = \ \frac{1}{\widetilde{c}} \cdot D_{\widetilde{c} \cdot \varphi_{0}}(\mathbf{Q}, \mathds{P}) + 1 - A.
\qquad \text{(cf. \eqref{brostu3:fo.divpow.RevKull1})}
\nonumber
\end{eqnarray}
\noindent
To proceed, let us fix an arbitrary constant $\widetilde{c} >0$.\\
(i) Case $\gamma \cdot (1-\gamma) \ne 0$. \\
(ia) Let us start with the subcase $\gamma \in \, ]-\infty,0[$.
From the first and the last line of \eqref{proof Lemma 11}, it is clear that
the corresponding $m-$infimum can not be achieved for $m \leq 0$;
since $H_{\gamma} > 0$ one gets the unique minimizer $m_{min} = \Big(\frac{H_{\gamma}}{A}\Big)^{1/(1-\gamma)} >0$
and the minimum 
$D_{\widetilde{c} \cdot \varphi_{\gamma}}(m_{min} \cdot \mathbf{Q},\mathds{P})
=\frac{\widetilde{c}}{\gamma} \cdot (1- \frac{H^{1/(1-\gamma)}}{A^{\gamma/(1-\gamma)}} )$.
Hence, \eqref{brostu3:fo.676} is established. 
The assertions \eqref{Inf in Lemma rate case finite 1} and 
\eqref{equiv infima in lemma rate finite case 1}
follow immediately by monotonicity inspection of 
$x \rightarrow \frac{\widetilde{c}}{\gamma } \cdot \left[ 1- 
\frac{1}{A^{\gamma/(1-\gamma)}} \cdot 
\left[ 1 + \gamma \cdot (A-1) + \frac{\gamma \cdot \left( \gamma -1\right)}{\widetilde{c}} 
\cdot x
\right] ^{-1/\left( \gamma -1\right) }\right]$
for $x \geq 0$ such that $1 + \gamma \cdot (A-1) + \frac{\gamma \cdot \left( \gamma -1\right)}{\widetilde{c}} 
\cdot x \geq 0$.\\
(ib) The subcase $\gamma \in \, ]0,1[$ (cf. the third line of \eqref{proof Lemma 11})
works analogously if $H_{\gamma} >0$;
furthermore, if $H_{\gamma} =0$ --- which can only appear when 
$\mathds{P}$, $\mathbf{Q}$ have disjoint supports (singularity) --- then 
$\inf_{m>0} D_{\widetilde{c} \cdot \varphi_{\gamma}}(m \cdot \mathbf{Q},\mathds{P}) = 
\frac{\widetilde{c}}{\gamma}$ which is (the corresponding special case of) \eqref{brostu3:fo.676}.\\
(ic) In the subcase $\gamma \in \, ]1,\infty[$ (cf. the fifth, sixth and seventh
line of \eqref{proof Lemma 11}) it is straightforward to see that the desired
infimum can not be achieved for $m < 0$. Hence, one can proceed analogously
to subcase (ia).\\
(id) The assertions \eqref{Inf in Lemma rate case finite 1b}
to \eqref{equiv infima in lemma rate finite case 1c} are straightforward. \\
(ii) Case $\gamma =1$.
From the fourth line of \eqref{proof Lemma 11},  
one obtains the unique minimizer $m_{min} = \exp\{-I/A\}$
and the minimum 
$D_{\widetilde{c} \cdot \varphi_{1}}(m_{min} \cdot \mathbf{Q},\mathds{P})
= \widetilde{c} \cdot (1- A \cdot m_{min})$, which leads to \eqref{brostu3:fo.677}. 
The monotonicity of
$x \rightarrow \widetilde{c} \cdot (1- \exp\{-x/\widetilde{c}\})$ for $x\geq0$
implies immediately \eqref{brostu3:fo.mmin-KL1} and 
\eqref{brostu3:fo.mmin-KL2}; moreover, \eqref{brostu3:fo.mmin-KL1b} and
\eqref{brostu3:fo.mmin-KL2b} are immediate.\\ 
(iii) Case $\gamma =0$.
The second line of \eqref{proof Lemma 11} 
implies the unique minimizer $m_{min} = 1/A$,
the minimum 
$D_{\widetilde{c} \cdot \varphi_{0}}(m_{min} \cdot \mathbf{Q},\mathds{P})
= \widetilde{c} \cdot (\widetilde{I} + \log A)$, and hence 
\eqref{brostu3:fo.678}.
The assertions \eqref{brostu3:fo.mmin-RKL1} to 
\eqref{brostu3:fo.mmin-RKL2b} are obvious.
\hspace{0.5cm} $\blacksquare$  \\

%
%

\section{Proofs --- Part 4}

\noindent
\textbf{Proof of Proposition \ref{ID1}.} 
The assertion follows straightforwardly from the following two facts:\\
(i) a moment generating function $MGF$ is infinitely divisible if and only if 
$MGF^{c}$ is a moment generating function for all $c >0$ 
(cf. e.g. (the MGF-version of) Prop. IV.2.5 of 
\cite{Ste:04}).\\ 
(ii) $z \mapsto MGF(z)$ is a moment generating function if and only if 
$z \mapsto MGF(\breve{c} \cdot z) =: MGF_{\breve{c}}(z)$ is a moment generating function for all $\breve{c} >0$.\\
Notice that for each $c >0$, $\breve{c} >0$ one has $int(dom(MGF)) = int(dom(MGF^{c}))$ and 
$int(dom(MGF_{\breve{c}})) = \frac{1}{\breve{c}} \cdot int(dom(MGF))$, and hence the light-tailedness
remains unchanged: $0 \in int(dom(MGF))$ if and only if $0 \in int(dom(MGF^{c}))$ if and only if 
$0 \in int(dom(MGF_{\breve{c}}))$.
Since  $\varphi \in \Upsilon (]a,b[)$, we have
\begin{equation}
\varphi (t)=
\sup_{z\in ]\lambda_{-}, \lambda_{+}[}\left( z \cdot t-\log \Big(\int_{\mathbb{R}
} e^{z\cdot y }\,  d\mathbb{\bbzeta} (y) \Big) \right)
, \quad t\in \, ]a,b[ \, , \  \
\label{Phi Legendre of mgf(W) multiple1}
\end{equation}
and thus for the exponential of its Fenchel-Legendre transform 
\begin{equation}
e^{\varphi_{*}(z)}  = \int_{\mathbb{R}
} e^{z \cdot y} d\mathbb{\bbzeta} (y) 
, \qquad z \in 
\, ]\lambda_{-}, \lambda_{+}[ . 
\label{Phi Legendre of mgf(W) multiple2}
\end{equation}
Now, let $\widetilde{\varphi} := \widetilde{c} \cdot \varphi \in \Upsilon (]a,b[)$
for arbitrarily fixed $\widetilde{c} >0$.
From the application of \eqref{Phi Legendre of mgf(W)} to $\widetilde{\varphi}$ we obtain
\begin{equation}
\widetilde{\varphi} (t)=\sup_{\widetilde{z} \in ]\widetilde{\lambda}_{-}, \widetilde{\lambda}_{+}[}
\left( \widetilde{z} \cdot t-\log \Big( \int_{\mathbb{R}
}e^{\widetilde{z} \cdot \widetilde{y}} d\widetilde{\mathbb{\bbzeta}}_{\widetilde{c}} (\widetilde{y})
\Big) \right), \qquad 
t\in \, ]a,b[ \, , \ \   
\label{Phi Legendre of mgf(W) multiple3}
\end{equation}
for some unique probability distribution $\widetilde{\mathbb{\bbzeta}}_{\widetilde{c}}$ on $\mathbb{R}$.
Here, we have used
$\widetilde{\lambda}_{-} := \inf_{t \in ]a,b[} \widetilde{\varphi}^{\prime}(t) = \widetilde{c} \cdot \lambda_{-}$ and
$\widetilde{\lambda}_{+} := \sup_{t \in ]a,b[} \widetilde{\varphi}^{\prime}(t) = \widetilde{c} \cdot \lambda_{+}$.
Dividing \eqref{Phi Legendre of mgf(W) multiple3} by $\widetilde{c}$, we arrive at
\begin{eqnarray}
\varphi(t) = \frac{\widetilde{\varphi} (t)}{\widetilde{c}} 
&=&\sup_{\widetilde{z} \in ]\widetilde{c} \cdot \lambda_{-}, \widetilde{c} \cdot \lambda_{+}[}
\left( \frac{\widetilde{z}}{\widetilde{c}} \cdot t-
 \log \Big( \Big(\int_{\mathbb{R}}
 e^{\frac{\widetilde{z}}{\widetilde{c}} \cdot \widetilde{y} \cdot \widetilde{c}} d\widetilde{\mathbb{\bbzeta}}_{\widetilde{c}} (\widetilde{y})\Big)^{1/\widetilde{c}} \, \Big) \right), 
\nonumber\\
&=& 
\sup_{z \in ]\lambda_{-}, \lambda_{+}[}
\left( z \cdot t-
 \log \Big( \Big(\int_{\mathbb{R}}
 e^{z \cdot \widetilde{y} \cdot \widetilde{c}} 
 d\widetilde{\mathbb{\bbzeta}}_{\widetilde{c}} (\widetilde{y})\Big)^{1/\widetilde{c}} \, \Big) \right),
\qquad 
t\in \, ]a,b[ \, , \ \ 
\nonumber 
\end{eqnarray}
and hence for the exponential of its Fenchel-Legendre transform 
\begin{equation}
e^{\varphi_{*}(z)}  =
\Big( \int_{\mathbb{R}} e^{z \cdot \widetilde{y} \cdot \widetilde{c}} 
d\widetilde{\mathbb{\bbzeta}}_{\widetilde{c}} (\widetilde{y})  \Big)^{1/\widetilde{c}}
, \qquad 
z \in \,  ]\lambda_{-}, \lambda_{+}[ .  
\label{Phi Legendre of mgf(W) multiple4}
\end{equation}
From \eqref{Phi Legendre of mgf(W) multiple2} and \eqref{Phi Legendre of mgf(W) multiple4} 
we deduce the relation 
$\Big(MGF_{\mathbb{\bbzeta}}(z)\Big)^{\widetilde{c}} = 
MGF_{\widetilde{\mathbb{\bbzeta}}_{\widetilde{c}}}(\widetilde{c} \cdot z)$
which (with the help of (i) and (ii)) implies the infinite
divisibility of $\mathbb{\bbzeta}$.\\
For the reverse direction, let us assume that $\varphi \in \Upsilon (]a,b[)$ and that
the corresponding $\mathbb{\bbzeta}$ is infinitely divisible. Recall that 
$]a,b[ \, = int(dom(\varphi))$.
Moreover,
we fix an arbitrary constant $\widetilde{c} >0$. 
Of course, there holds $\widetilde{c} \cdot \varphi \in \widetilde{\Upsilon}(]a,b[)$
and $dom(\widetilde{c} \cdot \varphi) = dom(\varphi)$. 
Furthermore, by multiplying \eqref{Phi Legendre of mgf(W) multiple1} with $\widetilde{c} >0$
and by employing (i), (ii)  we get
\begin{eqnarray}
\widetilde{c} \cdot \varphi (t) &=&
\sup_{z\in ]\lambda_{-}, \lambda_{+}[}\left( \widetilde{c} \cdot z \cdot t-\log \Big( \Big(\int_{\mathbb{R}
} e^{\widetilde{c} \cdot z\cdot \frac{y}{\widetilde{c}}} d\mathbb{\bbzeta} (y) \Big)^{\widetilde{c}} \, \Big) \right)
= \sup_{\widetilde{z} \in ]\widetilde{c} \cdot \lambda_{-}, \widetilde{c} \cdot \lambda_{+}[}
\left( \widetilde{z} \cdot t-\log \Big(\Big(\int_{\mathbb{R}} 
e^{\frac{\widetilde{z}}{\widetilde{c}} \cdot y} \,  d\mathbb{\bbzeta} (y) \Big)^{\widetilde{c}} \, \Big) \right)
\nonumber \\
&=& 
\sup_{\widetilde{z} \in ]\widetilde{c} \cdot \lambda_{-}, \widetilde{c} \cdot \lambda_{+}[}
\left( \widetilde{z} \cdot t-\log  \Big( \int_{\mathbb{R}} 
e^{\widetilde{z} \cdot y} d\mathbb{\bbzeta}_{\widetilde{c}} (y) \Big) \right)
, \quad t\in \, ]a,b[, \ \ 
\nonumber
\end{eqnarray}
for some probability distribution $\mathbb{\bbzeta}_{\widetilde{c}}$ on $\mathbb{R}$.
\hspace{0.5cm} $\blacksquare$  \\

\vspace{0.4cm}
\noindent
\textbf{Proof of Proposition \ref{ID2}.}
It is well known that a candidate function $M: \, ]-\infty,0[ \, \mapsto \, ]0,\infty[$
is the moment-generating function
of an infinitely divisible probability distribution
if and only if $(\log M)^{\prime}$ is absolutely monotone
(see e.g. Theorem 5.11 of \cite{Schi:12}).
By applying this to $M(z) :=  e^{-a\cdot z + \varphi^{*}(z)}$ 
respectively $M(z) := e^{b\cdot z + \varphi^{*}(- z)}$, one
gets straightforwardly the assertion (a) respectively (b);
notice that 
$b=\infty$ respectively $a =- \infty$ can be deduced from the fact
that the support of an infinitely divisible distribution is
always (one-sided or two-sided) unbounded. 
For the third case $a= - \infty$, $b= \infty$ one can use the 
assertion (cf. e.g. \cite{Morr:82}, p.73)
that a candidate function $M: \, ]\lambda_{-}, \lambda_{+}[ \, \mapsto \, ]0,\infty[$
is the moment-generating function of an infinitely divisible probability distribution
if the connected function $z \mapsto (\log M)^{\prime\prime}(z)/(\log M)^{\prime\prime}(0)$ 
is the moment-generating function of some auxiliary probability distribution; 
but the latter is equivalent to exponentially convexity (cf. Widder's theorem).
By applying this to $M(z) :=  e^{\varphi^{*}(z)}$, one ends up with (c). 
\hspace{0.5cm} $\blacksquare$  \\

%
%

\section{Proofs --- Part 5}

\noindent
\textbf{Proof of Theorem \ref{brostu3:thm.Wfind.new}.} \  
(i) \ Clearly, on $]\lambda_{-},\lambda_{+}[$ the function $\Lambda$ is
differentiable with strictly increasing derivative
\begin{equation}
\label{proof Ftheorem equ 1}
\Lambda^{\prime}(z) = F^{-1}(z+c) + 1- F^{-1}(c) ,  
\qquad z \in \, ]\lambda_{-},\lambda_{+}[.
\end{equation}
Hence, $\Lambda$ is strictly convex and smooth (because of the smoothness of $F^{-1}$), 
and satisfies $\Lambda(0) =0$ as well as $\Lambda^{\prime}(0) =1$.
Also, the corresponding extensions of $\Lambda$ to $z=\lambda_{-}$ and $z=\lambda_{+}$ are continuous.\\
\noindent
(ii) \  It is straightforward to see that on $]t_{-}^{sc},t_{+}^{sc}[$ the function $\varphi$ is
differentiable with strictly increasing derivative
\begin{equation}
\label{proof Ftheorem equ 2}
\varphi^{\prime}(t) = F(t+ F^{-1}(c) - 1) - c ,  
\qquad t \in \, ]t_{-}^{sc},t_{+}^{sc}[.
\end{equation}
Hence, $\varphi$ is strictly convex and smooth (because of the smoothness of $F$), 
and satisfies $\varphi(1) =0$ as well as $\varphi^{\prime}(1) =0$.
Also, the corresponding extensions of $\varphi$ to $t=t_{-}^{sc}$ and $t=t_{-}^{sc}$ are continuous.\\
To prove that $int(dom(\varphi)) = \, ]a,b[$,
let us first notice that
obviously there holds $a \leq t_{-}^{sc}$ and $t_{+}^{sc} \leq b$.
Moreover, the validity of $\varphi(t) < \infty$ for all $t \in ]t_{-}^{sc},t_{+}^{sc}[$
is clear from \eqref{brostu3:fo.Wfind2.new} since $t+F^{-1}(c)-1 \in \, ]a_{F},b_{F}[ \, = int(dom(F))$
and the involved integral over the continuous function $F^{-1}$ is taken over a compact interval.\\
\indent
For the subcase $t_{-}^{sc} =- \infty = a$ we have thus shown $dom(\varphi) \, \cap \, ]-\infty,1] 
= \, ]-\infty,1]
= \, ]a,1]$, whereas for the subcase $t_{+}^{sc} = \infty = b$ we have verified 
$dom(\varphi) \, \cap \, [1,\infty[ \, = [1,\infty[ \, = [1,b[$.\\
\indent
Let us next examine the subcase 
\textquotedblleft  $t_{-}^{sc} > - \infty$ and $\varphi(t_{-}^{sc}) < \infty$\textquotedblright :
if $\lambda_{-} > - \infty$ then $a = -\infty$ and
\eqref{brostu3:fo.Wfind2.new} implies
$\varphi(t) = \varphi(t_{-}^{sc}) + 
\lambda_{-} \cdot (t- t_{-}^{sc}) < \infty$
for all $t \in \, ]-\infty, t_{-}^{sc}] = \, ]a,t_{-}^{sc}]$, which leads to
$dom(\varphi) \, \cap \, ]-\infty,1] = \, ]-\infty,1] = \, ]a,1]$;
in contrast, if $\lambda_{-} = - \infty$ then $a = t_{-}^{sc}$ and
\eqref{brostu3:fo.Wfind2.new} implies
$\varphi(t) = \varphi(t_{-}^{sc}) + 
\lambda_{-} \cdot (t- t_{-}^{sc}) = \infty$
for all $t \in \, ]-\infty, t_{-}^{sc}[ \, = \, ]-\infty,a[$, which leads to
$dom(\varphi) \, \cap \, ]-\infty,1] = [a,1]$.\\
\indent
In the subcase 
\textquotedblleft  $t_{-}^{sc} > - \infty$ and $\varphi(t_{-}^{sc}) = \infty$\textquotedblright, 
due to the strict convexity of $\varphi$ one always has 
$\lim_{t \downarrow t_{-}^{sc}} \varphi^{\prime}(t) = - \infty$; this implies, 
by $\lim_{t \downarrow t_{-}^{sc}} \varphi^{\prime}(t) = \lambda_{-}$,
that $\lambda_{-} = - \infty$ and thus $a = t_{-}^{sc}$; from \eqref{brostu3:fo.Wfind2.new}
we derive $\varphi(t) = \varphi(t_{-}^{sc}) + 
\lambda_{-} \cdot (t- t_{-}^{sc}) = \infty$
for all $t \in \, ]-\infty, t_{-}^{sc}[ \, = \, ]-\infty,a[$, which leads to
$dom(\varphi) \, \cap \, ]-\infty,1] = \, ]a,1]$.\\
\indent
As a further step, we deal with the subcase 
\textquotedblleft  $t_{+}^{sc} < \infty$ and $\varphi(t_{+}^{sc}) < \infty$\textquotedblright :
if $\lambda_{+} < \infty$ then $b = \infty$ and
\eqref{brostu3:fo.Wfind2.new} implies
$\varphi(t) = \varphi(t_{+}^{sc}) + 
\lambda_{+} \cdot (t- t_{+}^{sc}) < \infty$
for all $t \in  [t_{+}^{sc},\infty[ \, = [t_{+}^{sc},b[$, which leads to
$dom(\varphi) \, \cap \, [1, \infty[ \, = [1, \infty[ \, = [1, b[$;
in contrast, if $\lambda_{+} = \infty$ then $b = t_{+}^{sc}$ and
\eqref{brostu3:fo.Wfind2.new} implies
$\varphi(t) = \varphi(t_{+}^{sc}) + 
\lambda_{+} \cdot (t- t_{+}^{sc}) = \infty$
for all $t \in \, ]t_{-}^{sc}, \infty[ \, = \, ]b,\infty[$, which leads to
$dom(\varphi) \cap [1,\infty[ \, = [1,b]$.\\
\indent
In the subcase 
\textquotedblleft  $t_{+}^{sc} < + \infty$ and $\varphi(t_{+}^{sc}) = \infty$\textquotedblright, 
due to the strict convexity of $\varphi$ one always gets 
$\lim_{t \uparrow t_{+}^{sc}} \varphi^{\prime}(t) = \infty$; this implies, 
by $\lim_{t \uparrow t_{+}^{sc}} \varphi^{\prime}(t) = \lambda_{+}$,
that $\lambda_{+} = \infty$ and thus $b = t_{+}^{sc}$; from \eqref{brostu3:fo.Wfind2.new}
we deduce $\varphi(t) = \varphi(t_{+}^{sc}) + 
\lambda_{+} \cdot (t- t_{+}^{sc}) = \infty$
for all $t \in \, ]t_{+}^{sc}, \infty[ \, = \, ]b, \infty[$, which leads to
$dom(\varphi) \, \cap \, [1,\infty[ \, = [1,b[$.

Putting things together, we have proved that $int(dom(\varphi)) = \, ]a,b[$.\\
\noindent
(iii) \  From \eqref{proof Ftheorem equ 1} and \eqref{proof Ftheorem equ 2}  
one gets easily
\begin{equation}
\Lambda^{\prime -1}(t) = F\left(t+F^{-1}(c)-1 \right) - c =  \varphi^{\prime}(t), \qquad t \in \, ]t_{-}^{sc},t_{+}^{sc}[,
\label{proof Ftheorem equ 3}
\end{equation} 
as well as $\Lambda^{\prime -1}(1)=0$. From this, we derive
\begin{eqnarray}
&& t \cdot \Lambda^{\prime -1}(t) -  \Lambda\left(\Lambda^{\prime -1}(t)\right)
\nonumber
\\
&& = t \cdot [ F\left(t+F^{-1}(c)-1 \right) - c \, ] 
+ [ F^{-1}(c)-1 \, ] \cdot [ F\left(t+F^{-1}(c)-1 \right) - c \, ]
\nonumber\\
& & -\int\displaylimits_{0}^{F\left(t+F^{-1}(c)-1 \right) - c} F^{-1}(u+c) \, du 
\nonumber \\
& & = \varphi(t), \qquad t \in \, ]t_{-}^{sc},t_{+}^{sc}[, 
\nonumber
\end{eqnarray}
and hence, 
\begin{equation}
\varphi(t) = \max_{z \in ]\lambda_{-},\lambda_{+}[} \left( z\cdot t - 
\Lambda(z)\right), \qquad t \in \, ]t_{-}^{sc},t_{+}^{sc}[ ,
\nonumber
\end{equation} 
i.e. on $]t_{-}^{sc},t_{+}^{sc}[$ the divergence generator $\varphi$ 
is the classical Legendre transform of the restriction of $\Lambda$ to $]\lambda_{-},\lambda_{+}[$.
If 
\textquotedblleft  $\lambda_{-} > -\infty$, $\Lambda(\lambda_{-}) \in \, ]-\infty,\infty[$
and $\Lambda^{\prime}(\lambda_{-}) \in \, ]-\infty,\infty[$\textquotedblright\ 
respectively 
\textquotedblleft  $\lambda_{+} < -\infty$, $\Lambda(\lambda_{+}) \in \, ]-\infty,\infty[$
and $\Lambda^{\prime}(\lambda_{+}) \in \, ]-\infty,\infty[$\textquotedblright,
then one can apply classical facts of Fenchel-Legendre transformation 
to get the corresponding left-hand respectively right-hand linear extensions of $\varphi$ on the complement
of $]t_{-}^{sc},t_{+}^{sc}[$, in order to obtain the desired
\begin{equation}
\varphi (t) = 
\sup_{z \in ]-\infty,\infty[} \left( z\cdot t - \Lambda(z)\right) ,
\qquad t \in \mathbb{R};
\nonumber
\end{equation} 
notice that
$t_{-}^{sc} =  \lim_{z \downarrow \lambda_{-}} \Lambda^{\prime}(z)$ and
$t_{+}^{sc} =  \lim_{z \uparrow \lambda_{+}} \Lambda^{\prime}(z)$.
\\
\noindent
(iv) \ This is just the reverse of (iii),
by applying standard Fenchel-Legendre-transformation theory.
\hspace{0.5cm} $\blacksquare$

%
%

\section{Further details and proofs for 
Subsection \ref{Subsect Estimators determ}
}

\noindent

\vspace{0.2cm}
\noindent
\textbf{Proof of Lemma \protect\ref{Lemma minim rate points}}. \ 
By Assumption (OM), one gets for all $\boldsymbol{\lambda} \in
cl(\mathbf{\Lambda})$ that 
$\{\mathbf{x} \in (dom(\widetilde{\varphi})^{n} : T(\mathbf{x}) = \boldsymbol{\lambda}\}
\, \cap \, ]t_{-}^{sc},t_{+}^{sc}[^{n} \ne \emptyset$. 
Moreover, for any $\mathbf{x}=\left( x_{1},..,x_{n}\right) $ in 
$\mathbb{R}^{n}$, by the independence of the components of $\mathbf{\widetilde{W}}$ 
we have
\begin{eqnarray}
& & I_{\mathbf{\widetilde{W}}}(\mathbf{x})
=\sup_{\mathbf{z}=\left( z_{1},\ldots,z_{n}\right) \in \mathbb{R}^{n}}
\left( \left\langle \mathbf{x},\mathbf{z}\right\rangle
-\sum_{i=1}^{n}\Lambda_{\widetilde{\mathbb{\bbzeta}}}(z_{i}) \right)
= \sup_{\mathbf{z}
\in ]\lambda_{-},\lambda_{+}[^{n}}
\left( \sum_{i=1}^{n} \left( x_{i} \cdot z_{i}
-\Lambda_{\widetilde{\mathbb{\bbzeta}}}(z_{i}) \right) \right)
\nonumber \\
& & 
= \sum_{i=1}^{n} \left( \sup_{z_{i}\in ]\lambda_{-},\lambda_{+}[}\left( x_{i} \cdot  z_{i} - \Lambda_{\widetilde{\mathbb{\bbzeta}}}(z_{i})\right) \right)
= \sum_{i=1}^{n} \widetilde{\varphi}(x_{i}) 
=\sum_{k=1}^{K}\sum_{i\in I_{k}^{(n)}} \widetilde{\varphi}(x_{i})
\nonumber 
\end{eqnarray}
which is finite if and only if $\mathbf{x} \in (dom(\widetilde{\varphi}))^{n}$
(recall that $\widetilde{\varphi}$ is a nonnegative function).
Hence, for each $\boldsymbol{\lambda} \in \mathbf{\Lambda}$
we obtain
\begin{eqnarray}
 I(\boldsymbol{\lambda}) &:=& 
\inf_{\mathbf{x} \in \mathbb{R}^{n}: \, T(\mathbf{x})=\boldsymbol{\lambda}} 
I_{\mathbf{\widetilde{W}}}(\mathbf{x})
= \inf_{\mathbf{x} \in (dom(\widetilde{\varphi}))^{n}: \, T(\mathbf{x})=\boldsymbol{\lambda}} 
I_{\mathbf{\widetilde{W}}}(\mathbf{x})
=
\inf_{\mathbf{x} \in (dom(\widetilde{\varphi}))^{n}: \, T(\mathbf{x})=\boldsymbol{\lambda}} \  
\sum_{k=1}^{K}\sum_{i\in I_{k}^{(n)}} \widetilde{\varphi}(x_{i})
\label{Imin1} 
\\
&=& \sum_{k=1}^{K} n_{k} \cdot \widetilde{\varphi}(\lambda_{k})
\ = \ n \cdot \sum_{k=1}^{K} \widetilde{p}_{k} \cdot \widetilde{\varphi}(\lambda_{k})
= \inf_{\mathbf{x} \in ]t_{-}^{sc},t_{+}^{sc}[^{n} \, : \, T(\mathbf{x})=\boldsymbol{\lambda}} 
I_{\mathbf{\widetilde{W}}}(\mathbf{x}) \, ;
\label{Imin2}
\end{eqnarray}
here, we have employed the following facts:
(i) the right-most infimum in \eqref{Imin1} is achieved by minimizing each of the $K$
terms $\sum_{i\in I_{k}^{(n)}}\widetilde{\varphi}(x_{i})$ under the linear constraint 
$\frac{1}{n_{k}} \cdot \sum_{i\in I_{k}^{(n)}} x_{i}= \lambda_{k}$, 
and by the strict convexity of $\widetilde{\varphi}$ on $]t_{-}^{sc},t_{+}^{sc}[$ 
the minimum of this generic term is attained when
all components $x_{i}$ are equal to $\lambda_{k}$, and 
(ii) the outcoming
minimum does not depend on the particular (generally non-unique) choice of the
$x_{i}$\textquoteright s.
Notice that we have used the relation $n_{k} = n \cdot \widetilde{p}_{k}$ as well.
To proceed, let $\underline{\boldsymbol{\lambda}}$ be a minimal rate point 
(mrp) of $\mathbf{\Lambda}$,
which means that $\underline{\boldsymbol{\lambda}} \in \partial \mathbf{\Lambda}$
and $I(\underline{\boldsymbol{\lambda}}) \leq I(\boldsymbol{\lambda})$ for all
$\boldsymbol{\lambda} \in \mathbf{\Lambda}$.
By Assumption (OM)
one can run all the steps in \eqref{Imin1} and \eqref{Imin2} with 
$\underline{\boldsymbol{\lambda}}$ instead of $\boldsymbol{\lambda}$, and hence
\begin{equation}
I(\underline{\boldsymbol{\lambda}}) \ = \  
\inf_{\mathbf{x} \in \mathbb{R}^{n}: \, T(\mathbf{x})=\underline{\boldsymbol{\lambda}}} 
I_{\mathbf{\widetilde{W}}}(\mathbf{x})
\ = \inf_{\mathbf{x} \in ]t_{-}^{sc},t_{+}^{sc}[^{n} \, : \, T(\mathbf{x})=\underline{\boldsymbol{\lambda}}} 
I_{\mathbf{\widetilde{W}}}(\mathbf{x})
\ = \ n \cdot \sum_{k=1}^{K} \widetilde{p}_{k} \cdot \widetilde{\varphi}(\underline{\lambda}_{k})
\ = \ n \cdot \sum_{k=1}^{K} \widetilde{p}_{k} \cdot \widetilde{\varphi}(\underline{\widetilde{q}}_{k}/\widetilde{p}_{k})
\nonumber 
\end{equation}
where for the last equality we have employed the vector
$\underline{\mathbf{\widetilde{Q}}} = \underline{\boldsymbol{\lambda}} \cdot \mathfrak{D}^{-1}
\in \partial \widetilde{\mathbf{\Omega}}$ 
which we have called the 
\textquotedblleft  
dominating point of $\widetilde{\mathbf{\Omega}}$\textquotedblright . 
Also we have proved 
\begin{equation}
I(\underline{\boldsymbol{\lambda}})
\ = \ n \cdot \inf_{\mathbf{\widetilde{Q}} \in \widetilde{\mathbf{\Omega}} 
}\sum_{k=1}^{K} \widetilde{p}_{k} \cdot \widetilde{\varphi}(
\widetilde{q}_{k}/\widetilde{p}_{k}).  
\nonumber
\end{equation}
The existence of a mrp $\underline{\boldsymbol{\lambda}}$ of $\mathbf{\Lambda}$ 
(or equivalently, of a mrp $\underline{\mathbf{\widetilde{Q}}}$ of $\widetilde{\mathbf{\Omega}}$) 
is a straightforward
consequence of the continuity of $D_{\widetilde{\varphi}}( \cdot , \widetilde{\mathds{P}})$,
the strict increasingness of 
$[0,1] \ni t \mapsto  
D_{\widetilde{\varphi}}((1-t) \cdot \widetilde{\mathds{P}} + t \cdot \widetilde{\mathbf{Q}}, \widetilde{\mathds{P}})$
for $\widetilde{\mathbf{Q}} \in int(\widetilde{\mathbf{\Omega}})$, 
and $\partial \widetilde{\mathbf{\Omega}} \ni \underline{\mathbf{\widetilde{Q}}} 
\cdot \mathfrak{D} = \underline{\boldsymbol{\lambda}} \in \partial
\mathbf{\Lambda}$.
$\blacksquare$

\vspace{0.4cm}

\noindent
\textbf{On the obtainment of proxies of minimal rate points by proxy method 2:}\\
For the rest of this section, besides (OM) we assume that 
$dom(\widetilde{\varphi})  = \, ]a,b[ \, = \, ]t_{-}^{sc},t_{+}^{sc}[$,
and that 
in case of $a=-\infty$ or $b=+\infty$ 
the divergence generator $\widetilde{\varphi}$ is regularly varying at $a$ or $b$ accordingly,
with positive index $\beta $, i.e. 
(with a slight abuse of notation)

\begin{itemize}
\item if $a=-\infty $, then for all $\lambda >0$ there holds
\[
\lim_{u\rightarrow -\infty }\frac{\widetilde{\varphi} \left( \lambda \cdot u\right) }{\widetilde{\varphi}
\left( u\right) }=\lambda ^{\beta }, 
\]

\item if $b=+\infty $, then for all $\lambda >0$ there holds
\[
\lim_{u\rightarrow +\infty }\frac{\widetilde{\varphi} \left( \lambda \cdot u\right) }{\widetilde{\varphi}
\left( u\right) }=\lambda ^{\beta }; 
\]
\end{itemize}
this assumption is denoted by (H$\widetilde{\varphi}$).\\
A proxy of $\underline{\mathbf{\widetilde{Q}}}$ can be obtained by sampling from a
distribution on $\mathbb{R}^{K}$ defined through  
\begin{equation}
f(\mathbf{\widetilde{Q}}):=C \cdot \exp\left( -\sum_{k=1}^{K} \widetilde{p}_{k} \cdot 
\widetilde{\varphi}(\widetilde{q}_{k}/\widetilde{p}_{k})\right) \ = \ 
C \cdot \exp\left( -D_{\widetilde{\varphi}}\left( \mathbf{\widetilde{Q}},
\widetilde{\mathds{P}}\right) 
\right)
\hspace{1.8cm} \textrm{(cf. \eqref{simul distr})}
\nonumber
\end{equation} 
where $C$ is a normalizing constant; strict convexity 
of $\widetilde{\varphi}$ together
with (H$\widetilde{\varphi}$) prove that $f$ is a well-defined (Lebesgue-) density for a 
random variable  $\mathbf{T}$ on $\mathbb{R}^{K}$. We denote by 
$\mathbb{F}(\cdot) := \mathbb{\Pi}[\mathbf{T} \in \cdot\, ]$
the corresponding distribution on $\mathbb{R}^{K}$ having density $f$.
The distribution of $\mathbf{T}$ 
given $\left( \mathbf{T} \in \widetilde{\mathbf{\Omega}} \right)$
concentrates on the points in 
$\widetilde{\mathbf{\Omega}}$ which minimize 
$D_{\widetilde{\varphi}}\left(\mathbf{\widetilde{Q}},\widetilde{\mathds{P}}\right) $ 
as $\mathbf{\widetilde{Q}}$ runs in $\widetilde{\mathbf{\Omega}}$, when 
$D_{\widetilde{\varphi}}(\widetilde{\mathbf{\Omega}},\widetilde{\mathds{P}})$
is large. This can be argued as follows.
We will consider the case when $\widetilde{\mathbf{\Omega}}$ is a compact subset in 
$\mathbb{R}_{> 0}^{K}$ and $\widetilde{\varphi}$ satisfies (H$\widetilde{\varphi}$)
with $b=+\infty$.
For the case when $\widetilde{\mathbf{\Omega}}$ is not compact, or belongs to 
$\mathbb{R}^{K}/\left\{ \mathbf{0}\right\}$, 
see the Remark \ref{Remark minim not standard bounds assumtions, reg var, etc}
hereunder. Consider a compact set $\mathbf{\Gamma}$ in $\widetilde{\mathbf{\Omega}}$ 
and let $\mathbf{\Gamma}_{t}$ be defined as deduced
from $\mathbf{\Gamma} $ in a way that makes $D_{\widetilde{\varphi}}\left( \mathbf{\Gamma} _{t},\widetilde{\mathds{P}}\right) $
increase with $t$ for sufficiently large $t$.
For instance, set 
\begin{equation}
\mathbf{\Gamma}_{t}:=t \cdot \mathbf{\Gamma} .  
\label{form of asymptotics for Gamma_t}
\end{equation}
Hence, in case of $b=+\infty$ the divergence
\[
D_{\widetilde{\varphi}}\left( \mathbf{\Gamma}_{t},\widetilde{\mathds{P}}\right) =
\inf_{\boldsymbol{g}_{t}\in \mathbf{\Gamma}_{t}}\sum_{k=1}^{K} \widetilde{p}_{k}
\cdot \widetilde{\varphi}\left( \frac{\left( 
\boldsymbol{g}_{t}\right)_{k}}{\widetilde{p}_{k}}
\right) =\inf_{\boldsymbol{g} \in \mathbf{\Gamma} }\sum_{k=1}^{K} \widetilde{p}_{k}
\cdot \widetilde{\varphi}\left( \frac{t \cdot g_{k}
}{\widetilde{p}_{k}}\right) 
\]
tends to infinity as $t \rightarrow \infty$; 
the case $a=-\infty $ works analogously with 
$t\rightarrow -\infty$. In case of $b < \infty$ we may consider 
\begin{equation}
\mathbf{\Gamma} _{t}:=\left\{ b-\boldsymbol{g} /t;\boldsymbol{g} \in \mathbf{\Gamma} \right\}
\label{form of asymptotics for Gamma_t bis}
\end{equation}
and indeed $D_{\widetilde{\varphi}}\left( \mathbf{\Gamma} _{t},\widetilde{\mathds{P}}\right) \rightarrow \infty $ as $
t\rightarrow \infty $, with a similar statement when $a>-\infty$.

\vspace{0.2cm}
\noindent
Assume that $\mathbf{\Gamma}$ has a dominating point $\underline{\boldsymbol{g}}$. 
Then $\mathbf{\Gamma}_{t}$ has dominating point
$\underline{\boldsymbol{g}}_{t} := t \cdot \underline{\boldsymbol{g}}$. 
We prove that $\mathbf{T}$ with distribution \eqref{simul distr}
cannot be too far
away (depending on $t$) from $\underline{\boldsymbol{g}}_{t}$ whenever $\mathbf{T}$ belongs
to $\mathbf{\Gamma} _{t}.$ This argument is valid in the present description of some
asymptotics which makes $\mathbf{\Gamma} _{t}$ as a model for $\widetilde{\mathbf{\Omega}}$ for large $t$; considering the case when 
$D_{\widetilde{\varphi} }\left( \widetilde{\mathbf{\Omega}} ,\widetilde{\mathds{P}}\right) $ 
is large is captured through the asymptotic statement 
\[
\lim_{t\rightarrow \infty }D_{\widetilde{\varphi} }\left( \mathbf{\Gamma} _{t},\widetilde{\mathds{P}}\right) =+\infty
. 
\]

\vspace{0.3cm}
\noindent
There holds the following

\vspace{0.3cm}

\begin{proposition}
\label{Prop simul mrp}
With the above notation and under condition (H$\widetilde{\varphi}$), 
denote by $\mathbf{B}$ a neighborhood of $\underline{\boldsymbol{g}}$ and $\mathbf{B}_{t}:=t \cdot \mathbf{B}$.
Then 
\[
\mathbb{F}[\mathbf{\Gamma}_{t}\cap \mathbf{B}_{t}^{c} \, \vert \, \mathbf{\Gamma}_{t} \, ] 
= \mathbb{\Pi }\left[ \left. \mathbf{T}\in \mathbf{\Gamma}_{t}\cap \mathbf{B}_{t}^{c}
\, \right\vert \, 
\mathbf{T}\in \mathbf{\Gamma}_{t}\right] \rightarrow 0 
\]
as $t\rightarrow \infty $, which proves that simulations under \eqref{simul distr}
produce proxies of the dominating points $\underline{\boldsymbol{g}}_{t}$
in $\mathbf{\Gamma}_{t}$.
\end{proposition}

\vspace{0.3cm}
\noindent
Before we start with the proof of Proposition \ref{Prop simul mrp},
we first quote the following

\vspace{0.3cm}

\begin{lemma}
\label{Lemma lilmit log integral =rate}
Let $\widetilde{\varphi} $ satisfy (H$\widetilde{\varphi}$)
with $b=+\infty $. \ Then for all $\mathbf{A}$ in $\mathbb{R}^{K}$ such that 
\[
\breve{\alpha} :=D_{\widetilde{\varphi}}(\mathbf{A},\widetilde{\mathds{P}}):=\inf_{\mathbf{v}\in \mathbf{A}}
\sum_{k=1}^{K} \widetilde{p}_{k} \cdot \widetilde{\varphi} \left( 
\frac{v_{k}}{\widetilde{p}_{k}}\right) 
\]
is finite there holds 
\[
\lim_{t\rightarrow \infty }\frac{1}{t} \cdot \log \int_{\mathbf{A}}
\exp\left(- \, t \cdot \sum_{k=1}^{K} \widetilde{p}_{k} \cdot \widetilde{\varphi}
\left( \frac{v_{k}}{\widetilde{p}_{k}}\right) \right) \,  dv_{1} \ldots dv_{k}
= - D_{\widetilde{\varphi}}\left(\mathbf{A},\widetilde{\mathds{P}}\right) . 
\]
\end{lemma}

\noindent
\textbf{Proof of Lemma \ref{Lemma lilmit log integral =rate}.}
Let us first remark that according to the geometry of the set $\mathbf{A}$, various
combinations for the asymptotics with \eqref{form of asymptotics for Gamma_t} or 
\eqref{form of asymptotics for Gamma_t bis} may occur; for sake of brevity, we
only handle the simplest ones, since all turn to be amenable through the
same arguments. Denote for positive $r$
\[
\mathbf{B}(r):=\left\{ \mathbf{v}\in \mathbb{R}^{K}:\sum_{k=1}^{K} \widetilde{p}_{k} \cdot \widetilde{\varphi} \left( \frac{
v_{k}}{\widetilde{p}_{k}}\right) >r\right\} . 
\]
It holds, by making the change of variable $r=t\cdot \breve{\alpha} +t \cdot s$,
\begin{eqnarray*}
& & \int_{\mathbf{A}}\exp\left( - t \cdot \sum_{k=1}^{K} \widetilde{p}_{k} \cdot 
\widetilde{\varphi} \left( \frac{v_{k}}{\widetilde{p}_{k}}\right) \right) \, 
dv_{1} \ldots dv_{k} 
=
\int \cdots \int 1_{\mathbb{R}^{+}}(r)
\cdot 1_{\mathbf{A}}(\mathbf{v})
\cdot 1_{\left]
t \cdot \sum_{k=1}^{K} \widetilde{p}_{k} \cdot \widetilde{\varphi} \left( \frac{v_{k}}{\widetilde{p}_{k}}\right) ,\infty \right[}(r)
\cdot e^{-r} \, dr \, dv_{1}\ldots dv_{K} \\
& & = t \cdot e^{-t \cdot \breve{\alpha} }
\int \cdots \int 
1_{]- \breve{\alpha},\infty[}(s) \cdot 
1_{\mathbf{A}}(\mathbf{v}) \cdot 1_{\mathbf{B}^{c}(\breve{\alpha}+s)}(\mathbf{v})
\cdot e^{-t \cdot s} \, ds \, dv_{1}\ldots dv_{K}
\ = \  
t \cdot e^{-t \cdot \breve{\alpha} }\int_{-\breve{\alpha}}^{\infty }
Vol\left( \mathbf{A}\cap \mathbf{B}^{c}(\breve{\alpha}+s)\right) \cdot 
e^{-t \cdot s} \, ds.
\end{eqnarray*}
Let $I_{t}:=t \cdot \int_{0}^{\infty }Vol\left( 
\mathbf{A}\cap \mathbf{B}^{c}(\breve{\alpha} +s)\right) \cdot e^{-t \cdot s}ds$. 
We prove that 
\begin{equation}
\lim_{t\rightarrow \infty }\frac{1}{t} \cdot \log I_{t}=0.  \label{claim Lemma I_t}
\end{equation}
When $a=-\infty$ or $b=+\infty$, since $\widetilde{\varphi}$ satisfies (H$\widetilde{\varphi}$) 
there exists a polynomial $P$ such that 
\[
Vol\left( \mathbf{A}\cap \mathbf{B}^{c}(\breve{\alpha} +s)\right) \leq P(s) \, ; 
\]
whence, assuming without loss of generality
that $dom(\widetilde{\varphi}) =\mathbb{R}^{+}$, we obtain
\[
\frac{1}{t} \cdot \log I_{t}\leq \frac{1}{t} \cdot \log \int_{0}^{\infty }P\left( \frac{u}{
t}\right) \cdot t \cdot e^{-u}du 
\]
which yields that for large $t$ 
\[
\frac{1}{t} \cdot \log I_{t}<0. 
\]
When dealing with a context where $a$ or $b$ has finite value 
and the corresponding sets $\mathbf{\Gamma}_{t}$ 
are \textquotedblleft  far away\textquotedblright\
from $\mathbf{\Gamma} $ in terms of 
the distance measure $D_{\widetilde{\varphi}}\left( \cdot ,\widetilde{\mathds{P}}\right)$,
then $Vol\left( \mathbf{A}\cap \mathbf{B}^{c}(\breve{\alpha} +s)\right) $ is bounded.\
Hence, $\lim \sup_{t\rightarrow \infty }\frac{1}{t} \cdot \log I_{t}\leq 0$. Now fix 
$\varepsilon >0.$ Then, since $Vol\left( \mathbf{A}\cap \mathbf{B}^{c}(a+s)\right) $ is
increasing in $s$, we get 
\begin{eqnarray*}
I_{t} &\geq & t \cdot \int_{\varepsilon }^{\infty }Vol\left( \mathbf{A}\cap \mathbf{B}^{c}(\breve{\alpha}
+s)\right) \cdot e^{-t \cdot s}ds \\
&\geq &Vol\left( \mathbf{A}\cap \mathbf{B}^{c}(\breve{\alpha} +\varepsilon )\right) 
\cdot e^{-t \cdot \varepsilon }.
\end{eqnarray*}
Hence 
\[
\frac{1}{t} \cdot \log I_{t}\geq \frac{1}{t}
\cdot \log Vol\left( \mathbf{A}\cap \mathbf{B}^{c}(\breve{\alpha}
+\varepsilon )\right) -\varepsilon 
\]
which yields $\lim \inf_{t\rightarrow \infty }\frac{1}{t} \cdot \log I_{t}\geq 0.$
Therefore (\ref{claim Lemma I_t}) holds, which concludes the proof.
\qquad $\blacksquare$

\vspace{0.3cm}
\noindent
We now turn to the 

\vspace{0.3cm}
\noindent
\textbf{Proof of Proposition \ref{Prop simul mrp}.}\ 
Without loss of generality, let $b=+\infty$, $\mathbf{\Gamma}_{t}$ as in 
\eqref{form of asymptotics for Gamma_t} and Condition (H$\widetilde{\varphi}$) hold.
Moreover, consider an arbitrary neighborhood $\mathbf{B}$ of $\underline{\boldsymbol{g}}$ 
and the corresponding neighborhoods $\mathbf{B}_{t} : =t \cdot \mathbf{B}$
of $\underline{\boldsymbol{g}}_{t} = t \cdot \underline{\boldsymbol{g}}$. There holds
\begin{eqnarray*}
\frac{1}{\widetilde{\varphi}(t)} \cdot \log \mathbb{\Pi}\left[ \mathbf{T}\in \mathbf{\Gamma}_{t}\right] &=&
\frac{C}{
\widetilde{\varphi}(t)} \cdot \log \int_{\mathbf{\Gamma} _{t}}\exp\left(-\sum_{k=1}^{K} 
\widetilde{p}_{k} \cdot \widetilde{\varphi}\left(\frac{w_{k}}{\widetilde{p}_{k}}\right) 
\right) \, dw_{1}\ldots dw_{K} \\
&\stackrel{(1)}{=}&\frac{C \cdot K}{\widetilde{\varphi}(t)} \cdot 
\log t+\frac{C}{\widetilde{\varphi}(t)} \cdot \log \int_{\mathbf{\Gamma}
}\exp \left(-t^{\beta } \cdot \sum_{k=1}^{K} 
\widetilde{p}_{k} \cdot \left( \widetilde{\varphi}\left( \frac{v_{k}}{\widetilde{p}_{k}}\right) \cdot
(1+o(1))\right)  \right) \,  dv_{1} \ldots dv_{K} \\
&\stackrel{(2)}{=}& \frac{C \cdot K}{\widetilde{\varphi}(t)} \cdot 
\log t+\frac{C}{\left( \widetilde{\varphi}(t)/t^{\beta
}\right) } \cdot \frac{1}{t^{\beta }} \cdot 
\log \left( (1+o(1)) \cdot \int_{\mathbf{\Gamma} }\exp\left( -t^{\beta} \cdot \sum_{k=1}^{K}
\widetilde{p}_{k} \cdot \widetilde{\varphi}\left( \frac{v_{k}}{\widetilde{p}_{k}}\right) \right) 
dv_{1} \ldots dv_{K} \right)
\\
&\stackrel{(3)}{=}& -\frac{C \cdot t^{\beta }}{\widetilde{\varphi}(t)}
\cdot D_{\widetilde{\varphi}}(\mathbf{\Gamma} ,\widetilde{\mathds{P}}) \cdot (1+o(1)) \\
&\stackrel{(4)}{=}& - \breve{\ell}(t) \cdot 
D_{\widetilde{\varphi}}(\mathbf{\Gamma} ,\widetilde{\mathds{P}}) \cdot (1+o(1))
\end{eqnarray*}
as $t$ tends to infinity. In the above display, $(1)$ follows from 
$\widetilde{\varphi}(t \cdot x)=\left( t \cdot x\right)^{\beta} \cdot \ell(t \cdot x)=
t^{\beta } \cdot x^{\beta} \cdot \ell(x) \cdot \frac{\ell(t \cdot x)}{\ell(x)} =
t^{\beta } \cdot \widetilde{\varphi} (x) \cdot \left( 1+o(1)\right) $ as $t$ tends
to infinity and $x$ lies in a compact subset of $]0,\infty[$,
where $\ell$ is a slowly varying function. The equality $(2)$ follows from compactness
of $\mathbf{\Gamma}$ together with the fact that $\widetilde{\varphi}$ is a regularly varying
function with index $\beta$, so that 
\[
\lim_{t\rightarrow \infty }\frac{\widetilde{\varphi}(t\cdot v)}{\widetilde{\varphi}(t)}=v^{\beta} 
\]
uniformly upon $v$ on compact sets in $]0,\infty[$.
The remaining equalities $(3)$ and $(4)$ follow from classical properties of 
regularly varying functions, where 
$\breve{\ell} := 1/\ell$ is a slowly varying function at infinity,
together with standard Laplace-Integral approximation. 

\vspace{0.3cm}
\noindent
In the same way we can show
\[
\frac{1}{\widetilde{\varphi}(t)} \cdot \log \mathbb{\Pi} \negthinspace 
\left[ \, \mathbf{T}\in \mathbf{\Gamma} _{t}\cap 
\mathbf{B}_{t}^{c} \, \right]
=- \breve{\ell}(t) \cdot D_{\widetilde{\varphi}}(\mathbf{\Gamma} 
\cap \mathbf{B}^{c},\widetilde{\mathds{P}}) \cdot (1+o(1)) 
\]
as $t$ tends to infinity.
Since $\mathbf{B}$ is a neighborhood of the unique 
dominating point $\underline{\boldsymbol{g}}$
of $\mathbf{\Gamma}$, one gets that 
$D_{\widetilde{\varphi}}(\mathbf{\Gamma} 
\cap \mathbf{B}^{c},\widetilde{\mathds{P}}) > D_{\widetilde{\varphi}}(\mathbf{\Gamma} ,\widetilde{\mathds{P}})$.
This implies that 
\[
\mathbb{\Pi} \negthinspace \left[ \, \mathbf{T}\in \mathbf{\Gamma}_{t}\cap \mathbf{B}_{t}^{c} \, \big\vert \, \mathbf{T}\in \mathbf{\Gamma}_{t} \, \right] \rightarrow 0 
\qquad \textrm{as $t\rightarrow \infty $. \qquad  $\blacksquare$}
\]

\vspace{0.3cm}

\begin{remark}
\label{Remark minim not standard bounds assumtions, reg var, etc}
Firstly, let
us quote that the case when $\widetilde{\mathbf{\Omega}}$ is an unbounded subset in 
$\mathbb{R}^{K}\backslash\{ \mathbf{0}\} $ is somewhat immaterial for applications.\
Anyhow, if compactness of $\mathbf{\Gamma} $ is lost, then in order to use the same
line of arguments as above, it is necessary to strengthen the assumptions (H$\widetilde{\varphi}$)
e.g. as follows: 
when $b=+\infty$ then $\widetilde{\varphi}$ has to be
asymptotically homogeneous with degree $\beta >0$, in the sense that
$\widetilde{\varphi}(t \cdot x)=t^{\beta } \cdot \widetilde{\varphi}(x)\cdot (1+o(1))$ as $t\rightarrow \infty $; 
for the subcase $a=-\infty$ one employs an analogous
assumption as $t\rightarrow -\infty$. 
The case when $\widetilde{\mathbf{\Omega}}$ is a compact set in 
$\mathbb{R}^{K}\backslash\{ \mathbf{0}\} $ can be treated as above, by combining the
asymptotics in $t$ in the neighborhood of $a$ and $b$ accordingly.
\end{remark}

%
%

\section{Proof for Subsection \ref{Subsect Estimators stoch}
}

\noindent
\textbf{Proof of Proposition \protect\ref{Proposition S}}.
Recall the weighted empirical measure
\[
\boldsymbol{\xi}_{n,\mathbf{X}}^{\mathbf{V}}
:=\left( \frac{1}{n} \sum_{i\in I_{1}^{(n)}} V_{i},\ldots,\frac{1}{n} \sum_{i\in
I_{K}^{(n)}} V_{i}\right)  
\]
which satisfies the $K$ linear constraints defined in \eqref{IS moment constraints}
through
\begin{equation*}
E_{S}[ \boldsymbol{\xi}_{n,\mathbf{X}}^{\mathbf{V}} ] =
\boldsymbol{\xi}_{M,\mathbf{X}}^{\mathbf{W}^{\ast}}
= \overline{W^{\ast}} \cdot \boldsymbol{\xi}_{M,\mathbf{X}}^{w\mathbf{W}^{\ast}}
\end{equation*}
where $\mathbf{Q}^{\ast}:= \left(q_{1}^{\ast}, \ldots, q_{K}^{\ast}\right)
= \boldsymbol{\xi}_{M,\mathbf{X}}^{w\mathbf{W}^{\ast}} 
\in int(\textrm{$\boldsymbol{\Omega}$\hspace{-0.23cm}$\boldsymbol{\Omega}$})$
and $\overline{W^{\ast}} = \frac{1}{M} \sum_{j=1}^{M}W_{j}^{\ast}$. 
The probability distribution $S$ defined on $\mathbb{R}^{n}$
is the Kullback-Leibler-divergence projection of $\mathbb{\bbzeta}^{\otimes n}$ 
on the class of all probability distributions on $\mathbb{R}^{n}$
which satisfy \eqref{IS moment constraints}.
We prove that
$\lim
\inf_{n\rightarrow \infty}
S\left[ 
\boldsymbol{\xi}_{n,\mathbf{X}}^{w\mathbf{V}} \in
\textrm{$\boldsymbol{\Omega}$\hspace{-0.23cm}$\boldsymbol{\Omega}$}
\right] 
> 0$. To start with, we define for strictly positive $\delta$ the set
\[
A_{n,\delta} := \left\{ \left\vert \frac{1}{n}\sum_{i=1}^{n}V_{i}-
\overline{W^{\ast }}\right\vert \leq \delta \right\} 
\]
and write 
\[
S\left[ 
\boldsymbol{\xi}_{n,\mathbf{X}}^{w\mathbf{V}} \in
\textrm{$\boldsymbol{\Omega}$\hspace{-0.23cm}$\boldsymbol{\Omega}$}
\right] 
=S\left[ 
\{ \boldsymbol{\xi}_{n,\mathbf{X}}^{w\mathbf{V}} \in
\textrm{$\boldsymbol{\Omega}$\hspace{-0.23cm}$\boldsymbol{\Omega}$} \}
\cap A_{n,\delta}
\right] 
+ S\left[ 
\{ \boldsymbol{\xi}_{n,\mathbf{X}}^{w\mathbf{V}} \in
\textrm{$\boldsymbol{\Omega}$\hspace{-0.23cm}$\boldsymbol{\Omega}$} \}
\cap A_{n,\delta}^{c}
\right]
=:I+II.
\]
By the law of large numbers, the second term $II$ tends to $0$ as $n$ tends to infinity. 
Moreover, one can rewrite
\[
I=S\bigg[ \dbigcup\limits_{m\in \left[ \overline{W^{\ast }}-\delta ,
\overline{W^{\ast }}+\delta \right] }
\left\{ \boldsymbol{\xi}_{n,\mathbf{X}}^{\mathbf{V}}\in 
m \cdot \textrm{$\boldsymbol{\Omega}$\hspace{-0.23cm}$\boldsymbol{\Omega}$} 
\right\} \bigg] 
\]
which entails 
\[
I \ \geq \ S\bigg[ \, \frac{1}{n_{k}} \sum_{i\in I_{k}^{(n)}}V_{i}\in 
\mathcal{V}_{\eta }\negthinspace\left( \overline{W^{\ast }} \cdot \frac{q_{k}^{\ast }}{p_{k}}
\right) \text{ for all }k \in \{1,\ldots,K\} \bigg] ,
\]
where $\mathcal{V}_{\eta }\negthinspace\left( \overline{W^{\ast }} \cdot \frac{
q_{k}^{\ast }}{p_{k}}\right) $ denotes a neighborhood of 
$\overline{W^{\ast }} \cdot \frac{q_{k}^{\ast }}{p_{k}}$ with radius $\eta$ being small 
when $\delta $ is small, for large enough $n$, making use of the a.s. convergence
of $n_{k}/n$ to $p_{k}$.  Now, for any $k \in \{1,\ldots,K\}$ one has
\begin{equation}
S\bigg[ \, \frac{1}{n_{k}} \sum_{i\in I_{k}^{(n)}} V_{i}\notin \mathcal{V}_{\eta}
\negthinspace\left( \overline{W^{\ast }} \cdot \frac{q_{k}^{\ast }}{p_{k}}\right) 
\bigg] 
\ \leq \ 
\exp\negthinspace\bigg(-n_{k} \cdot \inf_{x\in \mathcal{V}_{\eta }\left( 
\overline{W^{\ast }}\frac{q_{k}^{\ast }}{p_{k}}\right)^{\negthinspace c}} \, 
\varphi\left( x\right) \bigg)   
\label{ineg}
\end{equation}
since any margin of $S$ with index in $I_{k}^{(n)}$ is a corresponding
Kullback-Leibler-divergence projection of $\mathbb{\bbzeta}$ on the set of all distributions
on $\mathbb{R}$ with expectation $\overline{W^{\ast}} \cdot
\frac{q_{k}^{\ast }}{p_{M,k}^{emp}}$ ---
where $p_{M,k}^{emp}$ denotes the fraction of
the $X_{i}$\textquoteright s (within $X_{1},\ldots,X_{M}$) which are equal to $d_{k}$ 
(cf. \eqref{I^(n)_k for stat case}) ---  
and therefore has a
moment generating function which is finite in a non-void neighborhood of $0$, 
which yields \eqref{ineg} by the Markov Inequality. Note that the event 
$\left\{ \boldsymbol{\xi}_{M,\mathbf{X}}^{w\mathbf{W}^{\ast}}\in 
int\left( \textrm{$\boldsymbol{\Omega}$\hspace{-0.23cm}$\boldsymbol{\Omega}$} \right)
\right\}$ is regenerative, so that $M$ can be chosen large enough to make  
$p_{M,k}^{emp}$ close to $p_{k}$ for all $k \in \{1,\ldots,K\}$. This proves the claim.
\hspace{1.0cm} $\blacksquare$

\end{document}